\documentclass[amsmath,trackchanges,twocolumn]{aastex701}

\usepackage{graphicx}
\usepackage{dcolumn}
\usepackage{bm}
\usepackage{epsfig}
\usepackage{xspace}
\usepackage{hyperref}
\hypersetup{colorlinks,linkcolor=red,citecolor=blue,urlcolor=blue}  
\usepackage{color,xcolor}
\usepackage{verbatim}
\usepackage{float}
\usepackage{slashed}
\usepackage{rotating}
\usepackage[normalem]{ulem} 
\usepackage[caption=false]{subfig}
\usepackage{amsmath}
\usepackage[capitalize]{cleveref}
\usepackage{multirow}

\begin{document}

\title{On the maximum neutrino flux of blazars in the one-zone leptohadronic model}

\author{Wei-Jian Li}
\affiliation{Department of Physics, Zhejiang Normal University, Jinhua 321004, China}
\email{ruixue@zjnu.edu.cn} 

\author[orcid=0000-0003-1721-151X,gname=Rui,sname=Xue]{Rui Xue}
\affiliation{Department of Physics, Zhejiang Normal University, Jinhua 321004, China}
\email[show]{ruixue@zjnu.edu.cn} %

\author[orcid=0000-0002-3883-6669,gname=Ze-Rui,sname=Wang]{Ze-Rui Wang} 
\affiliation{College of Physics and Electronic Engineering, Qilu Normal University, Jinan 250200, China}
\affiliation{Shandong  Key Laboratory of Space Environment and Exploration Technology, China}
\email[show]{zerui\_wang62@163.com}

\author[0000-0002-6809-9575,gname=Dingrong,sname=Xiong]{Dingrong Xiong}
\affiliation{Yunnan Observatories, Chinese Academy of Sciences, 396 Yangfangwang, Guandu District, Kunming, 650216, China}
\affiliation{Center for Astronomical Mega-Science, Chinese Academy of Sciences, 20A Datun Road, Chaoyang District, Beijing, 100012, China}
\affiliation{Key Laboratory for the Structure and Evolution of Celestial Objects, Chinese Academy of Sciences, 396 Yangfangwang, Guandu District, Kunming, 650216, China}
\email[show]{\\xiongdingrong@ynao.ac.cn}

\begin{abstract}
The origin of extragalactic high-energy neutrinos remains a major mystery in astrophysics, with blazars as leading candidate sources. The widely adopted one-zone leptohadronic jet model, however, faces severe challenges from stringent X-ray observational constraints. In this work, we present an analytical approach that derives the maximum neutrino flux as a function of the observed X-ray flux and the corresponding physical parameters attainable within the one-zone leptohadronic framework. Applying this approach to a sample of neutrino candidate blazars, we further perform numerical modeling and find agreement between analytical and numerical results. 
Both approaches consistently show that the model-predicted neutrino fluxes do not significantly exceed those obtained in previous one-zone studies and remain below the flux levels inferred from IceCube observations, suggesting that the one-zone scenario alone is unlikely to fully account for high-energy neutrino-blazar associations. This highlights the importance of considering multi-zone models or alternative production sites (e.g., jet base, hot corona) to better explain high-energy neutrino origins in blazars.
\end{abstract}

\keywords{\uat{Active galactic nuclei}{16} --- \uat{Blazars}{164} ---\uat{High Energy astrophysics}{739} --- \uat{Neutrino astronomy}{1100}}


\section{Introduction} 
As one of the hottest topics in astroparticle physics, studying the origin of high-energy neutrinos offers unique insights into cosmic-ray origins \citep{2019PhR...801....1A, 2021PhRvL.126s1101R,2021ApJ...910..100D} and particle acceleration mechanisms \citep{2022A&A...668A.146D,2025MNRAS.tmp..237D}. The high-energy neutrino background discovered by the IceCube Observatory shows an isotropic distribution across the sky \citep{2013Sci...342E...1I}, uncorrelated with the Galactic plane, indicating its extragalactic origin \citep{2017ApJ...849...67A, 2024NatAs...8..241F}. As a subclass of active galactic nuclei (AGNs) with relativistic jets pointing to observers \citep{1995PASP..107..803U}, blazars have been widely discussed as one of the potential high-energy neutrino emitters \citep{2017PhRvD..96h2001A,2020ApJ...892...92A,2020PhRvL.124e1103A}. Depending on the equivalent width ($EW$) of emission lines, blazars are divided into flat spectrum radio quasars (FSRQs; $EW\geq5~\mathring{\rm A}$) and BL Lacertae objects (BL Lacs; $EW<5~\mathring{\rm A}$). The broadband spectral energy distributions (SEDs) of blazars are characterized by two bumps. It is generally accepted that the low-energy bump, from radio to optical/X-ray, originates from the synchrotron radiation of primary relativistic electrons in the jet. The origin of the high-energy bump, from X-ray to $\gamma$-ray, is still under debate. In leptonic models, inverse Compton (IC) radiation from relativistic electrons accounts for the high-energy bump. For FSRQs, which are characterized by abundant external photon fields, such as the broad-line region \citep[BLR;][]{1994ApJ...421..153S}, and the dusty torus \citep[DT;][]{2000ApJ...545..107B}, external Compton (EC) scattering plays a pivotal role when the emitting region is in close proximity to these ambient photon reservoirs. In contrast, the high-energy bump of BL Lacs is interpreted by the synchrotron self-Compton process \citep[SSC;][]{1985ApJ...298..114M}, where seed photons originate from co-spatial synchrotron radiation of primary relativistic electrons. In hadronic models, the high-energy bump could be contributed by the proton synchrotron radiation \citep{2000NewA....5..377A,2015MNRAS.448..910C, 2023PhRvD.107j3019X} and the emission from secondary particles produced in the hadronic interactions and internal $\gamma\gamma$ pair production \citep{2025ApJ...995..38L,2025ApJ...990..170X, 2013MNRAS.434.2684M, 2017MNRAS.464.2213P, 2024A&A...685A.110P, 2022PhRvD.106j3021X, 2024ApJS..271...10W, 2025ApJ...980...19O}. The abundant photon fields in blazar environments favor the photohadronic interactions, including the photopion interactions ($p\gamma$; $p + \gamma \rightarrow p/n + \pi^0 + \pi^\pm$) and Bethe-Heitler pair production (BH; $p + \gamma \rightarrow p + e^{\pm}$), although the hadronuclear interaction may also play an important role \citep{2022PhRvD.106j3021X, 2023ApJ...948...75M, 2024ApJS..271...10W, 2024ApJ...971..146X, 2025ApJ...980...19O}. In addition to producing neutrinos, these hadronic interactions are accompanied by $\gamma$-ray emission from $\pi^0$ decay, and the induced pair cascades generate electromagnetic radiation at lower energies as well \citep{2023ecnp.book..107H, 2025EPJC...85..779X}. Therefore, identifying the spatial and temporal associations between neutrinos and the multiwavelength (MWL) variability of astrophysical objects is a key approach to determining the potential neutrino sources.

In 2017, a high-energy (290 TeV) muon neutrino event, IC-170922A, was detected in both spatial and temporal coincidence with MWL flares of the blazar TXS 0506+056 at the significance level of $\sim 3\sigma$ for the first time \citep{2018Sci...361.1378I}. More interestingly, subsequent analysis of 9.5 years of archival IceCube data in the direction of TXS 0506+056 found a $\sim 3.5\sigma$ excess in the period between 2014 September and 2015 March, although no evidence of MWL activity was observed during this time \citep{2018Sci...361..147I}. These two associations have motivated extensive investigations into the possible neutrino-blazar connection \citep{2016NatPh..12..807K,2019ApJ...880..103G,2020ApJ...902...29P,2021ApJ...912...54R,2020ApJ...899..113P,2021JCAP...10..082O,2021NatAs...5..510S,2022ApJ...932L..25L,2022PhRvL.128v1101R,2023ApJ...958L...2F,2023MNRAS.519.1396S,2024ApJ...965L...2J,2024MNRAS.527.8746P, 2024ApJS..275...11Z}. Multi-wavelength behavior of neutrino-candidate blazars is intriguingly complex. Some cases show that variabilities in a specific band align with the arrival time of neutrinos \citep{2025ApJ...979....1J}, while in others, the MWL emission remains in a quiescent state, exhibiting only spatial coincidence with neutrinos \citep{2020ApJ...893..162F,2023Galax..11..117A}. Although the statistical significance of current associations remains limited, blazars are still considered a potential population of neutrino emitters that merit systematic investigation. For blazars, the relativistic jet, a powerful particle accelerator, is often considered the primary site for neutrino production. Due to the success of one-zone leptonic models in reproducing the SEDs and variability patterns of blazars, the one-zone framework has become the standard phenomenological setup for modeling blazar radiation. When extended to neutrino production, this framework is typically implemented in a one-zone leptohadronic scenario, in which relativistic protons interact with internal or external photon fields to generate high-energy neutrinos. However, in such one-zone leptohadronic models, hard X-ray observations substantially constrain the accompanying electromagnetic cascade emission. Since the generation of pair cascades is intrinsically tied to photohadronic interactions, the allowed neutrino flux is significantly constrained by hard X-ray observations \citep{2018ApJ...864...84K,2018ApJ...863L..10A, 2019NatAs...3...88G,2019MNRAS.483L..12C,2020ApJ...899..113P,2020ApJ...891..115P,2021JCAP...10..082O,2019ApJ...886...23X}. To alleviate this tension, multi-zone models, such as ``inner-outer blob model'' \citep{2019PhRvD..99f3008L,2019ApJ...886...23X,2021ApJ...906...51X} and ``stochastic dissipation model'' \citep{2024ApJ...962..142W, 2025PhRvD.112h3016W}, have been actively developed. These frameworks, while introducing additional parameters, provide richer physical interpretations and have been explored in parallel with alternative production sites, including hot coronae \citep{2019ApJ...880...40I, 2020ApJ...891L..33I, 2022ApJ...941L..17M, 2025ApJ...986..104F, 2025ApJ...995..166Y}, the accretion flow \citep{2025ApJ...980..255Y}, as well as dark matter origins \citep{2022PhRvL.128v1104W, 2023JCAP...05..057F,2023JCAP...07..039B, 2025PhRvD.112d3042D, 2025PhLB..87140015D, 2025PhRvD.112f3047X,2025JCAP...07..042Z}.

As previously mentioned, both multi-zone scenarios and one-zone models have been extensively studied. In observational contexts, the one-zone jet framework continues to be widely applied because its reduced parameter space enables clearer quantitative connections between high-energy neutrino signals and blazar emission \citep{2022MNRAS.510.2671P, 2024A&A...681A.119R, 2024A&A...689A.147R}. An open question is whether cascade emission in the one-zone leptohadronic framework can remain consistent with X-ray constraints while producing neutrino fluxes at a detectable level. In this work, we develop an analytical approach to delineate the permissible physical parameter space and the corresponding upper limit of the neutrino flux. We validate these results with numerical modeling, enabling an assessment of the one-zone leptohadronic model's ability to account for neutrino events. In Section \ref{sec:samples}, we list the sample of revisited blazars. In Section \ref{sec:methods}, we present analytical methods of deriving parameter space and estimating the maximum neutrino flux in the framework of the one-zone leptohadronic model. In Section \ref{sec:results}, we show the numerical modeling results. Finally, we end with discussion and conclusion in Section \ref{sec:discussion}. Throughout the paper, parameters with superscript ``obs'' are measured in the observer's frame, those with superscript ``AGN'' are measured in the AGN frame, whereas the parameters without the superscript are measured in the comoving frame, unless specified otherwise. We adopt the $\Lambda$CDM cosmological parameters $H_0=70~{\rm km~s^{-1}~Mpc^{-1}}$, $\Omega_{\rm m}=0.3$, $\Omega_{\Lambda}=0.7$ \citep{2014ApJ...794..135B}.

\section{Samples}\label{sec:samples}
In this section, we summarize a sample of neutrino-candidate blazars compiled from previous studies. The sample includes two masquerading BL Lacs \citep[intrinsically FSRQs with hidden broad lines and a standard accretion disk;][]{2013MNRAS.431.1914G}, four FSRQs, and two BL Lacs. The sources are selected based on the following considerations: (1) spatial coincidence with high-energy neutrino events, (2) temporal proximity between the neutrino arrival and enhanced MWL activity, and (3) the availability of quasi-simultaneous broadband SED data. The temporal information provides additional physical context for detailed emission modeling beyond purely positional associations, while the broadband SED coverage is essential for applying the analytical approach developed in this work and for reliably constraining the allowed parameter space within the one-zone leptohadronic framework. To the best of our knowledge, this sample includes all currently reported neutrino-candidate blazars that satisfy both temporal association and adequate broadband SED coverage. For this reason, we focus on this subset of sources rather than performing a sample selection from IceCat-1 based solely on spatial coincidence \citep{2023ApJS..269...25A}.
For GB6 J2113+1121, PKS 1502+106, and GB6 J1040+0617 lacking reported neutrino counts, we calculate the annual neutrino flux $\nu F_\nu (E_{\rm prob}^{\rm obs})$ at the most probable neutrino energy $E_{\rm prob}^{\rm obs}$ using the IceCube point-source effective area $A_{\rm eff}$ \citep{2019ICRC...36..851C} with
\begin{equation}
    \nu F_\nu (E_{\rm prob}^{\rm obs})=\frac{E_{{\rm prob}}}{A_{\rm eff}(E_{{\rm prob}},\delta_{\rm decl})\Delta T},
\end{equation}
where $\delta_{\rm decl}$ is the declination of the sources and $\Delta T=1~{\rm yr}$. For the other association events\footnote{For 5BZB J0630-2406, since $E_{\rm prob}^{\rm obs}$ is not given, we adopt the neutrino spectrum $\nu F_{\nu}(E_{\nu})\propto E_{\nu}^{-2},~100~{\rm TeV} \leq E_{\nu}\leq 40~{\rm PeV}$, as suggested by \cite{2023ApJ...958L...2F}.}, the detected neutrino energy and its corresponding flux $\nu F_{\nu,\rm orig}$ are reported in their original studies. To establish a uniform annual detection rate metric, the flux of each object is normalized to a one-year integration time using $\nu F_{\nu,\rm nor}=\nu F_{\nu,\rm orig}\times (\Delta T_{\rm orig}/1~\rm yr)^{-1}$, where $\Delta T_{\rm orig}$ denotes the observation duration specific to each study. The detailed information of these neutrino-candidate blazars is shown in Table \ref{table1}. More details of the sample are given in the following.


\begin{table*}[t]
    \centering
    \caption{Neutrino Source Candidates of Blazars.}
    \vspace{1mm}
    \label{table1}
    \resizebox{\textwidth}{!}{
    \hspace{-2.1cm}
    \begin{tabular}{ccccccc}
        \hline\hline
        Object & Class & $z$ & ${\rm Log}~M_{\rm BH}$ & $L_{\rm d}^{\rm AGN}$ & quasi-simultaneous MWL SED & Neutrino Event\\
        &  &  & ($M_{\odot}$) & (${\rm erg~s^{-1}}$) &  &  \\
        \hline
         &  &  &   &   & \cite{2018Sci...361.1378I} & IC-170922A \citep{2018Sci...361.1378I}\\
        TXS 0506+056 & {mas BL Lac} & 0.3365 & {8.477 \citep{2019MNRAS.484L.104P}} & {$5\times10^{44}$ \citep{2019MNRAS.484L.104P}}  &\cite{2024ApJ...962..142W} & GVD-210418CA \citep{2024MNRAS.527.8784A}\\
         & ~ & ~ & ~ & ~  & \cite{2024ApJ...962..142W} & IC-220918A \citep{2022ApJ...941L..25B}\\
        PKS 0735+178 & mas BL Lac & 0.45 & $8.8\pm0.4$ \citep{2023MNRAS.519.1396S} & $4\times10^{44}$ \citep{2023MNRAS.519.1396S}  & \cite{2023MNRAS.519.1396S} & IC-211208A \citep{2023MNRAS.519.1396S}\\
        GB6 J2113+1121 & FSRQ & 1.316 & 8.2 \citep{2022ApJ...932L..25L} & $1\times10^{46}$ \citep{2022ApJ...932L..25L} & \cite{2022ApJ...932L..25L} & IC-191001A \citep{2022ApJ...932L..25L}\\
        5BZB J0630-2406 & FSRQ & 1.239 & 10.362 \citep{2023ApJ...958L...2F} & $5.5\times10^{45}$ \citep{2012MNRAS.425.1371G} & \cite{2023ApJ...958L...2F} & IC J0630-2353 \citep{2023ApJ...958L...2F}\\
        PKS 1502+106 & FSRQ & 1.8385 & $9.64\pm0.44$ \citep{2011ApJS..194...45S} & $5\times10^{46}$ \citep{2021ApJ...912...54R} & \cite{2020ApJ...893..162F} & IC-190730A \citep{2020ApJ...893..162F}\\
        PKS B1424-418 & FSRQ & 1.522 & 9$^{\star}$ & $1\times10^{46}$ \citep{2013MNRAS.435L..24T} & \cite{2017ApJ...843..109G} & Big Bird \citep{2016NatPh..12..807K}\\
        GB6 J1040+0617 & LSP-BL Lac & 0.7351 & 9$^{\star}$ & - & \cite{2019ApJ...880..103G} & IC-141209A \citep{2019ApJ...880..103G}\\
        3HSP J095507.1+355101 & HSP-BL Lac & 0.557 & 8.477 \citep{2020MNRAS.495L.108P} & - & \cite{2020AA...640L...4G} & IC-200107A \citep{2020AA...640L...4G} \\
        \hline\hline
    \end{tabular}}
    \vspace{2mm}\\
    \raggedright
    \textbf{Notes.} Columns from left to right: the source name; the class of blazar; the redshift of the source; logarithm of the SMBH in units of the solar mass, $M_{\odot}$; the accretion disk luminosity in the AGN frame in units of ${\rm erg~s^{-1}}$; neutrino event associated with blazar; the reference of SED; the reference of neutrino. For TXS 0506+056, we select the latter three out of four reported neutrino association events, including the periods of 2017, 2021, and 2022. ``Mas BL Lac'' in this table represents the masquerading BL Lac.\\
    \footnotesize{$^{\star}$ In absence of an estimated black hole mass, we consider an average value of $10^9~M_{\odot}$ \citep{2022ApJ...925...40X}.} \\
\end{table*}

\subsection*{TXS 0506+056}
TXS 0506+056, classified as a masquerading BL Lac \citep{2019MNRAS.484L.104P} at redshift $z = 0.3365$ \citep{2018ApJ...854L..32P}, holds the distinction of being the first confirmed neutrino-emitting blazar and has been associated with four reported high-energy neutrino events to date. The first one is IC-170922A, an EHE muon neutrino event with an energy of $290~{\rm TeV}$. For the first time, a high-energy neutrino was detected in temporal and spatial coincidence with a MWL flare, with a significance of $\sim 3\sigma$ \citep{2018Sci...361.1378I}. From the direction of TXS 0506+056, a cascade neutrino event ($224\pm75~{\rm TeV}$, GVD-210418CA) and a {\tt Bronze} track-like neutrino event ($\sim170~{\rm TeV}$, IC-220918A) were detected by the Baikal-GVD neutrino telescope \citep{2024MNRAS.527.8784A} and the IceCube Collaboration \citep{2022ApJ...941L..25B}, respectively. However, TXS 0506+056 was in a low state in the ${\rm GeV}$ band during these two events \citep{2024ApJ...962..142W}. The last event, known as an orphan neutrino flare detected in the period between September 2014 and March 2015 \citep{2018Sci...361..147I}, has been excluded from this work due to the lack of quasi-simultaneous X-ray data.

\subsection*{PKS 0735+178}
PKS 0735+178 is known as a ``classical BL Lac'' owing to its featureless optical spectrum, complicating the determination of redshift \citep{1974ApJ...190L.101C,2012A&A...547A...1N,2021ATel15132....1F}. Here, we adopt $z=0.45$ consistent with \cite{2025A&A...695A.266O}. Recent analyses suggest it may instead be a ``masquerading'' BL Lac with potential external photon fields \citep{2023MNRAS.519.1396S}, motivating our consideration of the EC-dominated case. Taking into account the systematic error, PKS 0735+178 is located just inside the 90\% confidence region of IC-211208A \citep{2023MNRAS.519.1396S}, a {\tt Bronze} track-like event with an estimated energy of 172 TeV \citep{2021GCN.31191....1I}. In subsequent detections in this direction, Baikal-GVD and the Baksan Underground Scintillation Telescope reported a $43~{\rm TeV}$ cascade event 
\citep{2021ATel15112....1D} and a ${\rm GeV}$ muon neutrino 
\citep{2021ATel15143....1P}, respectively. Around the time of IC-211208A, the source was flaring in optical, X-ray, and $\gamma$-ray bands \citep{2025A&A...695A.266O}.

\subsection*{GB6 J2113+1121}
GB6 J2113+1121 is known as an FSRQ with $z=1.316$ \citep{2013ApJ...767...14P}. This blazar is associated with a track-like neutrino event ($\sim0.2~{\rm PeV}$, IC-191001A) under the {\tt Gold} alert streams \citep{2019GCN.25913....1I}, and it is located just outside the edge of the 90\% confidence region of the neutrino IC-191001A \citep{2022ApJ...932L..25L}. The radio-emitting tidal disruption event (AT2019dsg) is also proposed to be associated with the neutrino within the 90\% error region of IC-191001A, which is studied by \cite{2020PhRvD.102h3028L} and \cite{2021NatAs...5..510S}. A giant flare in long-term IR, optical, and $\gamma$-ray light curves of GB6 J2113+1121 emerged around the arrival time of the neutrino \citep{2022ApJ...932L..25L}.

\subsection*{5BZB J0630-2406}
5BZB J0630-2406 is classified as an FSRQ by \cite{2023ApJ...958L...2F} with $z\geq1.239$ \citep{2013ApJ...764..135S}. It has been proposed as one of the blazars associated with neutrino emission during the first $\sim7~{\rm yr}$ (2008-2015) of IceCube observations \citep{2022ApJ...933L..43B}. 5BZB J0630-2406 is located at an angular separation of $0^{\circ}\!.28$ from the hotspot IC J0630-2353, and this source is undergoing a long-term enhanced state in the more recent $\gamma$-ray data, though the $\gamma$-ray light curve shows a lower steady state in the first $\sim7~{\rm yr}$ \citep{2023ApJ...958L...2F}. Extending the observation time scale to 15 years, the optical band exhibits year-long modulations of the flux, and the overall variation trend is consistent with the $\gamma$-ray light curve.

\subsection*{PKS 1502+106}
PKS 1502+106 is an FSRQ at redshift $z=1.8385$ \citep{2008ApJS..175..297A}. It is located within the 50\% confidence region of the {\tt Gold} alert track-like muon neutrino event \citep[$\sim300~{\rm TeV}$, IC-190730A;][]{2019ATel12996....1K}. At the neutrino arrival time, PKS 1502+106 was in a low state, while the OVRO radio light curve reached the highest flux density in a long-term flare \citep{2020ApJ...893..162F}.

\subsection*{PKS B1424-418}
PKS B1424-418 is classified as an FSRQ with a redshift of $z=1.522$ \citep{1988ApJ...327..561W}. This blazar is associated with a cascade-like neutrino event ($2~{\rm PeV}$, Big Bird) spatially and temporally, while the probability of a chance coincidence is about 5\%. Along with the more than a year $\gamma$-ray outburst that begins around the neutrino event, the flux of X-ray, optical, and radio bands is also increased \citep{2016NatPh..12..807K}.

\subsection*{GB6 J1040+0617}
GB6 J1040+0617 is an LSP-BL Lac with redshift $z=0.7351$ \citep{2012ApJS..203...21A,2015Ap&SS.357..141M}. It is located within the 90\% confidence region of a HESE neutrino event ($97.4~{\rm TeV}$, IC-141209A) with a coincidence detection probability of just 30\%. The variability in both $\gamma$-ray and optical wavelengths coincided with the arrival time of the neutrino event \citep{2019ApJ...880..103G}. The MWL SED of this source is not contemporaneous, which is taken from \cite{2019ApJ...880..103G}.

\subsection*{3HSP J095507.1+355101}
3HSP J095507.1+355101 \citep[$z=0.557$;][]{2020MNRAS.495L.108P} is an HSP-BL Lac within the 90\% error region of the HESE ${\rm 0.33~PeV}$ neutrino event IC-200107A, and the source is located only $0^{\circ}\!.62$ away from the best-fit position of the neutrino event \citep{2020AA...640L...4G}. An X-ray flare exhibiting temporal coincidence with the neutrino event was detected commencing 1 day post-neutrino arrival, maintaining its maximum flux for the subsequent $\sim 30$ days \citep{2020AA...640L...4G}.

\section{Analytical Methods} \label{sec:methods}
Among the various frameworks proposed to interpret neutrino-blazar associations, the one-zone jet model, particularly in its leptohadronic form, has been widely adopted in both observational and phenomenological studies\footnote{As discussed in \cite{2019NatAs...3...88G}, no parameter space can be found in hadronic model and proton synchrotron model.}.
This hybrid model posits that the MWL SED is dominated by the leptonic emission from primary relativistic electrons (emission from secondary pairs may have a sub-dominant contribution at X-ray or TeV bands), while $p\gamma$ interactions are responsible for neutrino production \citep{2019NatAs...3...88G}. In the following, we will first employ an analytical method to determine the parameter space of primary leptonic emission required to explain MWL SEDs. Building upon this parameter space, we will then estimate the maximum neutrino flux analytically, and evaluate whether this one-zone leptohadronic framework can sufficiently interpret observations of high-energy neutrinos.

For the one-zone jet model, it is assumed that all of the nonthermal radiation of the observed jet comes from a single spherical region (hereafter referred to as the blob) of radius $R$, composed of a plasma of charged particles in a uniformly entangled magnetic field $B$ and moving with the bulk Lorentz factor $\Gamma=(1-\beta^2)^{-\frac{1}{2}}$ at a viewing angle $\theta^{\rm obs}$ to the observer's line of sight, where $\beta c$ is the speed of the blob and $c$ is the speed of light. Due to relativistic beaming effects, the observed flux is boosted by a factor of $\delta_{\rm D}^4$, where $\delta_{\rm D}=[\Gamma(1-\beta{\rm cos}\theta^{\rm obs})]^{-1}\approx\Gamma$ by assuming $\theta^{\rm obs}\lesssim1/\Gamma$.

\subsection{The physical parameter space}\label{sec:analytical method}
For the one-zone leptohadronic model, \cite{2017ApJ...843..109G} presented an analytical method to constrain the parameter space in the Thomson (TMS) regime when the high-energy peak is dominated by SSC emission (i.e., the SSC-dominated case), and applied it to the association between TXS 0506+056 and IC-170922A \citep{2019NatAs...3...88G}. In the following, we extend this analytical framework to the Klein-Nishina (KN) regime and to cases where the high-energy peak is dominated by SSC or EC emission (i.e., the EC-dominated case), thereby broadening its applicability to diverse blazar radiation scenarios.

When the primary leptonic emission dominates the MWL SED, the high-energy bump is explained by either the SSC emission or the EC emission, depending on whether broad emission lines are detected in specific blazars. Let us first derive the parameter space analytically in the SSC-dominated case. Following the analytical method investigated and applied in previous studies \citep{1998ApJ...509..608T, 2024MNRAS.528.7587H}, the correlations between blob physical parameters, including $R$, $B$, and $\delta_{\rm D}$, can be obtained through the peak frequencies and peak luminosities of SEDs' two bumps. In the TMS regime, we have
\begin{equation}\label{Bdelta1}
    B\delta_{\rm D}=(1+z)\frac{(\nu_{\rm s,p}^{\rm obs})^2}{2.8\times10^6 \nu_{\rm c,p}^{\rm obs}},
\end{equation}
where $z$ is the redshift, $\nu_{\rm s/c,p}^{\rm obs}$ represents peak frequencies of low- and high-energy bumps, and
\begin{equation}\label{Bdelta2}
    R=\left[\frac{2(L_{\rm s,p}^{\rm obs})^2f(\alpha_1,\alpha_2)}{L_{\rm c,p}^{\rm obs}c}\right]^{\frac{1}{2}}\frac{1}{B\delta_{\rm D}^2},
\end{equation}
where $L_{\rm s/c,p}^{\rm obs}$ represents the peak luminosities of low- and high-energy peaks, and $f(\alpha_1,\alpha_2)=\frac{1}{\alpha_2}-\frac{1}{\alpha_1}$ is a correction term \citep{1998ApJ...509..608T}, in which $\alpha_1$ and $\alpha_2$ are the spectra indexes below and above the synchrotron peak (i.e., $\nu F_\nu \propto \nu^{-\alpha}$), respectively. In the KN regime, high-energy spectrum will be softened, leading to lower $\nu_{\rm c,p}^{\rm obs}$ and $L_{\rm c,p}^{\rm obs}$ compared to those obtained under the TMS regime. If the KN effect is severe, the relation between $B$ and $\delta_{\rm D}$ satisfies \citep{2024MNRAS.528.7587H}
\begin{equation}\label{severe KN}
    B\delta_{\rm D}^3\leq\frac{(1+z)^3h^2(\nu_{\rm s,p}^{\rm obs})^3}{2.1\times10^6m_{\rm e}^2c^4},
\end{equation}
where $m_{\rm e}$ is the electron rest mass and $h$ is the Planck constant. Then relations of physical parameters are revised to \citep{1998ApJ...509..608T}
\begin{equation}\label{Bdelta3}
    B\delta_{\rm D}^{-1}=\frac{\nu_{\rm s,p}^{\rm obs}}{(\nu_{\rm c,p}^{\rm obs})^2}\left(\frac{m_{\rm e}c^2}{h}\right)^2\frac{g(\alpha_1,\alpha_2)^2}{3.7\times10^6}\frac{1}{1+z},
\end{equation}
where $g(\alpha_1,\alpha_2)={\rm exp}[\frac{1}{\alpha_1}+\frac{1}{2(\alpha_2-\alpha_1)}]\lesssim1$, and
\begin{equation}\label{Bdelta4}
    R=\left\{\frac{2(L_{\rm s,p}^{\rm obs})^2f(\alpha_1,\alpha_2)}{cL_{\rm c,p}^{\rm obs}\left[\frac{3}{4}\left(\frac{m_{\rm e}c^2}{h}\right)^2\frac{g(\alpha_1,\alpha_2)}{\nu_{\rm s,p}^{\rm obs}\nu_{\rm c,p}^{\rm obs}(1+z)}\right]^{\alpha_1}}\right\}^{\frac{1}{2}}\frac{1}{B\delta_{\rm D}^{(2+\alpha_1)}}.
\end{equation}
For the SSC-dominated case, by combining Eqs.~(\ref{Bdelta1}) and (\ref{Bdelta2}) for the TMS regime, and  Eqs.~(\ref{Bdelta3}) and (\ref{Bdelta4}) for the KN regime, the corresponding physical parameter space can be obtained, respectively.

For blazars having broad emission lines (i.e., FSRQs) or potential external photon fields \citep{2024MNRAS.528.7587H} (i.e., low-synchrotron-peaked BL Lacs, LSP-BL Lacs\footnote{According to the peak frequency of the low-energy synchrotron peak \citep{2010ApJ...716...30A}, blazars can be divided into low-synchrotron-peaked (LSP, $\nu_{\rm s,p}^{\rm obs}\lesssim10^{14}~{\rm Hz}$), intermediate-synchrotron-peaked (ISP, $10^{14}~\lesssim\nu_{\rm s,p}^{\rm obs}\lesssim10^{15}~{\rm Hz}$), and high-synchrotron-peaked (HSP, $\nu_{\rm s,p}^{\rm obs}\gtrsim10^{15}~{\rm Hz}$).}), the high-energy peak is EC-dominated. 
In general, external photons are considered to originate from the BLR and DT. The radiation from BLR and DT is taken as an isotropic greybody with peak frequencies at $\nu_{\rm BLR}^{\rm AGN}=2\times10^{15}~{\rm Hz}$ \citep{2008MNRAS.386..945T} and $\nu_{\rm DT}^{\rm AGN}=3\times10^{13}~{\rm Hz}$ \citep{2007ApJ...660..117C}, respectively, and their energy densities can be approximated by \citep{2012ApJ...754..114H}
\begin{equation}\label{U_BLR}
    U_{\rm BLR}(r)=\frac{\eta_{\rm BLR}\Gamma^2L_{\rm d}^{\rm AGN}}{4\pi(r_{\rm BLR}^{\rm AGN})^2c[1+(r/r_{\rm BLR}^{\rm AGN})^3]},
\end{equation}
and
\begin{equation}\label{U_DT}
    U_{\rm DT}(r)=\frac{\eta_{\rm DT}\Gamma^2L_{\rm d}^{\rm AGN}}{4\pi(r_{\rm DT}^{\rm AGN})^2c[1+(r/r_{\rm DT}^{\rm AGN})^4]},
\end{equation}
where $\eta_{\rm BLR}=\eta_{\rm DT}=0.1$ represent the fractions of the disk luminosity $L_{\rm d}^{\rm AGN}$ reprocessed into the BLR/DT radiation, $r$ is the distance between the blob and the central supermassive black hole (SMBH), $r_{\rm BLR}^{\rm AGN}=0.1(L_{\rm d}^{\rm AGN}/10^{46}~{\rm erg~s^{-1}})^\frac{1}{2}~{\rm pc}$ and $r_{\rm DT}^{\rm AGN}=2.5(L_{\rm d}^{\rm AGN}/10^{46}~{\rm erg~s^{-1}})^\frac{1}{2}~{\rm pc}$ are the characteristic distances of the BLR and DT, respectively \citep{2008MNRAS.387.1669G}.

Deriving the relations of blob physical parameters in EC-dominated situation is similar to the analytical method of SSC-dominated case. In the TMS regime, the peak frequency of high-energy bump is expressed as $\nu_{\rm c,p}^{\rm obs}\simeq\gamma_{\rm b}^2\delta_{\rm D}\Gamma\nu_{\rm BLR/DT}^{\rm AGN}$. From peak frequencies of two bumps, we have
\begin{equation}\label{Bdelta11}
    B\delta_{\rm D}^{-1}=\frac{1+z}{3.7\times10^6}\frac{\nu_{\rm s,p}^{\rm obs}\nu_{\rm BLR/DT}^{\rm AGN}}{\nu_{\rm c,p}^{\rm obs}}.
\end{equation}
From $L_{\rm c,p}^{\rm obs}/L_{\rm s,p}^{\rm obs}\approx (U_{\rm BLR}(r)+U_{\rm DT}(r))/U_{\rm B}$, where $U_{\rm B}=B^2/8\pi$ represents the energy density of magnetic field, the relation between $B$ and $\delta_{\rm D}$ can be derived as well, i.e.,
\begin{equation}\label{Bdelta22}
\begin{aligned}
    B\delta_{\rm D}^{-1}=&\left[\frac{2L_{\rm d}^{\rm AGN}}{c}\frac{L_{\rm s,p}^{\rm obs}}{L_{\rm c,p}^{\rm obs}}\left(\xi_{\rm BLR}+\xi_{\rm DT}\right)\right]^{\frac{1}{2}},
\end{aligned}
\end{equation}
where $\xi_{\rm BLR}=\frac{\eta_{\rm BLR}}{(r_{\rm BLR}^{\rm AGN})^2[1+(r/r_{\rm BLR}^{\rm AGN})^3]}$, $\xi_{\rm DT}=\frac{\eta_{\rm DT}}{(r_{\rm DT}^{\rm AGN})^2[1+(r/r_{\rm DT}^{\rm AGN})^4]}$.

Under the severe KN effect, we have $\nu_{\rm s,p}^{\rm obs}\geq3.7\times10^6\gamma_{\rm KN}^2B\delta_{\rm D}/(1+z)$, since $\gamma_{\rm b}\geq\gamma_{\rm KN}=3m_{\rm e}c^2/(4h\nu_{\rm BLR/DT}^{\rm AGN}\delta_{\rm D})$, then Eq.~(\ref{Bdelta11}) would be revised to 
\begin{equation}\label{Bdelta33}
    B\delta_{\rm D}^{-1}\leq4.8\times10^{-7}\nu_{\rm s,p}^{\rm obs}(1+z)\left(\frac{h\nu_{\rm BLR/DT}^{\rm AGN}}{m_{\rm e}c^2}\right)^2,
\end{equation}
and the peak luminosity constraint is revised to
\begin{equation}\label{Bdelta44}
    B\delta_{\rm D}^{-1}=\left\{\frac{2L_{\rm s,p}^{\rm obs}L_{\rm d}^{\rm AGN}\xi_{\rm BLR/DT}}{cL_{\rm c,p}^{\rm obs}\left[\frac{3}{4}\frac{m_{\rm e}c^2}{h\nu_{\rm BLR/DT}^{\rm AGN}}\left(\frac{\nu_{\rm BLR/DT}^{\rm AGN}}{\nu_{\rm c,p}^{\rm obs}}\right)^{\frac{1}{2}}\right]^{\beta}}\right\}^{\frac{1}{2}},
\end{equation}
where $\beta =3$ is the spectral index of external photons spectra. Therefore, in the EC-dominated scenarios, the parameter space can be constrained by combining Eqs.~(\ref{Bdelta11}) and (\ref{Bdelta22}) for the TMS regime, or Eqs.~(\ref{Bdelta33}) and (\ref{Bdelta44}) for the KN regime, respectively.

In addition to the above constraints, to ensure the predominance of EC components and leave a margin for hadronic radiation to obtain the maximum neutrino flux, the assumption $(U_{\rm BLR}(r)+U_{\rm DT}(r))\geq \chi U_{\rm syn}$ has been introduced when $r$ is determined, where $\chi$ is a value greater than unity and we choose $\chi=100$\footnote{The value of $\chi$ is directly proportional to the dominance of the EC process. By comparing the $\gamma$-ray and X-ray fluxes in \cref{TXS 17,TXS 21,TXS 22,PKS 0735+178,GB6 J2113+1121,5BZB J0630-2406,PKS 1502+106,PKS B1424-418}, it can be seen that 100 represents a relatively conservative lower bound, while a larger value would further restrict the effective parameter space (indicated by the black solid line), an excessively high setting may impose overly stringent constraints that lack physical justification.} in this work, and $U_{\rm syn}=L_{\rm syn}^{\rm obs}/(4\pi R^2c\delta_{\rm D}^4)$ is the energy density of synchrotron photons from primary electrons in which $L_{\rm syn}^{\rm obs}$ represents the integrated luminosity of low-energy bump. Combining with Eqs.~(\ref{U_BLR}) and (\ref{U_DT}), the EC dominant constraint is derived as
\begin{equation}\label{deltaECdomi}
    \delta_{\rm D}\geq\left[\frac{\chi}{R^2(\xi_{\rm BLR}+\xi_{\rm DT})}\frac{L_{\rm syn}^{\rm obs}}{L_{\rm d}^{\rm AGN}}\right]^{\frac{1}{6}}
\end{equation}
by giving a blob radius $R$.

In the one-zone leptohadronic scenario, the interaction efficiencies of $p\gamma$ interactions and $\gamma \gamma$ annihilation are inherently coupled through their shared soft photon fields \citep{2019ApJ...871...81X}. It implies that highly efficient $p\gamma$ processes would inevitably cause strong internal absorption of GeV--TeV photons, thereby inducing significant spectral softening.
Here, we define $E_{\rm h}^{\rm obs}$ to represent the critical energy at which the spectrum softening begins. If there is no such feature in the GeV--TeV spectrum, then $E_{\rm h}^{\rm obs}$ is equal to the highest energy of $\gamma$-ray data points. Using the $\delta$-approximation, the corresponding energy of the low-energy photon from synchrotron radiation or external photon fields can be deduced, i.e., $E_{\rm l}\approx\frac{2(m_{\rm e}c^2)^2\delta_{\rm D}}{E_{\rm h}^{\rm obs}(1+z)}$. For the internal $\gamma \gamma$ optical depth at $E_{\rm h}^{\rm obs}$, it needs to satisfy
\begin{equation}
    \tau_{\gamma\gamma}(E_{\rm h}^{\rm obs})\approx \frac{\sigma_{\gamma\gamma}E_{\rm h}^{\rm obs}U_{\rm l}(E_{\rm l})R(1+z)}{2m_{\rm e}^2c^4\delta_{\rm D}}\leq1,
\end{equation}
where $\sigma_{\gamma\gamma}\approx1.68\times10^{-25}~{\rm cm}^2$ is the peak cross section for $\gamma\gamma$ annihilation, and $U_{\rm l}(E_{\rm l})$ denotes the energy density of soft photons. For the SSC-dominated case, the energy density can be obtained by $U_{\rm l}(E_{\rm l})=U_{\rm syn}(E_{\rm l})$. For a specific blazar with a known SED, the lower limit of $\delta_{\rm D}$ can be derived, i.e.,
\begin{equation}\label{deltaSSC}
\begin{split}
    \delta_{\rm D}^{4-2\alpha}\geq& 7.7\times10^{-26}\times10^b\times2^\alpha D_{\rm L}^2(m_{\rm e}c^2)^{2\alpha-2}{E_{\rm h}^{\rm obs}}^{1-\alpha}\\
&\times (1+z)^{2-2\alpha} \frac{(\nu_{\rm s,p}^{\rm obs})^2}{\nu_{\rm c,p}^{\rm obs}}\left[\frac{L_{\rm c,p}^{\rm obs}}{2(L_{\rm s,p}^{\rm obs})^2f(\alpha_1,\alpha_2)}\right]^{\frac{1}{2}},
\end{split}
\end{equation}
where $\alpha$ represents either $\alpha_1$ or $\alpha_2$, depending on the energy of the target photons, $b$ means the corresponding intercepts of the power-law spectrum, and $D_{\rm L}$ is the luminosity distance.
In the case of EC-dominated, the target photons are mainly from external photon fields. The energy density distributions of BLR and DT follow the shape of the black-body radiation
\begin{equation}
    u(T,E)=\frac{8\pi E^3}{h^2c^3}\frac{1}{e^{E/k_{\rm B}T}-1},
\end{equation}
where $E$ is the photon energy of black-body radiation, $k_{\rm B}$ is the Boltzmann constant, and $T=h\nu_{\rm BLR/DT}^{\rm AGN}\delta_{\rm D}/(3.93k_{\rm B})$ is the characteristic temperature of BLR or DT, so the energy density distribution of external photon fields can be determined by $(U_{\rm BLR}(r,E)+U_{\rm DT}(r,E))=\frac{(U_{\rm BLR}(r)+U_{\rm DT}(r))}{\int u(T,E){\rm d}E}u(T,E)$ when $r$ is given. Thus, we can get the upper limit of $\delta_{\rm D}$ in the EC-dominated case, i.e.,
\begin{equation}\label{deltaEC}
    \delta_{\rm D}\leq\frac{2m_{\rm e}^2c^4}{U_{\rm ext}^{\rm AGN}(r,E_{\rm l}^{\rm AGN})\sigma_{\gamma\gamma}E_{\rm h}^{\rm obs}R(1+z)}.
\end{equation}
Overall, by giving the value of $R$, $\delta_{\rm D}$ can be further constrained by Eq.~(\ref{deltaSSC}) for SSC-dominated case, and by Eqs.~(\ref{deltaSSC}) and (\ref{deltaEC}) for EC-dominated case.

By applying the analytical method outlined above to the sample, we constrain the parameter spaces for each source, as illustrated in the left panels of \cref{TXS 17,TXS 21,TXS 22,PKS 0735+178,GB6 J2113+1121,5BZB J0630-2406,PKS 1502+106,PKS B1424-418,GB6 J1040+0617,3HSP J095507.1+355101}. Since the frequency obtained by the empirical function has some uncertainty, we consider an uncertainty of a factor of 3 in $\nu_{\rm c,p}^{\rm obs}$. For SSC-dominated cases (the upper left panels of \cref{TXS 17,TXS 21,TXS 22,PKS 0735+178,GB6 J2113+1121,5BZB J0630-2406,PKS 1502+106,PKS B1424-418,GB6 J1040+0617,3HSP J095507.1+355101}), the space below the red line represents the severe KN effect, while the space above this boundary remains under the TMS regime. As exemplified in \cref{TXS 17}, for a blob radius $R=1\times10^{16}~{\rm cm}$, the effective parameter space (white dashed contour with ratio 0) lies within the boundaries defined by the black and blue solid lines (the constraints of peak frequency and internal optical depth corresponding to Eqs.~\ref{Bdelta1} and \ref{deltaSSC}, respectively). No parameter space is found for $R=1\times10^{17}~{\rm cm}$. Overall, the SSC-dominated parameter spaces of most sources satisfy the TMS regime, except for an HSP (3HSP J095507.1+355101) under the severe KN effect. In EC-dominated scenarios, the frequency $\nu_{\rm ext}^{\rm AGN}$ of external soft photons corresponding to neutrinos with energy $E_{\rm prob}^{\rm obs}$ can be estimated using the $\delta$-approximation under the condition where the cross section of $p\gamma$ interaction peaks due to the $\Delta^+(1232)$ resonance \citep{1997PhRvL..78.2292W}
\begin{equation}\label{nudelta}
    \begin{aligned}
        \nu_{\rm ext}^{\rm AGN}&=\frac{0.3~{\rm GeV^2}}{20(1+z)hE_{\rm prob}^{\rm obs}}\\
        &\simeq1.8\times10^{16}~{\rm Hz}~\left(\frac{z}{1}\right)^{-1}\left(\frac{E_{\rm prob}^{\rm obs}}{100~{\rm TeV}}\right)^{-1}.
    \end{aligned}
\end{equation}
Given the proximity of $ \nu_{\rm ext}^{\rm AGN}$ to $\nu_{\rm BLR}^{\rm AGN}$, we first consider the location of blob $r$ within the BLR, so that a high $p\gamma$ interaction efficiency is expected. If there is no parameter space, the blob is placed between the BLR and the DT, and is prioritized to positions closer to the BLR. Since the peak luminosities of BLR and DT cannot be determined when both components contribute significantly to the EC peak, we adopt an uncertainty factor of 3 in $L_{\rm c,p}^{\rm obs}$. The results are shown in the lower left panels of \cref{TXS 17,TXS 21,TXS 22,PKS 0735+178,GB6 J2113+1121,5BZB J0630-2406,PKS 1502+106,PKS B1424-418}. When $r\leq r_{\rm BLR}^{\rm AGN}$ (corresponding to \cref{PKS 0735+178,GB6 J2113+1121,PKS 1502+106}), the effective parameter space (white dashed line) is within the peak frequency constraint (pink dashed lines corresponding to Eq.~\ref{Bdelta11}), the optical depth constraint (blue dotted and long-dashed lines corresponding to Eqs.~\ref{deltaSSC} and \ref{deltaEC}, respectively) and the constraint of predominant EC components (purple long-dashed line corresponding to Eq.~\ref{deltaECdomi}) under the TMS regime. For $r_{\rm BLR}^{\rm AGN}\leq r\leq r_{\rm DT}^{\rm AGN}$ (corresponding to \cref{TXS 17,TXS 21,TXS 22,5BZB J0630-2406,PKS B1424-418}), the effective parameter space is the area surrounded by the black solid lines. In general, since the high-energy peak frequency within the KN regime remains fixed \citep[$\nu_{\rm BLR}^{\rm obs}\simeq6\times10^{23}~{\rm Hz},~\nu_{\rm DT}^{\rm obs}\simeq5\times10^{25}~{\rm Hz}$, see][]{2024PASA...41...62D} when the external photon field serves as soft photon field, which in most cases cannot satisfy the spectral shape requirements of the IC peak, most sources are constrained within the TMS regime. A dual-component model (incorporating both BLR and DT photon fields) is required to explain the IC peak only when either the high-energy spectrum exhibits a flat shape or no parameter space exists for the blob within the BLR. In such cases, the BLR-related EC peak experiences significant KN suppression, which is consistent with the parameter spaces shown in the lower left panels of \cref{TXS 17,TXS 21,TXS 22,5BZB J0630-2406,PKS B1424-418}.

\subsection{The maximum model-predicted neutrino flux}\label{anaflux}
Previous studies have suggested that the neutrino flux predicted by the one-zone leptohadronic model falls below observations, as the X-ray data impose stringent constraints on the emission from secondary pairs, which are mainly from the BH process \citep[e.g., see the lower panel of Fig.~1 in][]{2019MNRAS.483L..12C}. Within the one-zone framework, a natural way to achieve a higher neutrino flux is to enhance the efficiency of the $p\gamma$ interactions $f_{p\gamma}$ while simultaneously suppressing that of the BH process $f_{\rm BH}$, i.e., to maximize the ratio $f_{p\gamma}/f_{\rm BH}$. By adjusting the injection of relativistic protons such that the BH-induced emission reaches the upper limit allowed by the X-ray data, the corresponding neutrino flux represents the maximum attainable value in the one-zone leptohadronic model.

Starting from the constraints imposed by X-ray observations on synchrotron emission from BH-induced pairs and the detection of neutrinos with energy $E_{\rm prob}^{\rm obs}$, one can, using the $\delta$-approximation, derive energies of soft photons $E_{\rm tar, BH/p\gamma}$ relevant to the BH and $p\gamma$ processes, and thereby further obtain the corresponding interaction efficiencies. In the SSC-dominated case, the corresponding number density of synchrotron photons in BH and $p\gamma$ processes is 
\begin{equation}
n_{\rm tar, p\gamma/BH}=\frac{L_{\rm tar}^{\rm obs}(E_{\rm tar, p\gamma/BH}^{\rm obs})}{4\pi R^2c\delta_{\rm D}^4E_{\rm tar, p\gamma/BH}}, 
\end{equation}
where $L_{\rm tar}^{\rm obs}(E_{\rm tar, p\gamma/BH}^{\rm obs})$ represents the target photons luminosity at
\begin{equation}\label{EBH}
E_{\rm tar, BH}\approx 0.04\rm~keV(E_{\rm keV}^{\rm obs}/10\rm~keV)^{-1/2}(\it{B}/\rm1~G)^{1/2}(\delta_{\rm D}/10)^{1/2}, 
\end{equation}
and
 \begin{equation}\label{Epg}
E_{\rm tar, p\gamma}\approx 1.5\rm~keV(E_{\rm prob}^{\rm obs}/100\rm~TeV)^{-1}(\delta_{\rm D}/10). 
\end{equation}
For BH process, we have  
\begin{equation}\label{fbh}
\begin{split}
f_{\rm BH} &\simeq n_{\rm tar, BH} \langle \sigma_{\rm BH} \kappa_{\rm BH}\rangle R \\
&\simeq 2.8\times 10^{-6} \big( \frac{L_{\rm tar}^{\rm obs}(E_{\rm tar, BH}^{\rm obs})}{10^{45}\rm~erg/s} \big) \big( \frac{E_{\rm keV}^{\rm obs}}{10\rm~keV} \big)^{\frac{1}{2}} \\
&\quad \big( \frac{R}{10^{15}\rm~cm} \big)^{-1} \big( \frac{B}{1\rm~G} \big)^{-\frac{1}{2}} \big( \frac{\delta_{\rm D}}{10} \big)^{-\frac{9}{2}},
\end{split}
\end{equation}
where $\langle \sigma_{\rm BH} \kappa_{\rm BH}\rangle\approx7\times10^{-31}\,\rm cm^2$ represents the BH cross section weighted by inelasticity.
For $p\gamma$ process, we have
\begin{equation}\label{fpg}
\begin{split}
f_{p\gamma} &\simeq \frac{1}{4}n_{\rm tar, p\gamma} \langle \sigma_{p\gamma} \kappa_{p\gamma}\rangle R \\
&\simeq 2.8\times 10^{-7} \big(\frac{L_{\rm tar}^{\rm obs}(E_{\rm tar,p\gamma}^{\rm obs})}{10^{44}\rm~erg/s}\big)\big(\frac{E_{\rm prob}^{\rm obs}}{100\rm~TeV} \big)\big(\frac{R}{10^{15}\rm~cm} \big)^{-1}\\
&\big(\frac{\delta_{\rm D}}{10} \big)^{-5},
\end{split}
\end{equation}
where $\langle \sigma_{p\gamma} \kappa_{p\gamma}\rangle\approx10^{-28}\,\rm cm^2$ represents the $p\gamma$ cross section weighted by inelasticity.
In SSC-dominated cases, target photons for $p\gamma$ and BH interactions are non-thermal emission, which follows a power law distribution, i.e.,
\begin{equation}\label{llratio}
\begin{split}
\frac{L_{\rm tar}^{\rm obs}(E_{\rm tar,p\gamma}^{\rm obs})}{L_{\rm tar}^{\rm obs}(E_{\rm tar,BH}^{\rm obs})}&= (\frac{E_{\rm tar,p\gamma}^{\rm obs}}{E_{\rm tar,BH}^{\rm obs}})^\alpha\\
& = 37.5^\alpha \big(\frac{E_{\rm prob}^{\rm obs}}{100\rm~TeV}\big)^{-\alpha} \big(\frac{E_{\rm keV}^{\rm obs}}{10\rm~keV} \big)^{\frac{\alpha}{2}}\\
&\times \big(\frac{B}{1\rm~G} \big)^{-\frac{\alpha}{2}} \big(\frac{\delta_{\rm D}}{10} \big)^{\frac{\alpha}{2}}.
\end{split}
\end{equation}
Due to the constraints imposed by the X-ray spectrum on the synchrotron emission of BH-induced pairs, and by combining Eqs.\,(\ref{fbh}--\ref{llratio}), the corresponding neutrino flux at $E_{\rm prob}^{\rm obs}$ is 
\begin{equation}\label{ffratio}
\begin{split}
&F_{\nu_{\mu}+\overline{\nu}_{\mu}}^{\rm obs}(E_{\rm prob}^{\rm obs})\simeq \big(\frac{E_{p,\rm p\gamma}}{E_{p,\rm BH}}\big)^{2-n_{\rm p}}\frac{f_{p\gamma}}{f_{\rm BH}}F_{\rm keV}^{\rm obs},\\
& \simeq 0.8^{2-n_p}\times37.5^\alpha\times10^{-12}\rm~erg~s^{-1}~cm^{-2}\big(\frac{\it{B}}{1\rm~G} \big)^{\frac{3-n_p-\it{\alpha}}{2}}\\
&\times \big(\frac{\delta_{\rm D}}{10} \big)^{\frac{\it{\alpha}+n_p-\rm 3}{2}}\big(\frac{F_{\rm keV}^{\rm obs}}{10^{-12}\rm~erg~s^{-1}~cm^{-2}}\big)\big(\frac{E_{\rm prob}^{\rm obs}}{100\rm~TeV} \big)^{3-n_p-\alpha}\\
&\times \big(\frac{E_{\rm keV}^{\rm obs}}{10\rm~keV} \big)^{\frac{\alpha+n_p-3}{2}},
\end{split}
\end{equation}
where $E_{p,\rm p\gamma}=20E_{\rm prob}^{\rm obs}/\delta_{\rm D}$ is the proton energy in $p\gamma$ interactions, $E_{p,\rm BH}=0.01~\rm GeV^2/E_{\rm tar, BH}$ is the proton energy in BH interactions, and $n_{\rm p}$ represents the spectral index of injection proton energy distribution.
It can be seen that, for a neutrino-associated blazar with detections extending below the hard keV range, the model-predicted neutrino flux depends on both $B$ and $\delta_{\rm D}$. For the SSC-dominated cases in TMS regime, substituting Eq.~(\ref{Bdelta1}) into Eq.~(\ref{ffratio}), we have
\begin{equation}\label{SSCTMS}
\begin{split}
&F_{\nu_{\mu}+\overline{\nu}_{\mu}}^{\rm obs}(E_{\rm prob}^{\rm obs})\simeq  37.5^\alpha \times10^{\alpha-1}\times 2.8^{\frac{\alpha-1}{2}}\times 0.05^{2-n_p}\\
&\times 10^{-12} \rm erg~s^{-1}~cm^{-2} \big(\frac{\delta_{\rm D}}{10} \big)^{n_p+\alpha-\rm 3} \big(\frac{\it{F}_{\rm keV}^{\rm obs}}{10^{-12}\rm~erg~s^{-1}~cm^{-2}}\big)\\ &\times \big(\frac{E_{\rm prob}^{\rm obs}}{100\rm~TeV} \big)^{3-n_p-\alpha} \big(\frac{E_{\rm keV}^{\rm obs}}{10\rm~keV} \big)^{\frac{n_p+\alpha-3}{2}}\times \big(\frac{\nu_{\rm s,p}^{\rm obs}}{10^{14}\rm~Hz} \big)^{3-n_p-\alpha}\\
&\times \big(\frac{\nu_{\rm c,p}^{\rm obs}}{10^{23}\rm~Hz} \big)^{\frac{\alpha+n_p-3}{2}}.
\end{split}
\end{equation}
Since acceleration mechanisms generally predict a spectral index below 3, and for all the SSC-dominated sources in the TMS regime within our sample we find $\alpha<0$, the maximum model-predicted neutrino flux can therefore be derived when a minimum $\delta_{\rm D}$ is imposed using Eq.~(\ref{deltaSSC}). Then the corresponding $R$ and $B$ can be obtained as well. As studied in Section~\ref{sec:analytical method}, the chosen parameter sets corresponding to the maximum neutrino flux are marked with red points in the effective parameter space in the upper left panels of \cref{TXS 17,TXS 21,TXS 22,PKS 0735+178,GB6 J2113+1121,5BZB J0630-2406,PKS 1502+106,PKS B1424-418,GB6 J1040+0617}. It can be seen that, within the framework of the one-zone SSC model, a larger emission region is more favorable for producing neutrino emission. Taking $n_p=2$ as a fiducial value\footnote{Generally, a harder spectrum leads to a moderate increase in the luminosity allocated to the relatively high-energy protons responsible for high-energy neutrino production. Utilizing Eq.\, (\ref{SSCTMS}) and adopting the fiducial values of $\alpha=-1$, $\delta_{\rm D}=5$, $E_{\rm prob}^{\rm obs}=290\,{\rm TeV}$, $E_{\rm keV}^{\rm obs}=5\,{\rm keV}$, $\nu_{\rm s,p}=10^{15}\,{\rm Hz}$, and $\nu_{\rm c,p}=5\times10^{23}\,{\rm Hz}$, the derived neutrino fluxes for $n_p = 1.5$, $2.0$, and $2.5$ differ by a factor of $\sim 1.5$ from one another.}, in line with the diffusive shock acceleration \citep{1983RPPh...46..973D}, the corresponding maximum neutrino fluxes are $3.7\times10^{-13}\rm~erg~s^{-1}~cm^{-2}$ for TXS 0506+056 (2017), $3.9\times10^{-13}\rm~erg~s^{-1}~cm^{-2}$ for TXS 0506+056 (2021), $3.8\times10^{-13}\rm~erg~s^{-1}~cm^{-2}$ for TXS 0506+056 (2022), $2.9\times10^{-12}\rm~erg~s^{-1}~cm^{-2}$ for PKS 0735+178, $6.7\times10^{-14}\rm~erg~s^{-1}~cm^{-2}$ for GB6 J2113+1121, $8.5\times10^{-15}\rm~erg~s^{-1}~cm^{-2}$ for 5BZB J0630-2406, $1.5\times10^{-13}\rm~erg~s^{-1}~cm^{-2}$ for PKS 1502+106, $1.9\times10^{-11}\rm~erg~s^{-1}~cm^{-2}$ for PKS B1424-418, $1.9\times10^{-14}\rm~erg~s^{-1}~cm^{-2}$ for GB6 J1040+0617, respectively. For SSC-dominated cases in KN regime, Eq.~(\ref{Bdelta3}) shows that $\delta_{\rm D}\propto B$. Substituting this relation into Eq.~(\ref{ffratio}), one finds that the maximum neutrino flux becomes independent of the physical parameters of the emission region and can be directly determined from the measured information of the SED and high-energy neutrino. It should be noted that the neutrino flux estimated using the $\delta$-approximation is in fact lower than the numerical result, since higher-energy interactions produce a broad spectral distribution \citep{2008PhRvD..78c4013K}. However, this effect has a minor impact on our estimation, as it is partially canceled out by taking the ratio $f_{p\gamma}/f_{\rm BH}$.

For EC-dominated cases, Eq.~(\ref{EBH}) shows that the target photon energy of the BH process $E_{\rm tar, BH}$ lies close to the peak energy of the BLR, at about $10\rm~eV\delta_{\rm D}$ \citep{2008MNRAS.386..945T}, and its photon number density is approximated as
\begin{equation}\label{necbh}
n_{\rm tar, BH}\approx \xi_{\rm BLR}\frac{\delta_{\rm D}^3L_{\rm d}^{\rm AGN}}{4\pi cE_{\rm tar, BH}^{\rm obs}}.
\end{equation}
Combining Eqs.\,(\ref{EBH}) and (\ref{necbh}), then the corresponding $f_{\rm BH}$ can be estimated through
\begin{equation}\label{ecfbh}
\begin{split}
f_{\rm BH}&\approx 1.5\times10^{-5}\big(\frac{\xi_{\rm BLR}}{5.3\times10^{-36}\rm~cm^{-2}} \big)\big(\frac{L_{\rm d}^{\rm AGN}}{10^{45}\rm~erg~s^{-1}} \big)\\
&\big(\frac{E_{\rm keV}^{\rm obs}}{10\rm~keV} \big)^{\frac{1}{2}}\big(\frac{B}{1\rm~G} \big)^{-\frac{1}{2}}\big(\frac{\delta_{\rm D}}{10} \big)^{\frac{3}{2}}\big(\frac{R}{10^{15}\rm~cm} \big).
\end{split}
\end{equation}
For $p\gamma$ interactions, Eq.~(\ref{nudelta}) indicates that the target photon energy in the jet comoving frame to produce the observed neutrinos is $\sim \rm1~keV$, nearly an order of magnitude higher than the characteristic energy of the BLR. To account for the impact of the broad distribution associated with higher-energy neutrino production \citep{2008PhRvD..78c4013K}, and thereby obtain analytical results more consistent with the numerical model, we firstly estimate the BLR photon number density at the characteristic energy as
\begin{equation}
n_{\rm tar, p\gamma}\approx \xi_{\rm BLR}\frac{\delta_{\rm D}L_{\rm d}^{\rm AGN}}{4\pi ch\nu_{\rm BLR}^{\rm AGN}},
\end{equation}
and apply the $\delta$-approximation to estimate a higher interaction efficiency than that at $E_{\rm prob}^{\rm obs}$, i.e.,
\begin{equation}\label{ecfpg}
\begin{split}
f_{p\gamma}&\approx 2.6\times10^{-4}\big(\frac{\xi_{\rm BLR}}{5.3\times10^{-36}\rm~cm^{-2}} \big)\big(\frac{L_{\rm d}^{\rm AGN}}{10^{45}\rm~erg~s^{-1}} \big)\\
&\big(\frac{\delta_{\rm D}}{10} \big)\big(\frac{R}{10^{15}\rm~cm} \big),
\end{split}
\end{equation}
and then rescale $\frac{f_{p\gamma}}{f_{\rm BH}}F_{\rm keV}^{\rm obs}$ according to the unified neutrino spectral shape to obtain the flux at $E_{\rm prob}^{\rm obs}$, i.e.,
\begin{equation}\label{ffecratio}
\begin{split}
&F_{\nu_{\mu}+\overline{\nu}_{\mu}}^{\rm obs}(E_{\rm prob}^{\rm obs})\simeq \zeta (\frac{E_{\rm prob}^{\rm obs}}{E_{\nu,\rm peak}^{\rm obs}})^k \big(\frac{E_{p,\rm p\gamma}}{E_{p,\rm BH}}\big)^{2-n_{\rm p}} \frac{f_{p\gamma}}{f_{\rm BH}}F_{\rm keV}^{\rm obs},\\
&\simeq  14.5^{2-n_p}\times 5.3\times10^{-14}\rm~erg~s^{-1}~cm^{-2}\zeta \big(\frac{\it{B}}{1\rm~G} \big)^{\frac{3-n_p}{2}}\\&\times \big(\frac{\delta_{\rm D}}{10} \big)^{\frac{n_p-3}{2}}\times \big(\frac{F_{\rm keV}^{\rm obs}}{10^{-12}\rm~erg~s^{-1}~cm^{-2}}\big)\big(\frac{E_{\rm prob}^{\rm obs}}{100\rm~TeV} \big)^2\\
&\big(\frac{E_{\rm keV}^{\rm obs}}{10\rm~keV} \big)^{\frac{n_p-3}{2}},
\end{split}
\end{equation}
where $E_{\nu,\rm peak}^{\rm obs}=\frac{0.3\rm~GeV^2}{20h\nu_{\rm BLR}^{\rm AGN}}\approx 1.8\rm~PeV$, $E_{p, p\gamma}=20E_{\nu,\rm peak}^{\rm obs}/\delta_{\rm D}$, $\zeta \approx4$ is the correction factor accounting for the curvature of the neutrino spectrum (with $\zeta=1$ when $E_{\rm prob}^{\rm obs}\approx E_{\nu,\rm peak}^{\rm obs}$), $k\approx2$ characterizes the spectral slope on the low-energy side of the unified neutrino spectral distribution. Our numerical tests find that both $\zeta$ and $k$ remain nearly unchanged for $n_{\rm p}$ in a physically reasonable range of 1.5--2.5.
In the TMS regime, substituting Eq.~(\ref{Bdelta11}) into Eq.~(\ref{ffecratio}) we have
\begin{equation}\label{fectms}
\begin{split}
F_{\nu_{\mu}+\overline{\nu}_{\mu}}^{\rm obs}&\simeq 33.7^{2-n_p}\times 1.3\times10^{-13}\rm~erg~s^{-1}~cm^{-2}\zeta \big(\frac{E_{\rm prob}^{\rm obs}}{100\rm~TeV} \big)^2\\
&\times \big(\frac{\nu_{\rm s,p}^{\rm obs}}{10^{14}\rm~Hz} \big)^{\frac{3-n_p}{2}}\big(\frac{\nu_{\rm c,p}^{\rm obs}}{10^{23}\rm~Hz} \big)^{\frac{n_p-3}{2}} \big(\frac{E_{\rm keV}^{\rm obs}}{10\rm~keV} \big)^{\frac{n_p-3}{2}}\\
&\times  \big(\frac{F_{\rm keV}^{\rm obs}}{10^{-12}\rm~erg~s^{-1}~cm^{-2}}\big),
\end{split}
\end{equation}
and in the KN regime, substituting Eq.~(\ref{Bdelta44}) into Eq.~(\ref{ffecratio}) we have
\begin{equation}\label{feckn}
\begin{split}
F_{\nu_{\mu}+\overline{\nu}_{\mu}}^{\rm obs}&\simeq 11.7^{2-n_p}\times 3.3\times10^{-14}\rm~erg~s^{-1}~cm^{-2}\zeta \big(\frac{\nu_{\rm c,p}^{\rm obs}}{10^{23}\rm~Hz} \big)^{\frac{9-3n_p}{8}}\\
&\times \big(\frac{F_{\rm keV}^{\rm obs}}{10^{-12}\rm~erg~s^{-1}~cm^{-2}}\big)\big(\frac{E_{\rm prob}^{\rm obs}}{100\rm~TeV} \big)^2\big(\frac{E_{\rm keV}^{\rm obs}}{10\rm~keV} \big)^{\frac{n_p-3}{2}}\\
&\times \big(\frac{L_{\rm d}^{\rm AGN}}{10^{45}\rm~erg~s^{-1}} \big)^{\frac{3-n_p}{4}} \big(\frac{\xi_{\rm BLR}}{5.3\times10^{-36}\rm~cm^{-2}} \big)^{\frac{3-n_p}{4}}\\
&\times \big(\frac{L_{\rm s,p}^{\rm obs}}{5\times10^{46}\rm~erg~s^{-1}} \big)^{\frac{3-n_p}{4}} \big(\frac{L_{\rm c,p}^{\rm obs}}{5\times10^{46}\rm~erg~s^{-1}} \big)^{\frac{n_p-3}{4}}.
\end{split}
\end{equation}
From Eqs.~(\ref{fectms}, \ref{feckn}), it can be seen that the maximum neutrino flux predicted in the EC-dominated case for a blazar with well-measured SED depends weakly on the model parameters, with the only remaining free parameter being $n_{\rm p}$, under the assumption of approximately isotropic external radiation fields.
It also means that the maximum neutrino flux can be derived by placing the emission region near or inside the BLR, and there is little room to further increase neutrino flux by fine-tuning free parameters because of the nearly constant ratio $f_{p\gamma}/f_{\rm BH}$. But if one expects a higher $f_{p\gamma}$, a large $\delta_{\rm D}$ should be adopted, as indicated by Eq.~(\ref{ecfpg}). As suggested by the lower left panels of \cref{TXS 17,TXS 21,TXS 22,PKS 0735+178,GB6 J2113+1121,5BZB J0630-2406,PKS 1502+106,PKS B1424-418}, effective parameter spaces are given in TMS or KN regime.  Similarly, setting $n_{\rm p}=2$ as the fiducial value\footnote{Using the fiducial values of $E_{\rm keV}^{\rm obs}=5\,{\rm keV}$, $\nu_{\rm s,p}=10^{15}\,{\rm Hz}$, $\nu_{\rm c,p}=5\times10^{23}\,{\rm Hz}$, $L_{\rm d}^{\rm AGN}=5\times10^{45}\,\rm erg\,s^{-1}$, $\xi_{\rm BLR}=2\times10^{-35}\,\rm cm^2$, and $L_{\rm s/c,p}^{\rm obs}=2\times10^{46}\,\rm erg\,s^{-1}$, the resulting neutrino fluxes for $n_p = 1.5$, $2.0$, and $2.5$ differ by a factor of $\sim 1.5$ from one another for the TMS scenario (Eq.\,\ref{fectms}). In contrast, for the KN scenario (Eq.\,\ref{feckn}), the fluxes differ by a larger factor of $\sim 5$.}, and employing Eqs.~(\ref{fectms}, \ref{feckn}), the maximum neutrino fluxes are $6.8\times10^{-13}\rm~erg~s^{-1}~cm^{-2}$ for TXS 0506+056 (2017), $5.8\times10^{-13}\rm~erg~s^{-1}~cm^{-2}$ for TXS 0506+056 (2021), $3.6\times10^{-13}\rm~erg~s^{-1}~cm^{-2}$ for TXS 0506+056 (2022), $1.6\times10^{-12}\rm~erg~s^{-1}~cm^{-2}$ for PKS 0735+178, $8.5\times10^{-13}\rm~erg~s^{-1}~cm^{-2}$ for GB6 J2113+1121, $1.1\times10^{-14}\rm~erg~s^{-1}~cm^{-2}$ for 5BZB J0630-2406, $7.1\times10^{-13}\rm~erg~s^{-1}~cm^{-2}$ for PKS 1502+106, $1.4\times10^{-11}\rm~erg~s^{-1}~cm^{-2}$ for PKS B1424-418, respectively.

In the above, we have provided analytical expressions for estimating the maximum neutrino flux of blazars in the SSC-dominated and EC-dominated cases, as well as under the TMS and KN regimes. A well-covered broadband SED is essential for the effective application of this analytical approach. For other neutrino-associated blazars that lack quasi-simultaneous broadband coverage, particularly in the X-ray band, the analytical approach proposed in this section is difficult to apply effectively, as the peak parameters of the two humps cannot be well constrained. Therefore, numerical simulations that explore the parameter space provide a more effective method, as exemplified by the orphan neutrino flare of TXS 0506+056 \citep{2019ApJ...874L..29R, 2021ApJ...906...51X}.

\section{Numerical modeling} \label{sec:results}
In this section, we employ numerical modeling to scan the parameter space identified by the analytical approach, determine the physical parameter combinations that yield the maximum neutrino flux, and contrast them with the analytical results.
\subsection{Model Description}\label{sec:model description}
In the modeling, primary relativistic electrons are assumed to be injected with a smooth broken power-law energy distribution at a constant rate given by \citep{Ghisellini2010}
\begin{equation}
\begin{aligned}
    \dot{Q}_{\rm e}^{\rm inj}(\gamma_{\rm e})=&\dot{Q}_{\rm e,0}\gamma_{\rm e}^{-n_{\rm e,1}}\left[1+\left(\frac{\gamma_{\rm e}}{\gamma_{\rm e,b}}\right)^{(n_{\rm e,2}-n_{\rm e,1})}\right]^{-1},\\
    &\gamma_{\rm e,min}<\gamma_{\rm e}<\gamma_{\rm e,max},
\end{aligned}
\end{equation}
where $\gamma_{\rm e,min/b/max}$ are the minimum, break, and maximum electron Lorentz factors, $n_{\rm e,1/2}$ are the electron spectral indices before and after $\gamma_{\rm e,b}$, and $\dot{Q}_{\rm e,0}$ is a normalization constant in units of ${\rm s^{-1}~cm^{-3}}$, which can be calculated by $\int\dot{Q}_{\rm e}^{\rm inj}\gamma_{\rm e}m_{\rm e}c^2{\rm d}\gamma_{\rm e}=3L_{\rm e,inj}/(4\pi R^3)$ by giving an electron injection luminosity $L_{\rm e,inj}$. When injection is balanced by radiative cooling and/or particle escape, the resulting steady-state electron energy distribution enables the calculation of synchrotron, SSC, and EC emissions \citep{2001A&A...367..809K}.

Relativistic protons are assumed to be injected with a power-law energy distribution at a constant rate, which is given by
\begin{equation}
    \dot{Q}_{\rm p}^{\rm inj}(\gamma_{\rm p})=\dot{Q}_{\rm p,0}\gamma_{\rm p}^{-n_{\rm p}},\hspace{0.7cm}\gamma_{\rm p,min}<\gamma_{\rm p}<\gamma_{\rm p,max},
\end{equation}
where $\dot{Q}_{\rm p,0}$ is the normalization in units of ${\rm s^{-1}~cm^{-3}}$, $\gamma_{\rm p,min/max}$ are the minimum and maximum proton Lorentz factors, and $n_{\rm p}$ is the spectral index. By giving a proton injection luminosity $L_{\rm p,inj}$, $\dot{Q}_{\rm p,0}$ can be determined by $\int\dot{Q}_{\rm p}^{\rm inj}\gamma_{\rm p}m_{\rm p}c^2{\rm d}\gamma_{\rm p}=3L_{\rm p,inj}/(4\pi R^3)$, where $m_{\rm p}$ is the rest mass of proton. When radiative cooling, including $p\gamma$ interactions, BH pair production, and the proton synchrotron process, and/or escape balance with injection, a steady-state proton energy distribution is achieved. With the steady-state proton energy distribution, the differential spectra of decayed $\gamma$-ray photons, electrons/positrons (pairs) and neutrinos produced in the $p\gamma$ and BH processes are calculated with analytical expressions developed in \cite{2008PhRvD..78c4013K}. 
Besides, the secondary pairs coming from internal $\gamma\gamma$ pair production is also calculated, which is given by \citep{1983Afz....19..323A}
\begin{equation}
    \begin{aligned}
        \dot{Q}^{\gamma \gamma}_{\rm e}(\gamma_{\rm e}&) = \frac{3\sigma_{\rm T} c}{16}\int_{\gamma_{\rm e}}^{\infty}{dE_{\gamma}\frac{n_{\rm h}(E_{\rm h})}{\epsilon_{\rm h}^3}} \int_{\frac{\epsilon_{\rm h}}{4\gamma_{\rm e}(\epsilon_{\rm h}-\gamma_{\rm e})}}^{\infty}{dE_{\rm l}} \\
        \times&\frac{n_{\rm l}(E_{\rm l})}{\epsilon_{\rm l}^2} \left[ \frac{4\epsilon_{\rm h}^2}{\gamma_{\rm e}(\epsilon_{\rm h}-\gamma_{\rm e})} \ln{\left( \frac{4\gamma_{\rm e} \epsilon_{\rm l} (\epsilon_{\rm h}-\gamma_{\rm e})}{\epsilon_{\rm h}} \right)} \right.\\
        -&8\epsilon_{\rm h} \epsilon_{\rm l}+\frac{2\epsilon_{\rm h}^2(2\epsilon_{\rm h}\epsilon_{\rm l}-1)}{\gamma_{\rm e}(\epsilon_{\rm h}-\gamma_{\rm e})} -\left(1-\frac{1}{\epsilon_{\rm h}\epsilon_{\rm l}}\right) \\
        \times&\left.\left( \frac{\epsilon_{\rm h}^2}{\gamma_{\rm e}(\epsilon_{\rm h}-\gamma_{\rm e})} \right)^2 \right],
    \end{aligned}
\end{equation}
where $\epsilon_{\rm l}$ and $\epsilon_{\rm h}$ are the dimensionless energies of low- and high-energy photons, and $n_{\rm l}(E_{\rm l})$ and $n_{\rm h}(E_{\rm h})$ are the number density distribution of low- and high-energy photons, respectively.
The generated secondary pairs then continue to emit $\gamma$-ray photons through leptonic emissions. If the energy of these photons remains above the annihilation threshold, they further collide with low-energy photons, generating new pairs and repeating the cycle. This process progressively transfers energy until photon energies drop below the $\gamma\gamma$ annihilation threshold. Here, we find that the contribution of the fourth-generation photons can be neglected. Therefore, we calculate the total photon flux from three generations of secondary pairs. For the $\gamma\gamma$ annihilation during propagation, we correct the GeV-TeV spectrum using the extragalactic background light model of \cite{2021MNRAS.507.5144S}. A more detailed description of numerical leptohadronic model can be found in \cite{2019ApJ...886...23X}.

\subsection{Modeling results}

\begin{table*}[t]
    \centering
    \caption{Neutrino Detection Rate of Blazars.}
    \vspace{1mm}
    \label{table2}
    \resizebox{\textwidth}{!}{
    \hspace{-2.1cm}
    \begin{tabular}{cccccc}
        \hline\hline
        Object & TXS 0506+056 & TXS 0506+056 & TXS 0506+056 & PKS 0735+178 & GB6 J2113+1121\\
        Neutrino Event & IC-170922A & GVD-210418CA & IC-220918A & IC-211208A & IC-191001A\\
        \hline
        SSC-dominated case & 0.0048 & 0.0025 & 0.0096 & 0.061 & 0.02\\
        EC-dominated case & 0.012 & 0.019 & 0.035 & 0.076 & 0.11\\
        {Reference} & SSC-dominated case: 0.0036 \citep{2019MNRAS.483L..12C} & {1.46$^{\star}$ \citep{2024ApJ...962..142W}} & {0.82$^{\star}$ \citep{2024ApJ...962..142W}} & {0.076 \citep{2023MNRAS.519.1396S}} & {0.011 \citep{2025ApJ...986..110J}}\\
         & EC-dominated case: 0.002 \citep{2018ApJ...864...84K}, 0.3$^{\star}$ \citep{2019ApJ...886...23X} &  &  &  & \\
        \hline\hline
        Object & 5BZB J0630-2406 & PKS 1502+106 & PKS B1424-418 & GB6 J1040+0617 & 3HSP J095507.1+355101\\
        Neutrino Event & IC J0630-2353 & IC-190730A & Big Bird & IC-141209A & IC-200107A\\
        \hline
        SSC-dominated case & 0.14 & 0.012 & 0.017 & 0.0048 & 0.0041\\
        EC-dominated case & 0.30 & 0.032 & 0.073 & - & -\\
        Reference & 0.68 \citep{2023ApJ...958L...2F} & 0.007 \citep{2021ApJ...912...54R} & 0.076 \citep{2017ApJ...843..109G} & 0.64$^{\star}$ \citep{2021RAA....21..305W} & 0.005 \citep{2020ApJ...899..113P}\\
        \hline\hline
    \end{tabular}}
    \vspace{2mm}\\
    \raggedright
    \textbf{Notes.} Columns from top to bottom: the source name; neutrino event associated with blazar; the predicted neutrino detection rate in SSC-dominated case; the predicted neutrino detection rate in EC-dominated case; the predicted neutrino detection rate from reference.\\
    \footnotesize{$^{\star}$ The predicted neutrino detection rates from multi-zone models.}\\
\end{table*}

In the modeling, to better reconcile neutrino detections, we depart from the conventional Eddington luminosity as the upper limit of the proton injection luminosity. Instead, we mainly rely on observations from the keV band to constrain the proton injection luminosity. Because the neutrino flux $\nu F_{\nu}(E_{\rm prob}^{\rm obs})$ would not be overly influenced when the maximum energy of proton is greater than the proton energy corresponding to the observed neutrino, i.e., $E_{\rm p,max}^{\rm obs}>20E_{\rm prob}^{\rm obs}$, the maximum proton Lorentz factor is fixed at $\gamma_{\rm p,max}=1\times10^8$. In addition, we set $\gamma_{\rm p,min}=1$ and $n_{\rm p}=2$ for simplicity.

\begin{figure*}[htbp]
    \centering
    \begin{minipage}{0.49\linewidth}
        \centering
        \includegraphics[width=\linewidth, trim=40 15 90 30,clip]{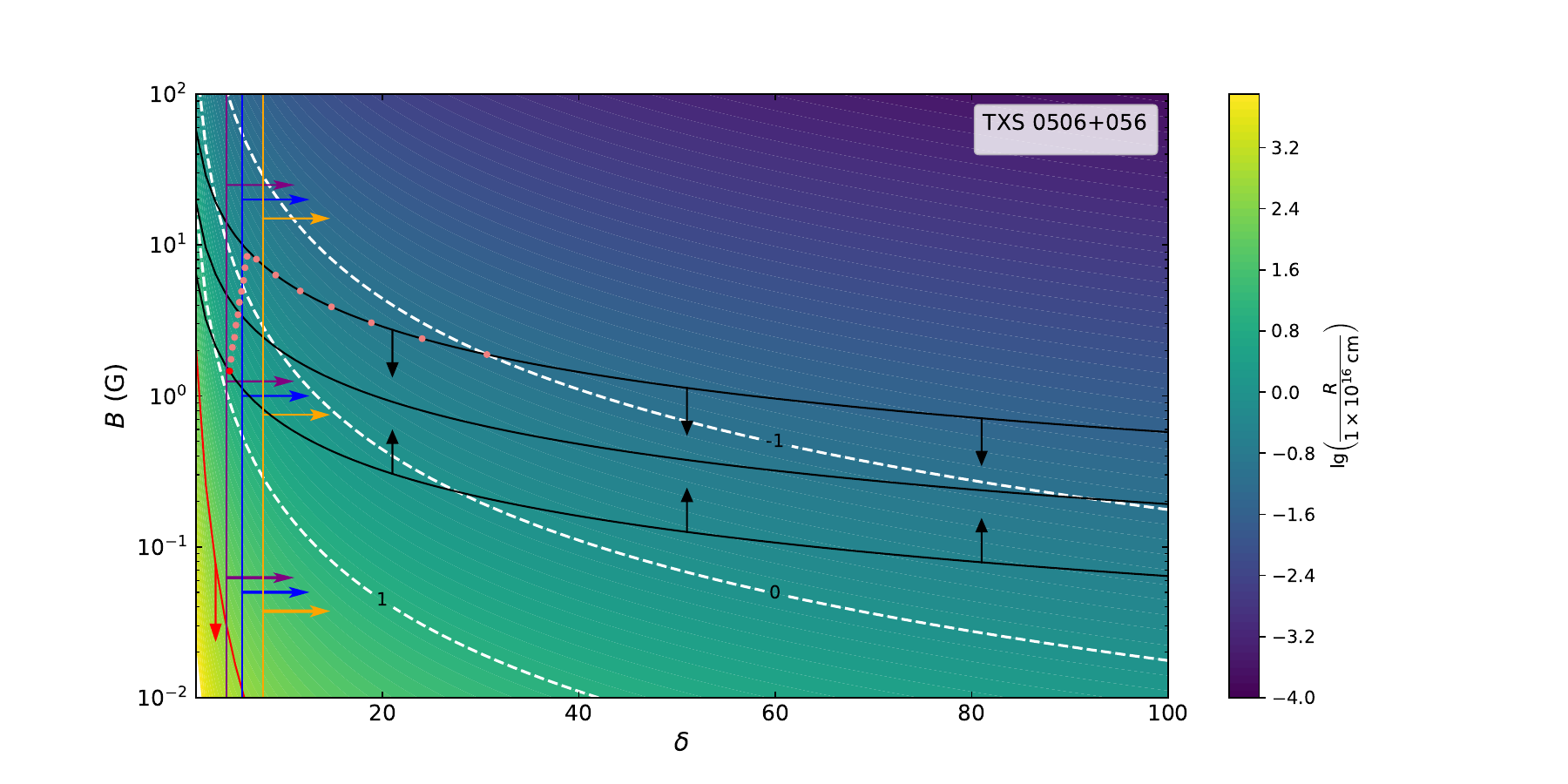}
        \label{TXS 17 SSC space}
    \end{minipage}\hspace{-4mm}
    \begin{minipage}{0.49\linewidth}
        \centering
        \includegraphics[width=\linewidth, trim=10 15 60 10,clip]{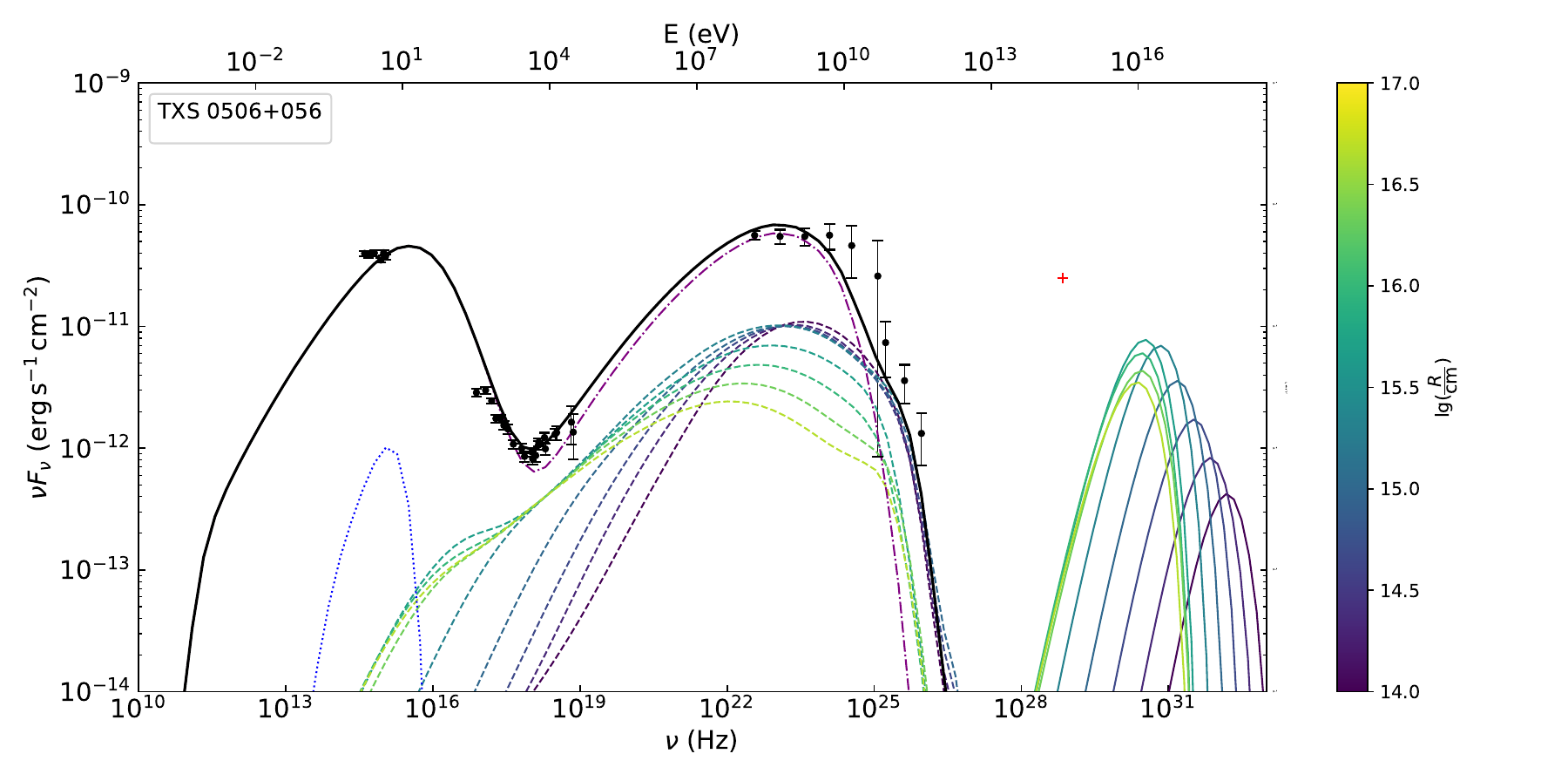}
        \label{TXS 17 SSC variable R}
    \end{minipage}

    \begin{minipage}{0.49\linewidth}
        \centering
        \includegraphics[width=\linewidth, trim=40 15 90 30,clip]{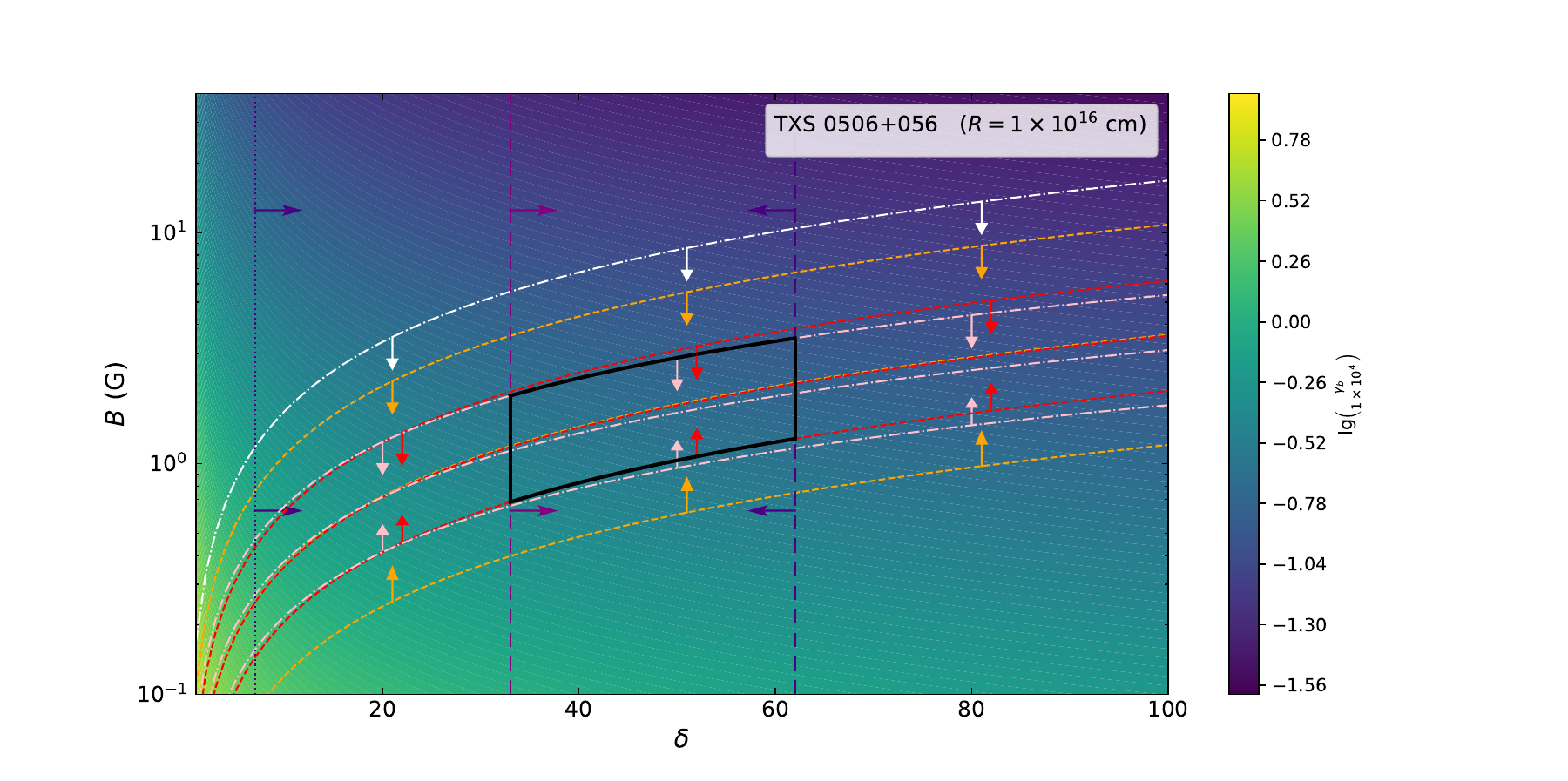}
        \label{TXS 17 EC space}
    \end{minipage}\hspace{-4mm}
    \begin{minipage}{0.49\linewidth}
        \centering
        \includegraphics[width=\linewidth, trim=10 15 60 10,clip]{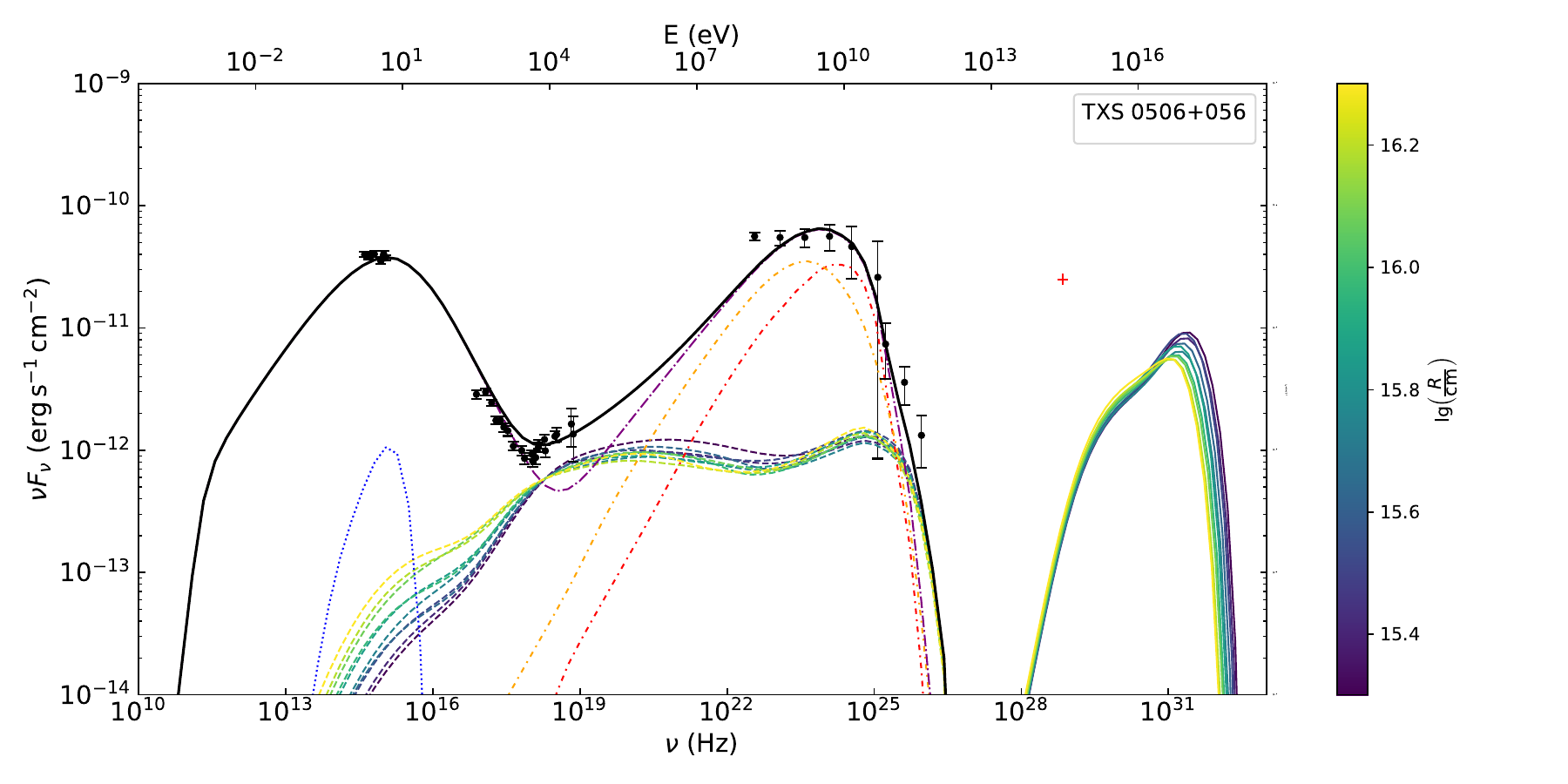}
        \label{TXS 17 EC variable R}
    \end{minipage}
    \caption{TXS 0506+056 associated with IC-170922A. Upper panels: the parameter space (left panel) and the fitting result of the SED for different radii of blob (right panel) under the SSC-dominated case. In the upper left panel, the red line with arrows represents the parameter space under the severe KN effect corresponding to Eq.~(\ref{severe KN}). The area within the black lines with arrows represents the peak frequency constraint, corresponding to Eq.~(\ref{Bdelta1}). The white dashed contours denote specific values of log$\left(\frac{R}{1\times10^{16}~{\rm cm}}\right)$ associated with the color bar that corresponds to Eq.~(\ref{Bdelta2}). The orange, blue and purple lines with arrows represent the internal optical depth constraint corresponding to Eq.~(\ref{deltaSSC}) for $R=1\times10^{15}~{\rm cm}$, $R=1\times10^{16}~{\rm cm}$ and $R=1\times10^{17}~{\rm cm}$, respectively. In the upper right panel, the black data points are quasi-simultaneous data and the red cross represents the neutrino flux at the energy of $290~{\rm TeV}$ during 1 year observation taken from \cite{2018Sci...361.1378I}. The blue dotted line represents the thermal radiation of the accretion disk. The purple dash-dotted line represents the leptonic emission from primary relativistic electrons. The colored dashed lines represent the secondary pair cascade emission, in which secondary pairs are from $\pi^{\pm}$ decay, BH pair production and internal $\gamma \gamma$ pair production for different radii of blob. The colored solid curves are the neutrino spectrum. The black solid line is the total emission with a certain radius of the blob. Lower panels: the parameter space for $R=1\times10^{16}~{\rm cm}$ (left panel) and the fitting result of the SED for different radii of blob (right panel) under the EC-dominated case. In the lower left panel, the white and pink dashed-dotted lines with arrows respectively represent the peak frequency constraint and peak luminosity constraint of BLR, corresponding to Eqs.~(\ref{Bdelta33}) and (\ref{Bdelta44}). The orange and red dashed lines with arrows respectively represent the peak frequency constraint and peak luminosity constraint of DT, corresponding to Eqs.~(\ref{Bdelta11}) and (\ref{Bdelta22}). The blue dotted and long-dashed lines with arrows are the optical depth constraint corresponding to Eqs.~(\ref{deltaSSC}) and (\ref{deltaEC}), respectively. The purple long-dashed line with arrows ensure the predominance of EC components corresponding to Eq.~(\ref{deltaECdomi}). The area surrounded by the black solid lines is the effective parameter space. In the lower right panel, the red and orange dashed-dotted lines respectively represent the EC emissions for BLR and DT photon fields. Other points and lines are the same as the SSC-dominated case.}
    \label{TXS 17}
\end{figure*}

\begin{figure*}[htbp]
    \centering
    \begin{minipage}[t]{0.49\linewidth}
        \centering
        \includegraphics[width=\linewidth, trim=40 15 90 30,clip]{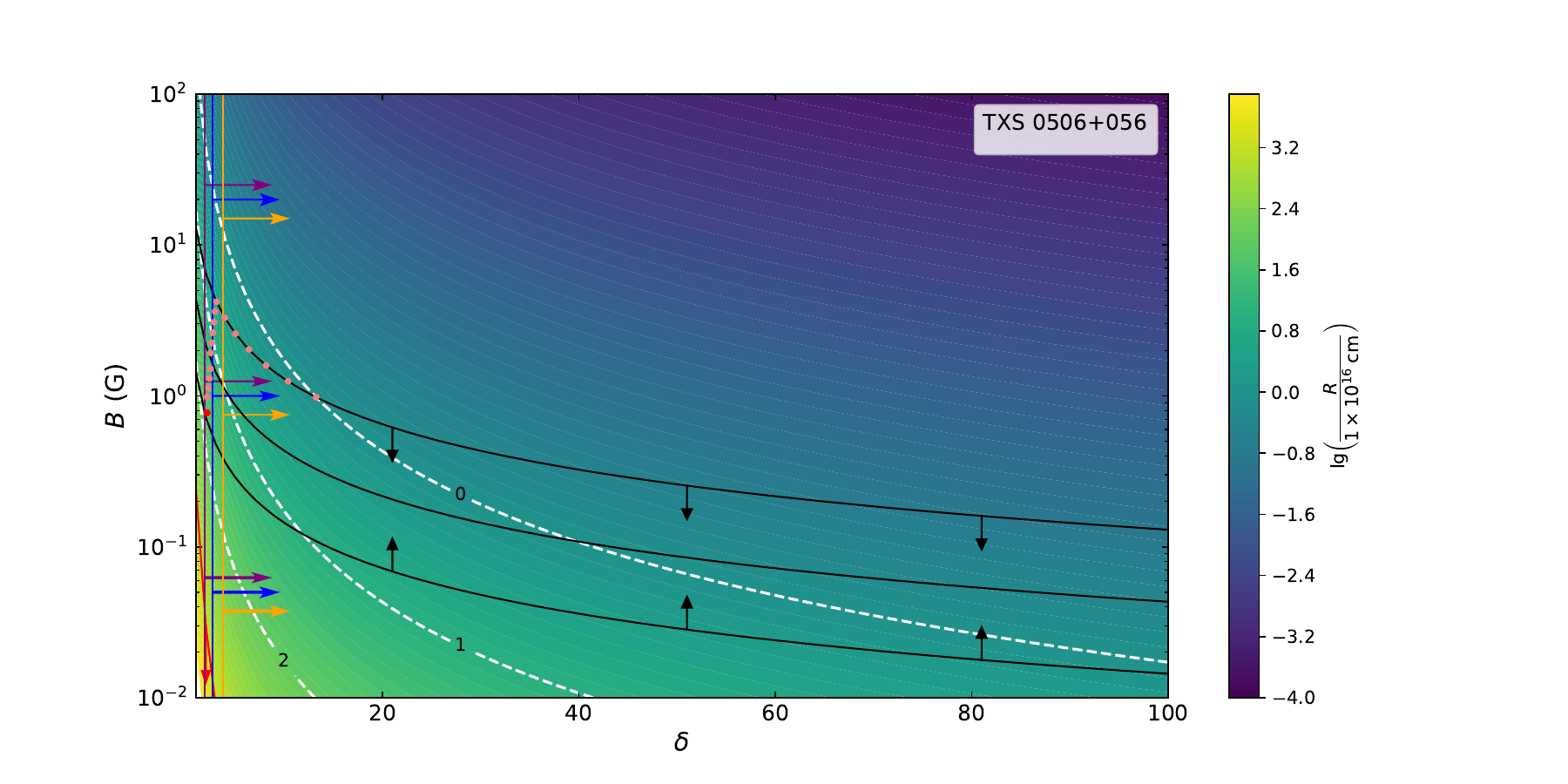}
        \label{TXS 21 SSC space}
    \end{minipage}\hspace{-4mm}
    \begin{minipage}[t]{0.49\linewidth}
        \centering
        \includegraphics[width=\linewidth, trim=10 15 55 10,clip]{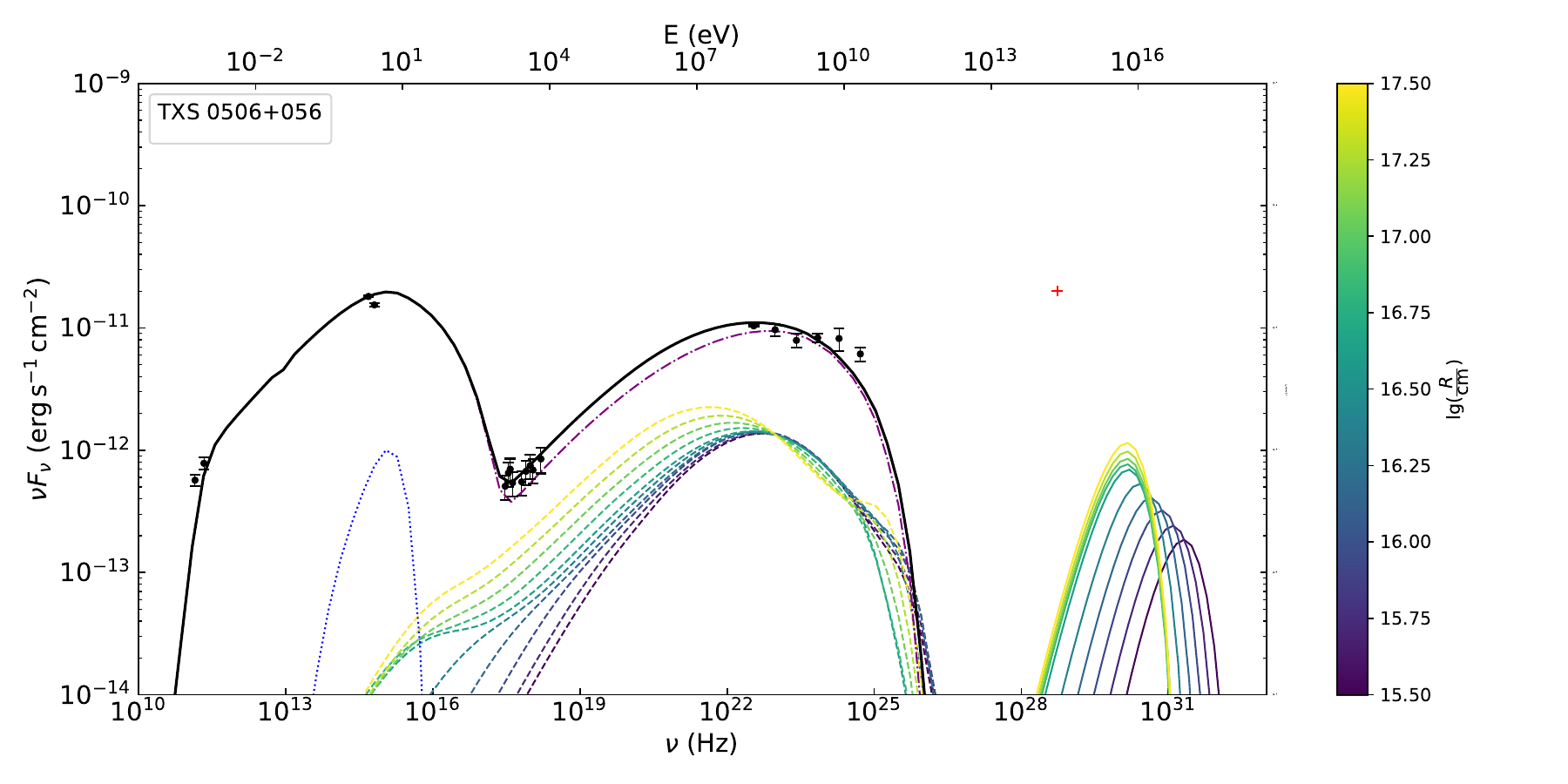}
        \label{TXS 21 SSC variable R}
    \end{minipage}
    
    \begin{minipage}{0.49\linewidth}
        \centering
        \includegraphics[width=\linewidth, trim=40 15 90 30,clip]{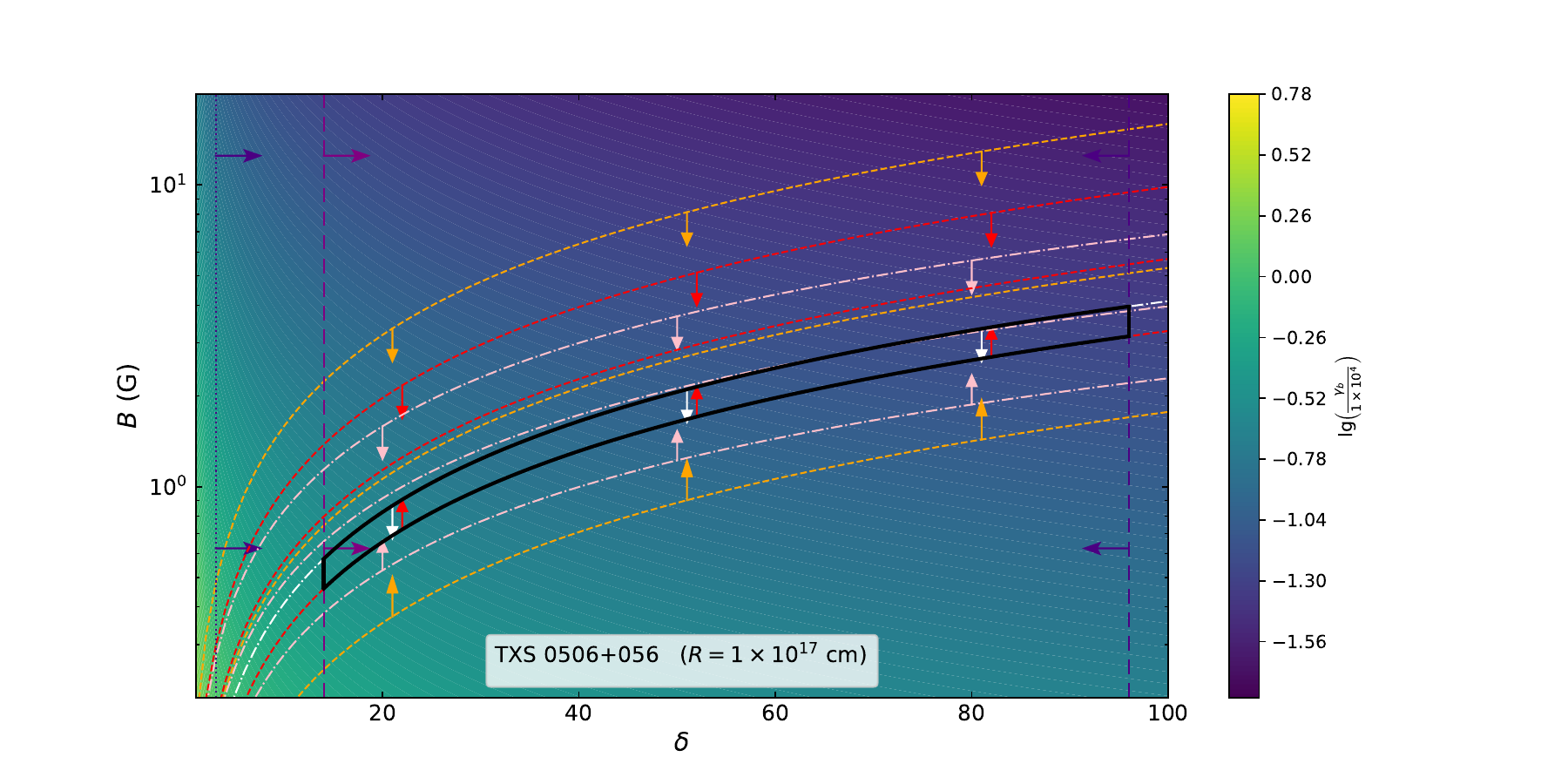}
        \label{TXS 21 EC space}
    \end{minipage}\hspace{-4mm}
    \begin{minipage}{0.49\linewidth}
        \centering
        \includegraphics[width=\linewidth, trim=10 15 55 10,clip]{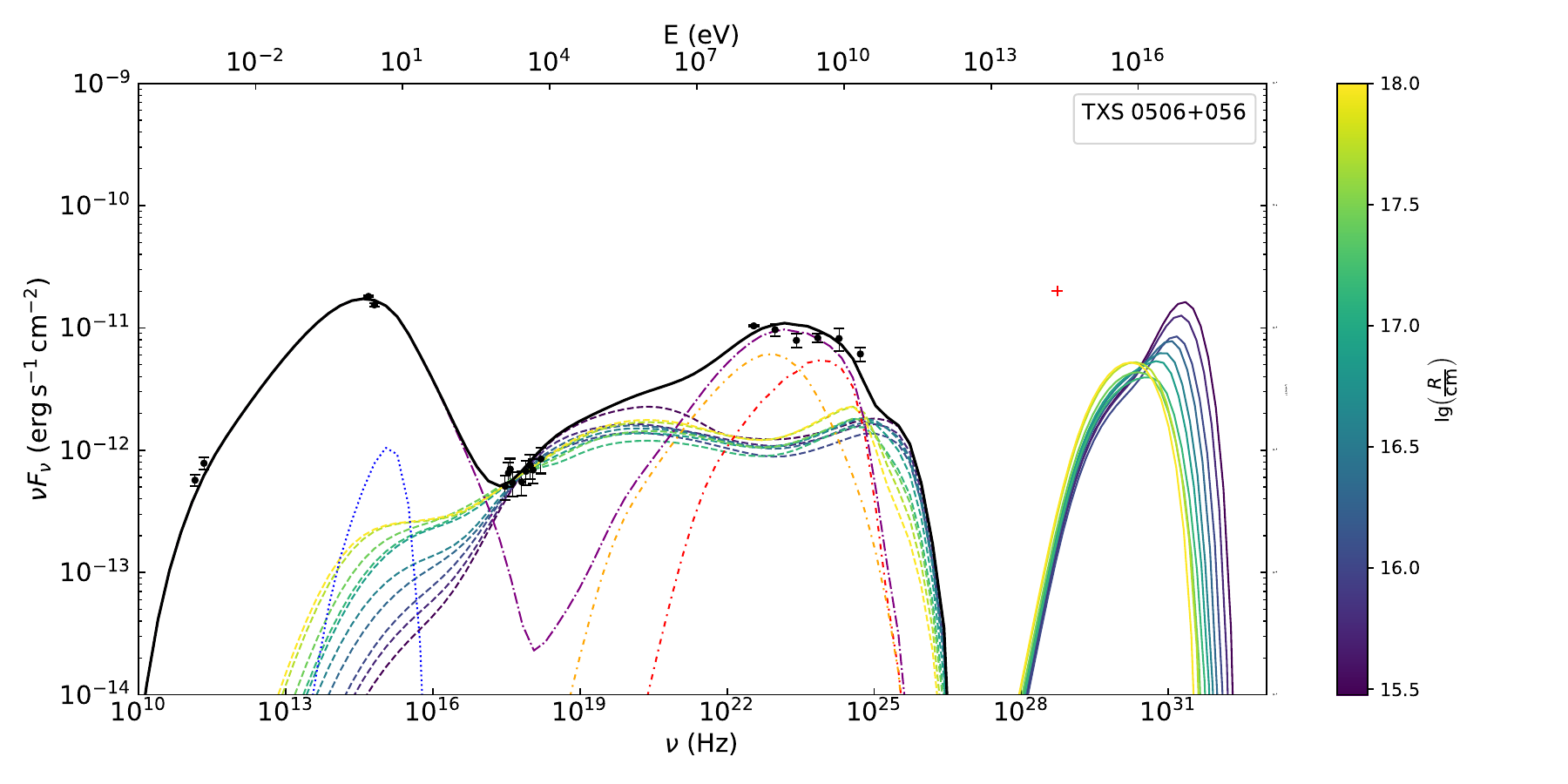}
        \label{TXS 21 EC variable R}
    \end{minipage}
    \caption{TXS 0506+056 associated with GVD-210418CA. Upper panels: the parameter space (left panel) and the fitting result of the SED for different radii of blob (right panel) under the SSC-dominated case. The line styles in upper panels have the same meaning as in Fig.~\ref{TXS 17}, expect orange, blue and purple lines with arrows represent the internal optical depth constraint corresponding to Eq.~(\ref{deltaSSC}) for $R=1\times10^{16}~{\rm cm}$, $R=1\times10^{17}~{\rm cm}$ and $R=1\times10^{18}~{\rm cm}$, respectively. In the upper right panel, the black data points are quasi-simultaneous data taken from \cite{2024ApJ...962..142W} and the red cross represents the neutrino flux at the energy of $224~{\rm TeV}$ during 1 year observation taken from \cite{2024MNRAS.527.8784A}. Lower panels: the parameter space for $R=1\times10^{17}~{\rm cm}$ (left panel) and the fitting result of the SED for different radii of blob (right panel) under the EC-dominated case. The line styles in lower panels have the same meaning as in Fig.~\ref{TXS 17}.}
    \label{TXS 21}
\end{figure*}

\begin{figure*}[htbp]
    \centering
    \begin{minipage}{0.49\linewidth}
        \centering
        \includegraphics[width=\linewidth, trim=40 15 90 30,clip]{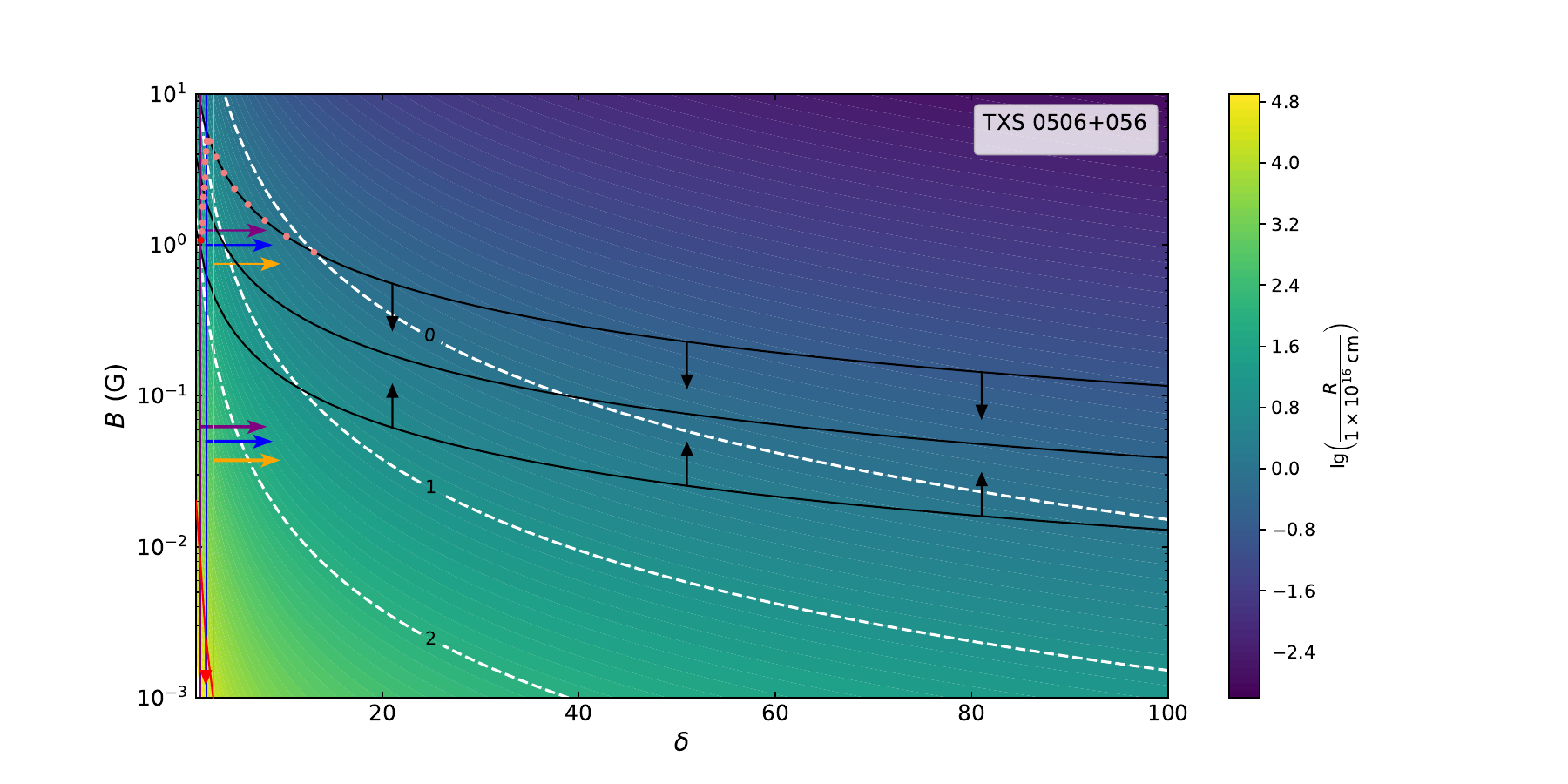}
        \label{TXS 22 SSC space}
    \end{minipage}\hspace{-4mm}
    \begin{minipage}{0.49\linewidth}
        \centering
        \includegraphics[width=\linewidth, trim=10 15 60 10,clip]{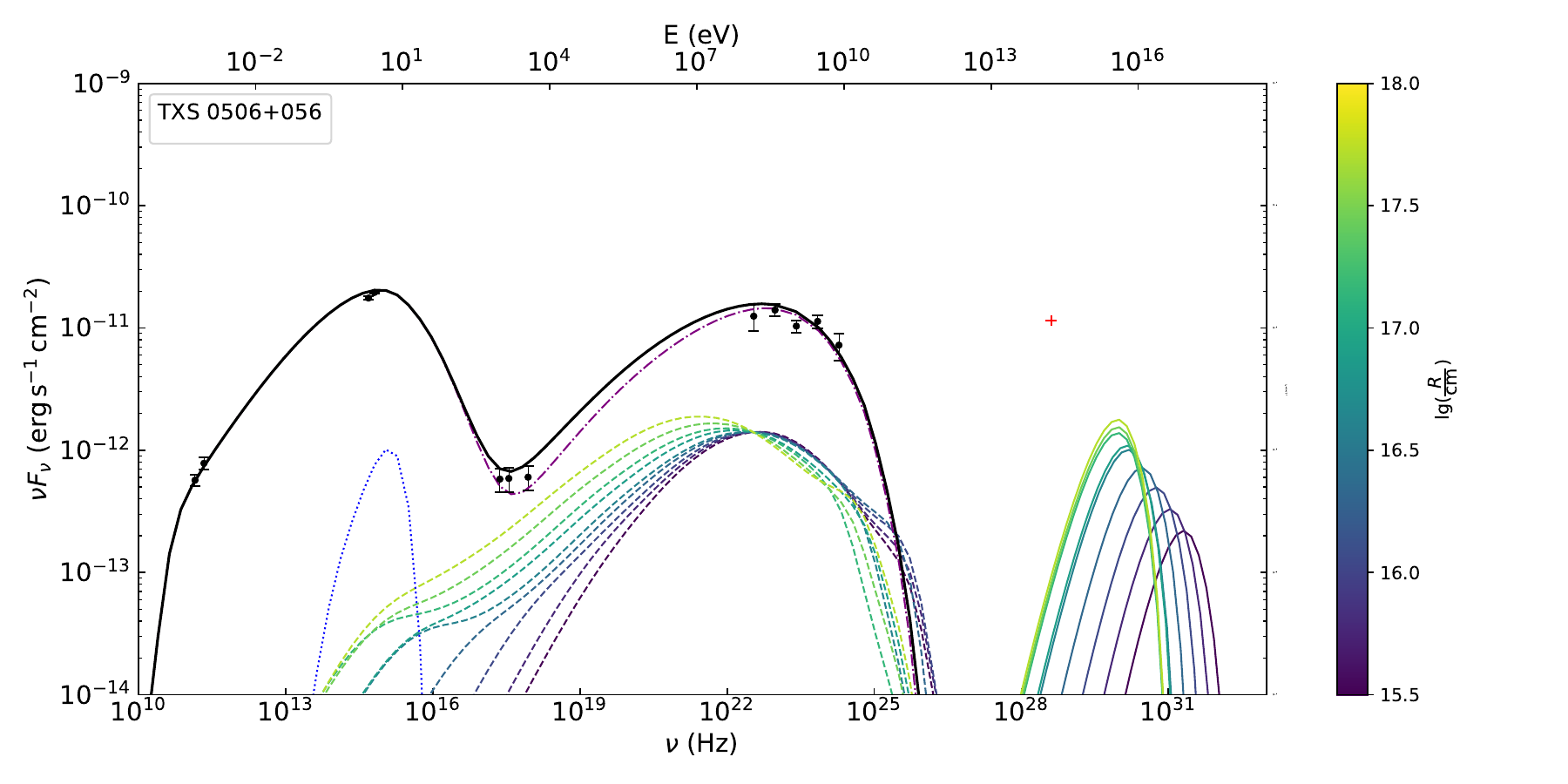}
        \label{TXS 22 SSC variable R}
    \end{minipage}

    \begin{minipage}{0.49\linewidth}
        \centering
        \includegraphics[width=\linewidth, trim=40 15 90 30,clip]{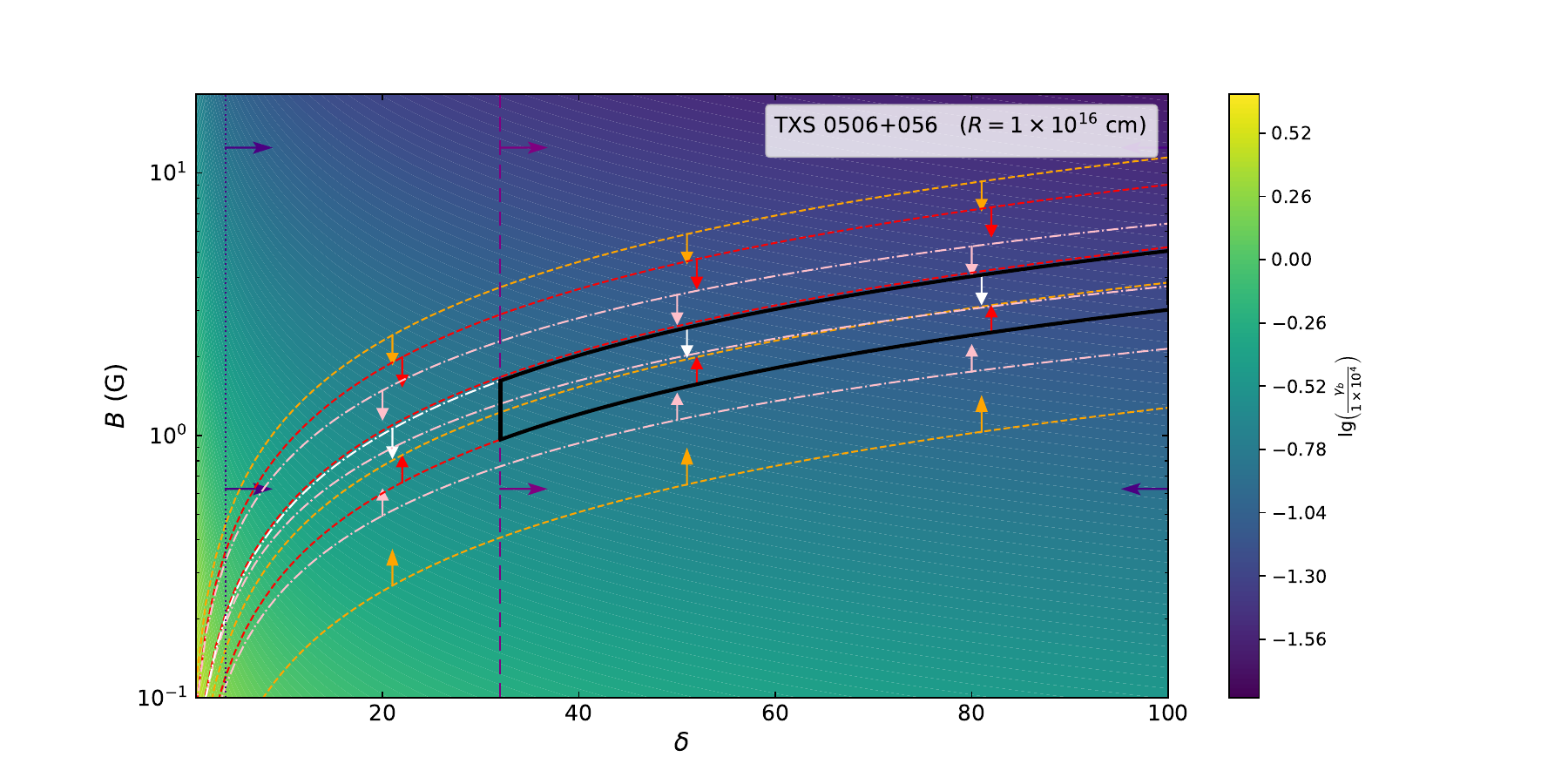}
        \label{TXS 22 EC space}
    \end{minipage}\hspace{-4mm}
    \begin{minipage}{0.49\linewidth}
        \centering
        \includegraphics[width=\linewidth, trim=10 15 60 10,clip]{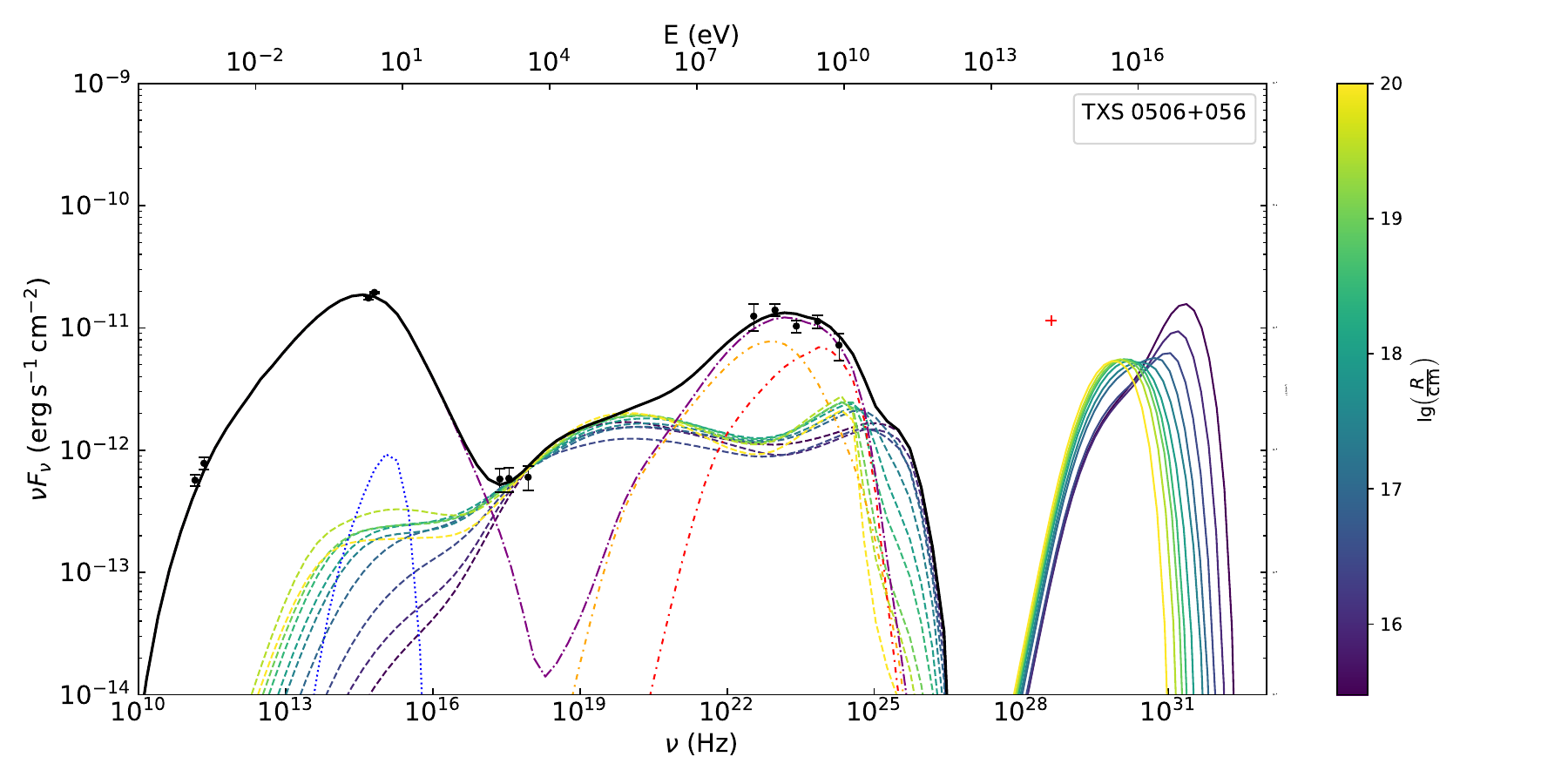}
        \label{TXS 22 EC variable R}
    \end{minipage}
    \caption{TXS 0506+056 associated with IC-220918A. Upper panels: the parameter space (left panel) and the fitting result of the SED for different radii of blob (right panel) under the SSC-dominated case. The line styles in upper panels have the same meaning as in Fig.~\ref{TXS 21}. In the upper right panel, the black data points are quasi-simultaneous data taken from \cite{2024ApJ...962..142W} and the red cross represents the neutrino flux at the energy of $170~{\rm TeV}$ during 1 year observation taken from \cite{2022ApJ...941L..25B}. Lower panels: the parameter space for $R=1\times10^{16}~{\rm cm}$ (left panel) and the fitting result of the SED for different radii of blob (right panel) under the EC-dominated case. The line styles in lower panels have the same meaning as in Fig.~\ref{TXS 17}.}
    \label{TXS 22}
\end{figure*}

\begin{figure*}[htbp]
    \centering
    \begin{minipage}{0.49\linewidth}
        \centering
        \includegraphics[width=\linewidth, trim=40 15 90 30,clip]{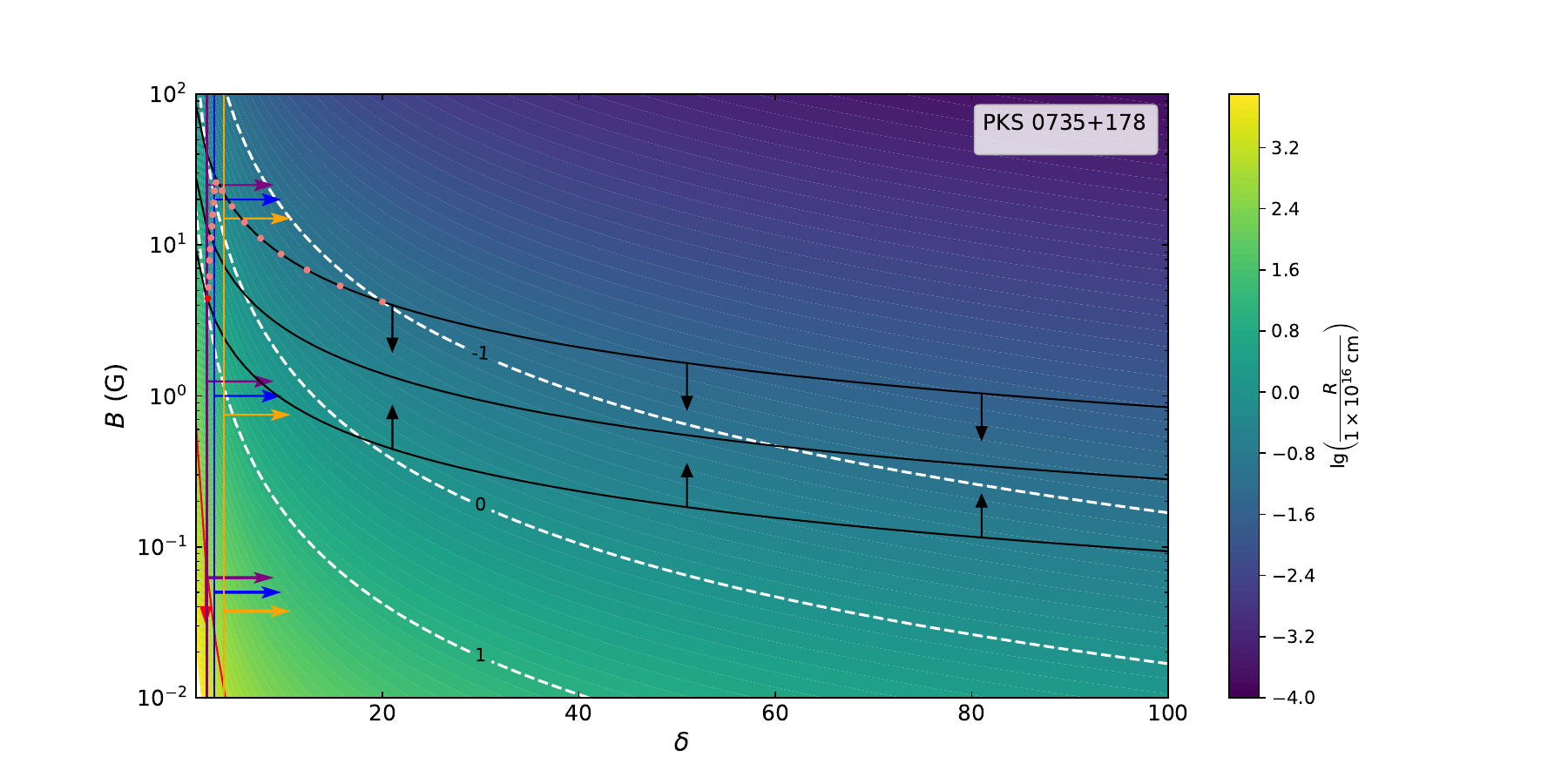}
        \label{PKS 0735+178 SSC space}
    \end{minipage}\hspace{-4mm}
    \begin{minipage}{0.49\linewidth}
        \centering
        \includegraphics[width=\linewidth, trim=10 15 60 10,clip]{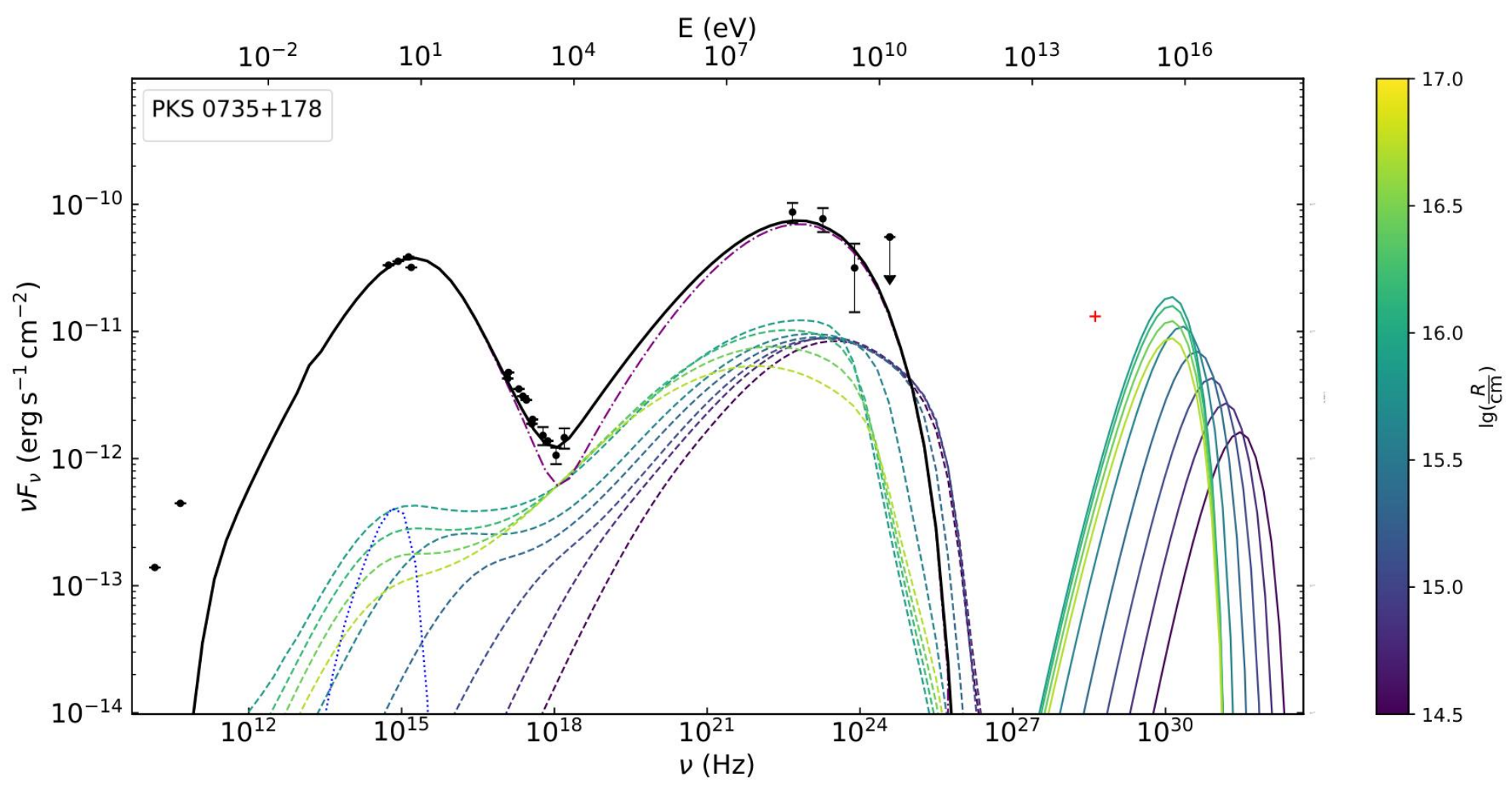}
        \label{PKS 0735+178 SSC variable R}
    \end{minipage}

    \begin{minipage}{0.49\linewidth}
        \centering
        \includegraphics[width=\linewidth, trim=40 15 90 30,clip]{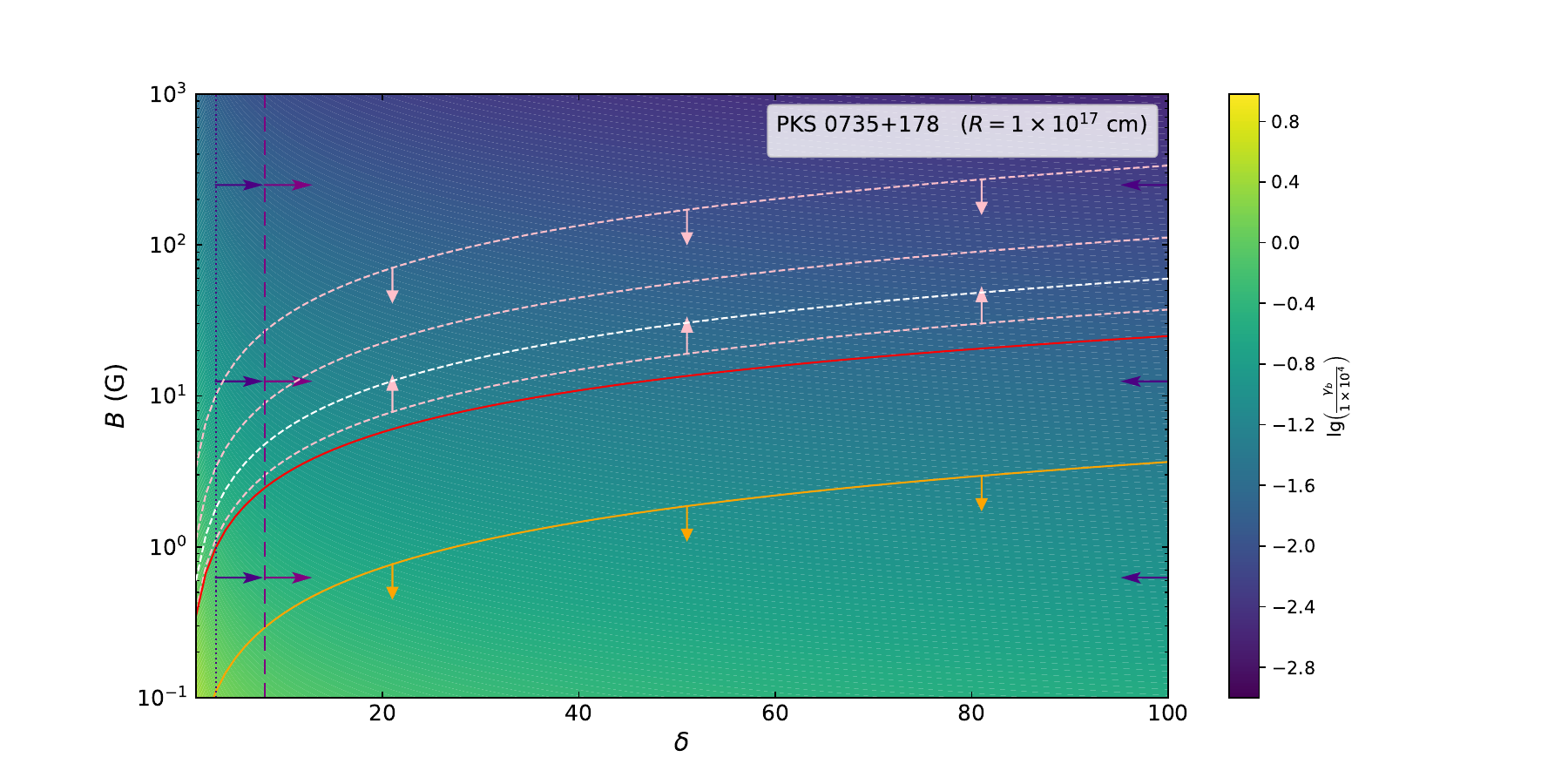}
        \label{PKS 0735+178 EC space}
    \end{minipage}\hspace{-4mm}
    \begin{minipage}{0.49\linewidth}
        \centering
        \includegraphics[width=\linewidth, trim=10 15 60 10,clip]{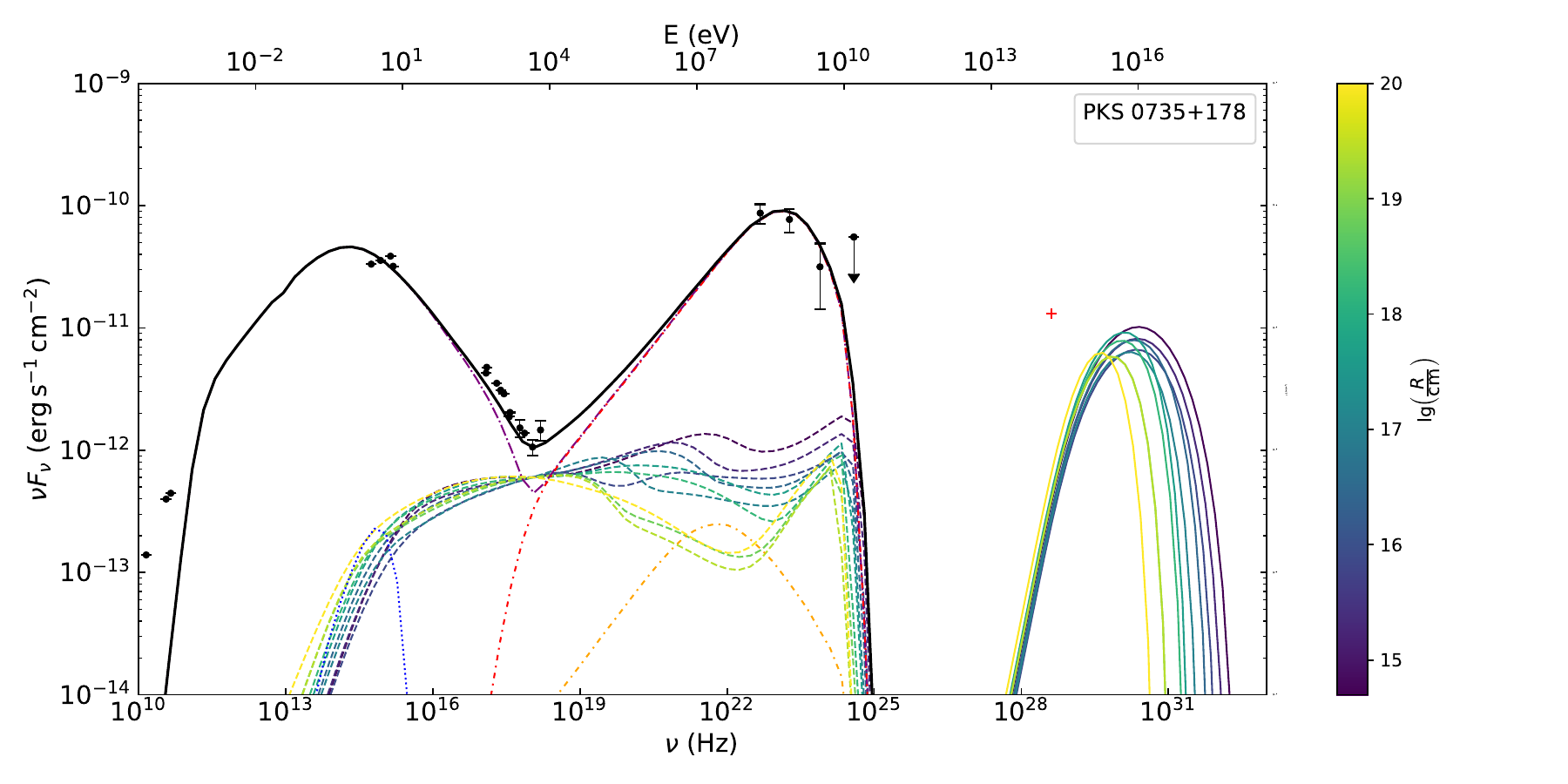}
        \label{PKS 0735+178 EC variable R}
    \end{minipage}
    \caption{PKS 0735+178 associated with IC-211208A. Upper panels: the parameter space (left panel) and the fitting result of the SED for different radii of blob (right panel) under the SSC-dominated case. The line styles in upper panels have the same meaning as in Fig.~\ref{TXS 17}. In the upper right panel, the black data points are quasi-simultaneous data and the red cross represents the neutrino flux at the energy of $172~{\rm TeV}$ during 1 year observation taken from \cite{2023MNRAS.519.1396S}. Lower panels: the parameter space for $R=1\times10^{17}~{\rm cm}$ (left panel) and the fitting result of the SED for different radii of blob (right panel) under the EC-dominated case. In the lower left panel, the pink and white dashed lines respectively represent the peak frequency constraint and peak luminosity constraint of BLR under the TMS regime, corresponding to Eqs.~\ref{Bdelta11} and \ref{Bdelta22}. The orange and red solid lines respectively represent the peak frequency constraint and peak luminosity constraint of BLR under the KN regime, corresponding to Eqs.~\ref{Bdelta33} and \ref{Bdelta44}. The other line styles in lower panels have the same meaning as in Fig.~\ref{TXS 17}.}
    \label{PKS 0735+178}
\end{figure*}

\begin{figure*}[htbp]
    \centering
    \begin{minipage}{0.49\linewidth}
        \centering
        \includegraphics[width=\linewidth, trim=40 15 90 30,clip]{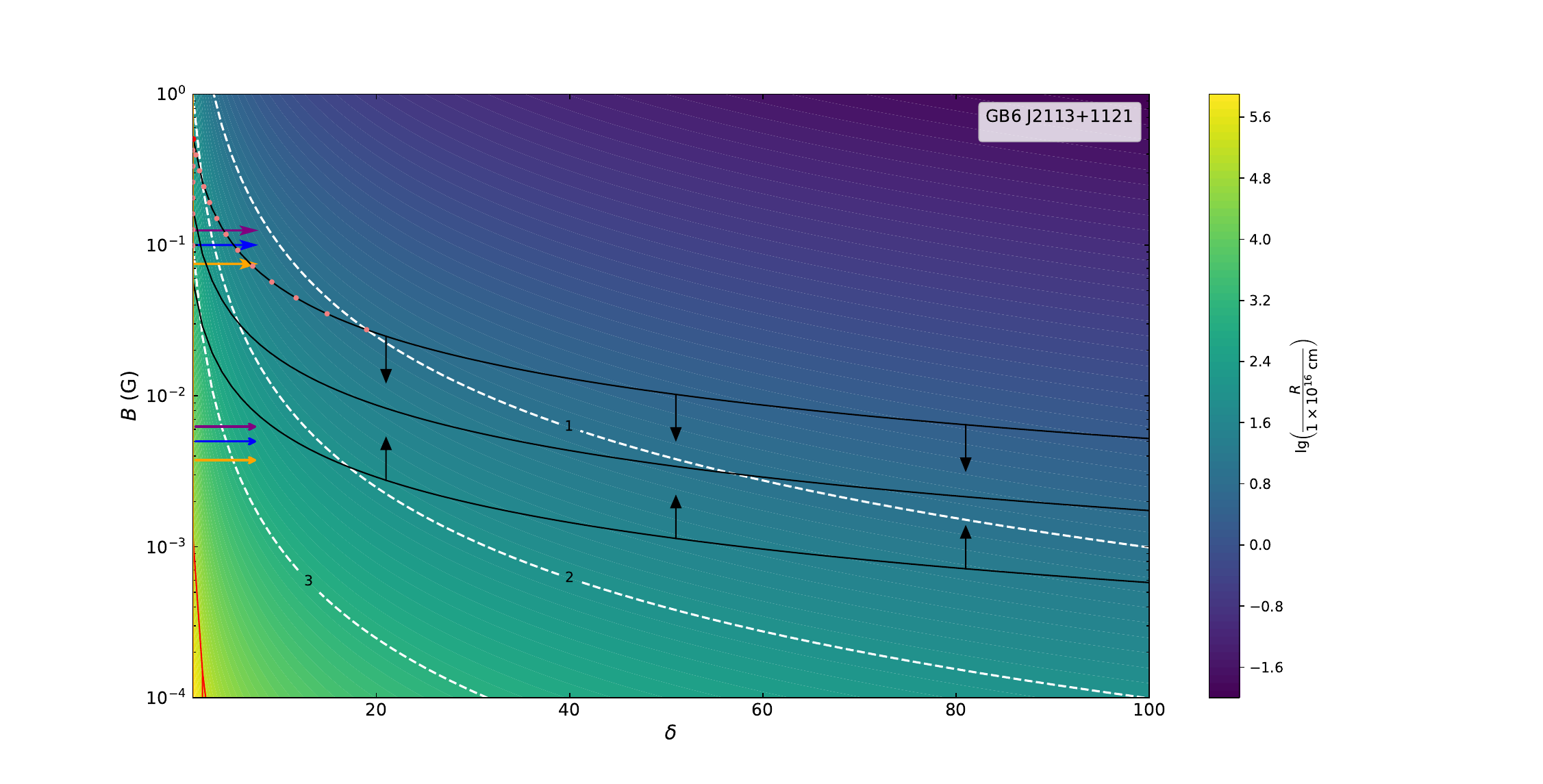}
        \label{GB6 J2113+1121 SSC space}
    \end{minipage}\hspace{-4mm}
    \begin{minipage}{0.49\linewidth}
        \centering
        \includegraphics[width=\linewidth, trim=10 15 55 10,clip]{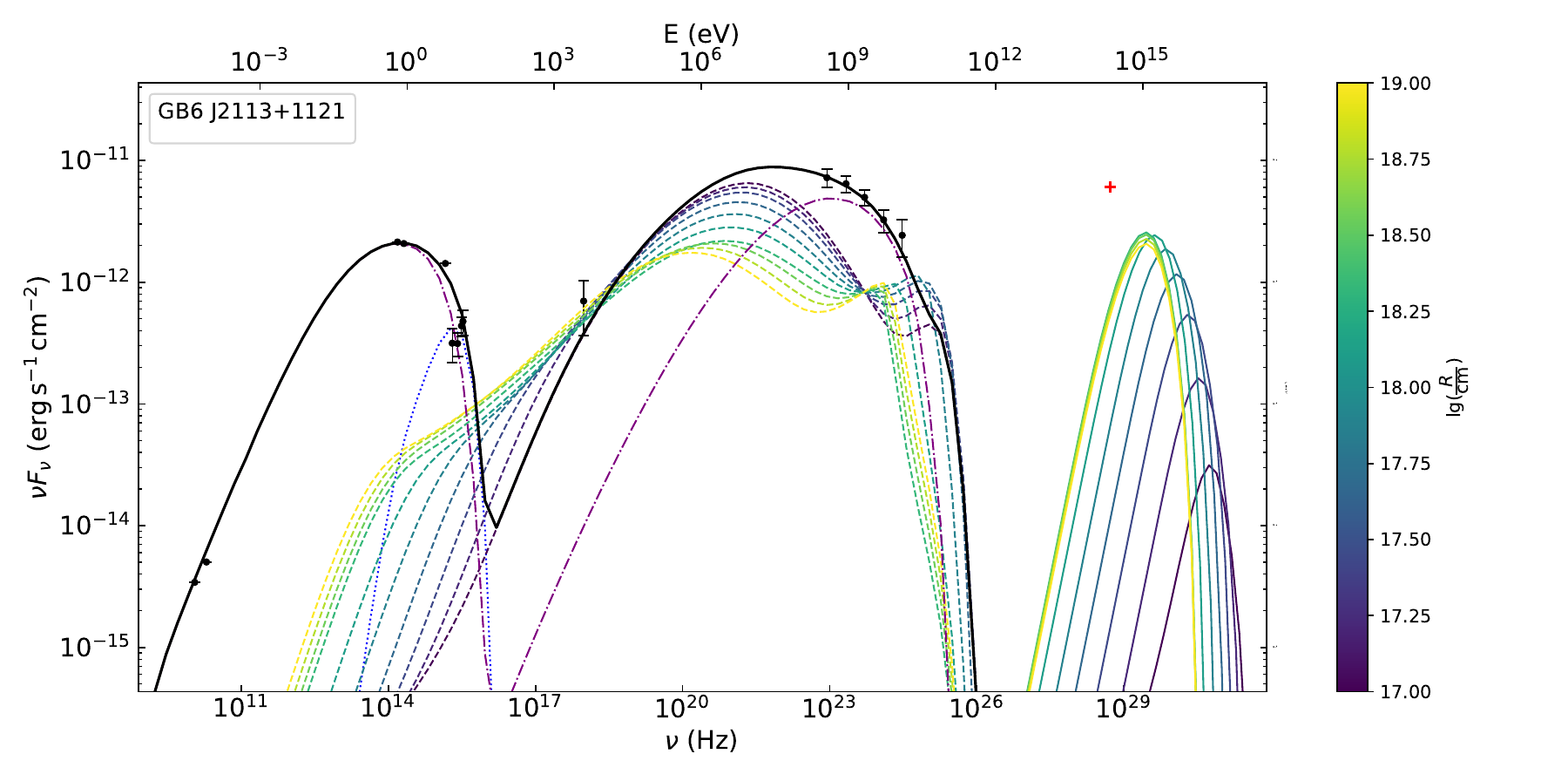}
        \label{GB6 J2113+1121 SSC variable R}
    \end{minipage}

    \begin{minipage}{0.49\linewidth}
        \centering
        \includegraphics[width=\linewidth, trim=40 15 90 30,clip]{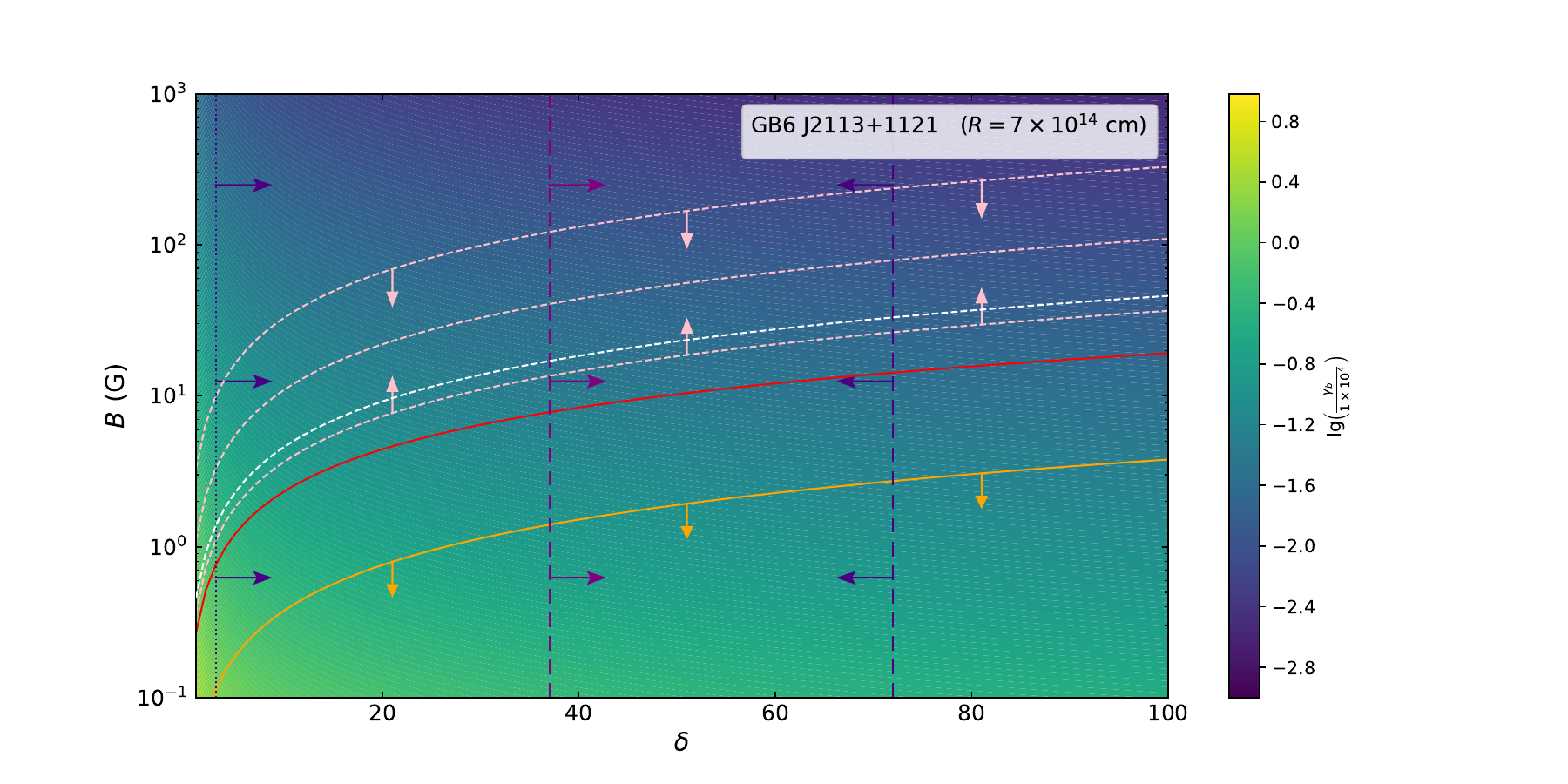}
        \label{GB6 J2113+1121 EC space}
    \end{minipage}\hspace{-4mm}
    \begin{minipage}{0.49\linewidth}
        \centering
        \includegraphics[width=\linewidth, trim=10 15 55 10,clip]{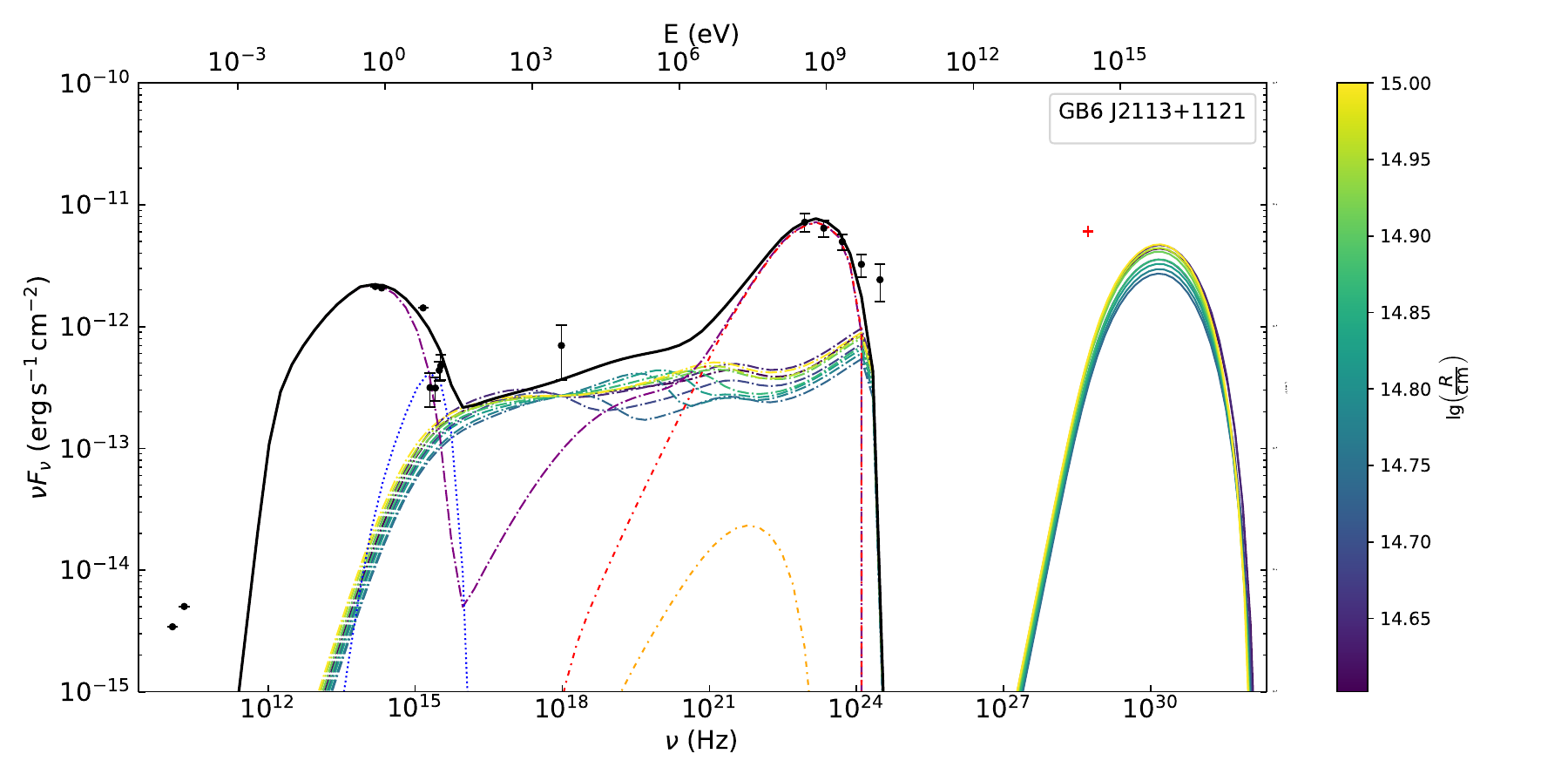}
        \label{GB6 J2113+1121 EC variable R}
    \end{minipage}
    \caption{GB6 J2113+1121 associated with IC-191001A. Upper panels: the parameter space (left panel) and the fitting result of the SED for different radii of blob (right panel) under the SSC-dominated case. The line styles in upper panels have the same meaning as in Fig.~\ref{TXS 17}, expect orange, blue and purple lines with arrows represent the internal optical depth constraint corresponding to Eq.~(\ref{deltaSSC}) for $R=1\times10^{17}~{\rm cm}$, $R=1\times10^{18}~{\rm cm}$ and $R=1\times10^{19}~{\rm cm}$, respectively. In the upper right panel, the black data points are quasi-simultaneous data and the red cross represents the neutrino flux at the energy of $0.2~{\rm PeV}$ during 1 year observation taken from \cite{2022ApJ...932L..25L}. Lower panels: the parameter space for $R=7\times10^{14}~{\rm cm}$ (left panel) and the fitting result of the SED for different radii of blob (right panel) under the EC-dominated case. The line styles in lower panels have the same meaning as in Fig.~\ref{PKS 0735+178}.}
    \label{GB6 J2113+1121}
\end{figure*}

\begin{figure*}[htbp]
    \centering
    \begin{minipage}{0.49\linewidth}
        \centering
        \includegraphics[width=\linewidth, trim=40 15 90 30,clip]{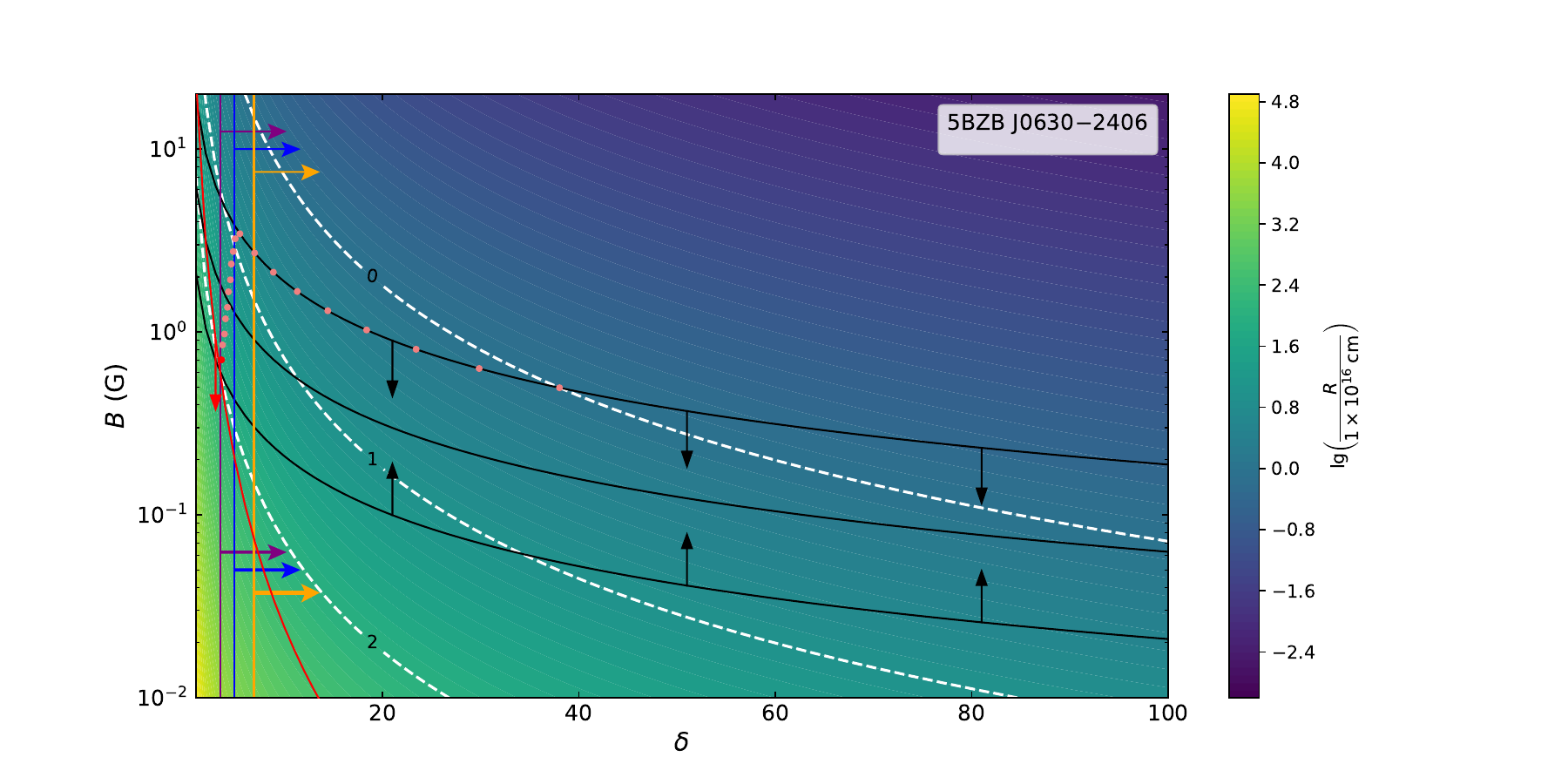}
        \label{5BZB J0630-2406 SSC space}
    \end{minipage}\hspace{-4mm}
    \begin{minipage}{0.49\linewidth}
        \centering
        \includegraphics[width=\linewidth, trim=10 15 55 10,clip]{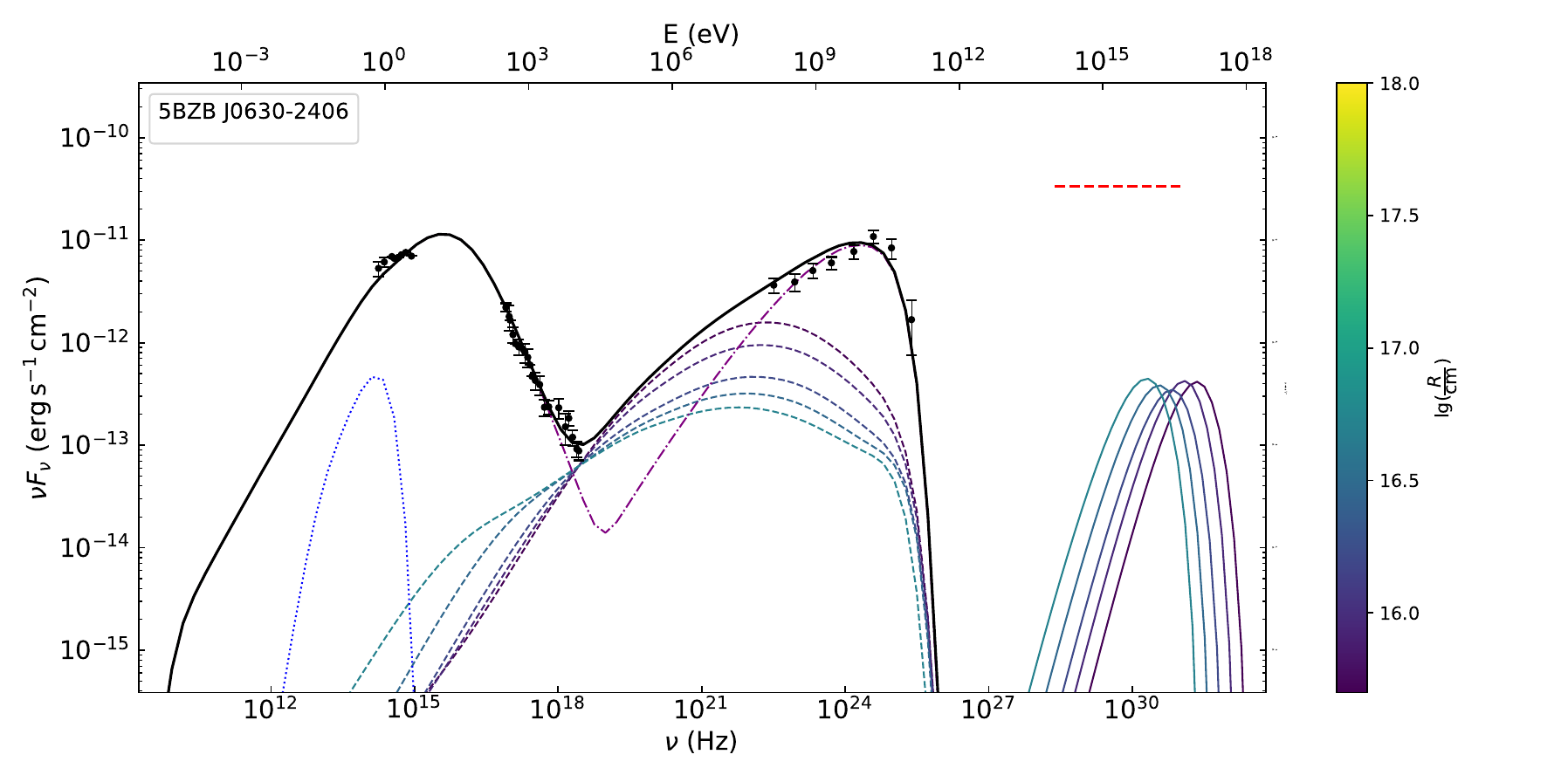}
        \label{5BZB J0630-2406 SSC variable R}
    \end{minipage}

    \begin{minipage}{0.49\linewidth}
        \centering
        \includegraphics[width=\linewidth, trim=40 15 90 30,clip]{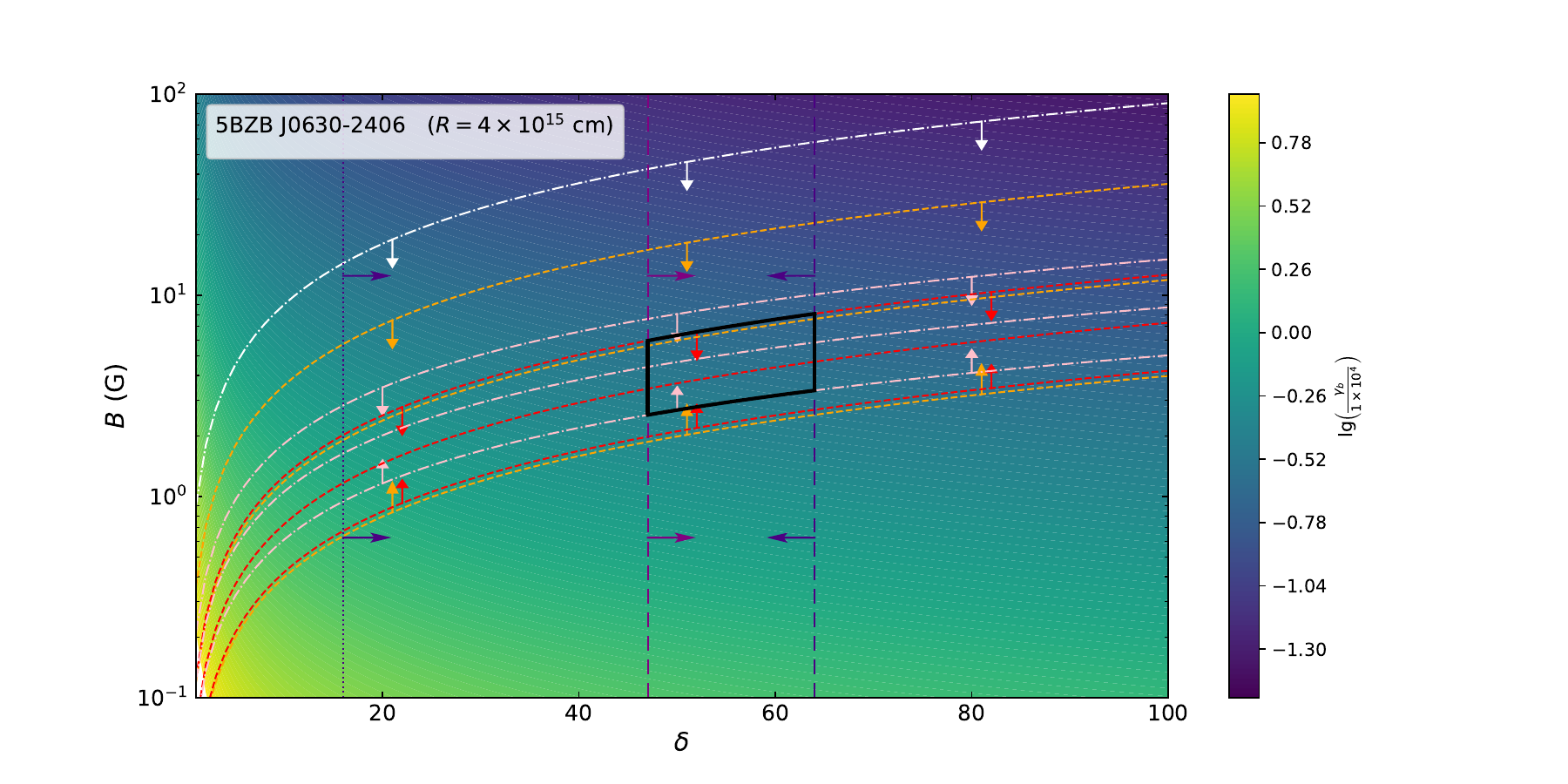}
        \label{5BZB J0630-2406 EC space}
    \end{minipage}\hspace{-4mm}
    \begin{minipage}{0.49\linewidth}
        \centering
        \includegraphics[width=\linewidth, trim=10 15 55 10,clip]{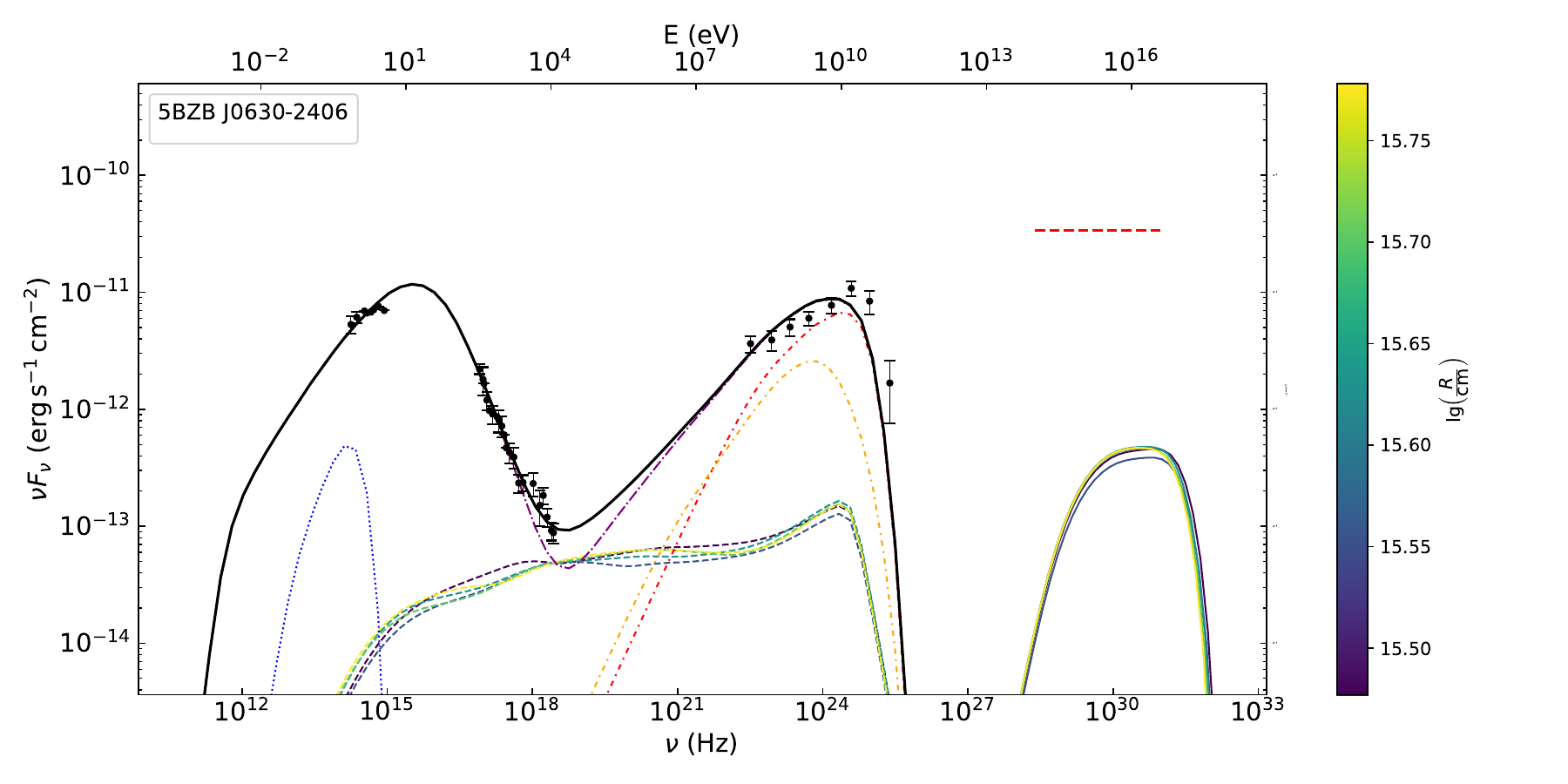}
        \label{5BZB J0630-2406 EC variable R}
    \end{minipage}
    \caption{5BZB J0630-2406 associated with IC J0630-2353. Upper panels: the parameter space (left panel) and the fitting result of the SED for different radii of blob (right panel) under the SSC-dominated case. The line styles in upper panels have the same meaning as in Fig.~\ref{TXS 21}. In the upper right panel, the black data points are quasi-simultaneous data and the horizontal red dashed line represents the neutrino flux taken from \cite{2023ApJ...958L...2F}. Lower panels: the parameter space for $R=4\times10^{15}~{\rm cm}$ (left panel) and the fitting result of the SED for different radii of blob (right panel) under the EC-dominated case. The line styles in lower panels have the same meaning as in Fig.~\ref{TXS 17}.}
    \label{5BZB J0630-2406}
\end{figure*}

\begin{figure*}[htbp]
    \centering
    \begin{minipage}{0.49\linewidth}
        \centering
        \includegraphics[width=\linewidth, trim=40 15 90 30,clip]{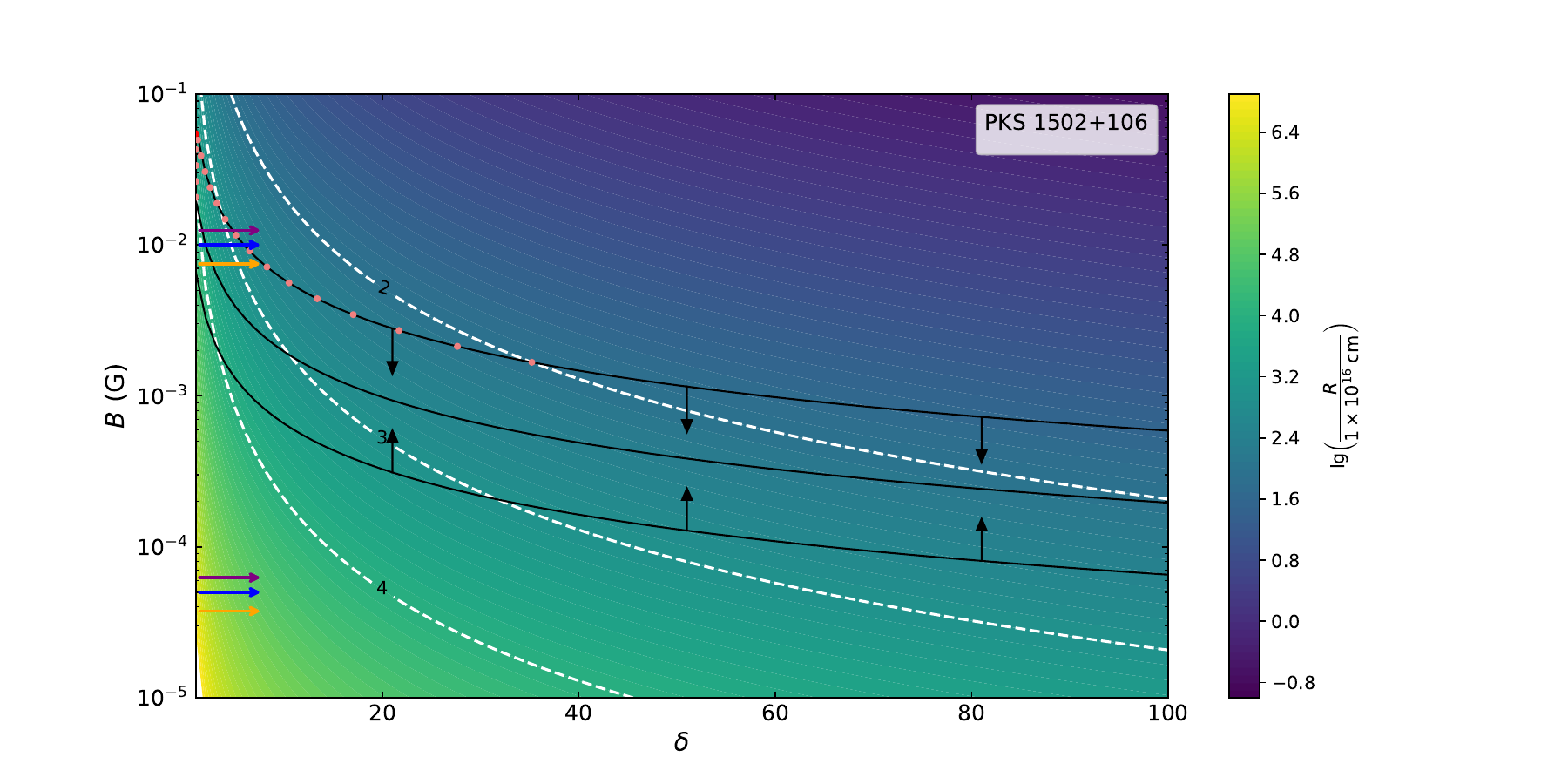}
        \label{PKS 1502+106 SSC space}
    \end{minipage}\hspace{-4mm}
    \begin{minipage}{0.49\linewidth}
        \centering
        \includegraphics[width=\linewidth, trim=10 15 60 10,clip]{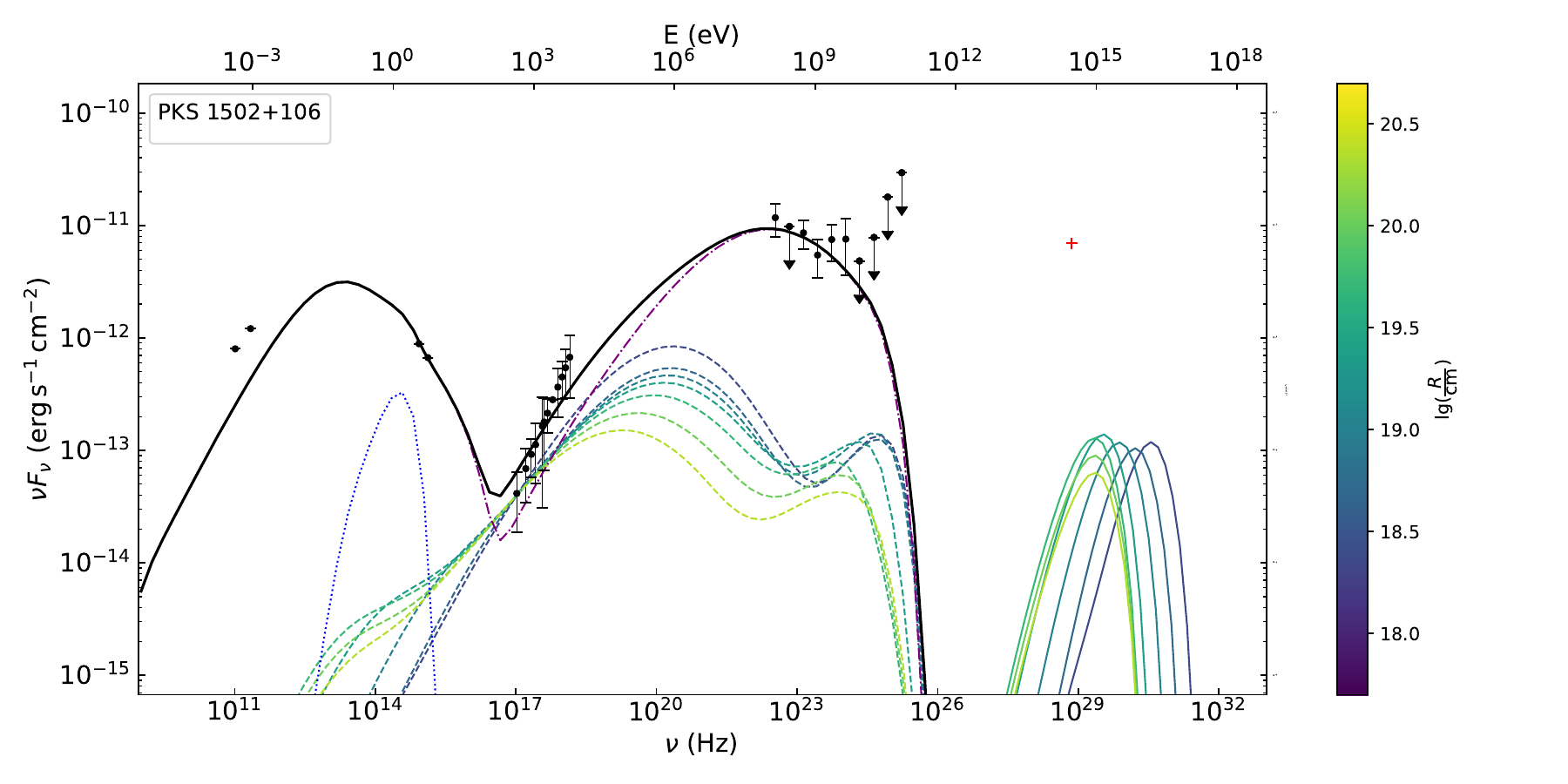}
        \label{PKS 1502+106 SSC variable R}
    \end{minipage}

    \begin{minipage}{0.49\linewidth}
        \centering
        \includegraphics[width=\linewidth, trim=40 15 90 30,clip]{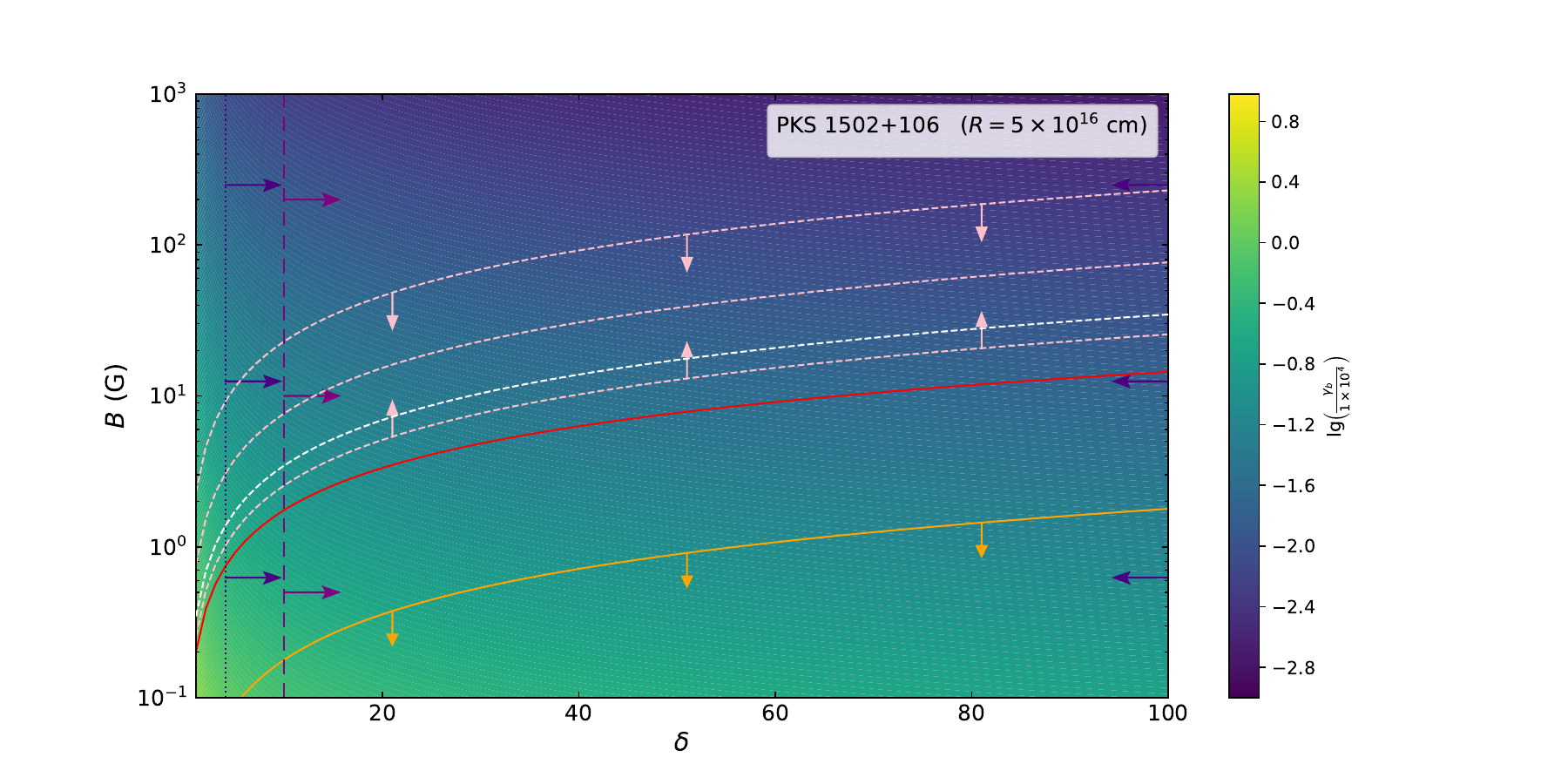}
        \label{PKS 1502+106 EC space}
    \end{minipage}\hspace{-4mm}
    \begin{minipage}{0.49\linewidth}
        \centering
        \includegraphics[width=\linewidth, trim=10 15 60 10,clip]{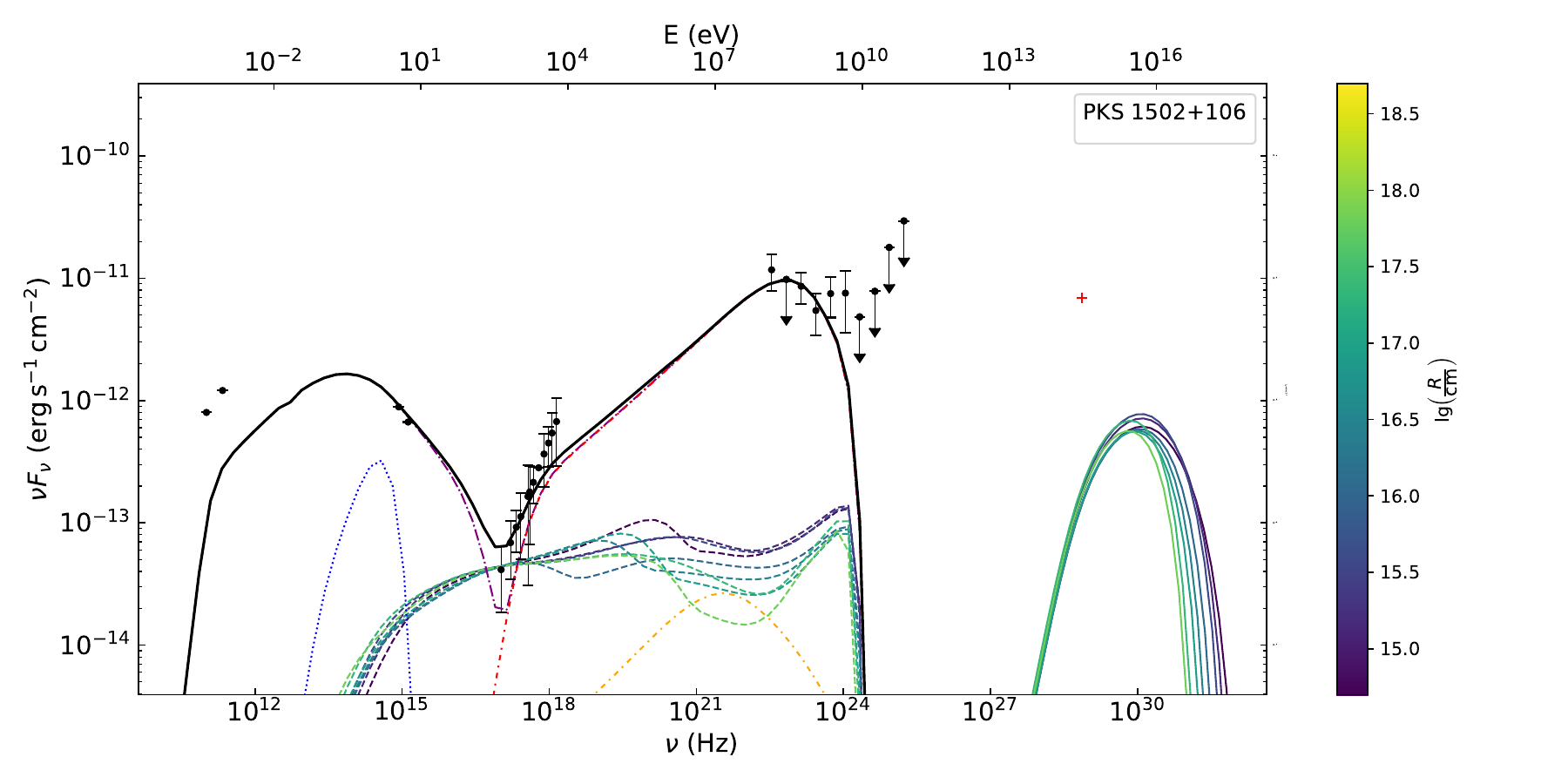}
        \label{PKS 1502+106 EC variable R}
    \end{minipage}
    \caption{PKS 1502+106 associated with IC-190730A. Upper panels: the parameter space (left panel) and the fitting result of the SED for different radii of blob (right panel) under the SSC-dominated case. The line styles in upper panels have the same meaning as in Fig.~\ref{TXS 17}, expect orange, blue and purple lines with arrows represent the internal optical depth constraint corresponding to Eq.~(\ref{deltaSSC}) for $R=1\times10^{18}~{\rm cm}$, $R=1\times10^{19}~{\rm cm}$ and $R=1\times10^{20}~{\rm cm}$, respectively. In the upper right panel, the black data points are quasi-simultaneous data and the red cross represents the neutrino flux at the energy of $300~{\rm TeV}$ during 1 year observation taken from \cite{2020ApJ...893..162F}. Lower panels: the parameter space for $R=5\times10^{16}~{\rm cm}$ (left panel) and the fitting result of the SED for different radii of blob (right panel) under the EC-dominated case. The line styles in lower panels have the same meaning as in Fig.~\ref{PKS 0735+178}.}
    \label{PKS 1502+106}
\end{figure*}

\begin{figure*}[htbp]
    \centering
    \begin{minipage}{0.49\linewidth}
        \centering
        \includegraphics[width=\linewidth, trim=40 15 90 30,clip]{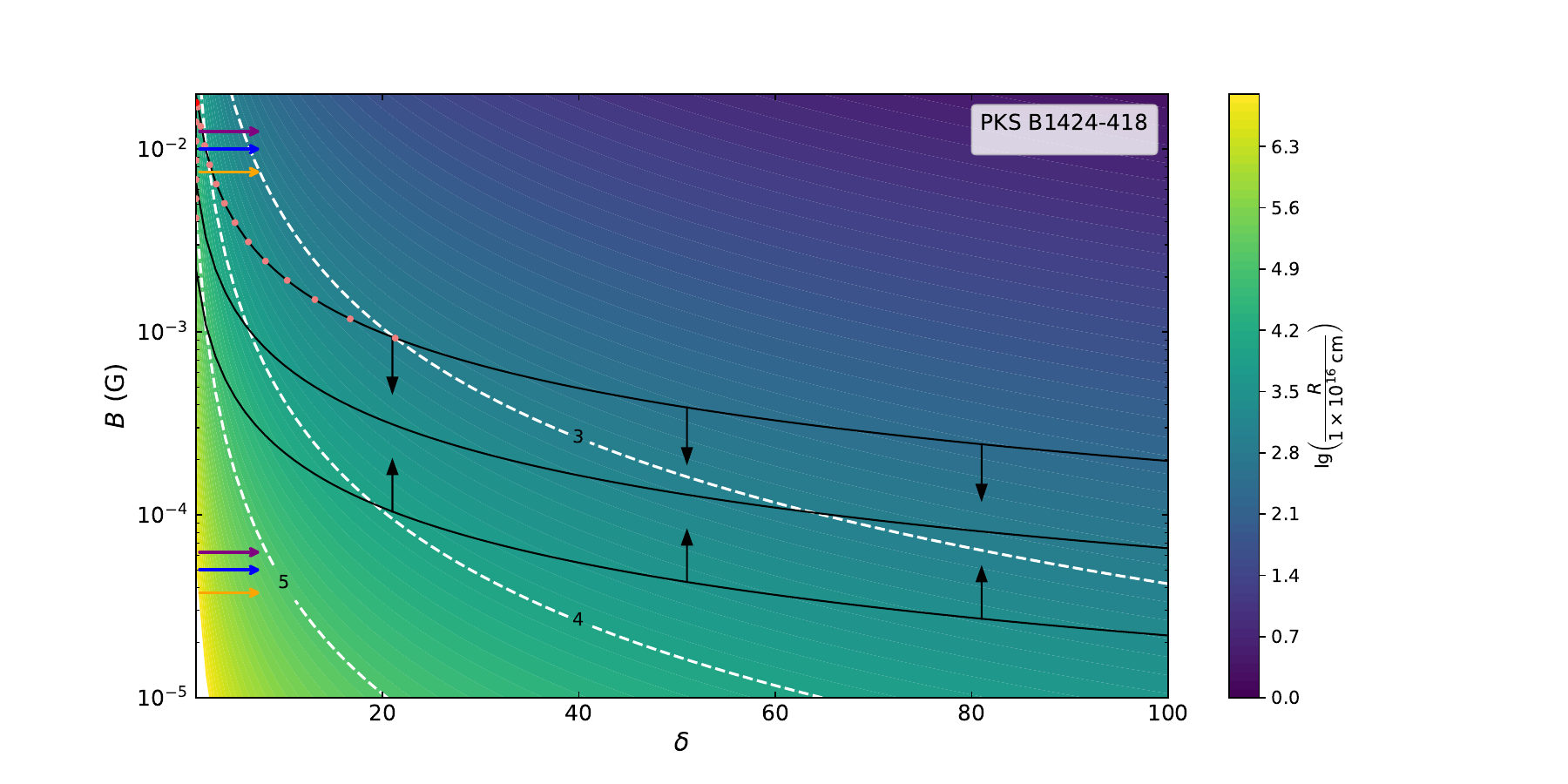}
        \label{PKS B1424-418 SSC space}
    \end{minipage}\hspace{-4mm}
    \begin{minipage}{0.49\linewidth}
        \centering
        \includegraphics[width=\linewidth, trim=10 15 55 10,clip]{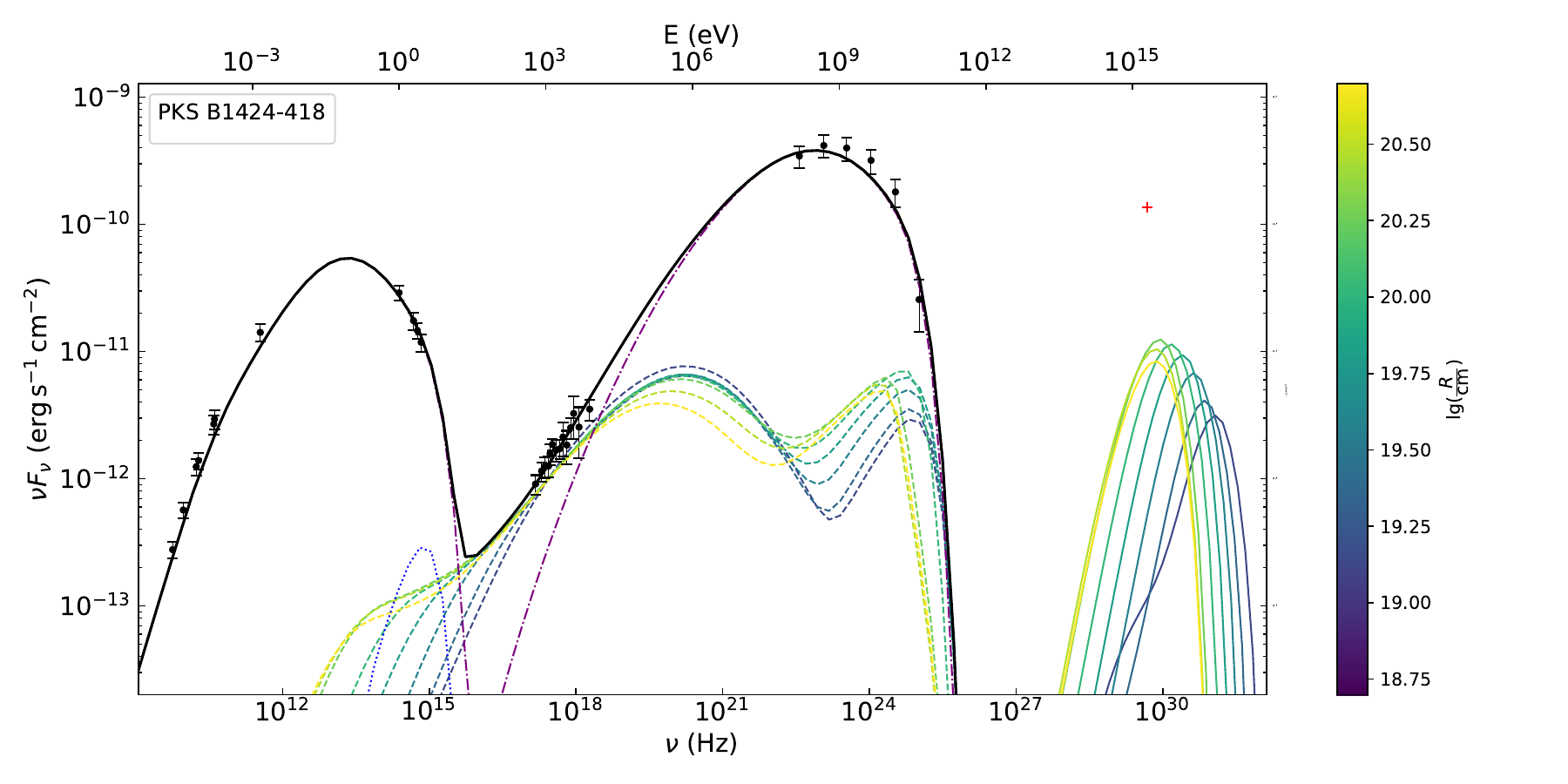}
        \label{PKS B1424-418 SSC variable R}
    \end{minipage}

    \begin{minipage}{0.49\linewidth}
        \centering
        \includegraphics[width=\linewidth, trim=40 15 90 30,clip]{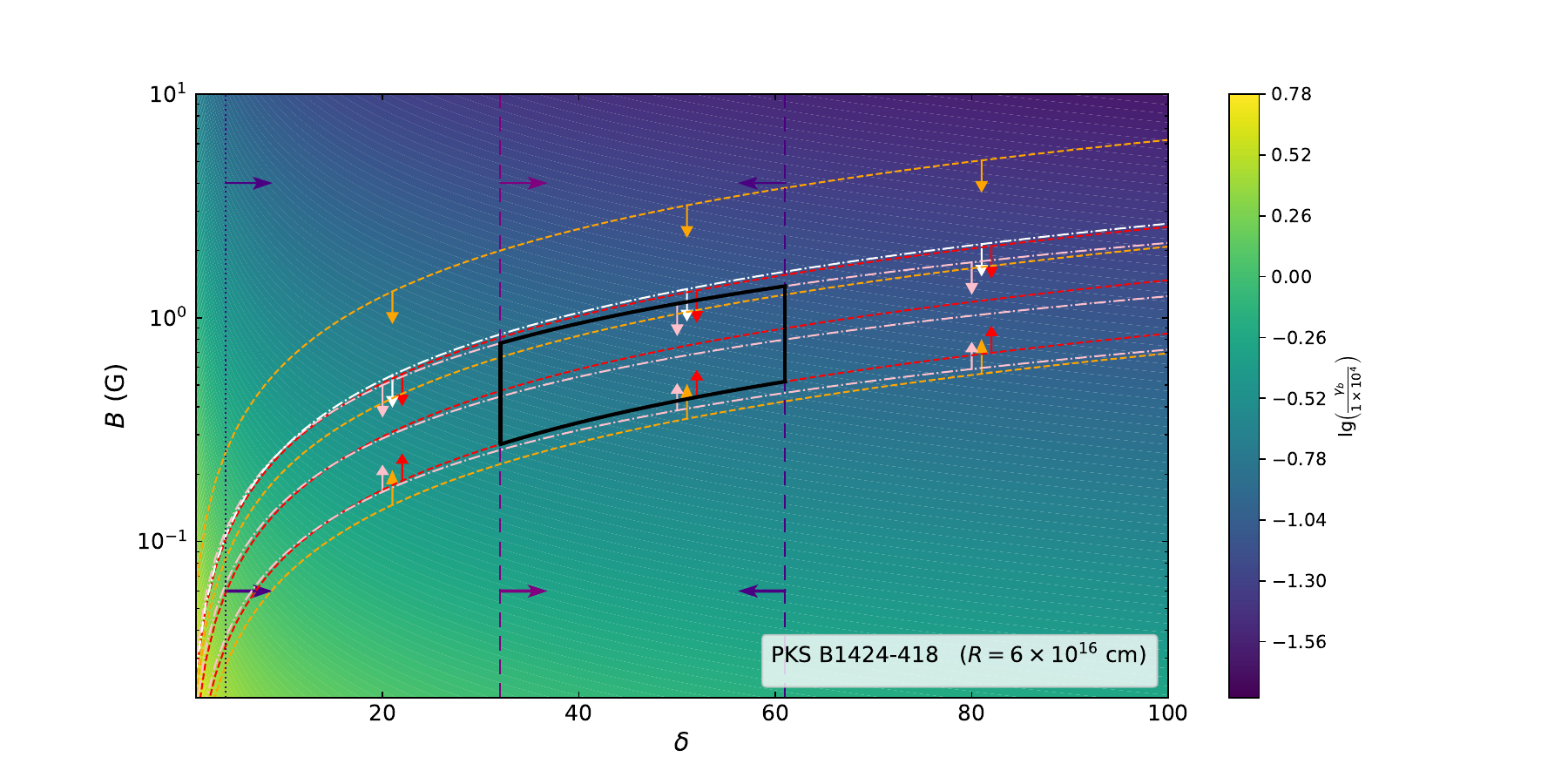}
        \label{PKS B1424-418 EC space}
    \end{minipage}\hspace{-4mm}
    \begin{minipage}{0.49\linewidth}
        \centering
        \includegraphics[width=\linewidth, trim=10 15 55 10,clip]{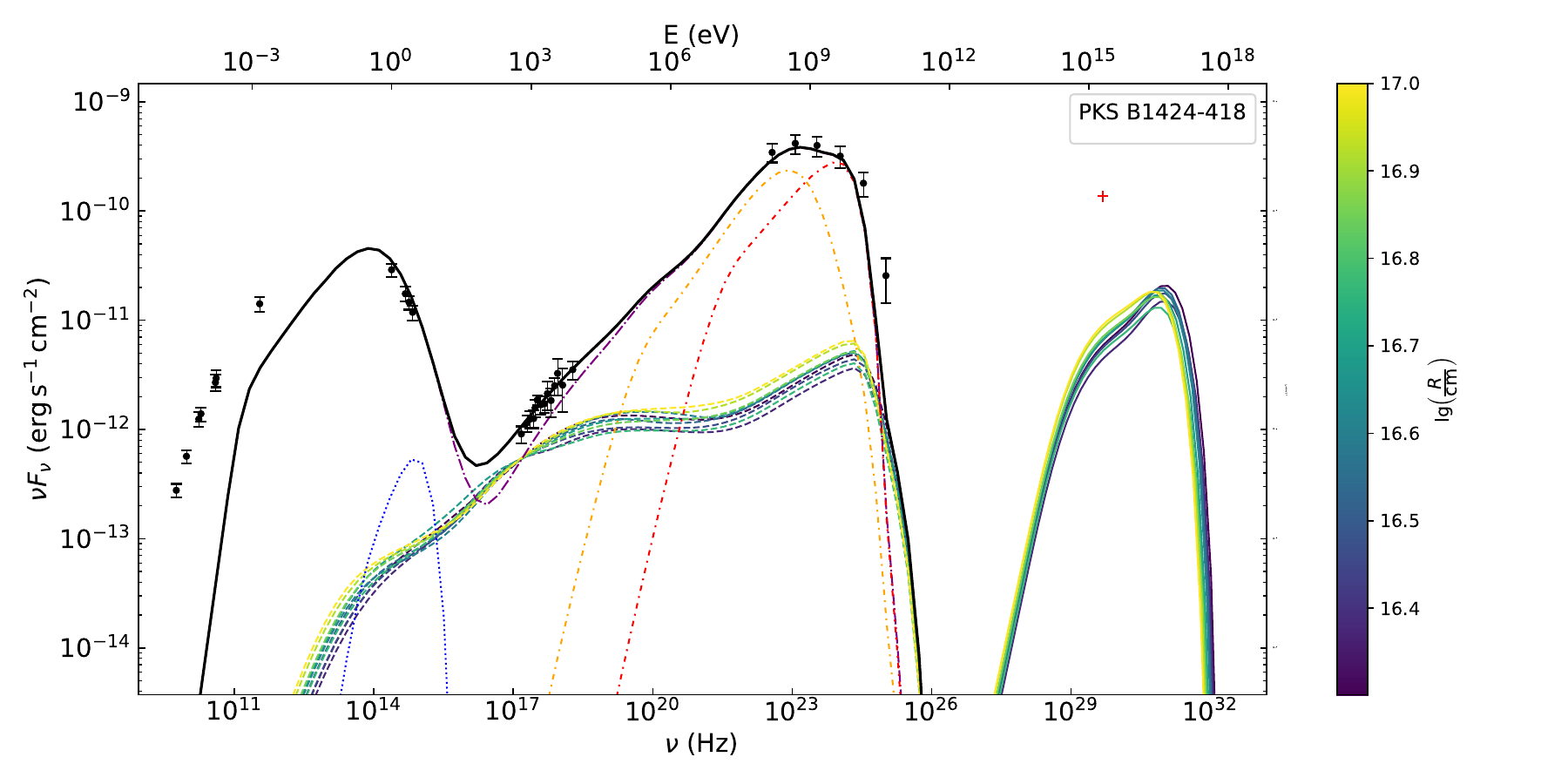}
        \label{PKS B1424-418 EC variable R}
    \end{minipage}
    \caption{PKS B1424-418 associated with Big Bird. Upper panels: the parameter space (left panel) and the fitting result of the SED for different radii of blob (right panel) under the SSC-dominated case. The line styles in upper panels have the same meaning as in Fig.~\ref{TXS 17}, expect orange, blue and purple lines with arrows represent the internal optical depth constraint corresponding to Eq.~(\ref{deltaSSC}) for $R=1\times10^{19}~{\rm cm}$, $R=1\times10^{20}~{\rm cm}$ and $R=1\times10^{21}~{\rm cm}$, respectively. In the upper right panel, the black data points are quasi-simultaneous data taken from \cite{2017ApJ...843..109G} and the red cross represents the neutrino flux at the energy of $2~{\rm PeV}$ during 1 year observation taken from \cite{2016NatPh..12..807K}. Lower panels: the parameter space for $R=6\times10^{16}~{\rm cm}$ (left panel) and the fitting result of the SED for different radii of blob (right panel) under the EC-dominated case. The line styles in lower panels have the same meaning as in Fig.~\ref{TXS 17}.}
    \label{PKS B1424-418}
\end{figure*}

\begin{figure*}[htbp]
    \centering
    \begin{minipage}{0.49\linewidth}
        \centering
        \includegraphics[width=\linewidth, trim=40 15 90 30,clip]{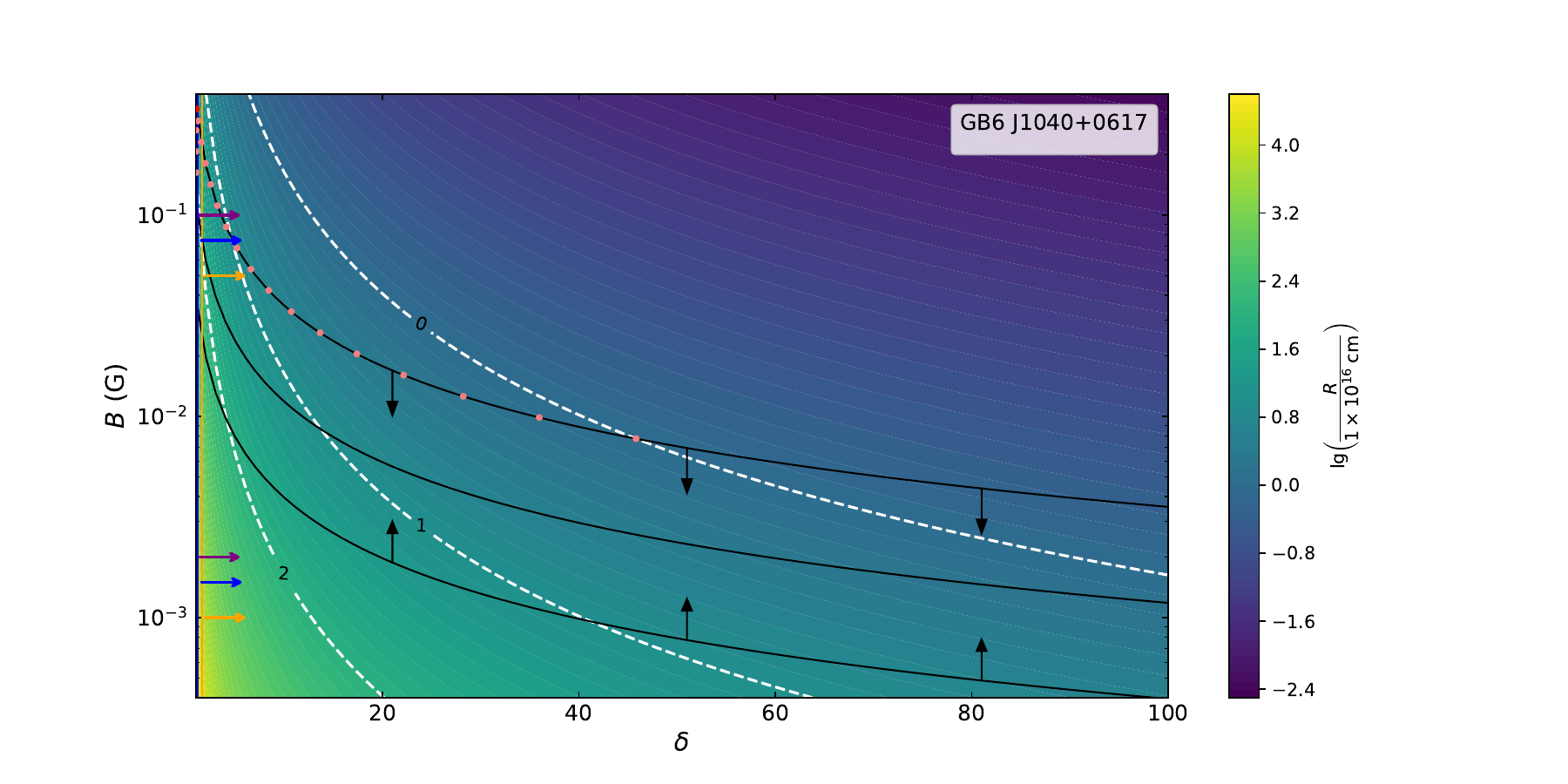}
        \label{GB6 J1040+0617 SSC space}
    \end{minipage}\hspace{-4mm}
    \begin{minipage}{0.49\linewidth}
        \centering
        \includegraphics[width=\linewidth, trim=10 15 55 10,clip]{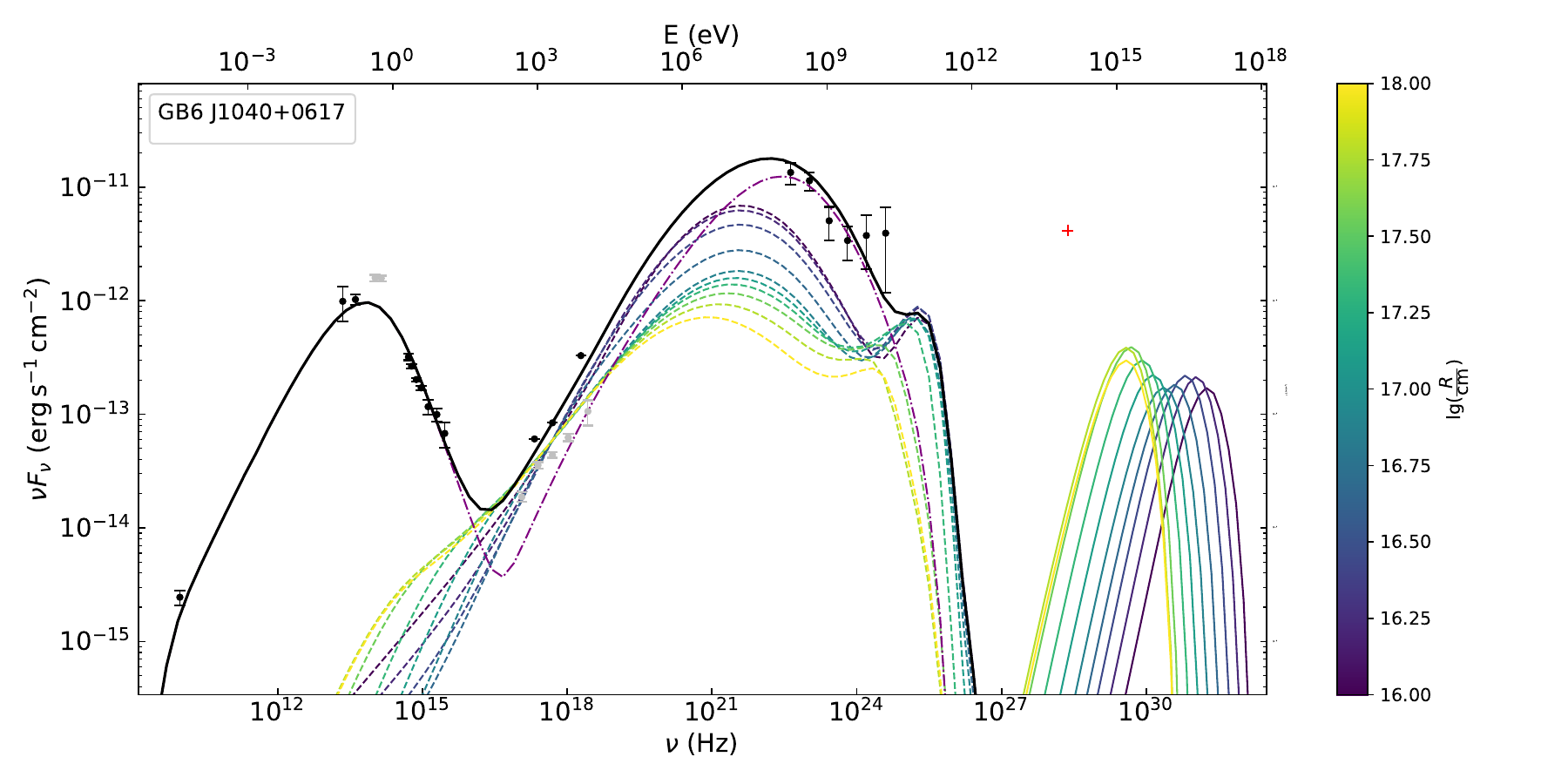}
        \label{GB6 J1040+0617 SSC variable R}
    \end{minipage}
    \caption{GB6 J1040+0617 associated with IC-141209A. The parameter space (left panel) and the fitting result of the SED for different radii of blob (right panel) under the SSC-dominated case. The line styles have the same meaning as in Fig.~\ref{TXS 22}. In the right panel, the black data points are quasi-simultaneous data and the red cross represents the neutrino flux at the energy of $97.4~{\rm TeV}$ during 1 year observation taken from Fig.~\cite{2019ApJ...880..103G}.}
    \label{GB6 J1040+0617}
\end{figure*}

\begin{figure*}[htbp]
    \centering
    \begin{minipage}{0.49\linewidth}
        \centering
        \includegraphics[width=\linewidth, trim=40 15 90 30,clip]{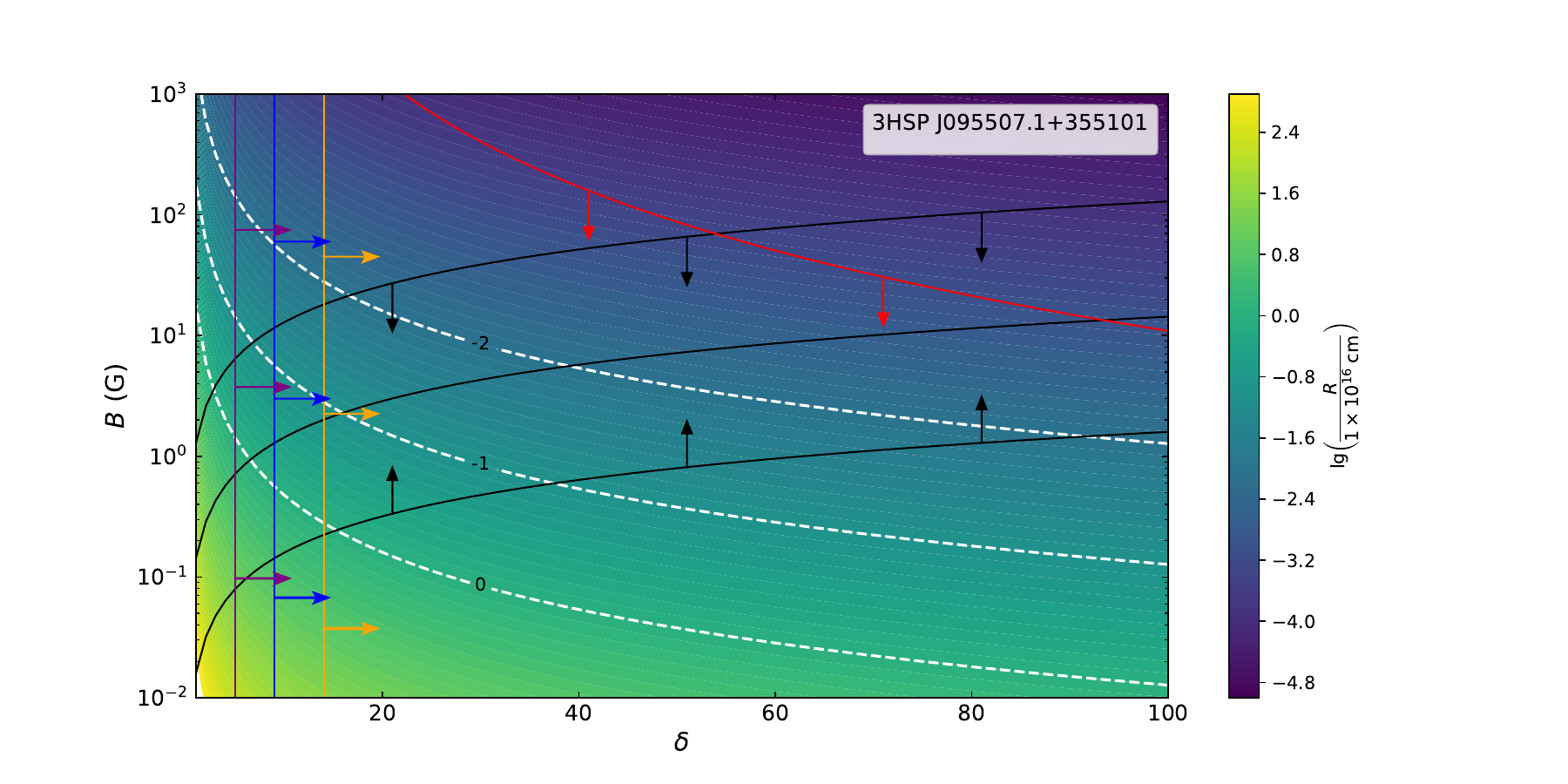}
        \label{3HSP J095507.1+355101 SSC space}
    \end{minipage}\hspace{-4mm}
    \begin{minipage}{0.49\linewidth}
        \centering
        \includegraphics[width=\linewidth, trim=10 15 60 10,clip]{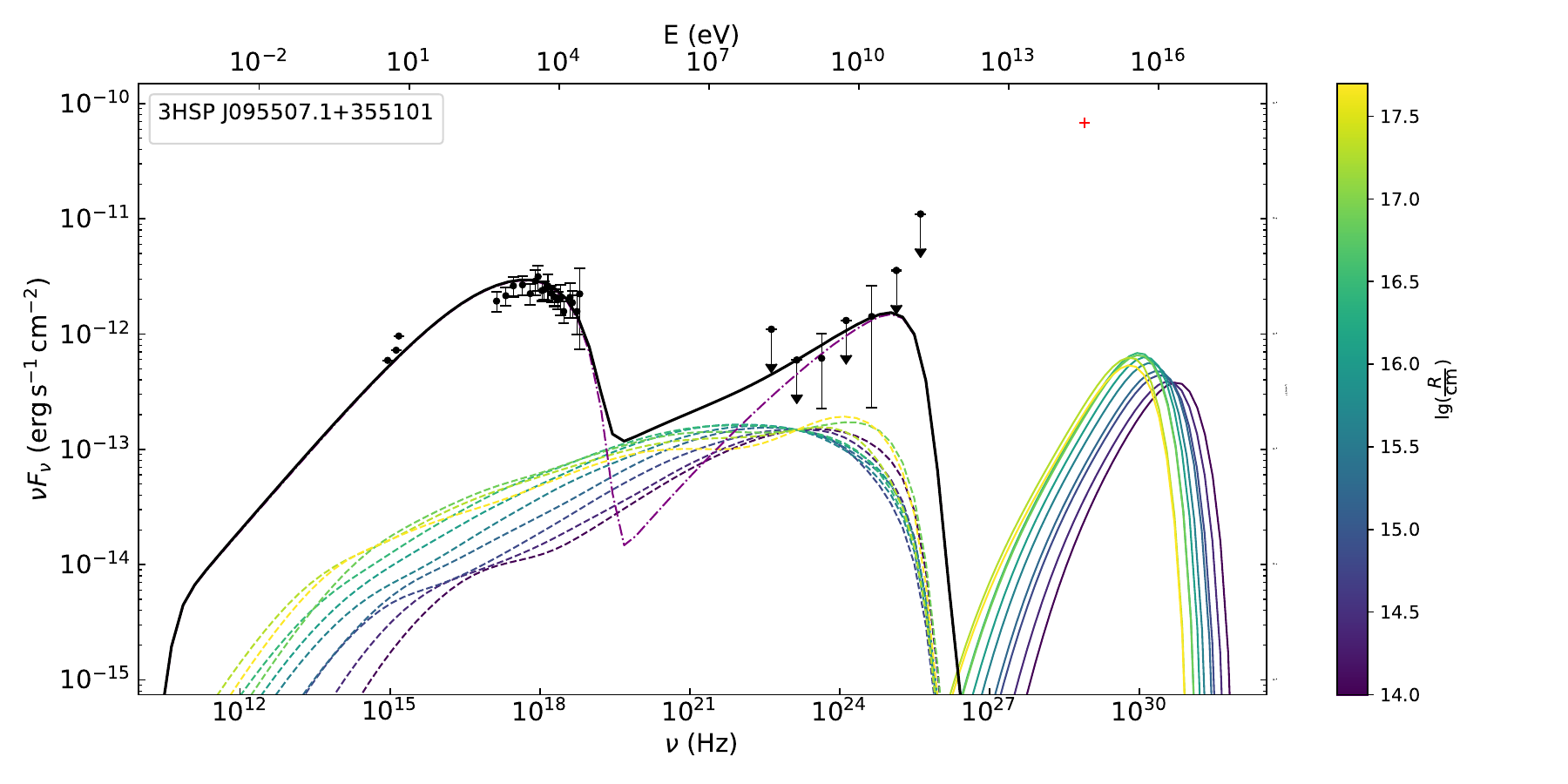}
        \label{3HSP J095507.1+355101 SSC variable R}
    \end{minipage}
    \caption{3HSP J095507.1+355101 associated with IC-200107A. The parameter space (left panel) and the fitting result of the SED for different radii of blob (right panel) under the SSC-dominated case. In the left panel, the red line with arrows represents the parameter space under the severe KN effect corresponding to Eq.~(\ref{severe KN}). The area within the black lines with arrows represents the peak frequency constraint, corresponding to Eq.~(\ref{Bdelta3}). The white dashed contours denote specific values of log$\left(\frac{R}{1\times10^{16}~{\rm cm}}\right)$ associated with the color bar that corresponds to Eq.~(\ref{Bdelta4}). The orange, blue and purple lines with arrows represent the internal optical depth constraint corresponding to Eq.~(\ref{deltaSSC}) for $R=1\times10^{14}~{\rm cm}$, $R=1\times10^{15}~{\rm cm}$ and $R=1\times10^{16}~{\rm cm}$, respectively. In the right panel, the black data points are quasi-simultaneous data and the red cross represents the neutrino flux at the energy of $0.33~{\rm PeV}$ during 1 year observation taken from \cite{2020AA...640L...4G}. The other line styles have the same meaning as in Fig.~\ref{TXS 17}.}
    \label{3HSP J095507.1+355101}
\end{figure*}

Employing the one-zone leptohadronic model, we perform a parameter-space scan at each blob radius to determine the combinations that yield the maximum neutrino flux. Taking the left panels of Fig.~\ref{TXS 17} as an example, the allowed parameter space for the SSC-dominated case (upper left panel) is delineated by the black arrowed boundary. Within this region, we sample parameter combinations along representative contours corresponding to different values of $R$ (as indicated by the color bar) and identify the parameter set that yields the maximum neutrino flux. For the EC-dominated case (lower left panel), the allowed parameter space for a given $R$ is enclosed within the black box. We systematically vary the parameter combinations within this region and evaluate the maximum neutrino flux. We verify that increasing the sampling density within the allowed parameter space does not significantly change the derived maximum neutrino flux, indicating that the parameter-space scan has reached convergence and that the reported maxima are robust against sampling resolution. The modeling results corresponding to the maximum neutrino flux for each radius are summarized in the right panels of \cref{TXS 17,TXS 21,TXS 22,PKS 0735+178,GB6 J2113+1121,5BZB J0630-2406,PKS 1502+106,PKS B1424-418,GB6 J1040+0617,3HSP J095507.1+355101} (see Appendix \ref{appendix} for supplementary modeling results). For SSC-dominated cases in the TMS regime (upper panels of \cref{TXS 17,TXS 21,TXS 22,PKS 0735+178,GB6 J2113+1121,5BZB J0630-2406,PKS 1502+106,PKS B1424-418,GB6 J1040+0617}), the physical parameters of the best results for each radius obtained from numerical modeling are marked with pink points, and the parameter set with maximum neutrino flux at $E_{\rm prob}^{\rm obs}$ is given by red points, in agreement with those derived analytically in Section~\ref{anaflux}. By comparison, we find that the maximum neutrino flux obtained by the analytical method is $<4$ times that obtained by the numerical modeling; for EC-dominated cases (lower panels of \cref{TXS 17,TXS 21,TXS 22,PKS 0735+178,GB6 J2113+1121,5BZB J0630-2406,PKS 1502+106,PKS B1424-418}), the maximum neutrino flux at $E_{\rm prob}^{\rm obs}$ obtained from the numerical modeling is approximately a factor of $<2$ lower than that derived from the analytical result. These discrepancies arise mainly from the contributions of leptonic emission from primary electrons and secondary pairs produced via internal $\gamma \gamma$ annihilation, both of which exert varying degrees of influence. Nevertheless, the flux predicted by the analytical approach can be regarded as an upper limit achievable within the one-zone model. The parameter set found by analytical methods is in good agreement with that given by the parameter-space scan using numerical modeling. These agreements indicate that our analytical estimates and numerical modeling are consistent within the expected uncertainties. An exception arises in the case of the HSP-BL Lac object 3HSP J095507.1+355101. For this source, the GeV spectrum provides a stronger constraint than the X-ray data located at the synchrotron peak, and thus the analytical approach becomes invalid. In this situation, exploring the parameter space through numerical modeling is important. Nevertheless, since the dominant contribution to the GeV band originates from $\pi^{\pm}$ decay secondaries, the resulting neutrino flux cannot exceed the observed GeV spectral flux, as illustrated in the right panel of \cref{3HSP J095507.1+355101}.

For SSC-dominated cases (the upper right panels of \cref{TXS 17,TXS 21,TXS 22,PKS 0735+178,GB6 J2113+1121,5BZB J0630-2406,PKS 1502+106,PKS B1424-418,GB6 J1040+0617,3HSP J095507.1+355101}), it can be seen that the peak energy of the neutrino spectrum has two changing trends as the blob radius increases: (i) first decreases and (ii) then remains almost unchanged\footnote{Taking the upper right panel of \cref{TXS 17} (TXS 0506+056 and IC-170922A association event) as an example, when the blob is compact ($R<3\times10^{15}~{\rm cm}$ in this case), the peak energy of neutrino spectrum decreases with increasing radius. When the blob is sufficiently large ($R\geq3\times10^{15}~{\rm cm}$), the peak energy of neutrino spectrum remains nearly unchanged despite a further increase in blob radius.}. When the blob is compact, increasing in the blob radius reduces the synchrotron photon energy density, necessitating a lower Doppler factor to maintain the flux of the IC peak, i.e., $R\propto \delta_{\rm D}^{-1}$, as indicated by combining Eqs.~(\ref{Bdelta1}) and (\ref{Bdelta2}). Due to the decrease in the Doppler factor caused by the increasing radius, the peak energy of the neutrino energy spectrum in the observer frame progressively decreases; When $R$ is sufficiently large, the Doppler factor changes slightly due to the optical depth constraint, resulting in a tiny variation in the peak frequencies of neutrino spectra across different radii. Compared to the situation found in previous one-zone studies where the peak of pair cascade spectrum is close to the X-ray data \citep[e.g.,][]{2019MNRAS.483L..12C,2023MNRAS.519.1396S}, we find that this peak can be moved away from the X-ray band through optimized parameter selection, especially for a compact blob with $R\lesssim10^{16}\rm~cm$. However, for small radius emission regions, in order to avoid internal $\gamma\gamma$ absorption, a relatively large $\delta_{\rm D}$ must be assumed. The analytical results indicate that this implies a lower $f_{p\gamma}/f_{\rm BH}$ at small radii compared to larger ones.
Consequently, the maximum neutrino flux with smaller emitting region will be lower than that given by the large emitting region. 
The required proton injection luminosities for these SSC-dominated maximum-flux solutions span a broad range, from sub-Eddington to highly super-Eddington values. The corresponding ratios of proton injection luminosity in the AGN frame to the Eddington luminosity ($L_{p, \rm inj}\delta_{\rm D}^2/L_{\rm Edd}$) are approximately 5--300 for TXS\,0506+056 (2017), 3--40 for TXS\,0506+056 (2021/2022), 1.8--15 for PKS\,0735+178, $7\times10^3$--$2.5\times10^5$ for GB6\,J2113+1121, 0.2--7.4 for 5BZB\,J0630-2406, 40--$2.7\times10^3$ for PKS\,1502+106, $1.4\times10^3$--$2.9\times10^4$ for PKS\,B1424-418, 32--$7.1\times10^4$ for GB6\,J1040+0617, and 0.6--100 for 3HSP\,J095507.1+355101. For larger blob radii, which correspond to lower $\delta_{\rm D}$, it generally requires lower $L_{p, \rm inj}$. While some solutions for 5BZB J0630-2406 and 3HSP J095507.1+355101 are sub-Eddington, compact emission regions often necessitate extremely high, potentially unphysical proton injection luminosities.
In EC-dominated cases (the lower right panels of \cref{TXS 17,TXS 21,TXS 22,PKS 0735+178,GB6 J2113+1121,5BZB J0630-2406,PKS 1502+106,PKS B1424-418}), the peak energy of neutrino spectrum exhibits a regular decrease with increasing blob radius, caused by the Doppler factor. More specifically, larger blob radii correlate with improved EC scattering efficiency when $r$ is fixed. This permits a smaller Doppler factor to satisfy the condition for EC to dominate the high-energy bump (Eq.~\ref{deltaECdomi}), which correspondingly manifests in the neutrino spectrum as a downward shift in peak energy with increasing radius. For the lower right panels of \cref{TXS 17,TXS 21,TXS 22,5BZB J0630-2406,PKS B1424-418} in compact blob radii, there is a concave structure in the neutrino spectra. This phenomenon primarily arises because a higher Doppler factor enables relativistic protons to interact concurrently with photons from both the DT and the BLR, resulting in a merged double-peak structure in the neutrino spectra.  
Similar to the SSC-dominated cases, the optimal neutrino flux also cannot be reconciled with the neutrino observations by the one-zone EC model.
In contrast to the SSC-dominated scenarios, the EC-dominated solutions instead exhibit a much narrower range of required proton injection luminosities. The corresponding proton-to-Eddington luminosity ratios are approximately 3--4 for TXS\,0506+056 (2017), 7--16 for TXS\,0506+056 (2021/2022), 0.18--0.71 for PKS\,0735+178, 13--23 for GB6\,J2113+1121, 0.02--0.03 for 5BZB\,J0630-2406, 0.14--0.25 for PKS\,1502+106, and 13--27 for PKS\,B1424-418. As can be seen, the required injection luminosities range from sub-Eddington to at most 27 times the Eddington luminosity. Unlike the SSC scenarios, which in certain cases necessitate extremely super-Eddington proton powers, the EC-dominated solutions avoid such extreme requirements and are therefore physically more plausible.

The maximum annual neutrino detection rates derived from numerical modeling under SSC-dominated and EC-dominated scenarios, along with previous one-zone leptohadronic modeling results, are presented in Table~\ref{table2}. For FSRQs and masquerading BL Lacs, the neutrino detection rates have been calculated for both SSC- and EC-dominated cases. A comparison indicates that when external photon fields such as the BLR are taken into account, the neutrino flux is indeed enhanced, although the overall improvement remains modest. This suggests that external photon fields may play a potentially important role in neutrino production, particularly in the case of masquerading BL Lacs \citep{2022MNRAS.510.2671P, 2024A&A...689A.147R}.
Generally, our results are consistent with previous one-zone leptohadronic model studies that employed publicly available hadronic codes \citep{2026ApJS..282...22C}, and our values lie within or slightly above the ranges reported there. Such agreement indicates that our calculations represent a reliable application of established numerical models, and that further enhancement of the event rate within the one-zone framework is unlikely.
While it may be possible to explain the detection of one neutrino as a consequence of the Eddington bias \citep{2019A&A...622L...9S}, if a higher neutrino flux and hence a more favorable detection probability are assumed, considering multi-zone models remains a natural and more compelling approach.
For comparison, Table~\ref{table2} also includes results from multi-zone scenarios, which stand out by providing significantly higher detection probabilities. These multi-zone predictions underscore that more complex source structures offer a more plausible explanation for the observed neutrino events.

\section{Discussion and Conclusion} \label{sec:discussion}

\subsection{Can $\gamma$-ray and neutrino flares be correlated?}
Previous studies \citep{2018ApJ...863L..10A,2019NatAs...3...88G,2019MNRAS.483L..12C,2021ApJ...906...51X,2023MNRAS.519.1396S} have revealed that the peak of secondary pair cascade predominantly resides in the keV--MeV range, suggesting a potential correlation between X-ray flares and neutrino events. Constrained by the X-ray data, the predicted annual neutrino detection rates remain insufficient to explain observed neutrino events in the framework of one-zone leptohadronic jet model. Beyond the keV--MeV range, whether $\gamma$-ray flares are correlated with the neutrino events is also of significant interest. Since high-energy neutrinos and secondary pairs generated via $p\gamma$ interactions carry comparable fluxes, it implies that the cascade originated $\gamma$-ray flares and neutrino events might be correlated. As shown in our previous modeling, we find that the peak of secondary pair cascade emission shifts away from the keV band in some cases of the SSC-dominated scenario (see the upper right panels of \cref{TXS 17,TXS 21,TXS 22,PKS 0735+178,GB6 J2113+1121,5BZB J0630-2406,PKS 1502+106,PKS B1424-418}, \cref{GB6 J1040+0617,3HSP J095507.1+355101}). 
Such spectral shifts of the cascade emission have also been analyzed in detail by \cite{2024JCAP...07..006K}, which investigated that BH pair injection can produce cascade spectra with peaks moving to $\gamma$-ray band under certain conditions. Our findings are consistent with their results, as both studies highlight that cascade emission may depart from the X-ray band.
This provides an opportunity to re-examine the possible correlation between the $\gamma$-ray flare and the neutrino event.

\begin{figure*}[htbp]
    \centering
    \begin{minipage}{0.49\linewidth}
        \centering
        \includegraphics[width=\linewidth, trim=40 15 90 30,clip]{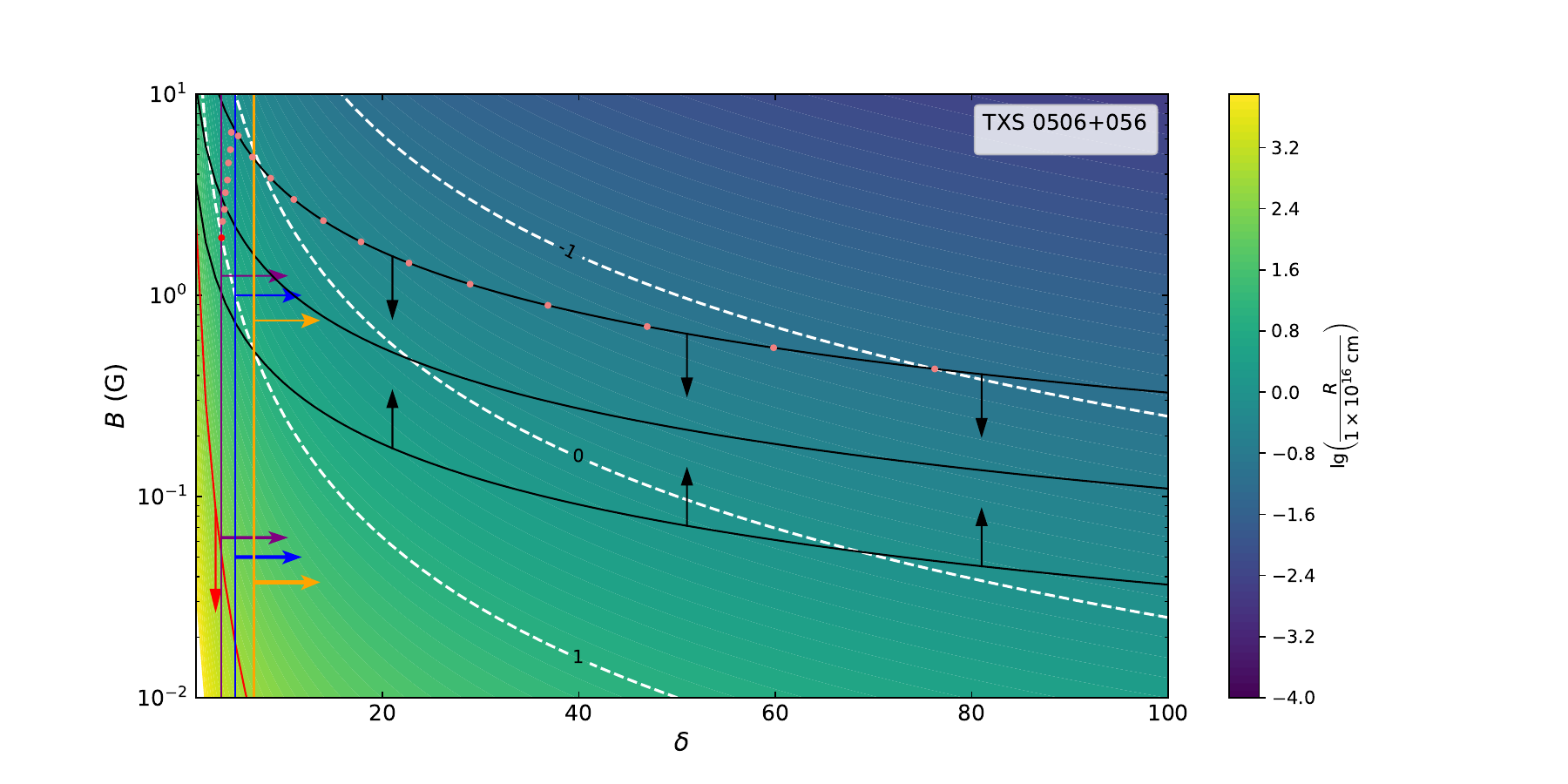}
    \end{minipage}\hspace{-4mm}
    \begin{minipage}{0.49\linewidth}
        \centering
        \includegraphics[width=\linewidth, trim=10 15 55 10,clip]{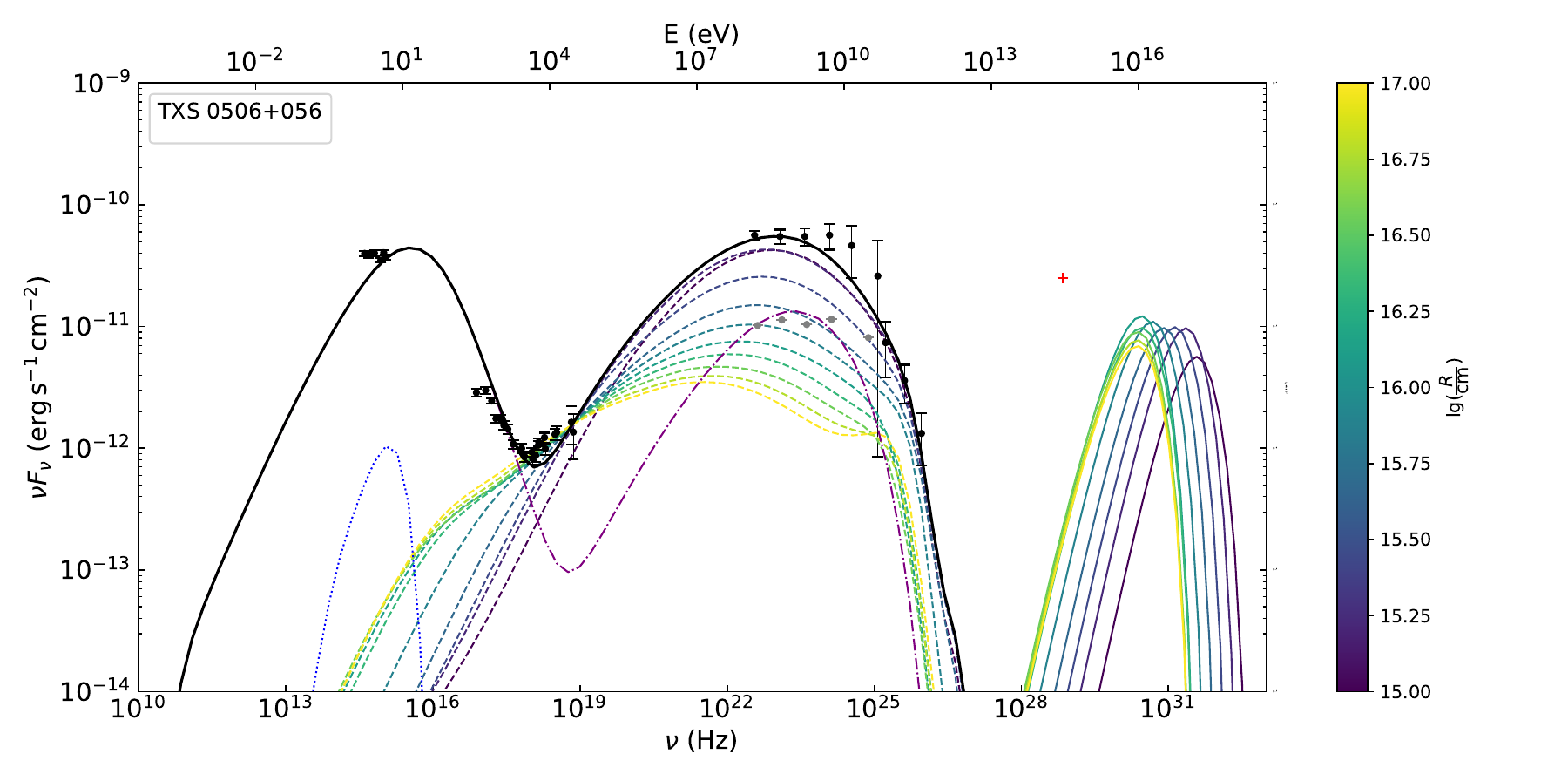}
    \end{minipage}
    \caption{TXS 0506+056 associated with IC-170922A. The parameter space (left panel) and the fitting result of the SED for different radii of blob (right panel) under the SSC-dominated case. The line styles have the same meaning as in Fig.~\ref{TXS 17}. In the right panel, the black and gray data points are respectively quasi-simultaneous and archival data, and the red cross represents the neutrino flux at the energy of $290~{\rm TeV}$ during 1 year observation taken from \cite{2018Sci...361.1378I}.}
    \label{TXS 17 discussion}
\end{figure*}

Here we take TXS 0506+056 as a representative case, in which a $\gamma$-ray flare is coincident with the neutrino event IC-170922A. If the $\gamma$-ray flare is triggered by the secondary pair cascades, this would suggest a potential physical correlation between the $\gamma$-ray flare and the neutrino emission. To constrain the parameter space, we assume that the archival GeV data from \cite{2018Sci...361.1378I} originate from the SSC emission from primary electrons (the left panel of \cref{TXS 17 discussion}). Following the procedure outlined in Section~\ref{anaflux} and Section~\ref{sec:results}, we obtain optimal results for different blob radii (the right panel of \cref{TXS 17 discussion}) by requiring that the quasi-simultaneous GeV data be predominantly explained by secondary pair cascade emission. It can be seen that the secondary pairs can indeed dominate the high-energy bump when the blob is compact. However, in cases with larger blob radii, the spectrum remains constrained by X-ray data. This discrepancy primarily stems from the reduced Doppler factor at larger radii shifting the peak of secondary pair emission toward lower energy bands in the observer's frame. Therefore, a correlation between the $\gamma$-ray flare and the neutrino event is only viable in scenarios where the blob maintains a sufficiently compact radius. Even so, the peak of neutrino spectrum shifts to higher energies in such compact radii, indicating that the model predicted neutrino detection rate is still too low to account for the observation. For the neutrino best-fit case, the annual neutrino detection rate is 0.0089. Compared to the optimal result of the SSC-dominated scenario in \cref{TXS 17}, the neutrino detection rate increases by a factor of 2.5. 
This level of event rate suggests that the association between the $\gamma$-ray flare and the neutrino event might be interpreted as a statistical fluctuation, consistent with the Eddington bias \citep{2019A&A...622L...9S}; otherwise, invoking multi-zone models provides a more natural explanation for such a correlation.

\subsection{Comparison with previous one-zone leptohadronic models}
Previous studies have investigated the maximum neutrino flux from blazars using one-zone leptohadronic models. \cite{2019NatAs...3...88G} and \cite{2019MNRAS.483L..12C} performed extensive numerical simulations, scanning over model parameters to find configurations that satisfy MWL constraints. However, their results differ noticeably, partly due to the specific choices made during parameter selection. For example, \cite{2019MNRAS.483L..12C} focused on $\delta_{\rm D}>30$ and $R\approx10^{16}\rm~cm$, leaving out other regions of parameter space that may also be relevant. This narrow sampling introduces bias into the inferred neutrino flux.

\cite{2024A&A...689A.147R} presented a more recent study, applying a one-zone leptohadronic model to a sample of 32 BL Lacs associated with IceCube events. Their modeling is based on a new statistical method that compares model-predicted neutrino spectra directly with IceCube point-source data, rather than assuming a power-law signal. This approach allows them to find consistency with IceCube fluxes at the 68\% confidence level for many masquerading BL Lacs. The underlying model predictions, such as the required proton power and the neutrino peak energy, remain similar to those in earlier works. In particular, the predicted neutrino spectra peak above the PeV range, and the flux is limited by X-ray observations, which continues to suppress neutrino flux in the one-zone scenario. The neutrino spectrum of TXS 0506+056 with significant external photon fields is similar to that given in \cite{2018ApJ...864...84K}. By comparison, the neutrino spectrum in \cite{2024A&A...689A.147R} exhibits higher fluxes above the PeV range, which may be more favorable for interpreting the observation, especially if the true neutrino energy exceeds 290 TeV, the median expected energy for a power-law spectrum.

Our analytical method offers a complementary perspective. Rather than scanning the full parameter space, it derives the maximum neutrino flux directly from X-ray constraints and SED peak properties. This approach is straightforward and helps avoid biases from incomplete parameter coverage. It also makes it easier to apply the method to a larger sample of sources, which is especially useful for population studies. While our method does not aim to reproduce detailed SEDs or time-dependent behavior, it provides a clear upper limit on the neutrino flux and helps assess whether the one-zone leptohadronic model can meet observational requirements. The consistency between our analytical estimates and numerical results supports the reliability of this approach. Overall, our findings suggest that the one-zone leptohadronic model shows limited ability to explain the observed neutrino fluxes, suggesting that more complex scenarios such as multi-zone models or alternative production sites may be needed.

\subsection{Conclusion}
In this work, we have proposed an analytical method to derive upper limits on the neutrino flux and the corresponding physical parameters for both SSC- and EC-dominated scenarios in the framework of the one-zone leptohadronic model. These analytical estimates, based on the X-ray flux constraints, were further tested with numerical modeling for a sample of candidate neutrino-emitting blazars, and the two approaches show consistent results. Both the analytical and numerical studies indicate that the neutrino flux predicted within the one-zone framework tends to fall short of the levels inferred from current detections, although statistical fluctuations may still render occasional events consistent with this scenario. This outcome reflects the intrinsic coupling between $p\gamma$ interactions and secondary cascades: the same photon fields that enable neutrino production also generate cascade emission that is tightly constrained by X-ray data. As a result, the one-zone scenario may not fully reproduce the observed MWL spectra and account for the observed neutrino flux simultaneously.

These findings suggest that the one-zone framework, while useful as a first-order description, has limited explanatory power for the current neutrino-blazar associations. To make further progress, it is necessary to consider more complex scenarios, such as multi-zone jet models, or to explore alternative production sites including the jet base, hot corona, or accretion flow. Such environments may provide the conditions required for efficient neutrino generation without violating electromagnetic constraints.

In summary, the analytical approach developed here provides a practical tool to assess the maximum neutrino yield of blazars under the one-zone assumption, explicitly tied to X-ray flux measurements. At the same time, the results highlight the need to move beyond this simplified picture and to develop more sophisticated models in order to understand the origin of high-energy neutrinos in blazars.

\begin{acknowledgments}
This work is supported by the National Key R\&D Program of China (2023YFB4503305), the National Natural Science Foundation of China  (NSFC) under the grant No. 12203043, No. 12473020, No. 12203024, No. 12473042, No. 12373109, No. 12473042, and No. 12373109, the Department of Science \& Technology of Shandong Province under Grant No. ZR2022QA071, the CAS ``Light of West China'' Program and the
Yunnan Province Youth Top Talent Project (grant No. YNWR-QNBJ-2020-116).
\end{acknowledgments}

\clearpage

\software{NumPy \citep{2020Natur.585..357H}, astropy \citep{2013A&A...558A..33A,2018AJ....156..123A,2022ApJ...935..167A}
          }


\appendix

\section{Results of different parameter combinations under the specific radius}\label{appendix}
\begin{figure*}[htbp]
    \centering
    \begin{minipage}{0.49\linewidth}
        \centering
        \includegraphics[width=\linewidth, trim=10 15 35 10,clip]{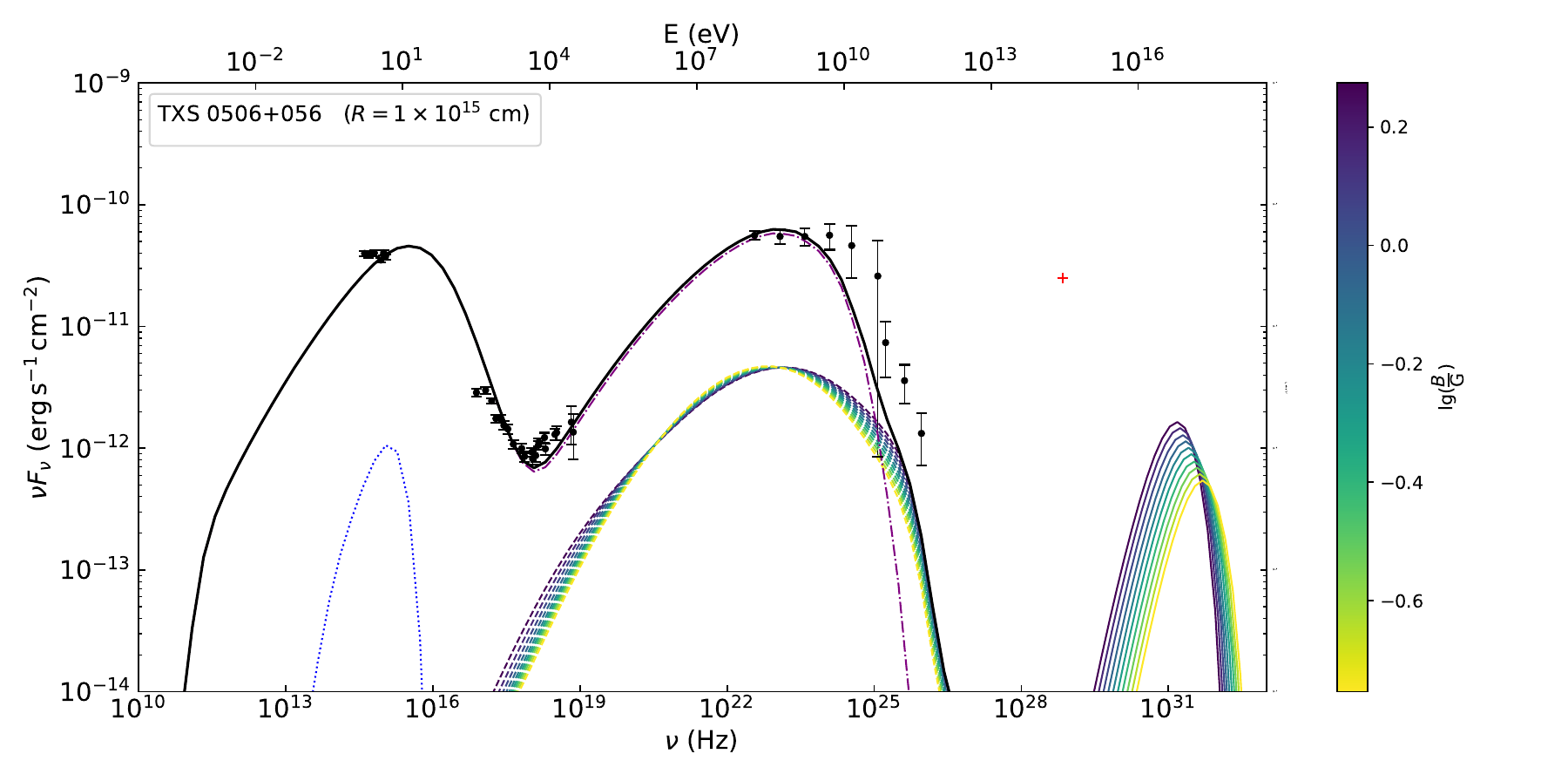}
        \label{TXS 17 SSC R=1e15}
    \end{minipage}\hspace{-4mm}
    \begin{minipage}{0.49\linewidth}
        \centering
        \includegraphics[width=\linewidth, trim=10 15 35 10,clip]{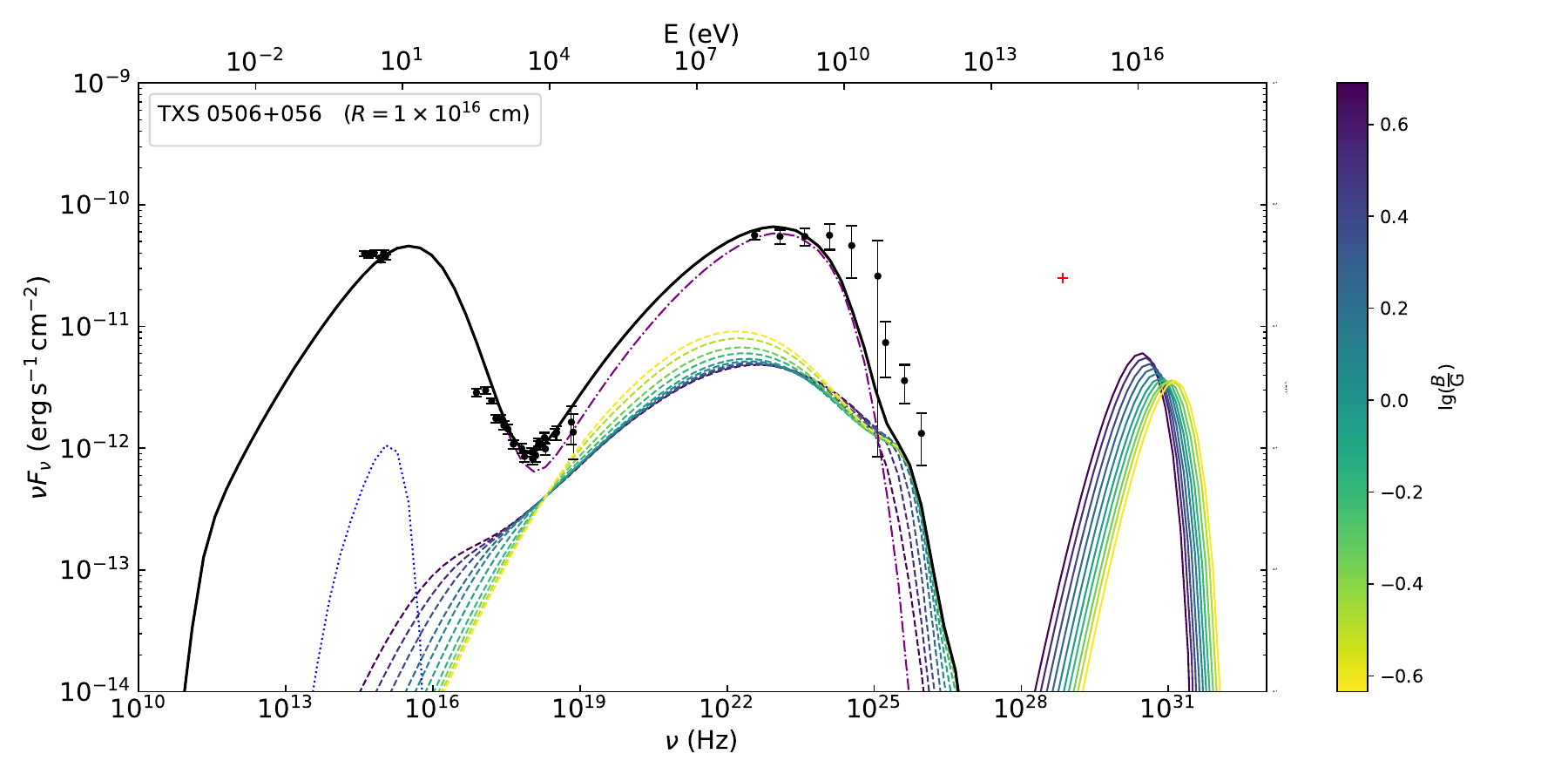}
        \label{TXS 17 SSC R=1e16}
    \end{minipage}

    \begin{minipage}{0.49\linewidth}
        \centering
        \includegraphics[width=\linewidth, trim=35 15 90 30,clip]{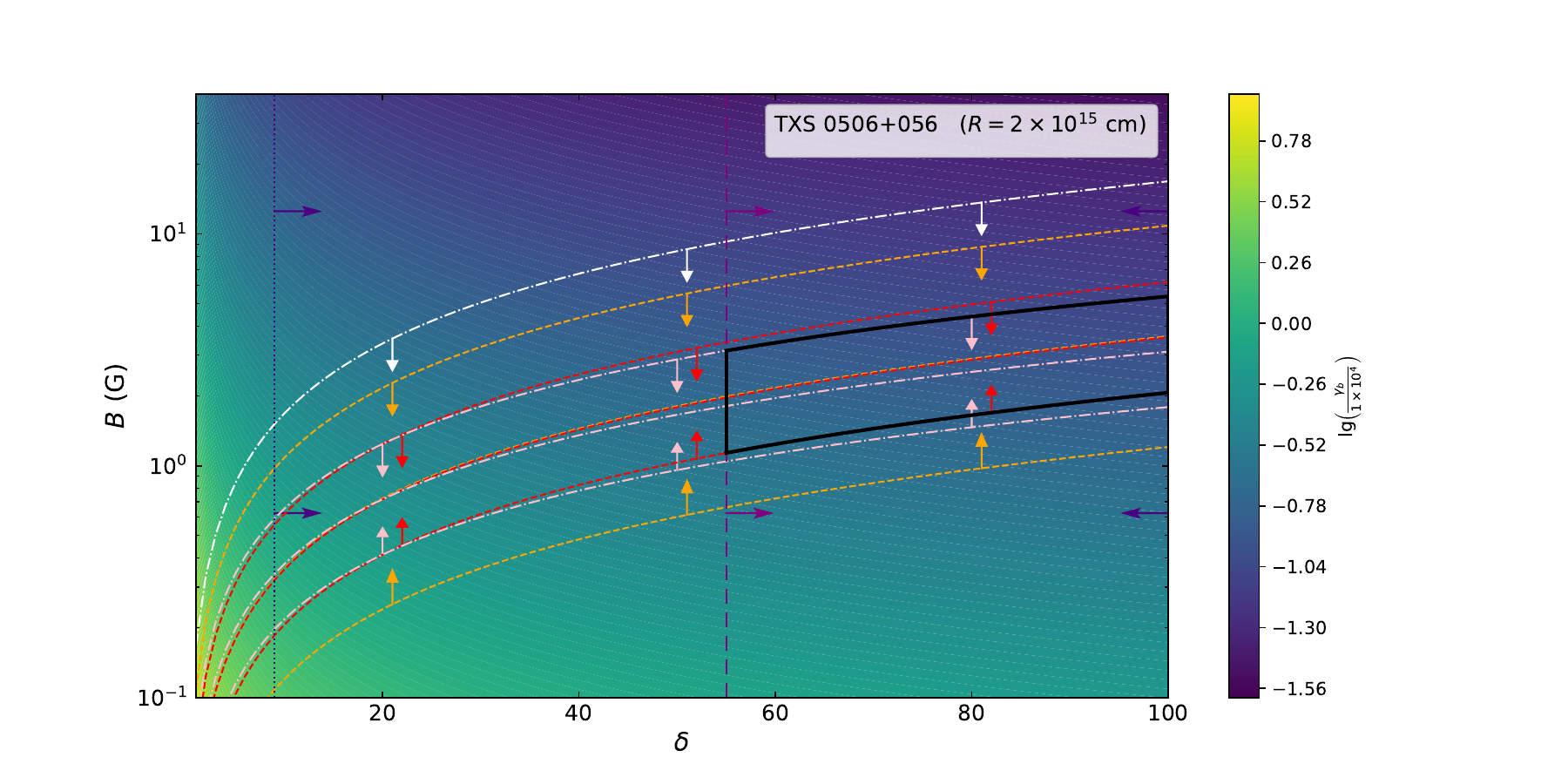}
        \label{TXS 17 EC R=2e15 space}
    \end{minipage}\hspace{-4mm}
    \begin{minipage}{0.49\linewidth}
        \centering
        \includegraphics[width=\linewidth, trim=35 15 90 30,clip]{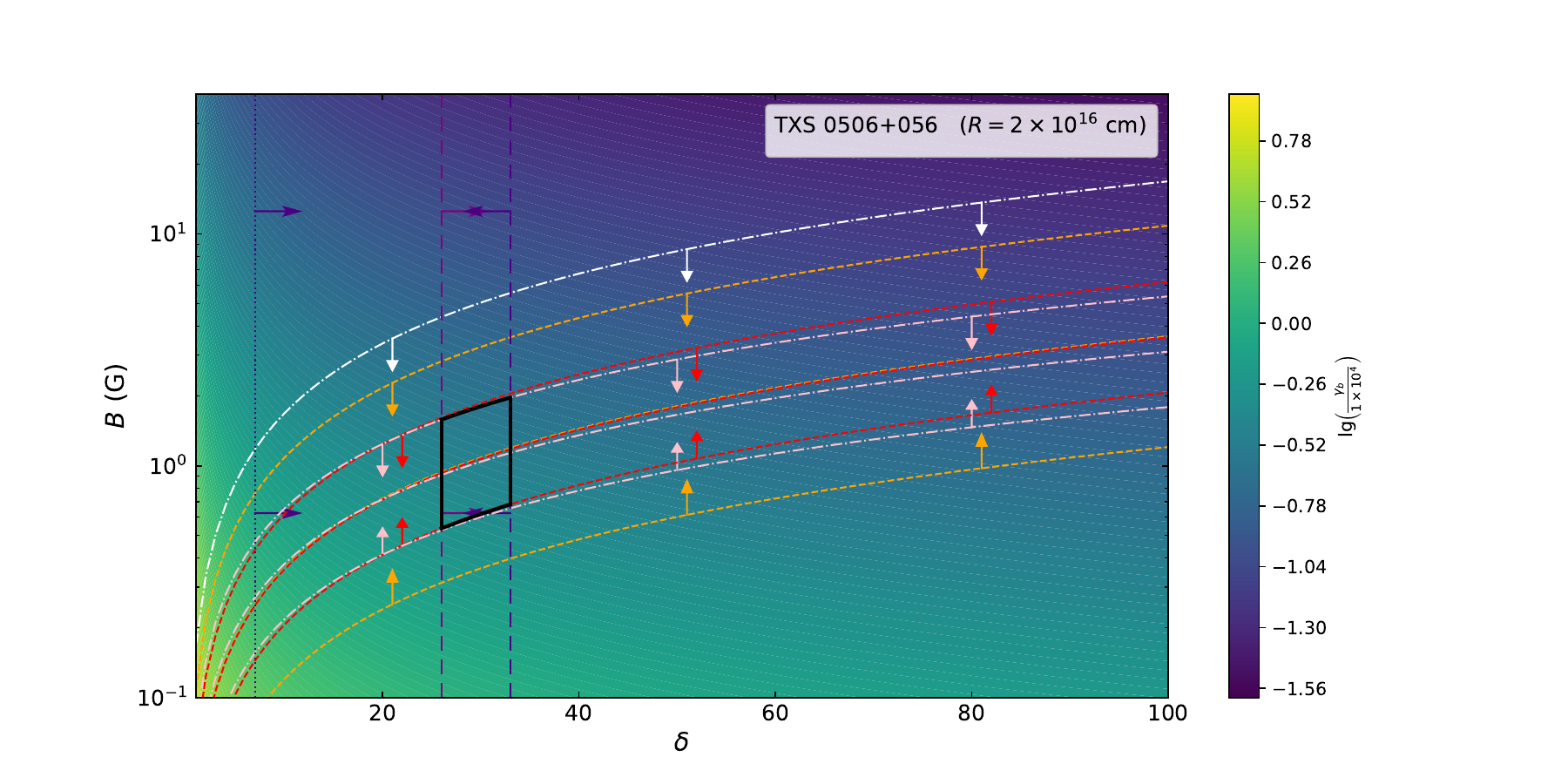}
        \label{TXS 17 EC R=2e16 space}
    \end{minipage}

    \begin{minipage}{0.49\linewidth}
        \centering
        \includegraphics[width=\linewidth, trim=10 15 30 10,clip]{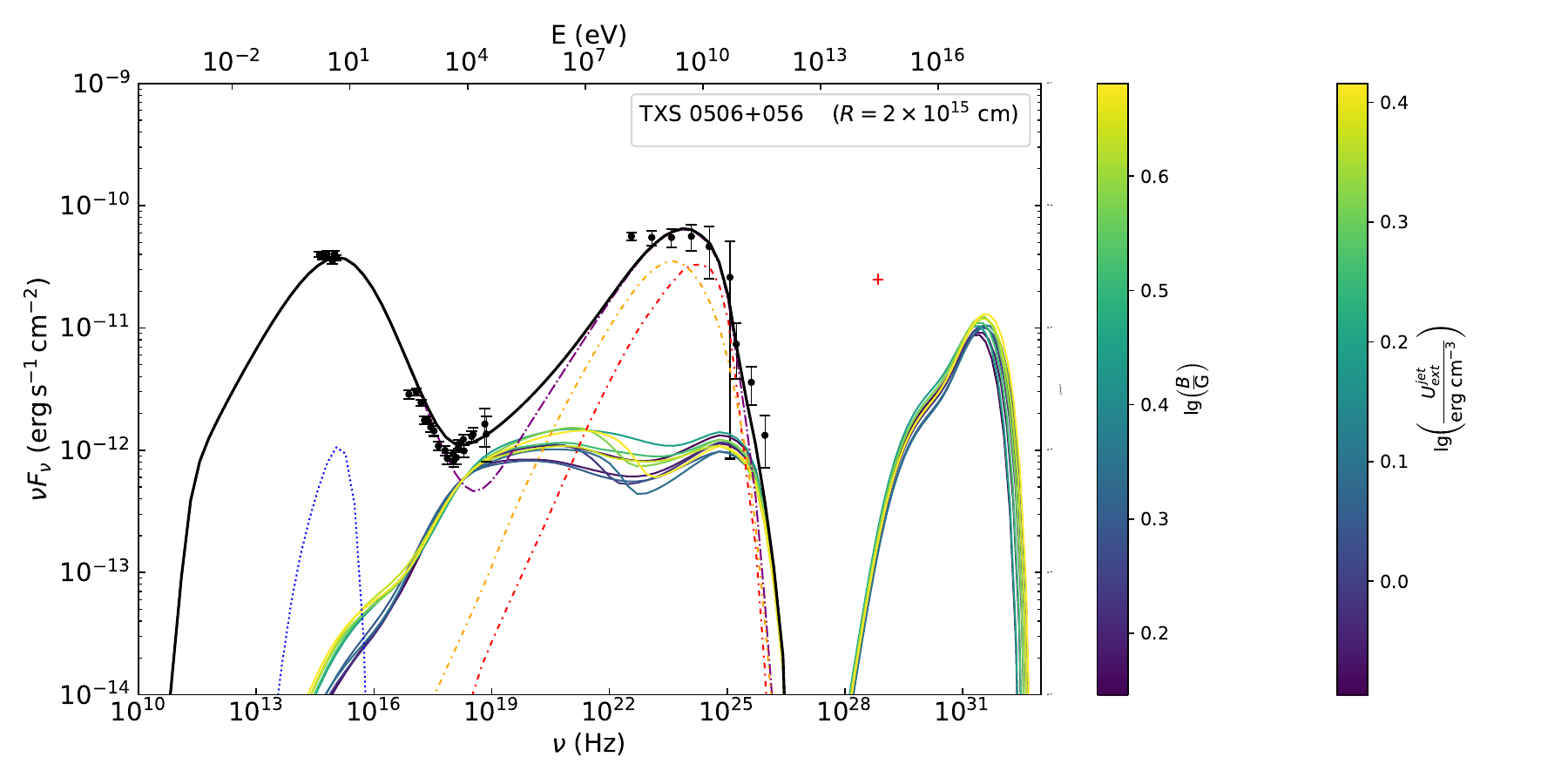}
        \label{TXS 17 EC R=2e15}
    \end{minipage}\hspace{-4mm}
    \begin{minipage}{0.49\linewidth}
        \centering
        \includegraphics[width=\linewidth, trim=10 15 30 10,clip]{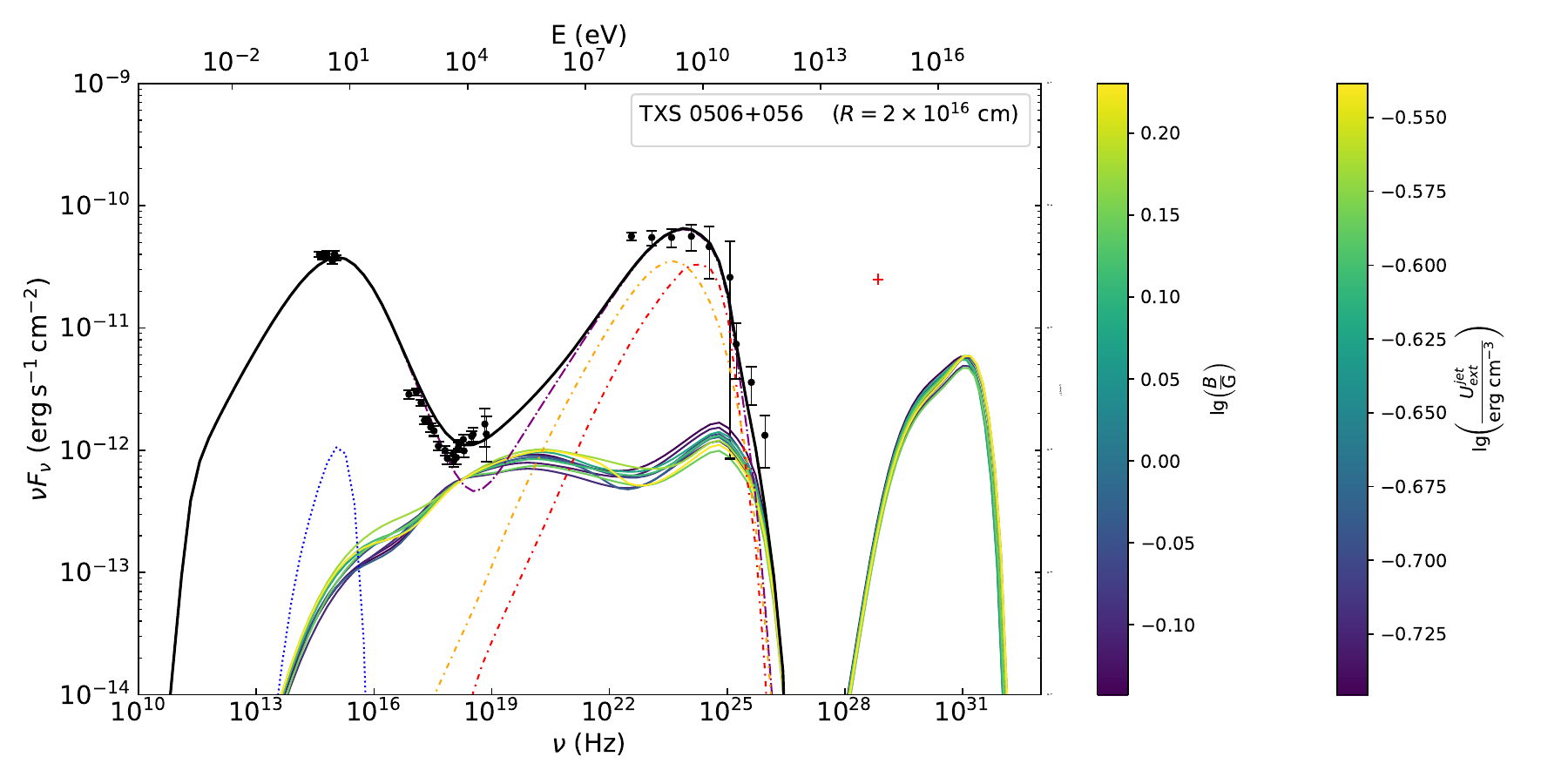}
        \label{TXS 17 EC R=2e16}
    \end{minipage}
    \caption{TXS 0506+056 associated with IC-170922A. Upper panels: the fitting results of the SED for $R=1\times10^{15}~{\rm cm}$ (left panel) and $R=1\times10^{16}~{\rm cm}$ (right panel) under the SSC-dominated case. Middle panels: the parameter space for $R=2\times10^{15}~{\rm cm}$ (left panel) and $R=2\times10^{16}~{\rm cm}$ (right panel) under the EC-dominated case. Lower panels: the fitting results of the SED for $R=2\times10^{15}~{\rm cm}$ (left panel) and $R=2\times10^{16}~{\rm cm}$ (right panel) under the EC-dominated case. In upper and lower panels, the colored dashed and solid lines respectively represent the secondary pair cascade emission and the neutrino spectrum for different parameter combinations, which correspond to the color bar. The quasi-simultaneous data, neutrino data and other line styles in all panels have the same meaning as in Fig.~\ref{TXS 17}.}
    \label{TXS 17 appendix}
\end{figure*}

\begin{figure*}[htbp]
    \centering
    \begin{minipage}{0.49\linewidth}
        \centering
        \includegraphics[width=\linewidth, trim=10 15 35 10,clip]{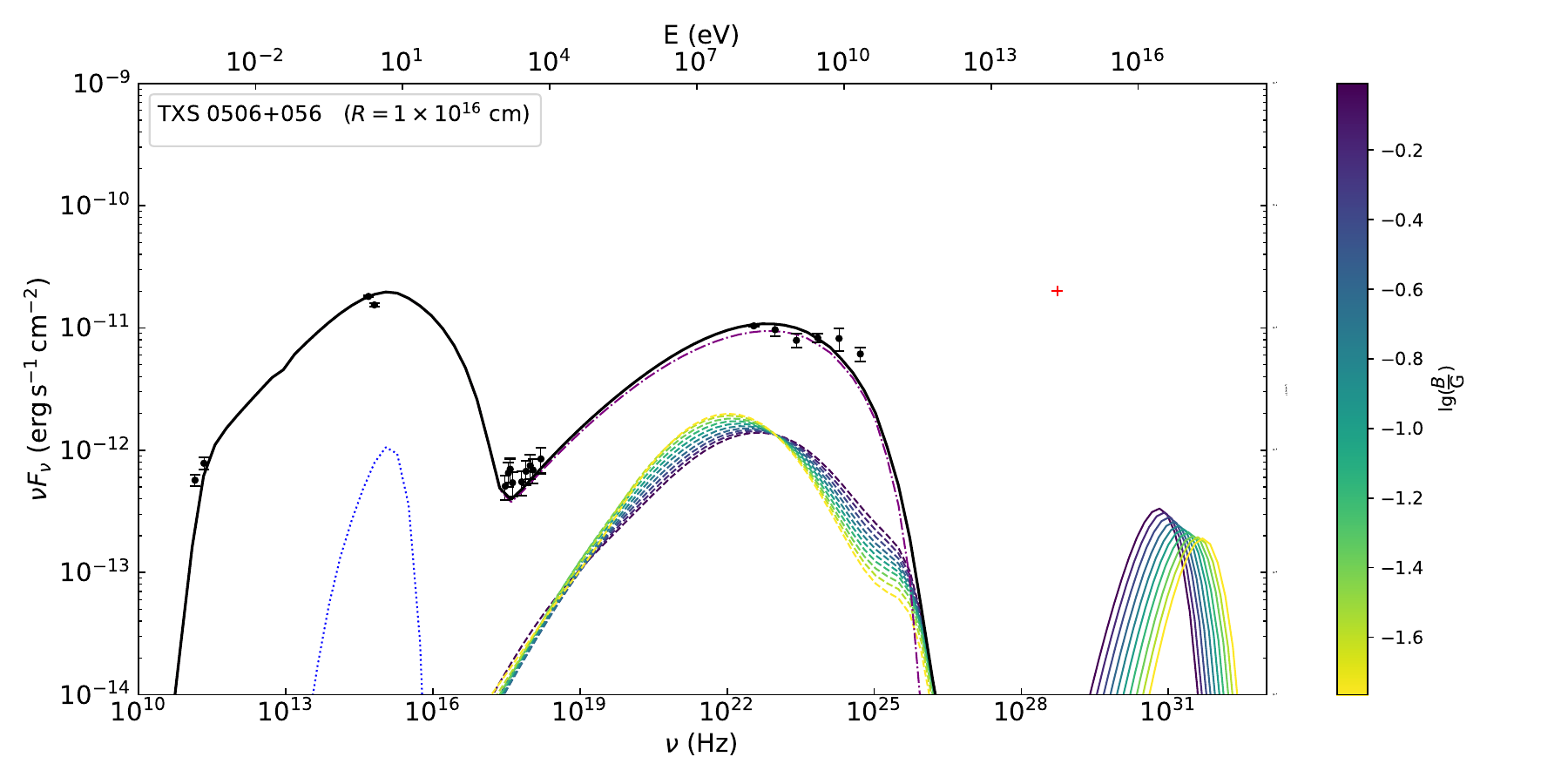}
        \label{TXS 21 SSC R=1e16}
    \end{minipage}\hspace{-4mm}
    \begin{minipage}{0.49\linewidth}
        \centering
        \includegraphics[width=\linewidth, trim=10 15 35 10,clip]{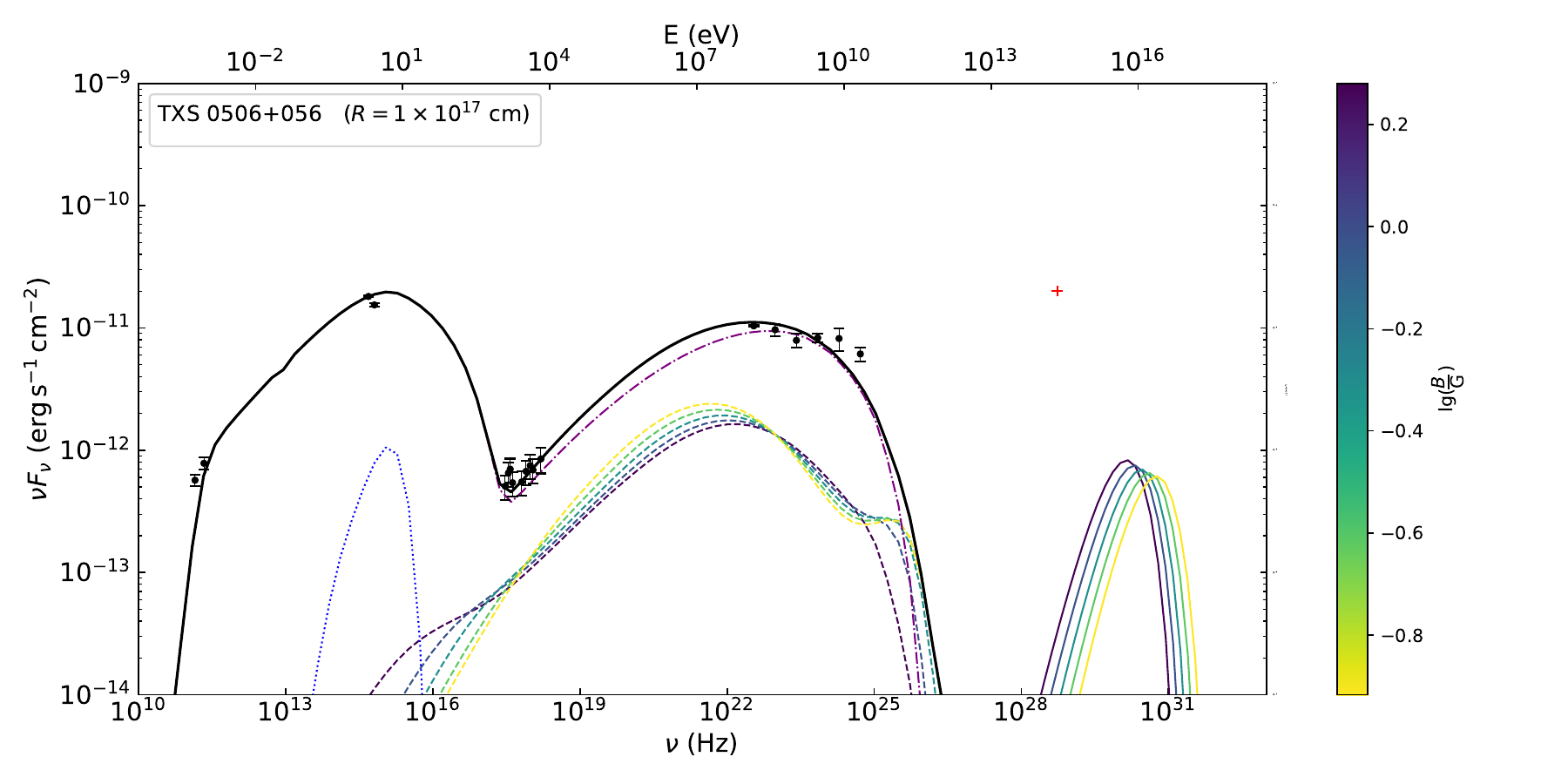}
        \label{TXS 21 SSC R=1e17}
    \end{minipage}

    \begin{minipage}{0.49\linewidth}
        \centering
        \includegraphics[width=\linewidth, trim=35 15 90 30,clip]{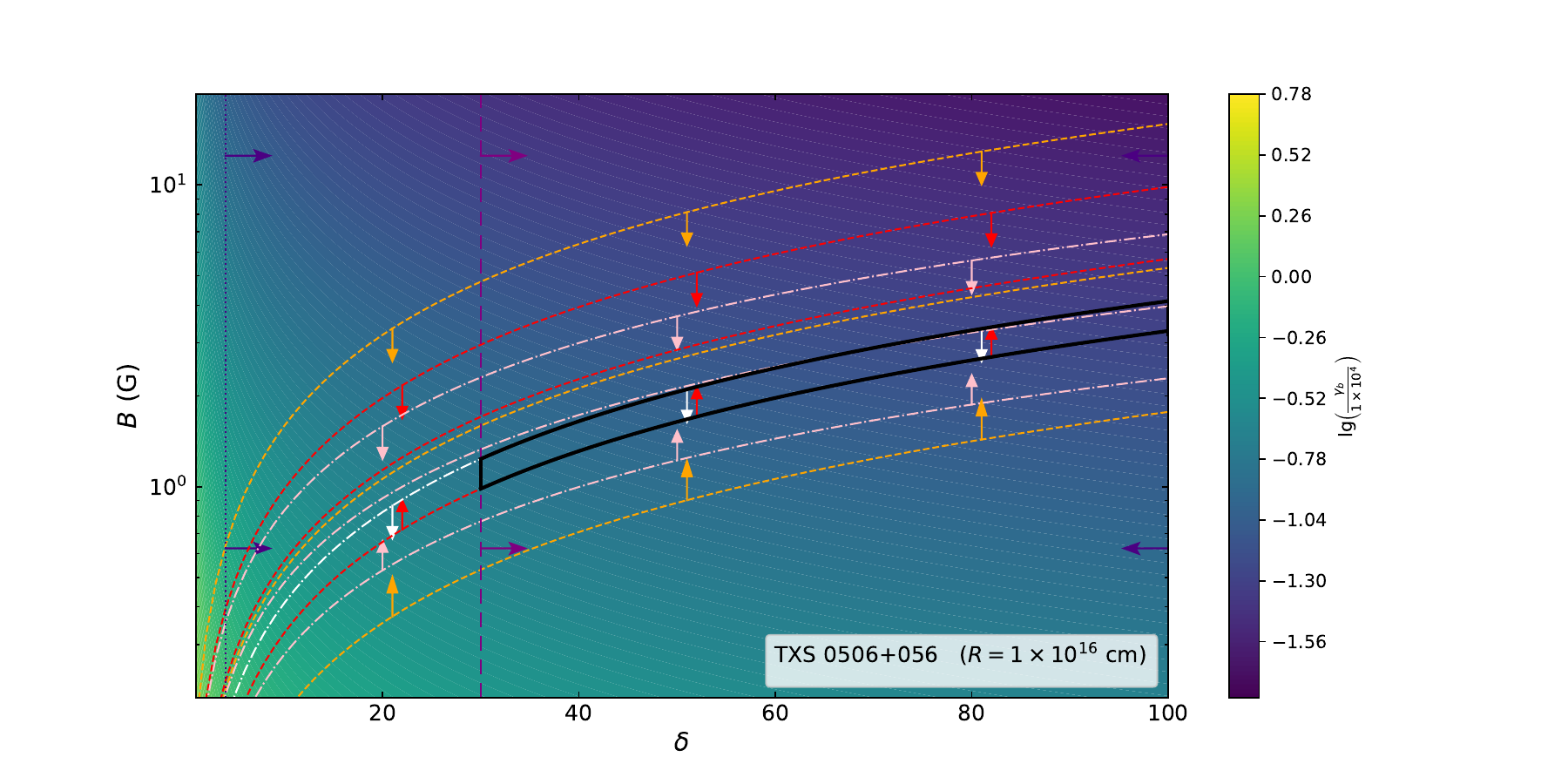}
        \label{TXS 21 EC R=1e16 space}
    \end{minipage}\hspace{-4mm}
    \begin{minipage}{0.49\linewidth}
        \centering
        \includegraphics[width=\linewidth, trim=35 15 90 30,clip]{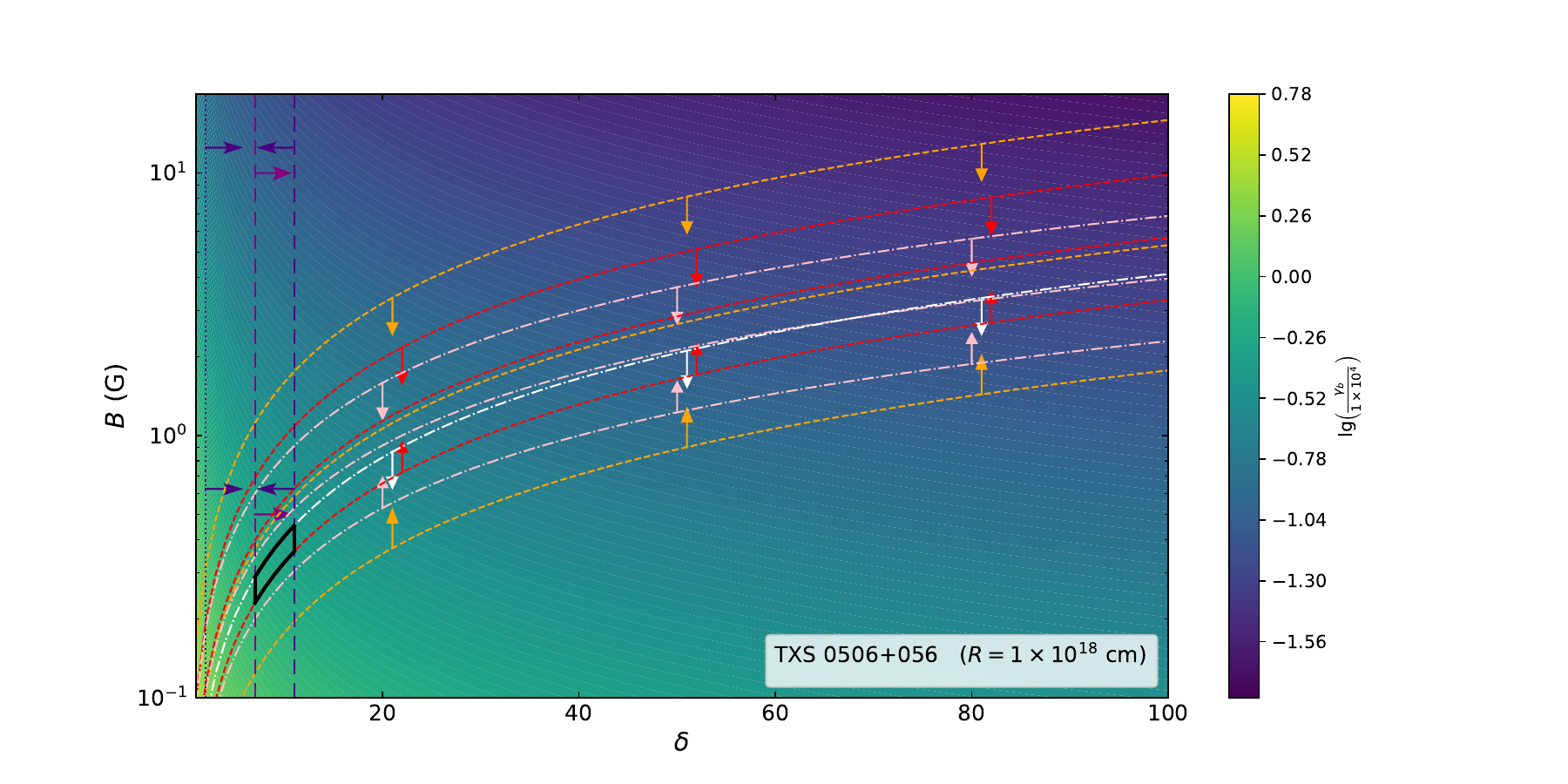}
        \label{TXS 21 EC R=1e18 space}
    \end{minipage}

    \begin{minipage}{0.49\linewidth}
        \centering
        \includegraphics[width=\linewidth, trim=10 15 25 10,clip]{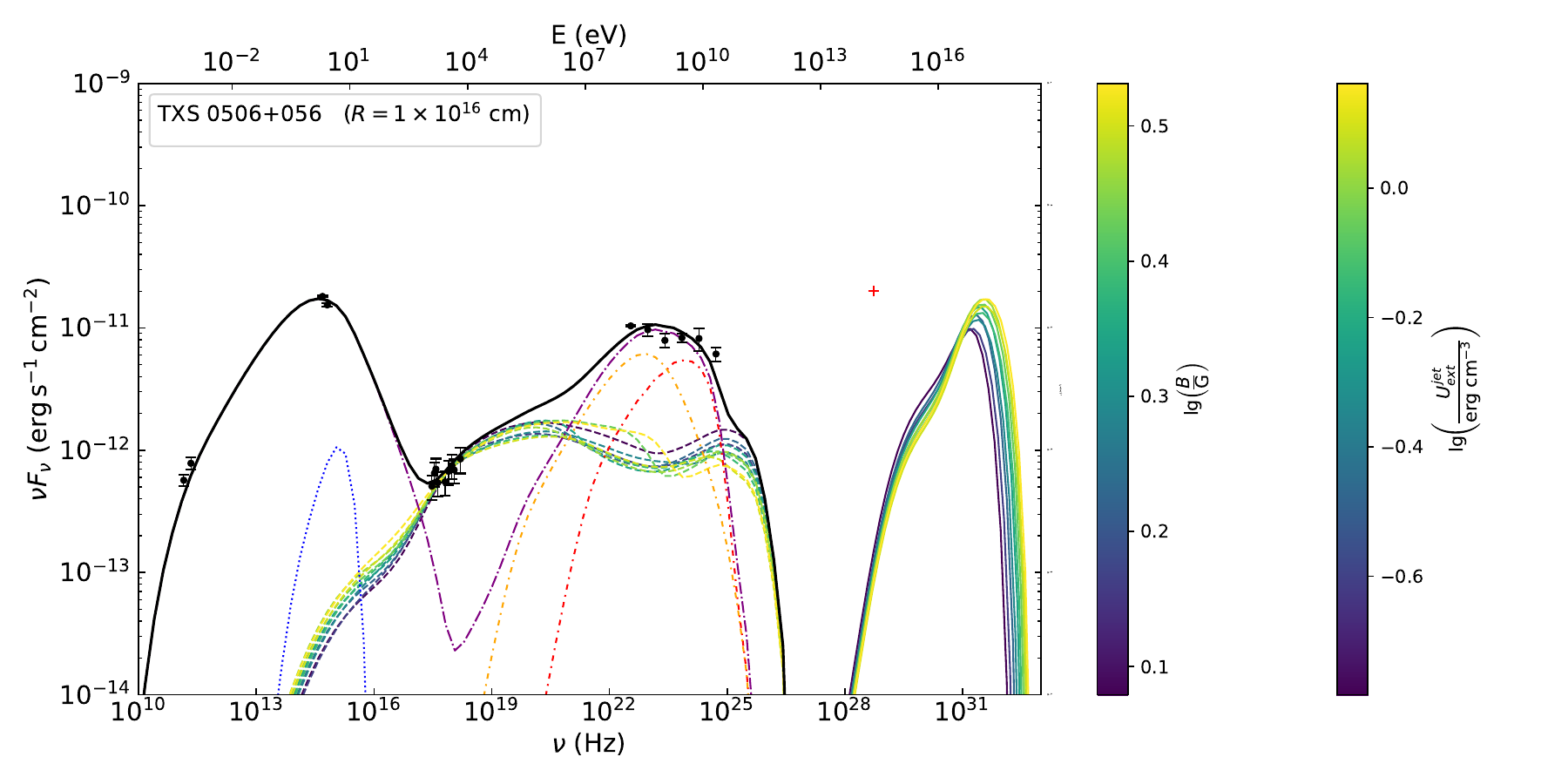}
        \label{TXS 21 EC R=1e16}
    \end{minipage}\hspace{-4mm}
    \begin{minipage}{0.49\linewidth}
        \centering
        \includegraphics[width=\linewidth, trim=10 15 25 10,clip]{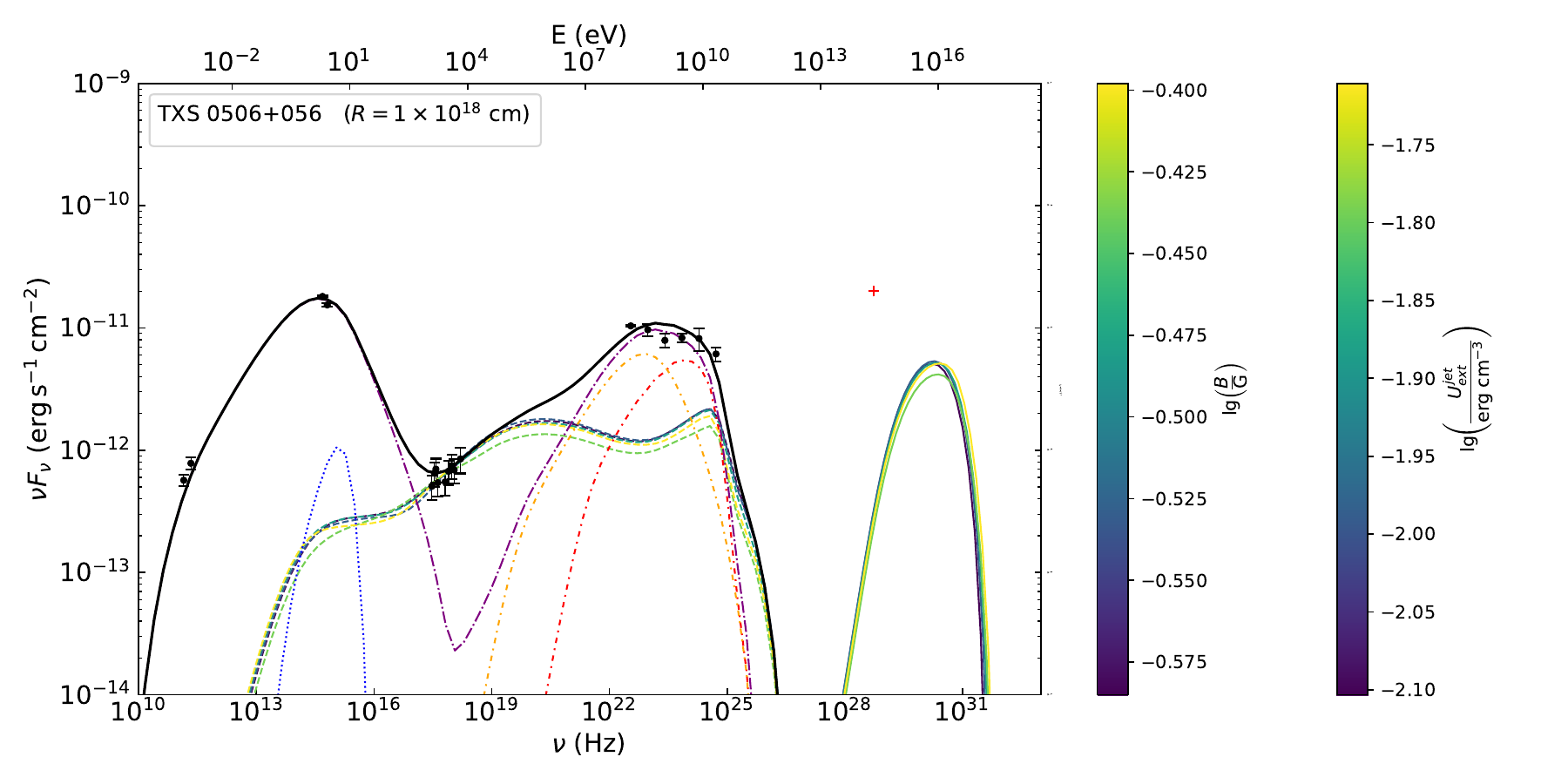}
        \label{TXS 21 EC R=1e18}
    \end{minipage}
    \caption{TXS 0506+056 associated with GVD-210418CA. Upper panels: the fitting results of the SED for $R=1\times10^{16}~{\rm cm}$ (left panel) and $R=1\times10^{17}~{\rm cm}$ (right panel) under the SSC-dominated case. Middle panels: the parameter space for $R=1\times10^{16}~{\rm cm}$ (left panel) and $R=1\times10^{18}~{\rm cm}$ (right panel) under the EC-dominated case. Lower panels: the fitting results of the SED for $R=1\times10^{16}~{\rm cm}$ (left panel) and $R=1\times10^{18}~{\rm cm}$ (right panel) under the EC-dominated case. In upper and lower panels, the colored dashed and solid lines respectively represent the secondary pair cascade emission and the neutrino spectrum for different parameter combinations, which correspond to the color bar. The quasi-simultaneous data, neutrino data and other line styles in all panels have the same meaning as in Fig.~\ref{TXS 21}.}
    \label{TXS 21 appendix}
\end{figure*}

\begin{figure*}[htbp]
    \centering
    \begin{minipage}{0.49\linewidth}
        \centering
        \includegraphics[width=\linewidth, trim=10 15 30 10,clip]{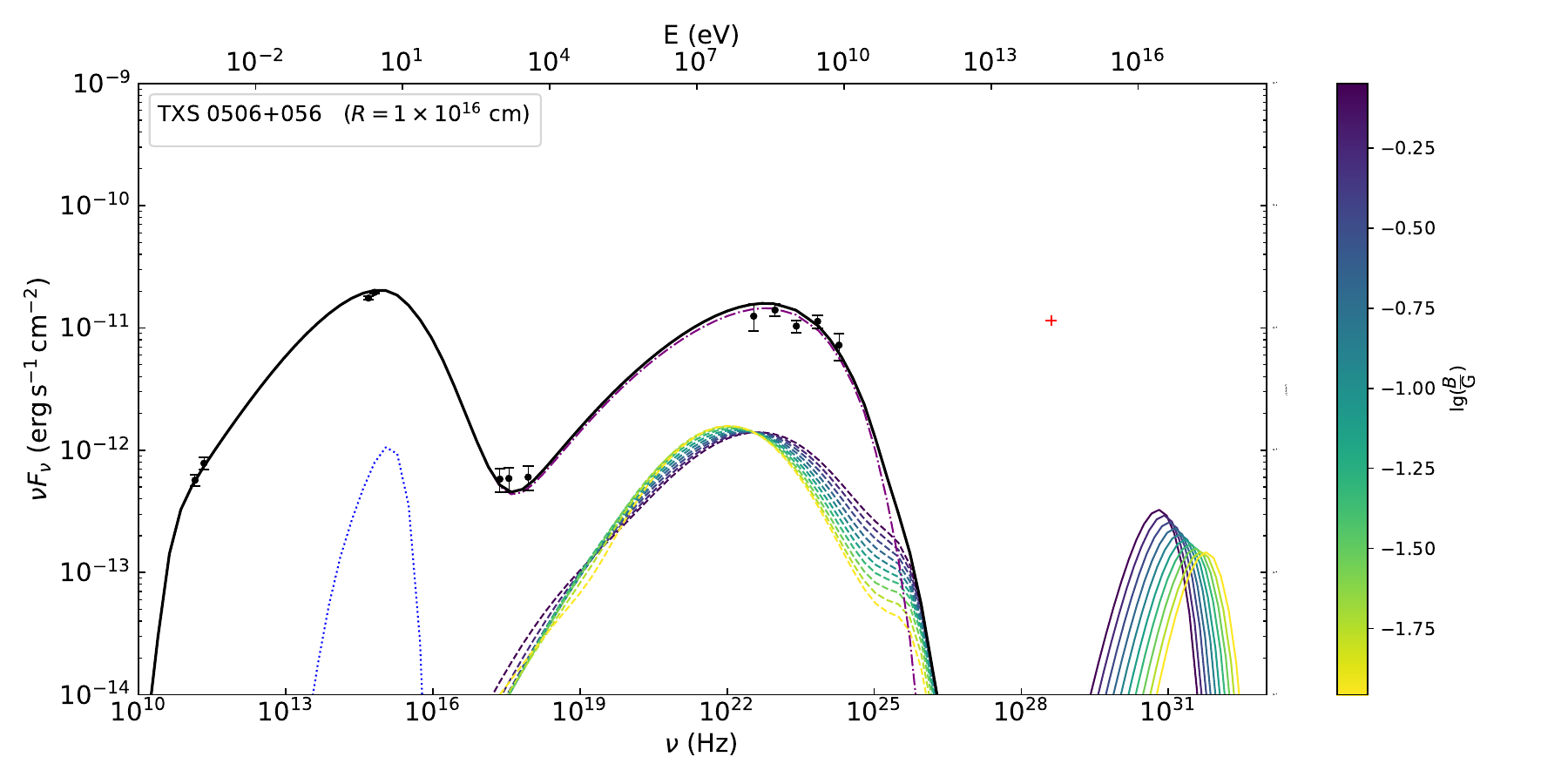}
        \label{TXS 22 SSC R=1e16}
    \end{minipage}\hspace{-4mm}
    \begin{minipage}{0.49\linewidth}
        \centering
        \includegraphics[width=\linewidth, trim=10 15 30 10,clip]{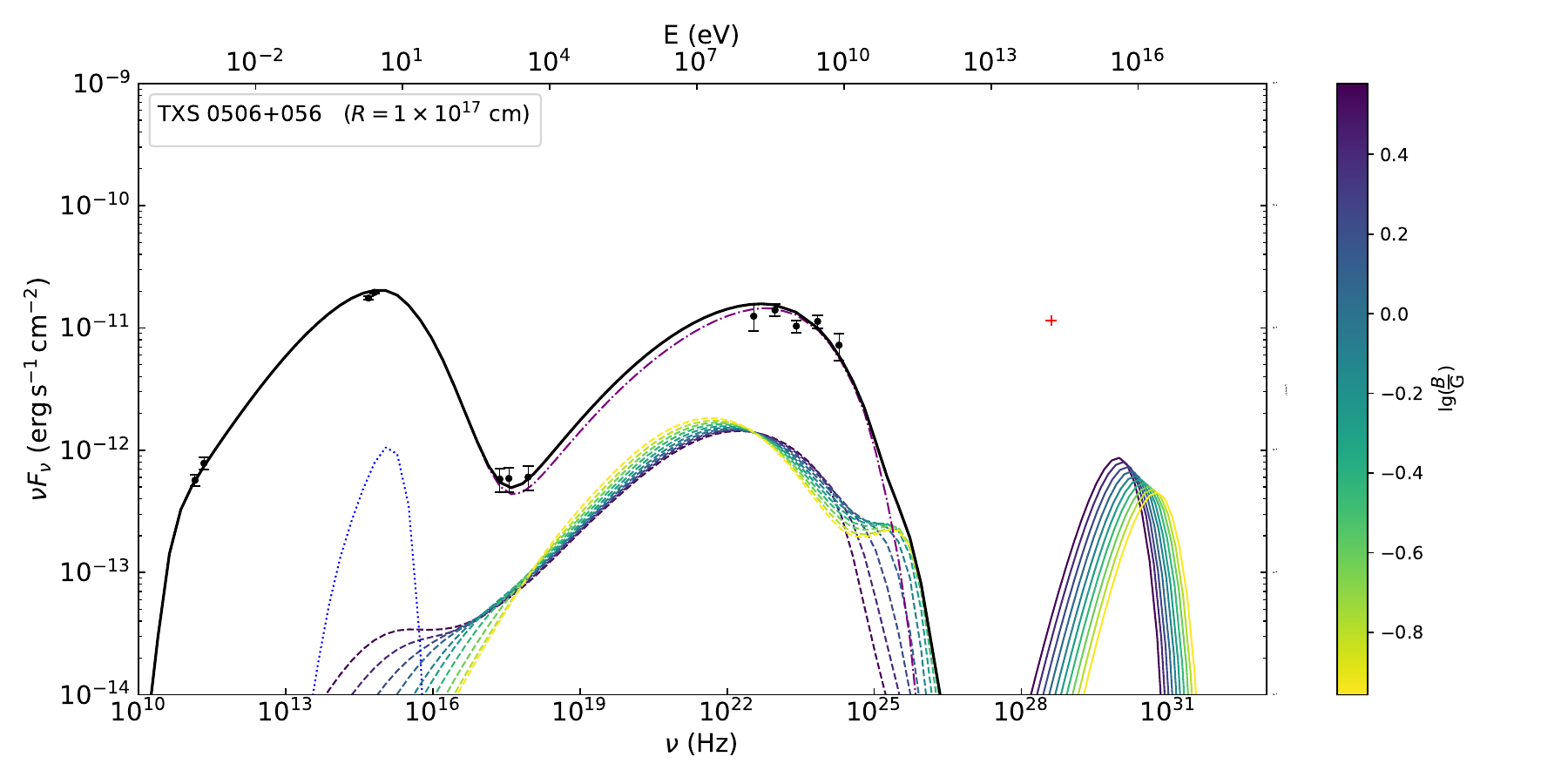}
        \label{TXS 22 SSC R=1e17}
    \end{minipage}

    \begin{minipage}{0.49\linewidth}
        \centering
        \includegraphics[width=\linewidth, trim=35 15 90 30,clip]{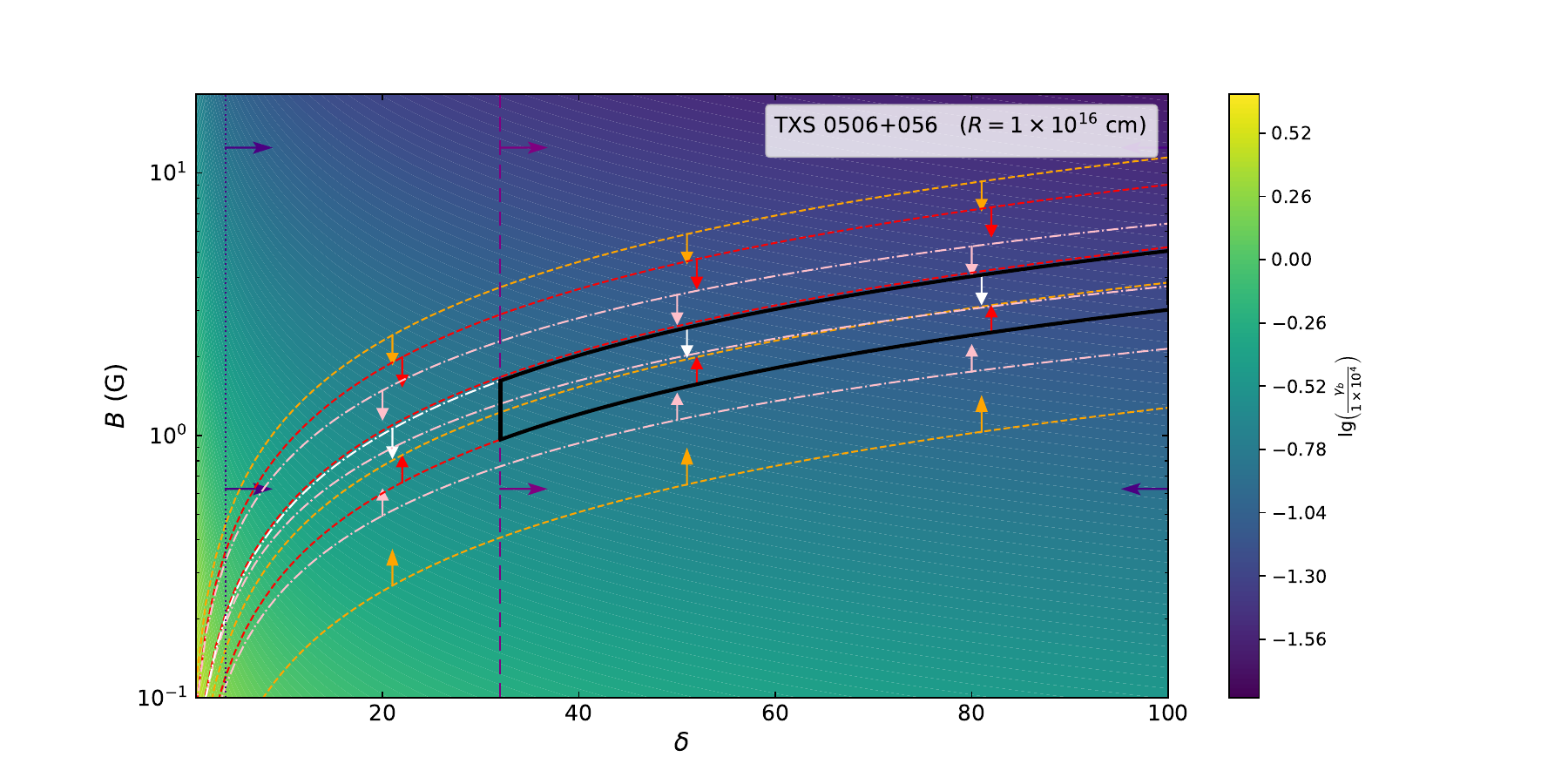}
        \label{TXS 22 EC R=1e16 space}
    \end{minipage}\hspace{-4mm}
    \begin{minipage}{0.49\linewidth}
        \centering
        \includegraphics[width=\linewidth, trim=35 15 90 30,clip]{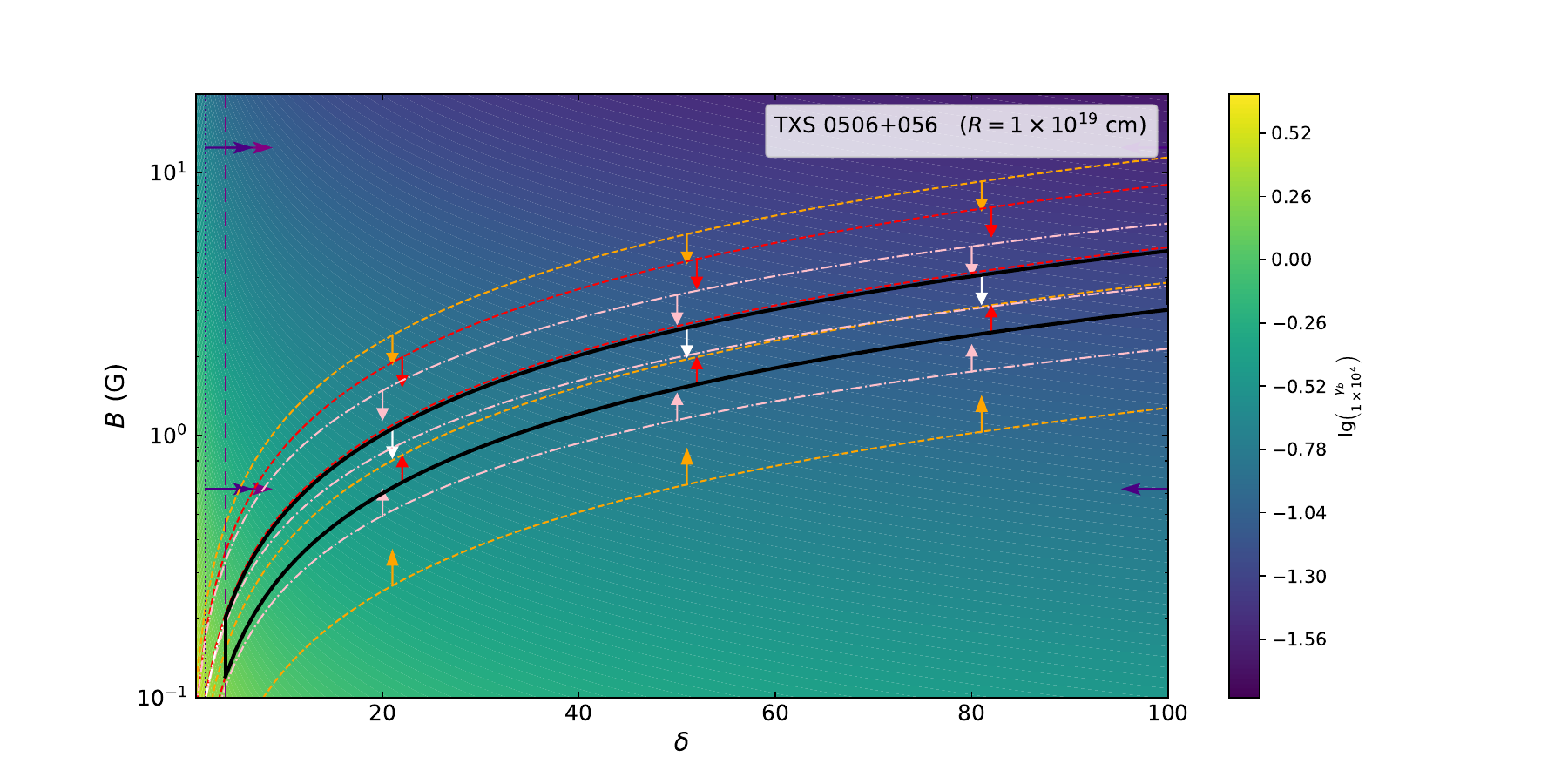}
        \label{TXS 22 EC R=1e19 space}
    \end{minipage}

    \begin{minipage}{0.49\linewidth}
        \centering
        \includegraphics[width=\linewidth, trim=10 15 25 10,clip]{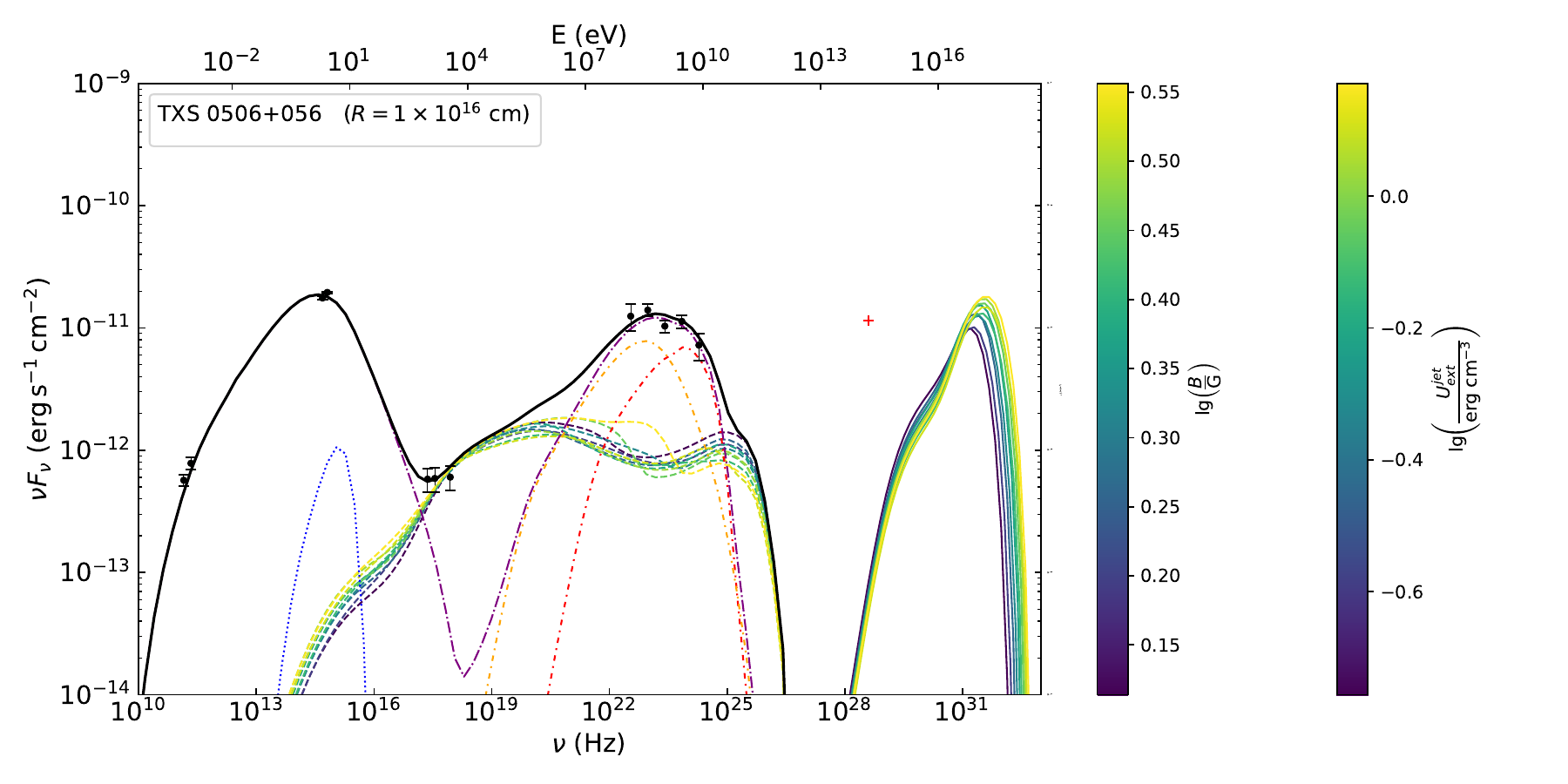}
        \label{TXS 22 EC R=1e16}
    \end{minipage}\hspace{-4mm}
    \begin{minipage}{0.49\linewidth}
        \centering
        \includegraphics[width=\linewidth, trim=10 15 25 10,clip]{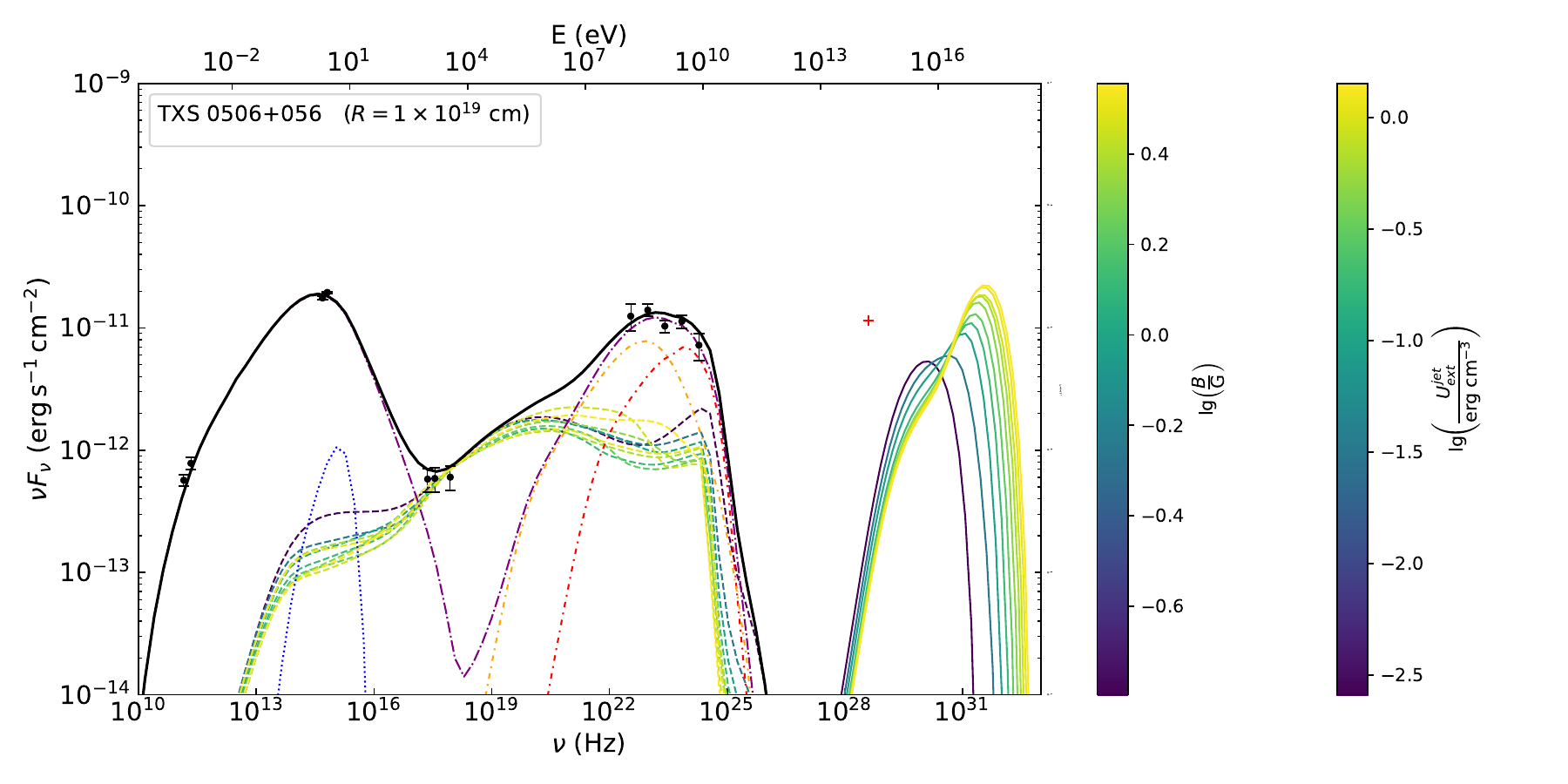}
        \label{TXS 22 EC R=1e19}
    \end{minipage}
    \caption{TXS 0506+056 associated with IC-220918A. Upper panels: the fitting results of the SED for $R=1\times10^{16}~{\rm cm}$ (left panel) and $R=1\times10^{17}~{\rm cm}$ (right panel) under the SSC-dominated case. Middle panels: the parameter space for $R=1\times10^{16}~{\rm cm}$ (left panel) and $R=1\times10^{19}~{\rm cm}$ (right panel) under the EC-dominated case. Lower panels: the fitting results of the SED for $R=1\times10^{16}~{\rm cm}$ (left panel) and $R=1\times10^{19}~{\rm cm}$ (right panel) under the EC-dominated case. In upper and lower panels, the colored dashed and solid lines respectively represent the secondary pair cascade emission and the neutrino spectrum for different parameter combinations, which correspond to the color bar. The quasi-simultaneous data, neutrino data and other line styles in all panels have the same meaning as in Fig.~\ref{TXS 22}.}
    \label{TXS 22 appendix}
\end{figure*}

\begin{figure*}[htbp]
    \centering
    \begin{minipage}{0.49\linewidth}
        \centering
        \includegraphics[width=\linewidth, trim=10 15 30 10,clip]{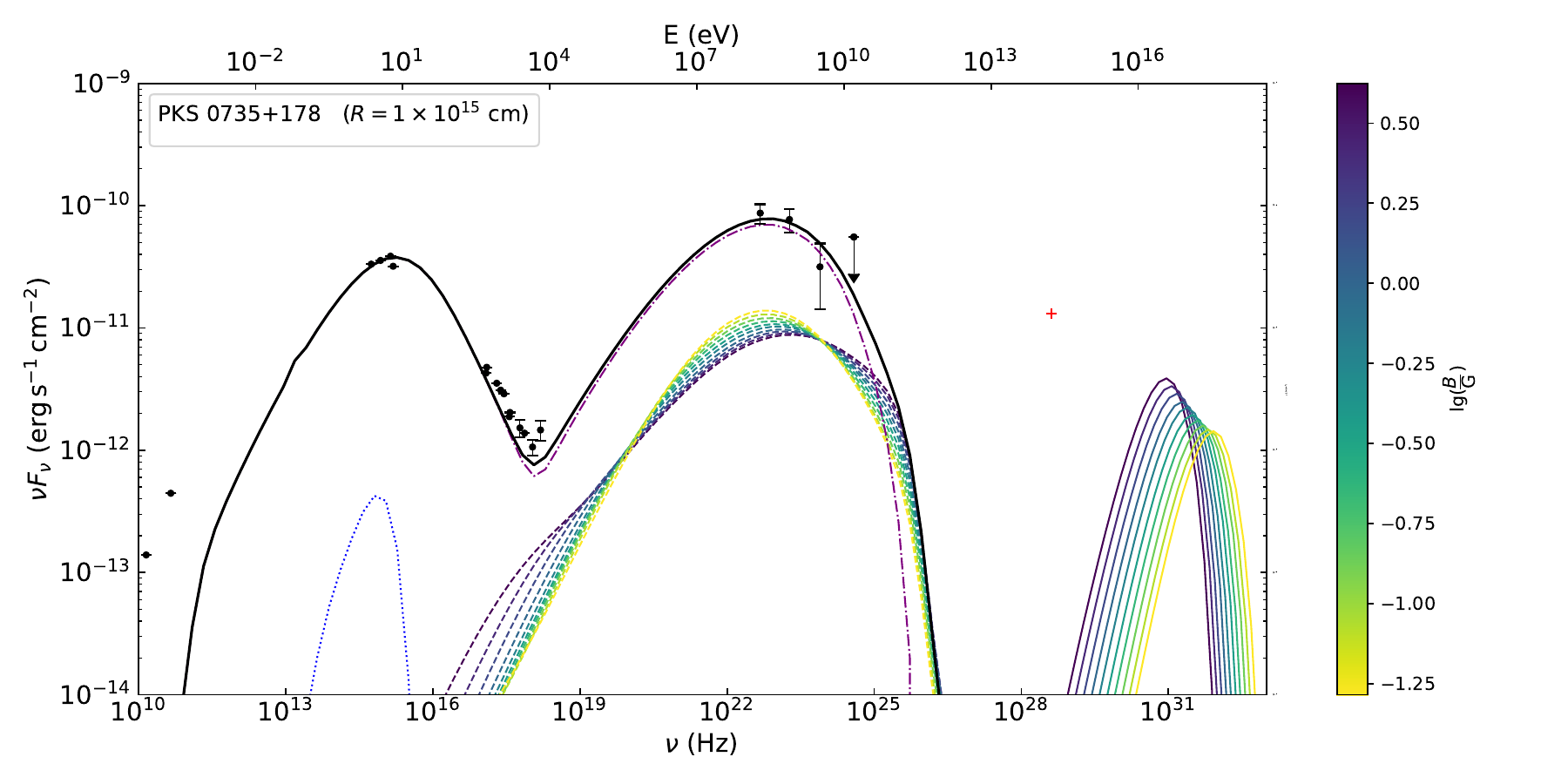}
        \label{PKS 0735+178 SSC R=1e15}
    \end{minipage}\hspace{-4mm}
    \begin{minipage}{0.49\linewidth}
        \centering
        \includegraphics[width=\linewidth, trim=10 15 30 10,clip]{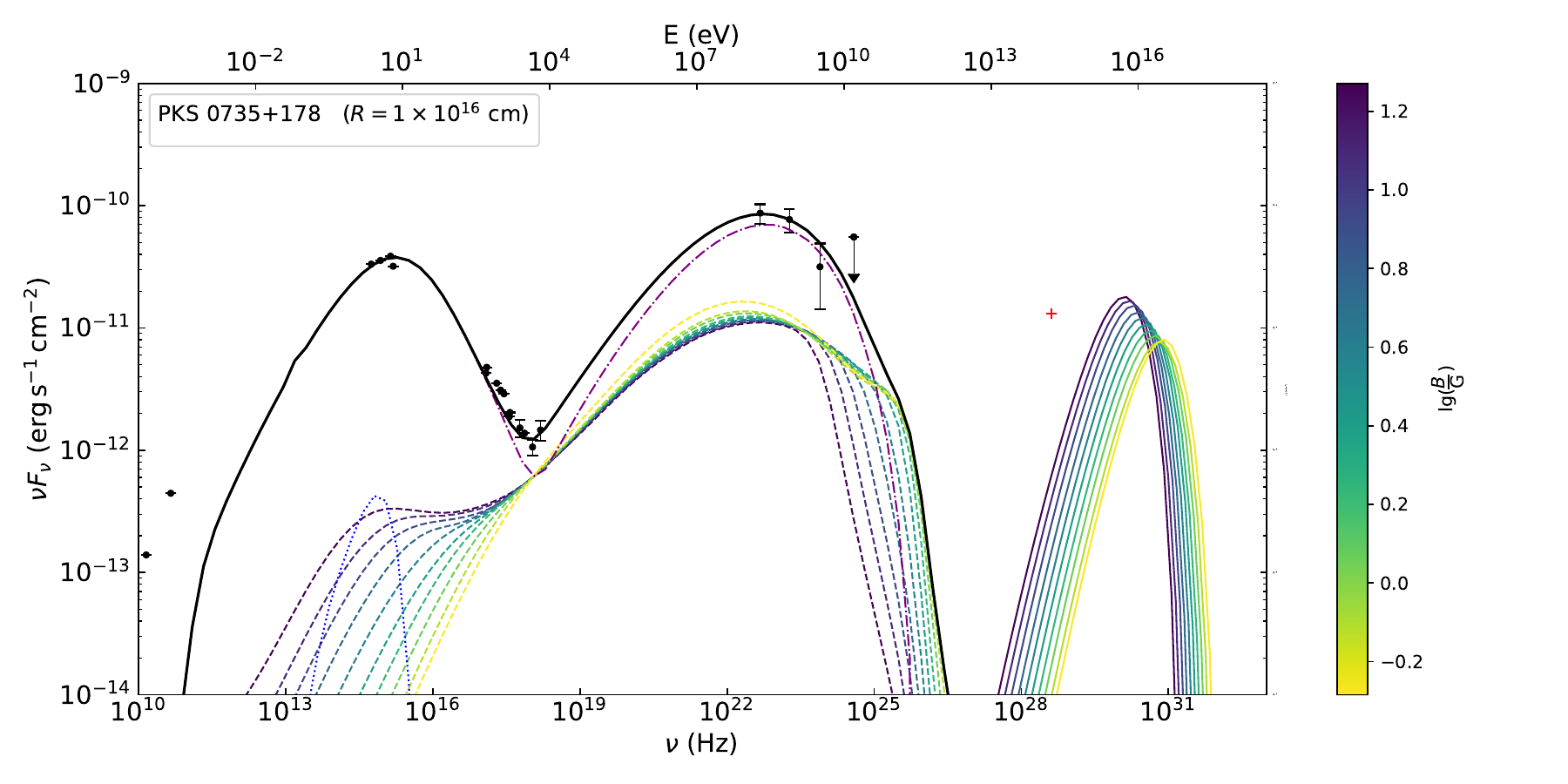}
        \label{PKS 0735+178 SSC R=1e16}
    \end{minipage}

    \begin{minipage}{0.49\linewidth}
        \centering
        \includegraphics[width=\linewidth, trim=35 15 90 30,clip]{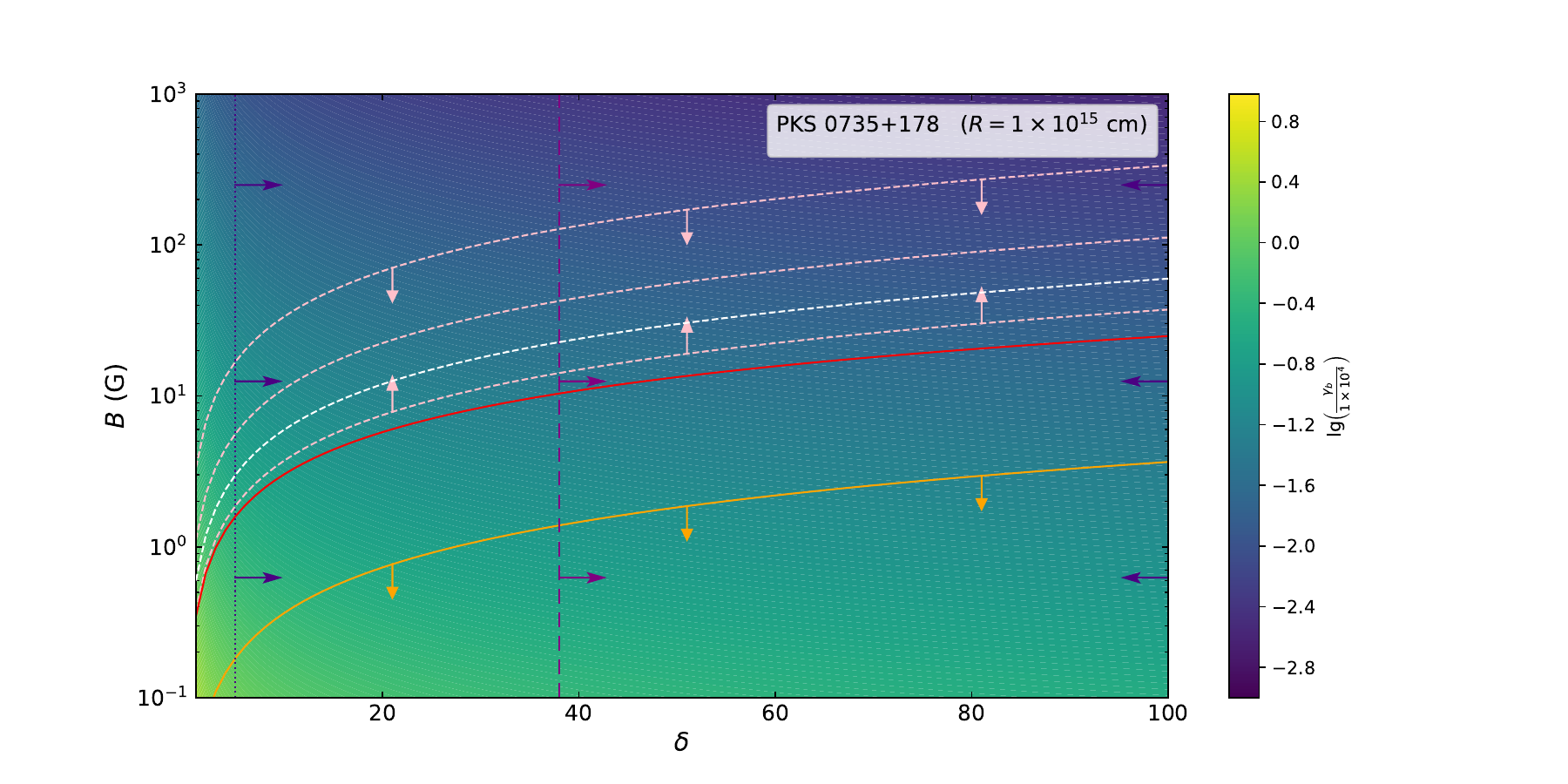}
        \label{PKS 0735+178 EC R=1e15 space}
    \end{minipage}\hspace{-4mm}
    \begin{minipage}{0.49\linewidth}
        \centering
        \includegraphics[width=\linewidth, trim=35 15 90 30,clip]{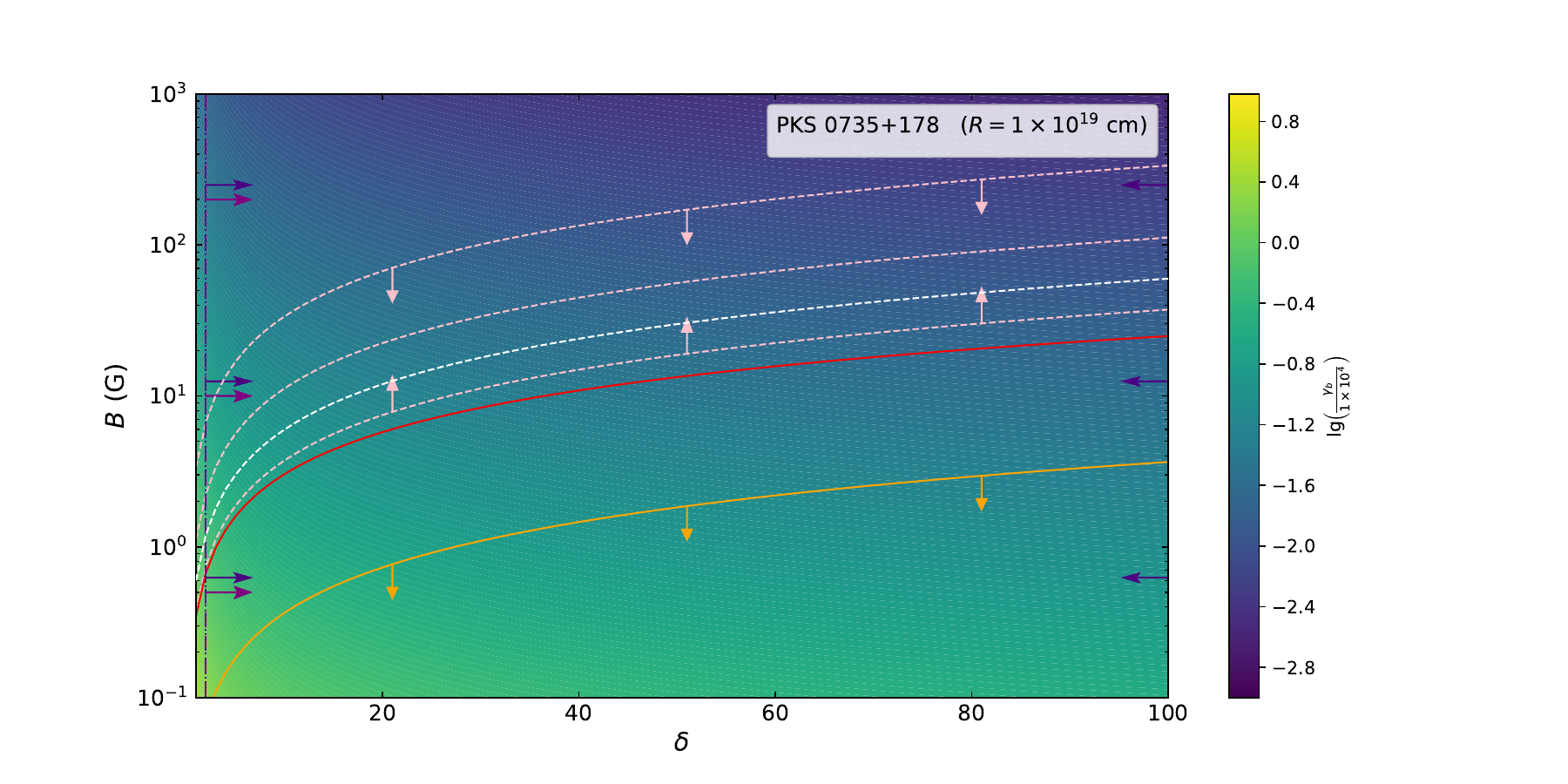}
        \label{PKS 0735+178 EC R=1e19 space}
    \end{minipage}

    \begin{minipage}{0.49\linewidth}
        \centering
        \includegraphics[width=\linewidth, trim=10 15 25 10,clip]{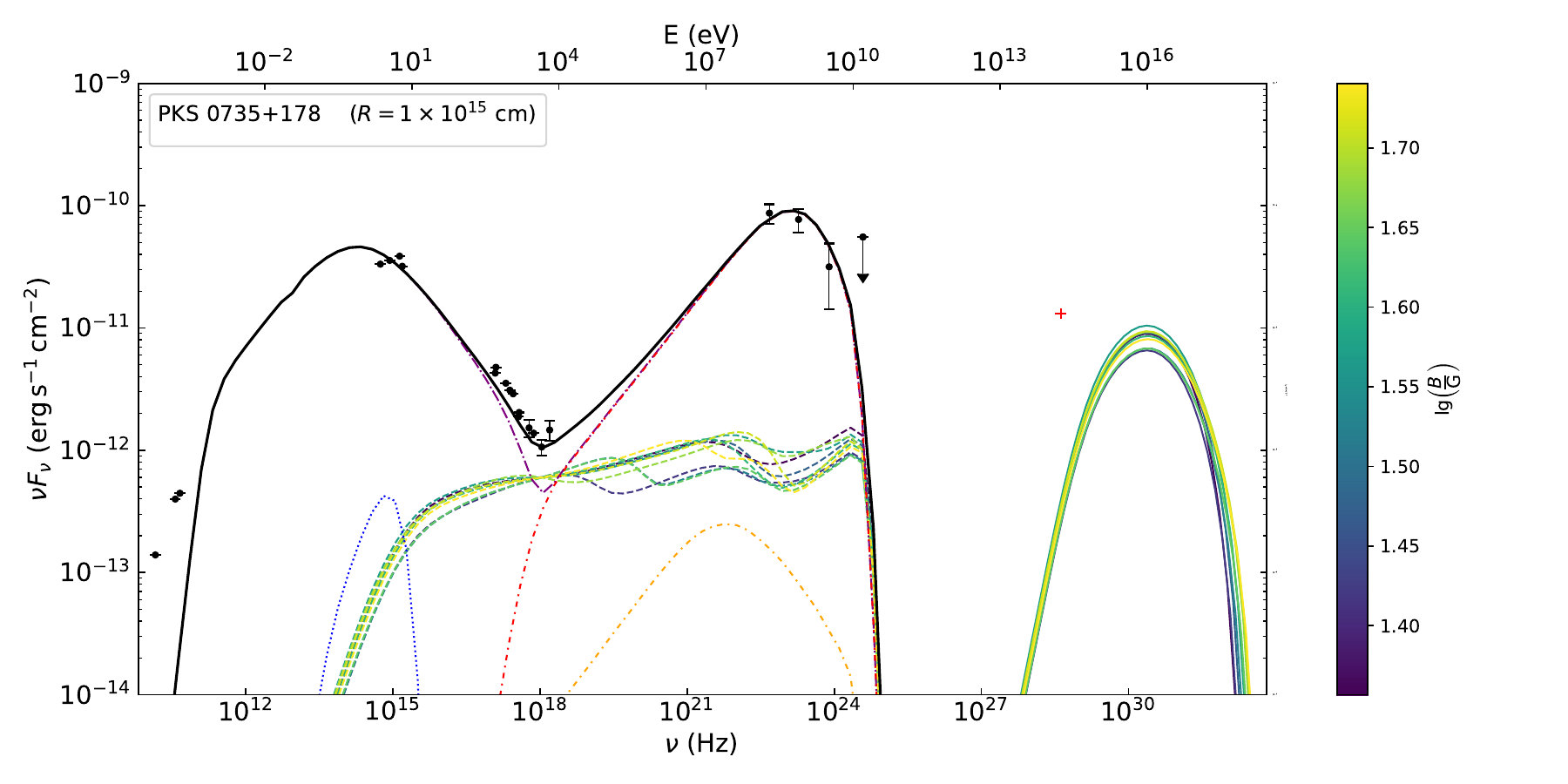}
        \label{PKS 0735+178 EC R=1e15}
    \end{minipage}\hspace{-4mm}
    \begin{minipage}{0.49\linewidth}
        \centering
        \includegraphics[width=\linewidth, trim=10 15 25 10,clip]{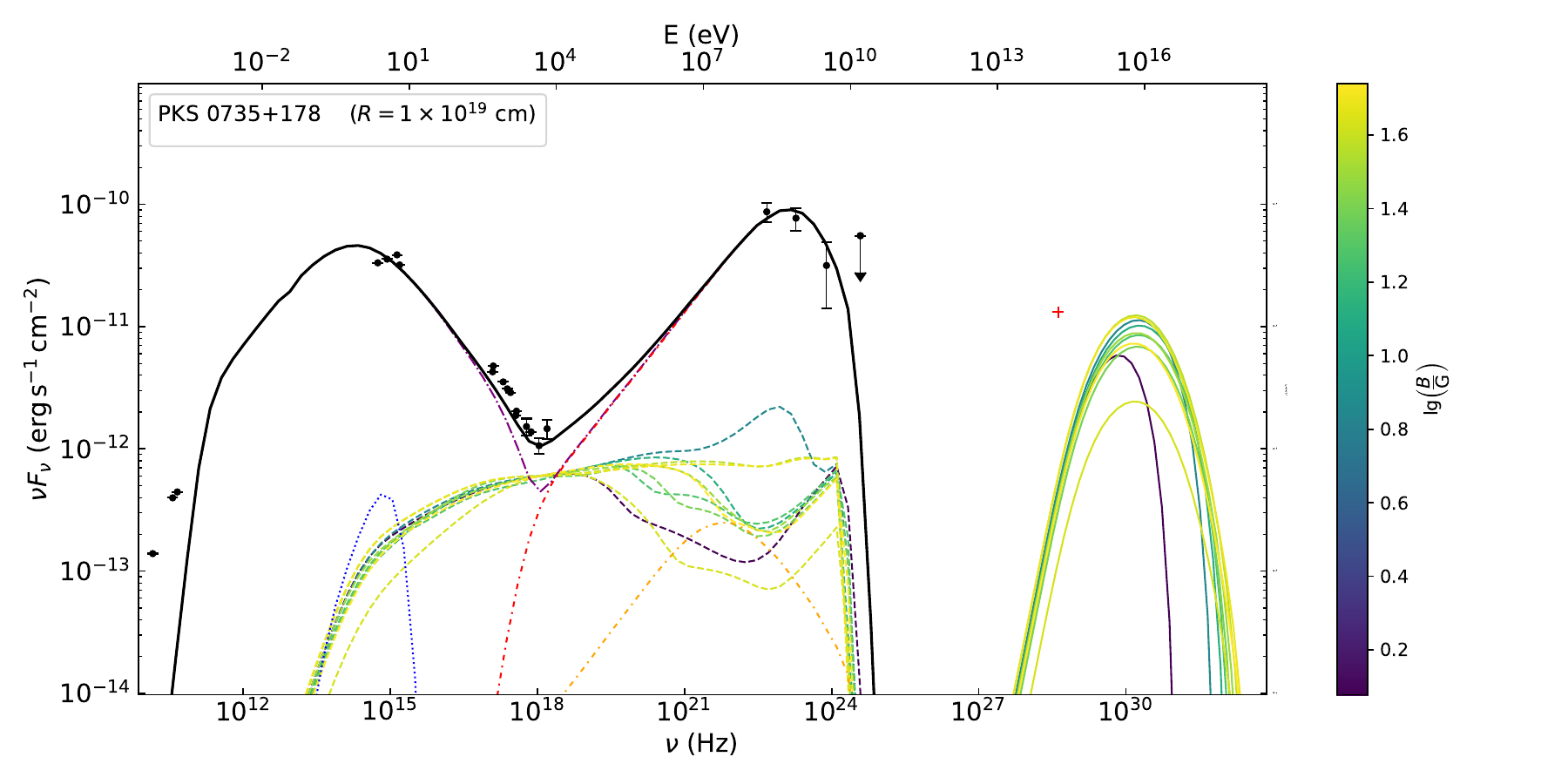}
        \label{PKS 0735+178 EC R=1e19}
    \end{minipage}
    \caption{PKS 0735+178 associated with IC-211208A. Upper panels: the fitting results of the SED for $R=1\times10^{15}~{\rm cm}$ (left panel) and $R=1\times10^{16}~{\rm cm}$ (right panel) under the SSC-dominated case. Middle panels: the parameter space for $R=1\times10^{15}~{\rm cm}$ (left panel) and $R=1\times10^{19}~{\rm cm}$ (right panel) under the EC-dominated case. Lower panels: the fitting results of the SED for $R=1\times10^{15}~{\rm cm}$ (left panel) and $R=1\times10^{19}~{\rm cm}$ (right panel) under the EC-dominated case. In upper and lower panels, the colored dashed and solid lines respectively represent the secondary pair cascade emission and the neutrino spectrum for different parameter combinations, which correspond to the color bar. The quasi-simultaneous data, neutrino data and other line styles in all panels have the same meaning as in Fig.~\ref{PKS 0735+178}.}
    \label{PKS 0735+178 appendix}
\end{figure*}

\begin{figure*}[htbp]
    \centering
    \begin{minipage}{0.49\linewidth}
        \centering
        \includegraphics[width=\linewidth, trim=10 15 30 10,clip]{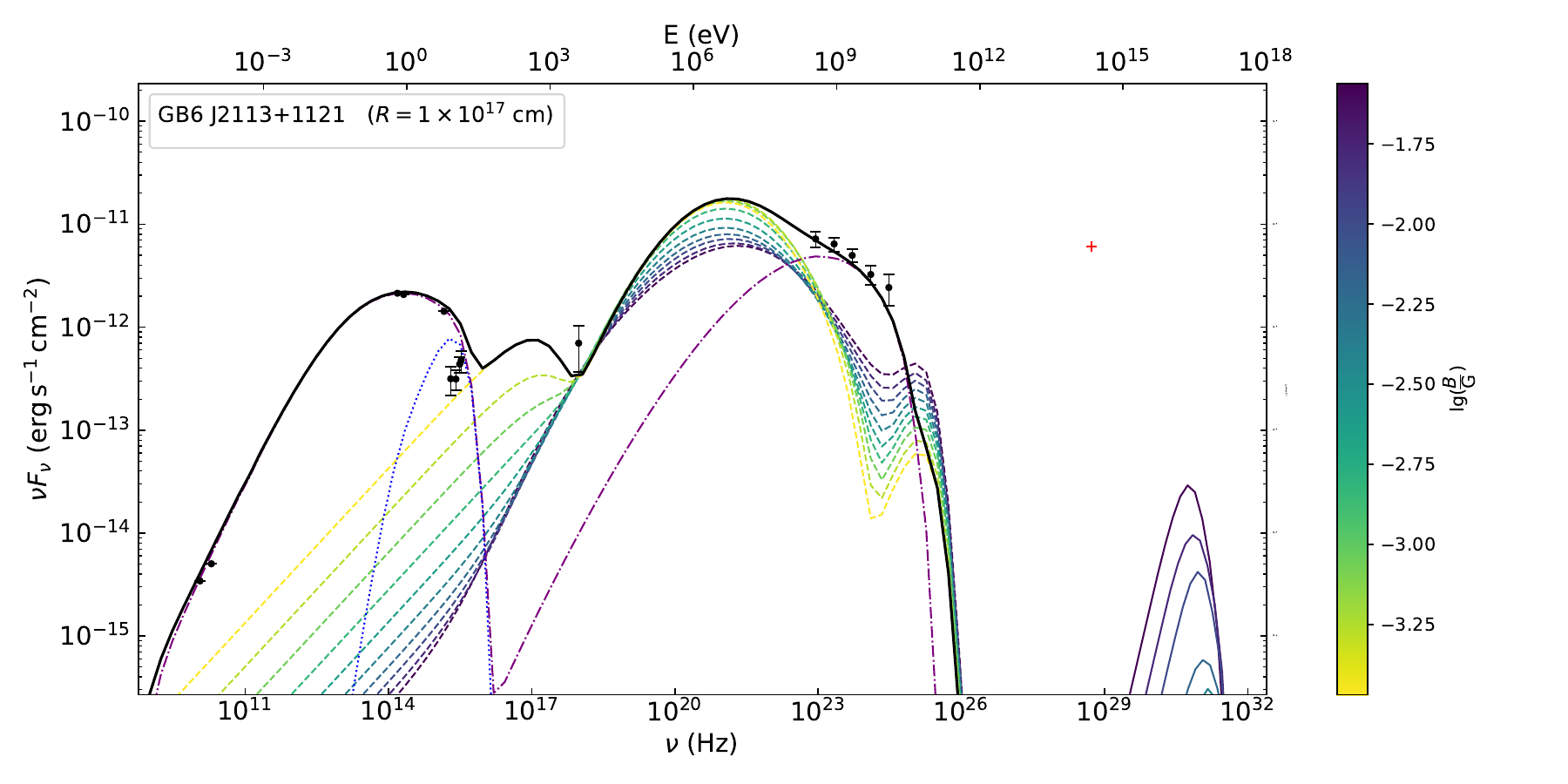}
        \label{GB6 J2113+1121 SSC R=1e17}
    \end{minipage}\hspace{-4mm}
    \begin{minipage}{0.49\linewidth}
        \centering
        \includegraphics[width=\linewidth, trim=10 15 30 10,clip]{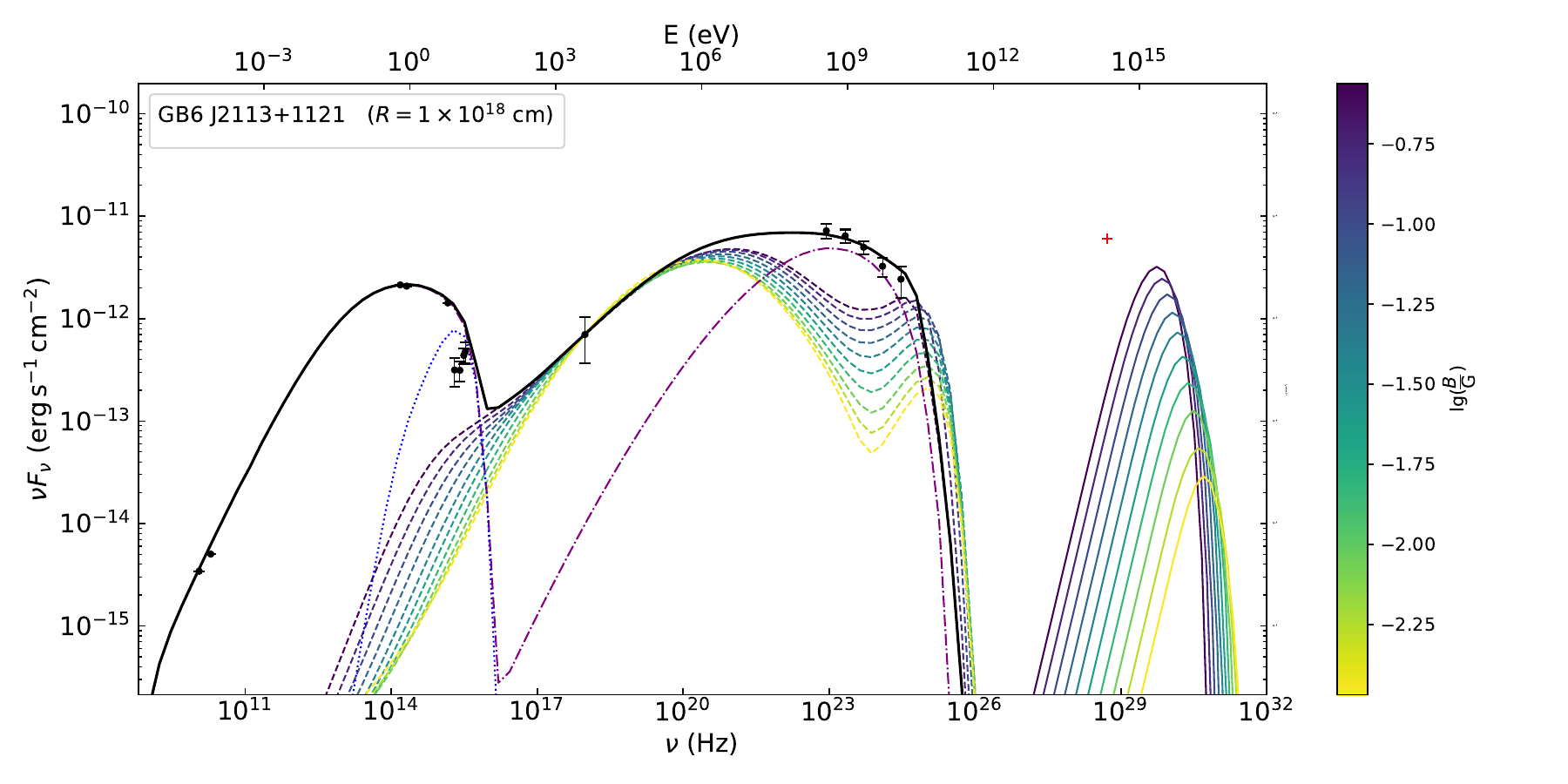}
        \label{GB6 J2113+1121 SSC R=1e18}
    \end{minipage}

    \begin{minipage}{0.49\linewidth}
        \centering
        \includegraphics[width=\linewidth, trim=35 15 90 30,clip]{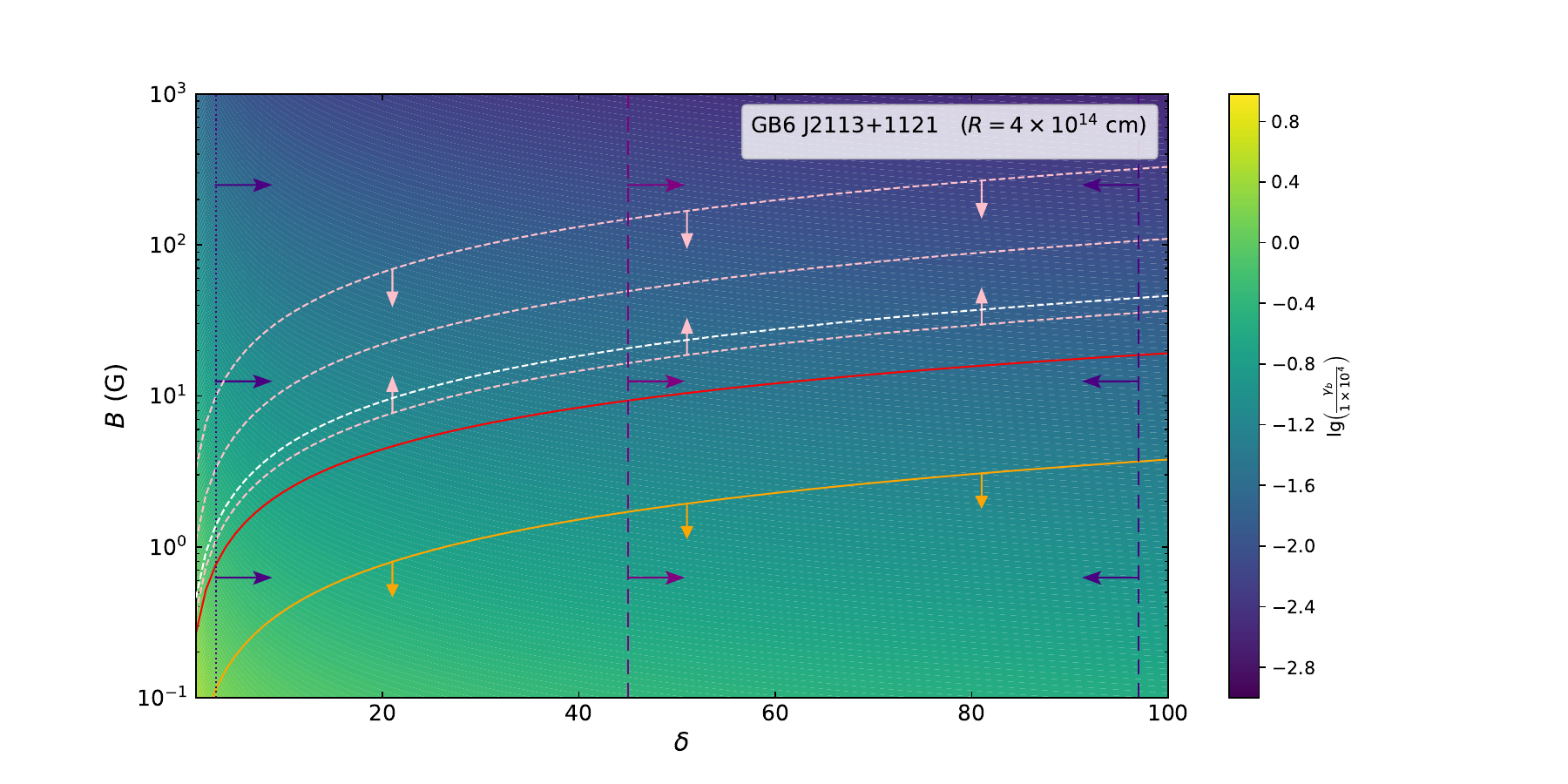}
        \label{GB6 J2113+1121 EC R=4e14 space}
    \end{minipage}\hspace{-4mm}
    \begin{minipage}{0.49\linewidth}
        \centering
        \includegraphics[width=\linewidth, trim=35 15 90 30,clip]{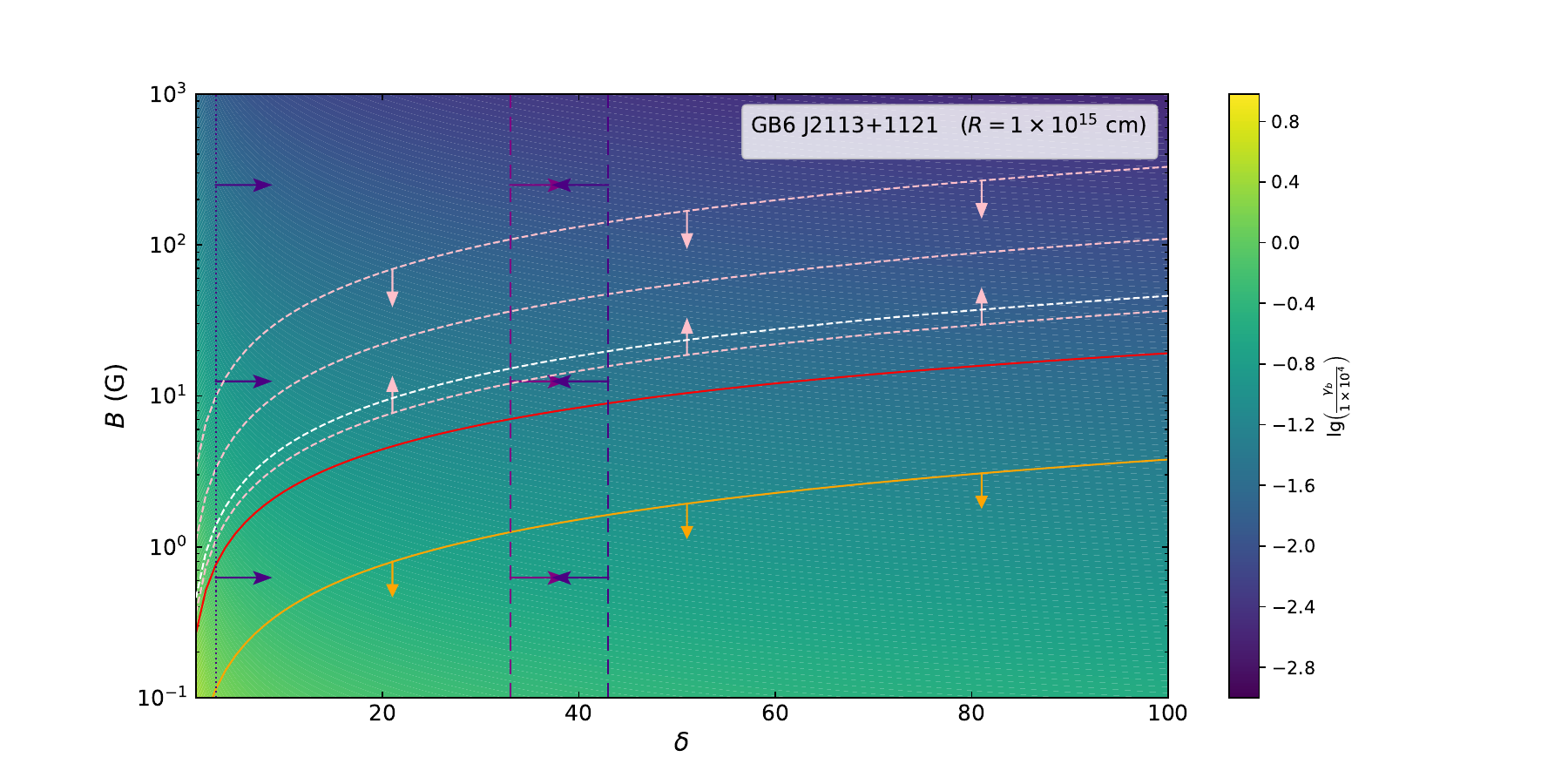}
        \label{GB6 J2113+1121 EC R=1e15 space}
    \end{minipage}

    \begin{minipage}{0.49\linewidth}
        \centering
        \includegraphics[width=\linewidth, trim=10 15 25 10,clip]{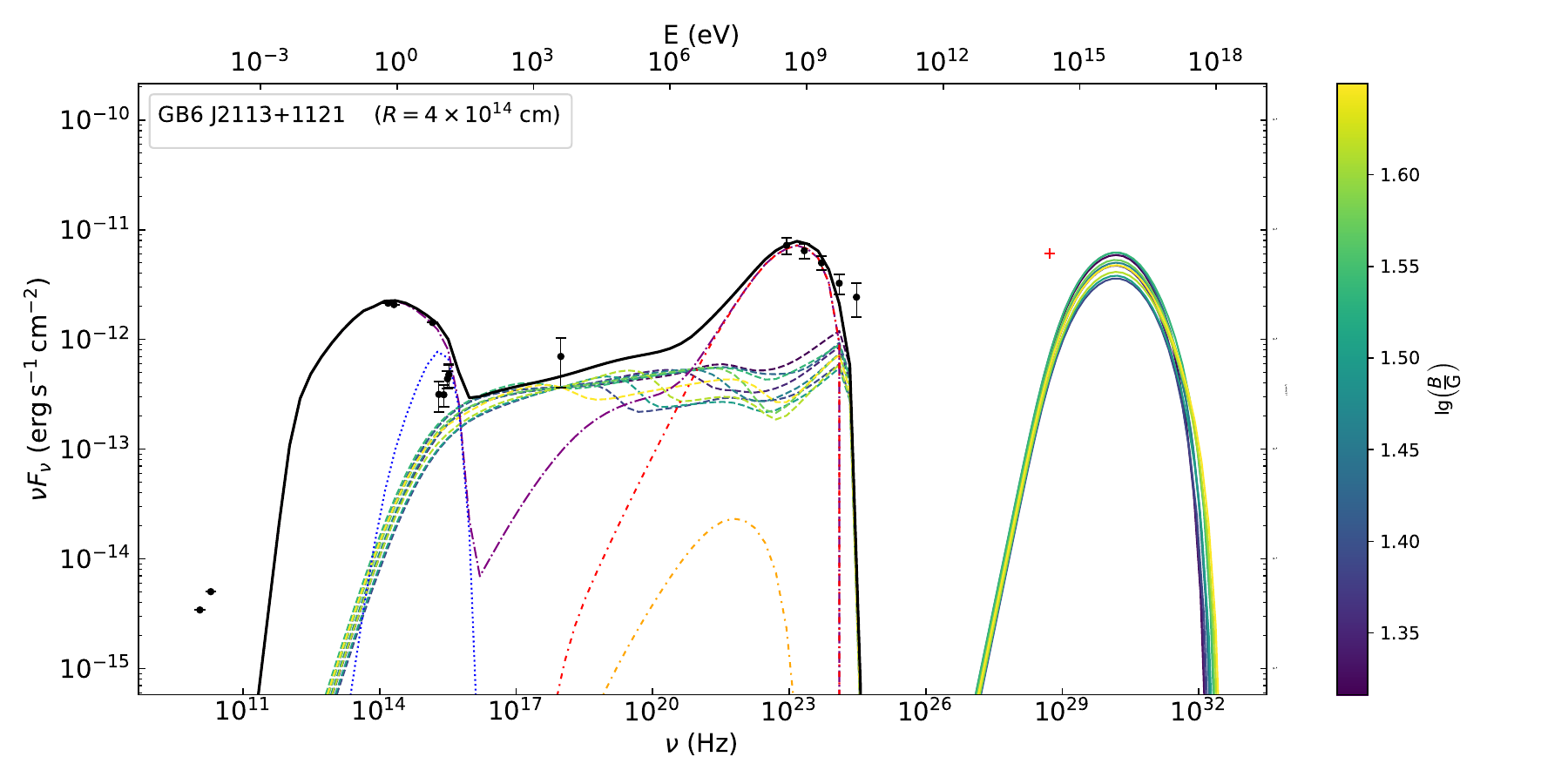}
        \label{GB6 J2113+1121 EC R=4e14}
    \end{minipage}\hspace{-4mm}
    \begin{minipage}{0.49\linewidth}
        \centering
        \includegraphics[width=\linewidth, trim=10 15 25 10,clip]{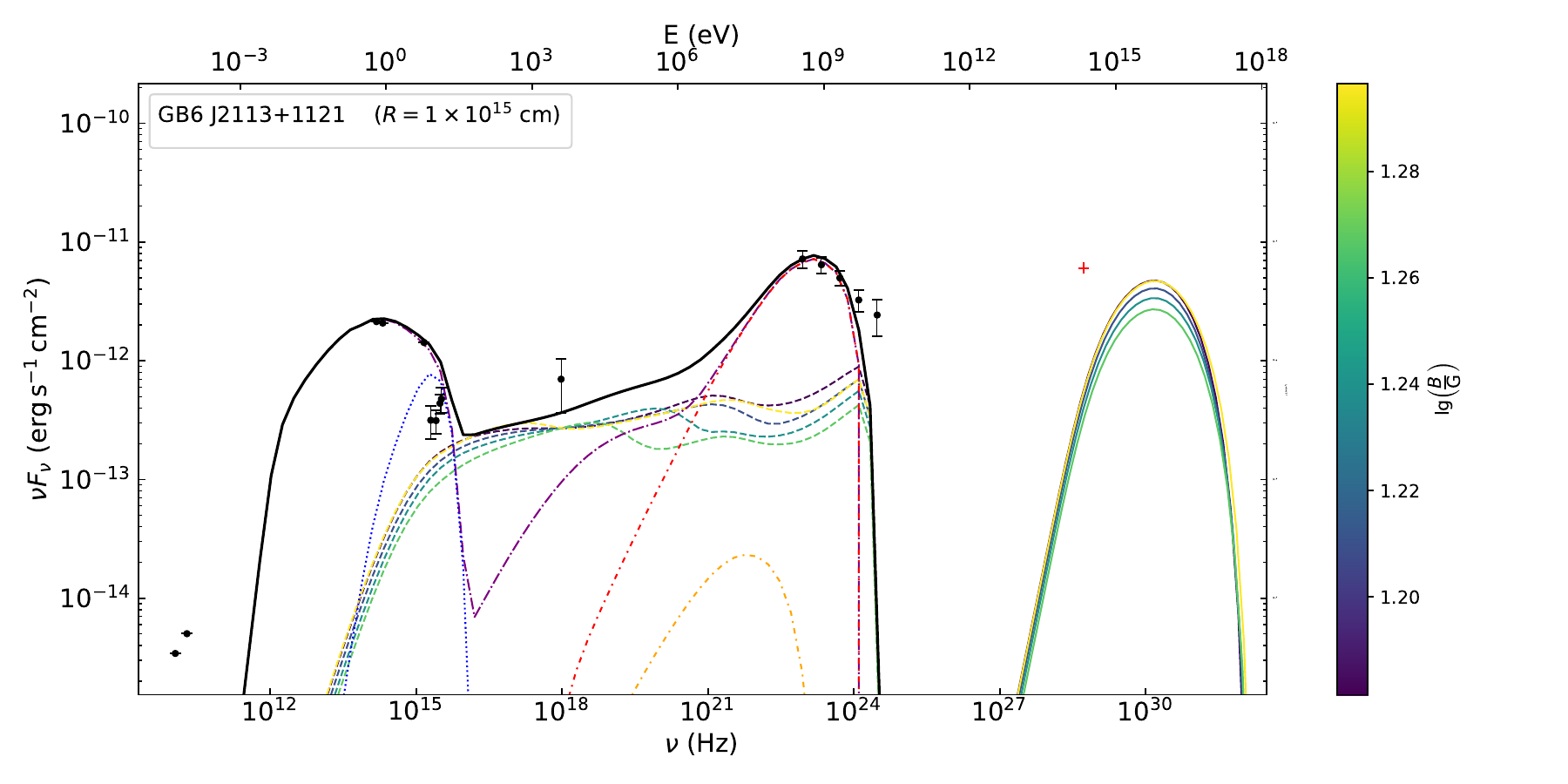}
        \label{GB6 J2113+1121 EC R=1e15}
    \end{minipage}
    \caption{GB6 J2113+1121 associated with IC-191001A. Upper panels: the fitting results of the SED for $R=1\times10^{17}~{\rm cm}$ (left panel) and $R=1\times10^{18}~{\rm cm}$ (right panel) under the SSC-dominated case. Middle panels: the parameter space for $R=4\times10^{14}~{\rm cm}$ (left panel) and $R=1\times10^{15}~{\rm cm}$ (right panel) under the EC-dominated case. Lower panels: the fitting results of the SED for $R=4\times10^{14}~{\rm cm}$ (left panel) and $R=1\times10^{15}~{\rm cm}$ (right panel) under the EC-dominated case. In upper and lower panels, the colored dashed and solid lines respectively represent the secondary pair cascade emission and the neutrino spectrum for different parameter combinations, which correspond to the color bar. The quasi-simultaneous data, neutrino data and other line styles in all panels have the same meaning as in Fig.~\ref{GB6 J2113+1121}.}
    \label{GB6 J2113+1121 appendix}
\end{figure*}

\begin{figure*}[htbp]
    \centering
    \begin{minipage}{0.49\linewidth}
        \centering
        \includegraphics[width=\linewidth, trim=10 15 30 10,clip]{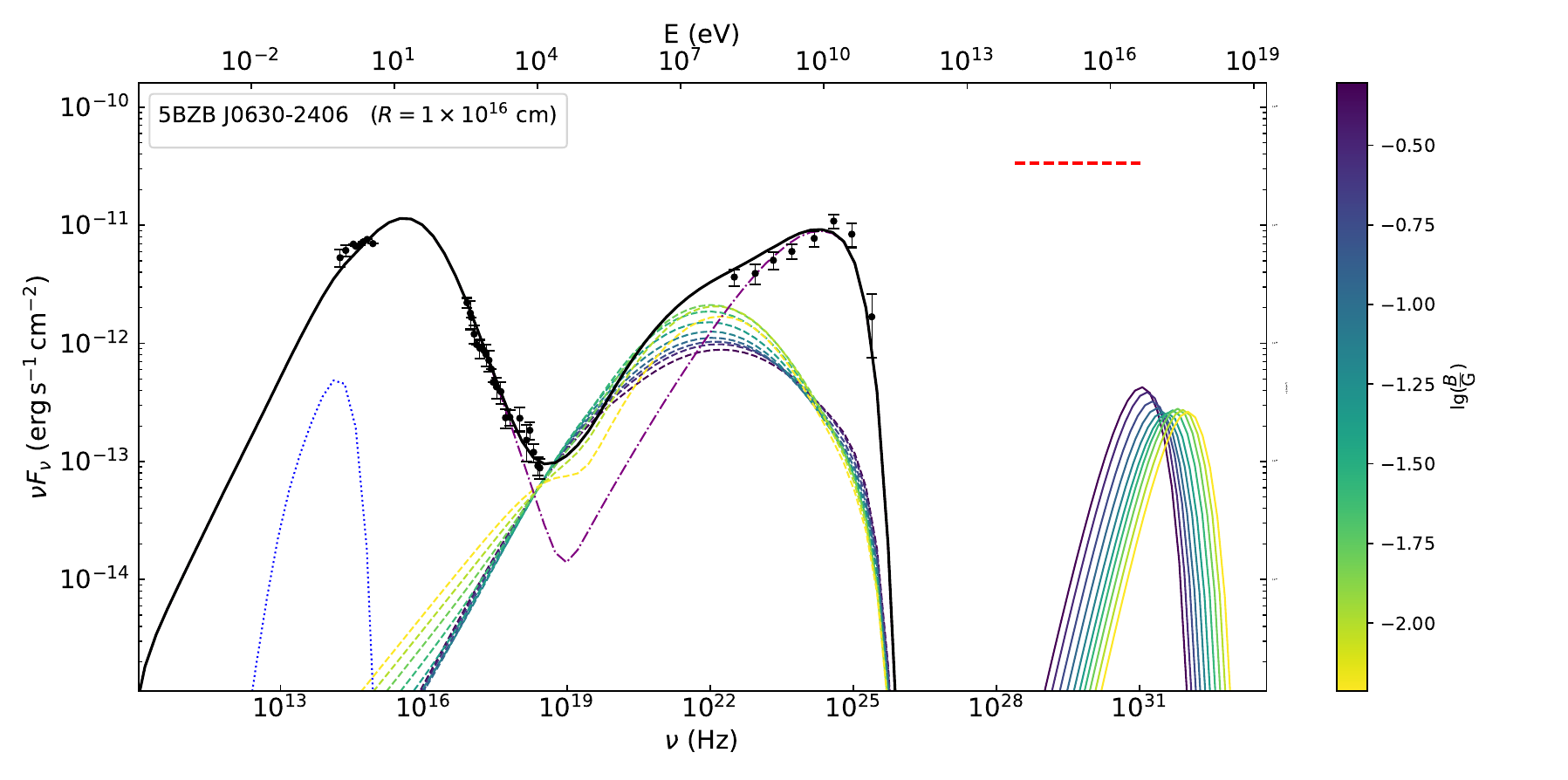}
        \label{5BZB J0630-2406 SSC R=1e16}
    \end{minipage}\hspace{-4mm}
    \begin{minipage}{0.49\linewidth}
        \centering
        \includegraphics[width=\linewidth, trim=10 15 30 10,clip]{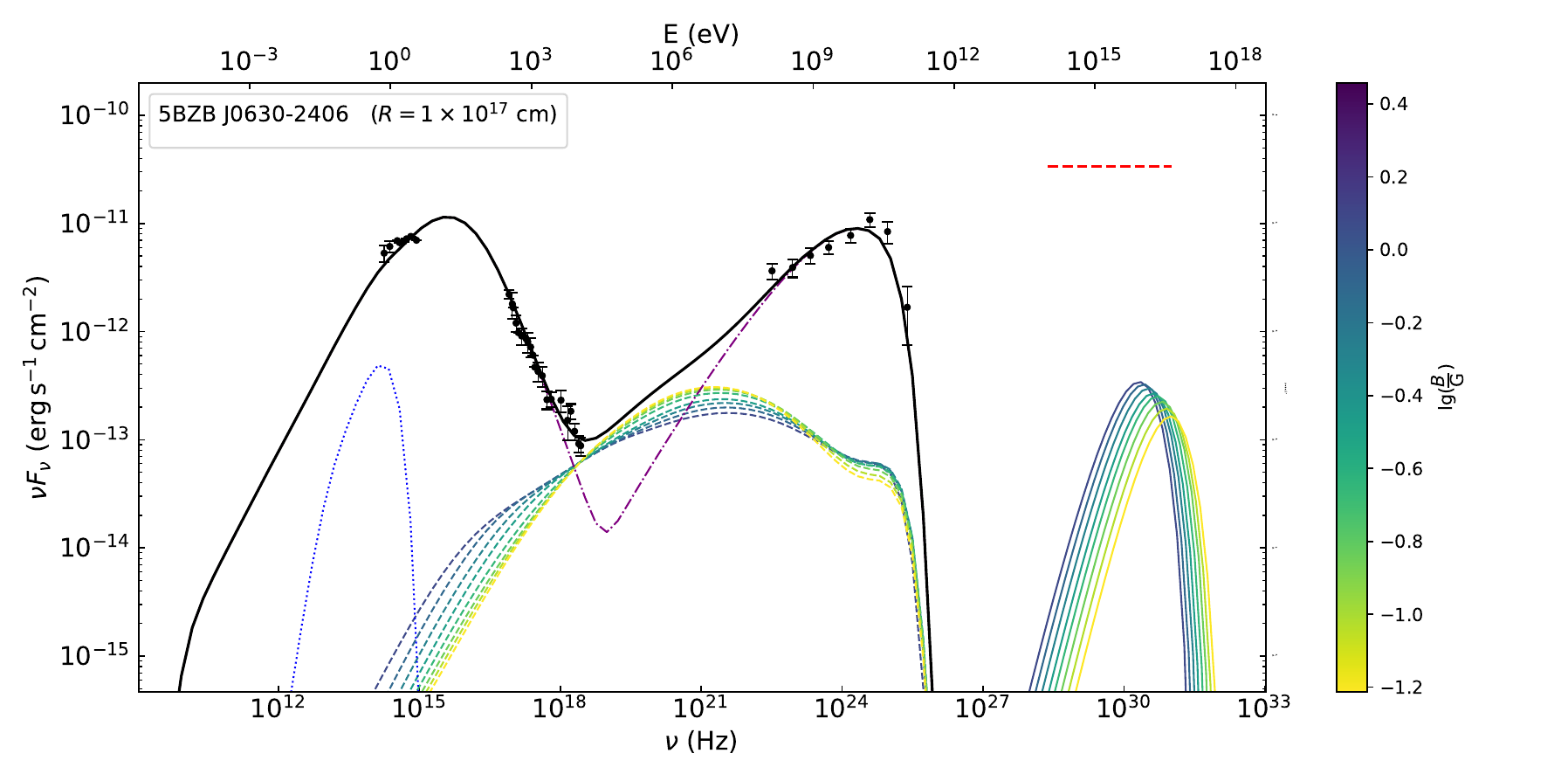}
        \label{5BZB J0630-2406 SSC R=1e17}
    \end{minipage}

    \begin{minipage}{0.49\linewidth}
        \centering
        \includegraphics[width=\linewidth, trim=35 15 90 30,clip]{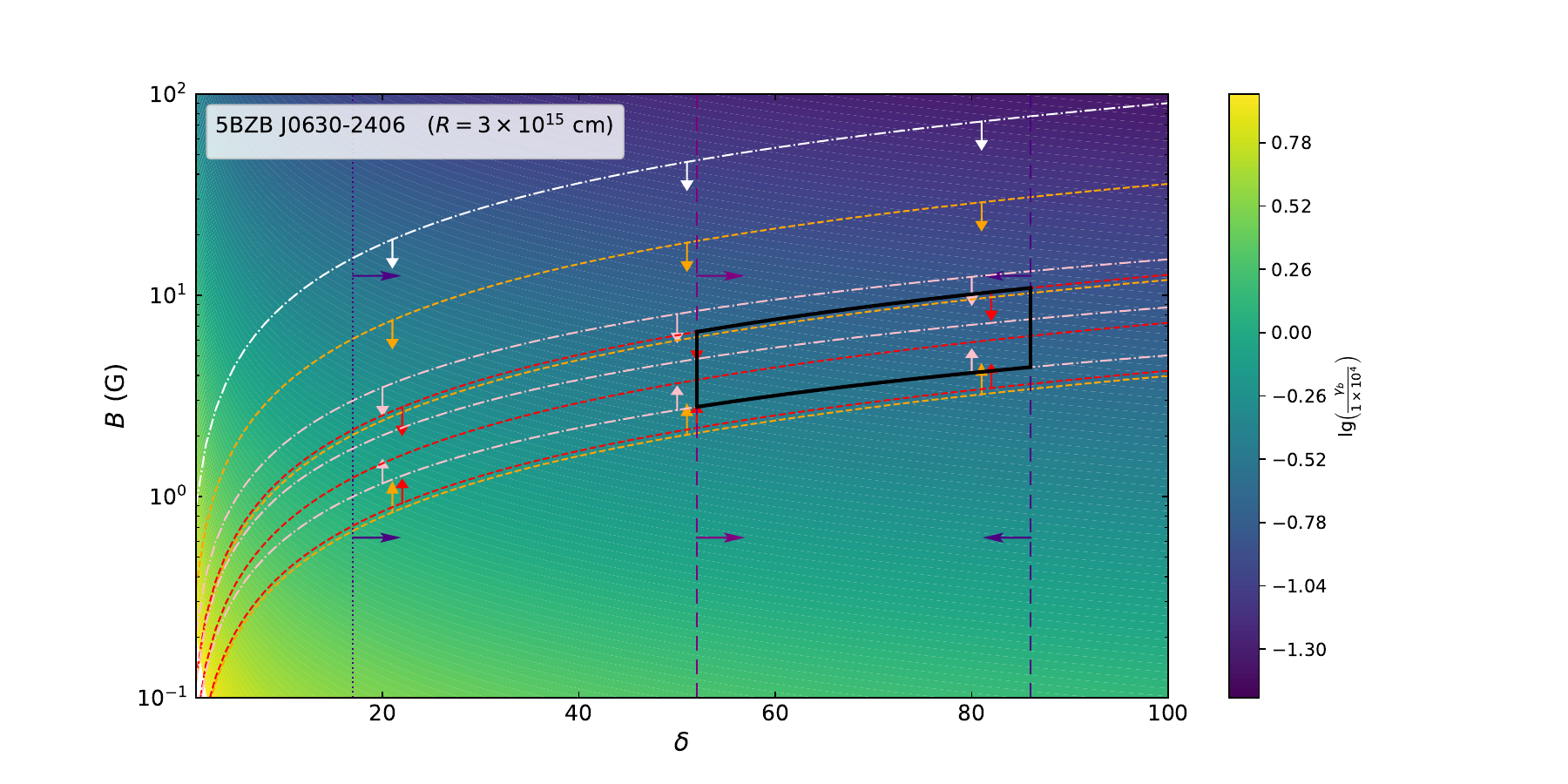}
        \label{5BZB J0630-2406 EC R=3e15 space}
    \end{minipage}\hspace{-4mm}
    \begin{minipage}{0.49\linewidth}
        \centering
        \includegraphics[width=\linewidth, trim=35 15 90 30,clip]{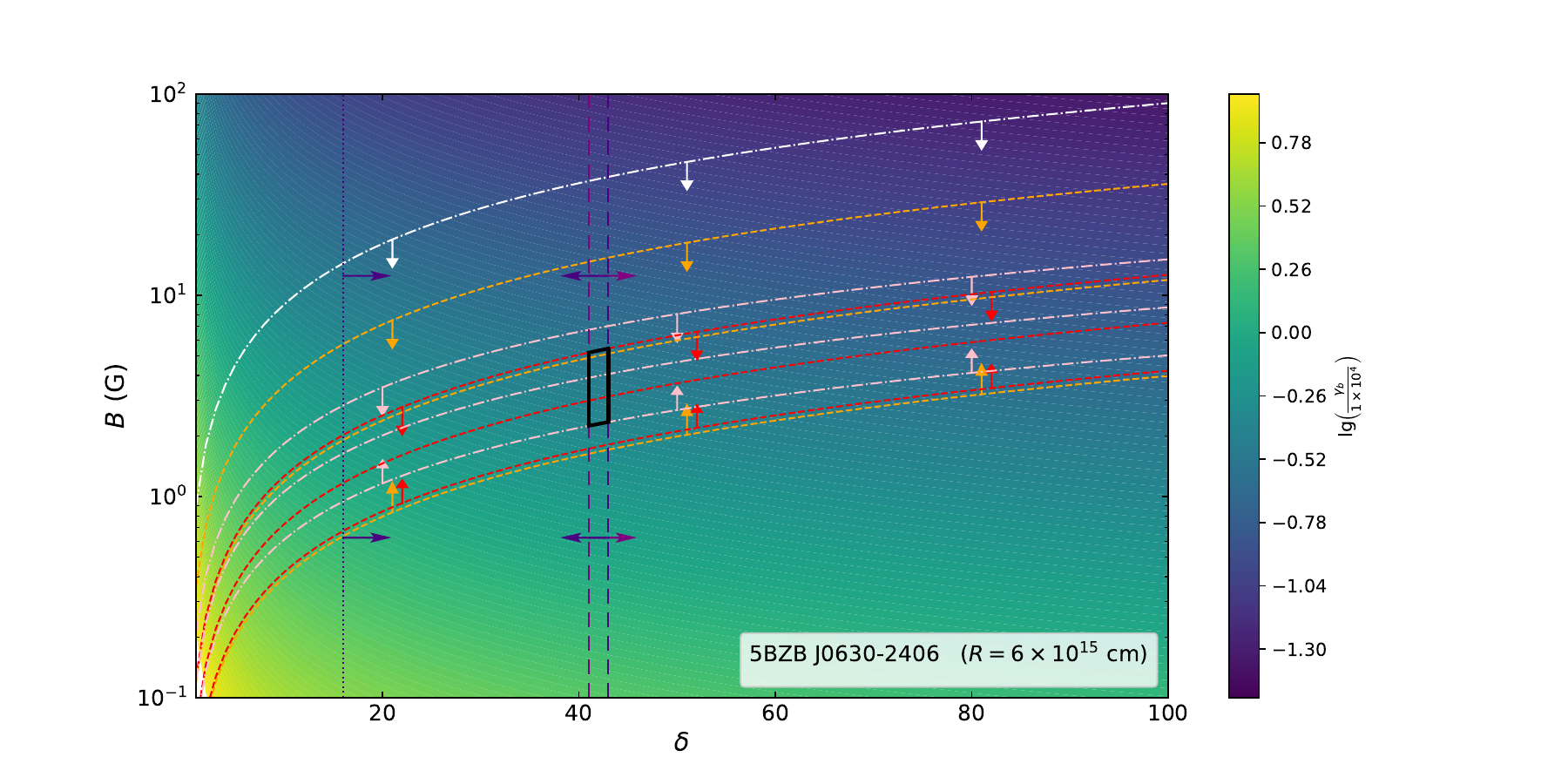}
        \label{5BZB J0630-2406 EC R=6e15 space}
    \end{minipage}

    \begin{minipage}{0.49\linewidth}
        \centering
        \includegraphics[width=\linewidth, trim=10 15 25 10,clip]{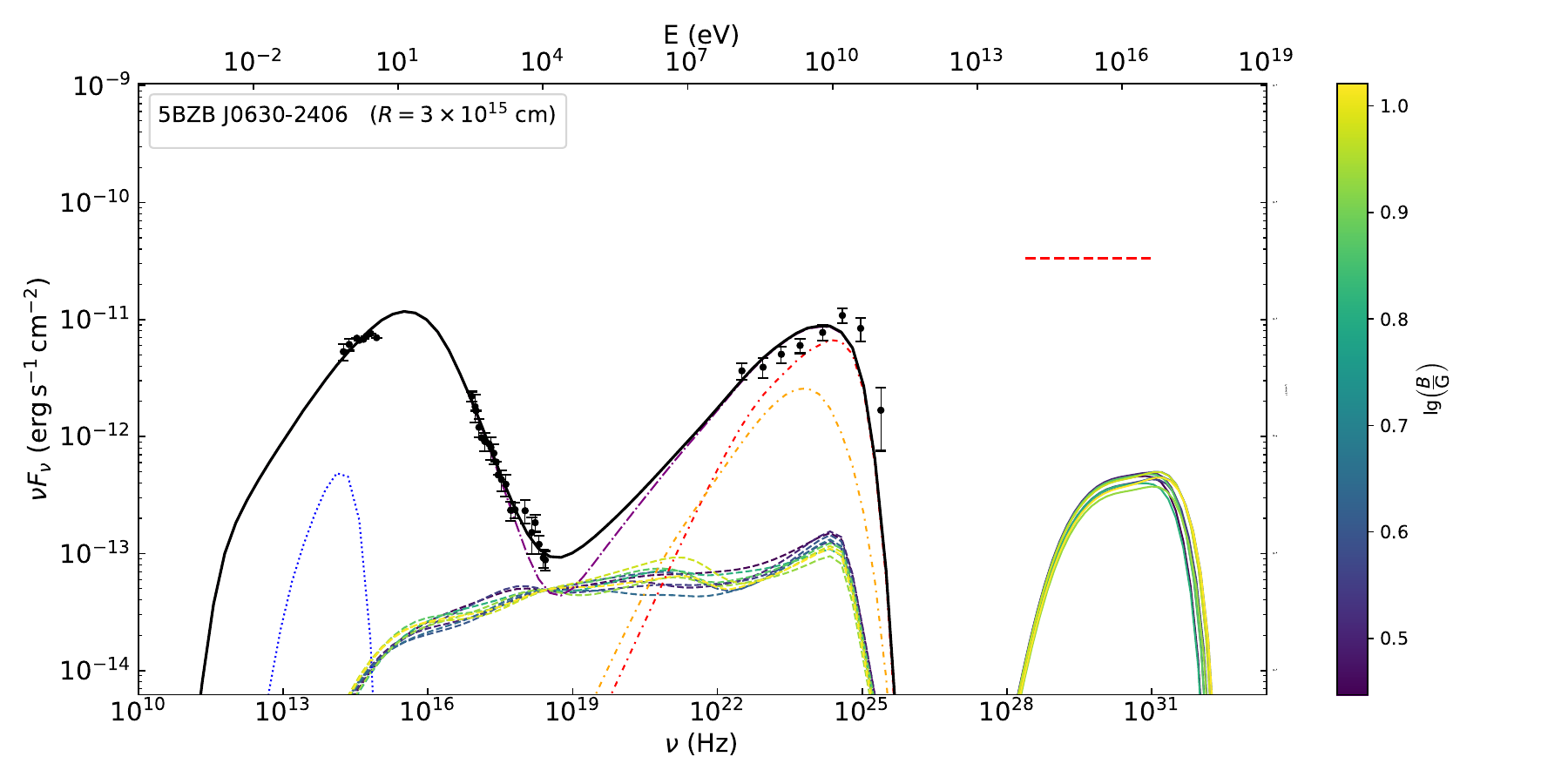}
        \label{5BZB J0630-2406 EC R=3e15}
    \end{minipage}\hspace{-4mm}
    \begin{minipage}{0.49\linewidth}
        \centering
        \includegraphics[width=\linewidth, trim=10 15 25 10,clip]{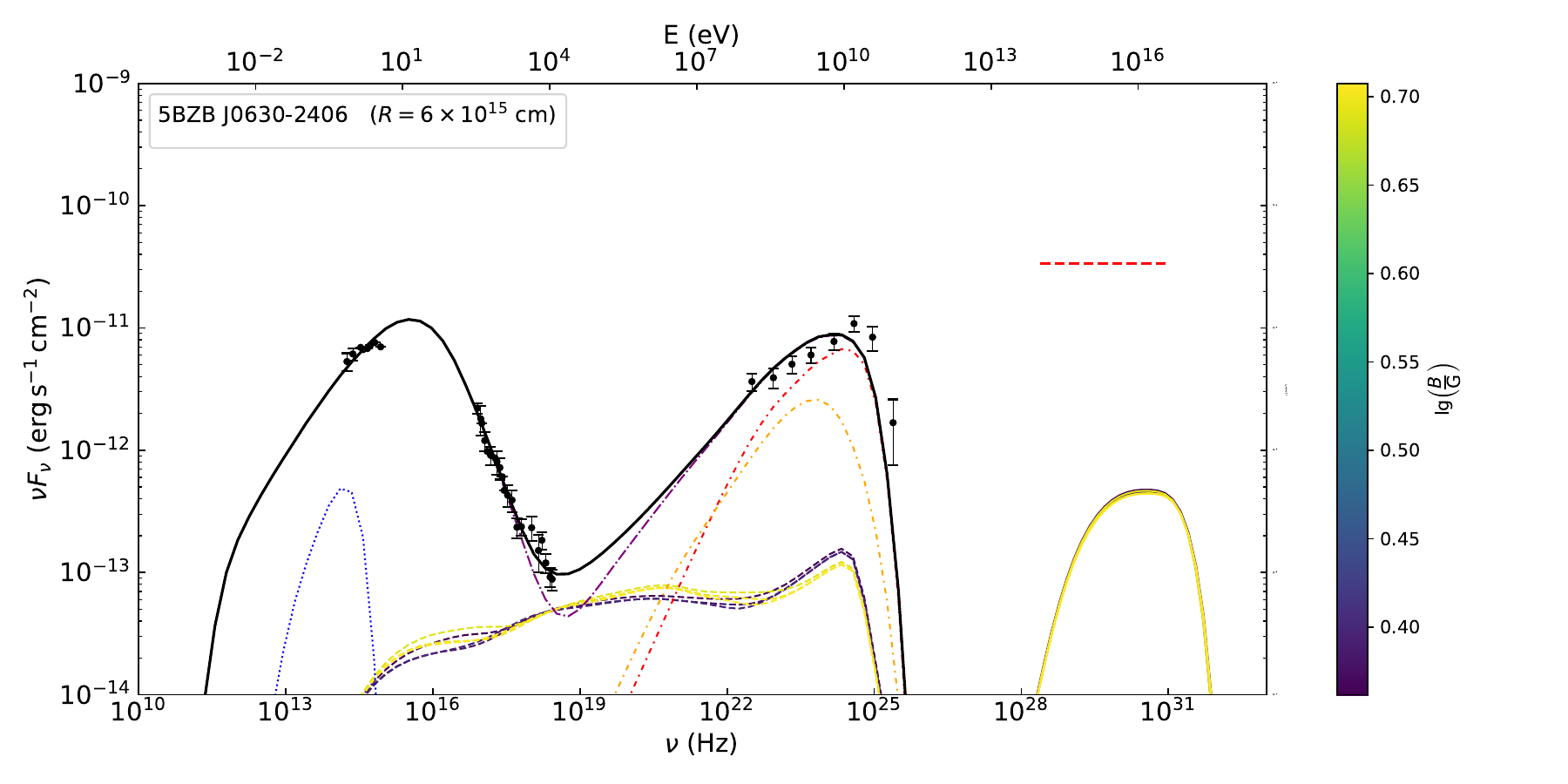}
        \label{5BZB_J0630-2406 EC R=6e15}
    \end{minipage}
    \caption{5BZB J0630-2406 associated with IC J0630-2353. Upper panels: the fitting results of the SED for $R=1\times10^{16}~{\rm cm}$ (left panel) and $R=1\times10^{17}~{\rm cm}$ (right panel) under the SSC-dominated case. Middle panels: the parameter space for $R=3\times10^{15}~{\rm cm}$ (left panel) and $R=6\times10^{15}~{\rm cm}$ (right panel) under the EC-dominated case. Lower panels: the fitting results of the SED for $R=3\times10^{15}~{\rm cm}$ (left panel) and $R=6\times10^{15}~{\rm cm}$ (right panel) under the EC-dominated case. In upper and lower panels, the colored dashed and solid lines respectively represent the secondary pair cascade emission and the neutrino spectrum for different parameter combinations, which correspond to the color bar. The quasi-simultaneous data, neutrino data and other line styles in all panels have the same meaning as in Fig.~\ref{5BZB J0630-2406}.}
    \label{5BZB J0630-2406 appendix}
\end{figure*}

\begin{figure*}[htbp]
    \centering
    \begin{minipage}{0.49\linewidth}
        \centering
        \includegraphics[width=\linewidth, trim=10 15 30 10,clip]{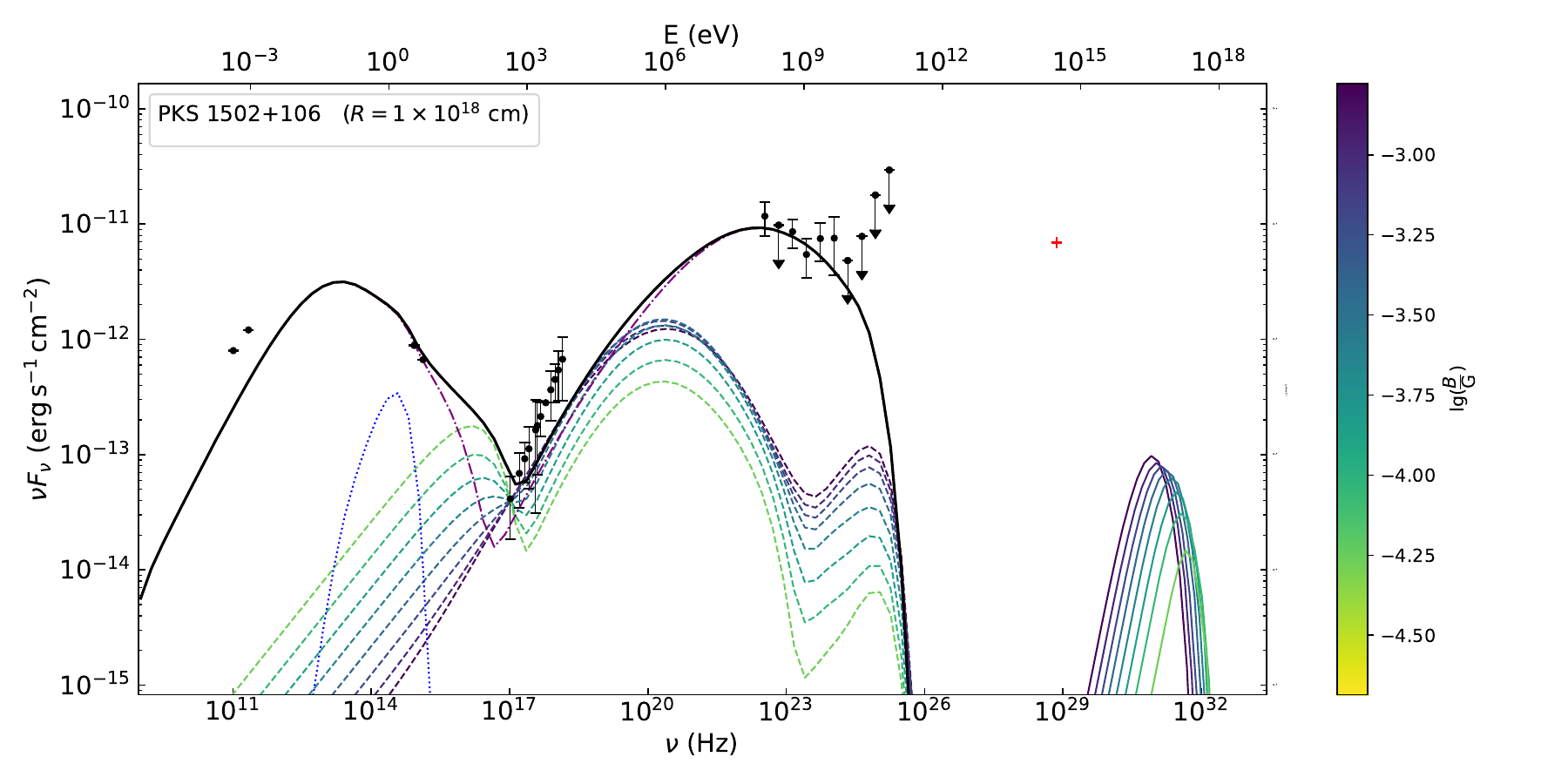}
        \label{PKS 1502+106 SSC R=1e18}
    \end{minipage}\hspace{-4mm}
    \begin{minipage}{0.49\linewidth}
        \centering
        \includegraphics[width=\linewidth, trim=10 15 30 10,clip]{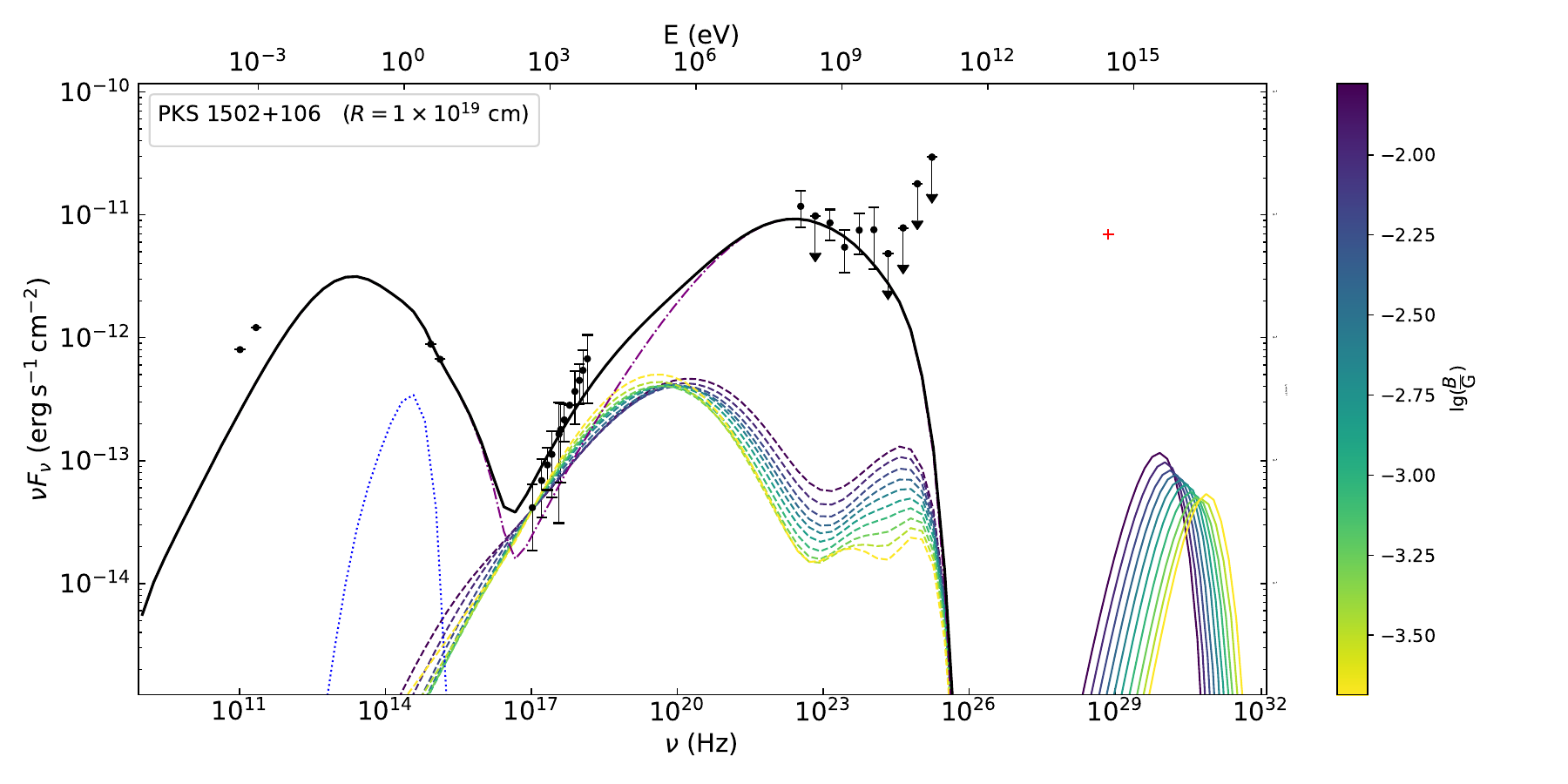}
        \label{PKS 1502+106 SSC R=1e19}
    \end{minipage}

    \begin{minipage}{0.49\linewidth}
        \centering
        \includegraphics[width=\linewidth, trim=35 15 90 30,clip]{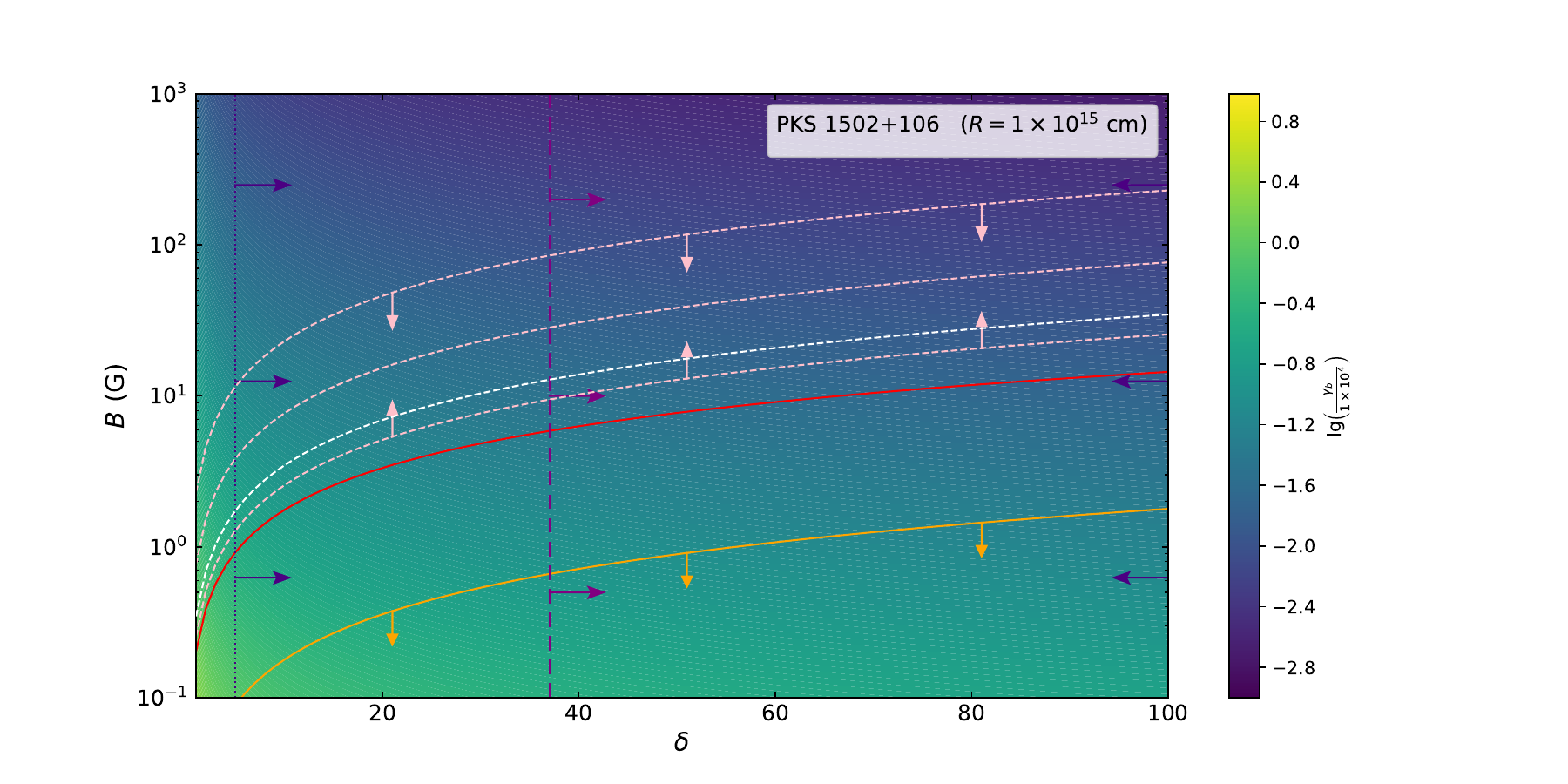}
        \label{PKS 1502+106 EC R=1e15 space}
    \end{minipage}\hspace{-4mm}
    \begin{minipage}{0.49\linewidth}
        \centering
        \includegraphics[width=\linewidth, trim=35 15 90 30,clip]{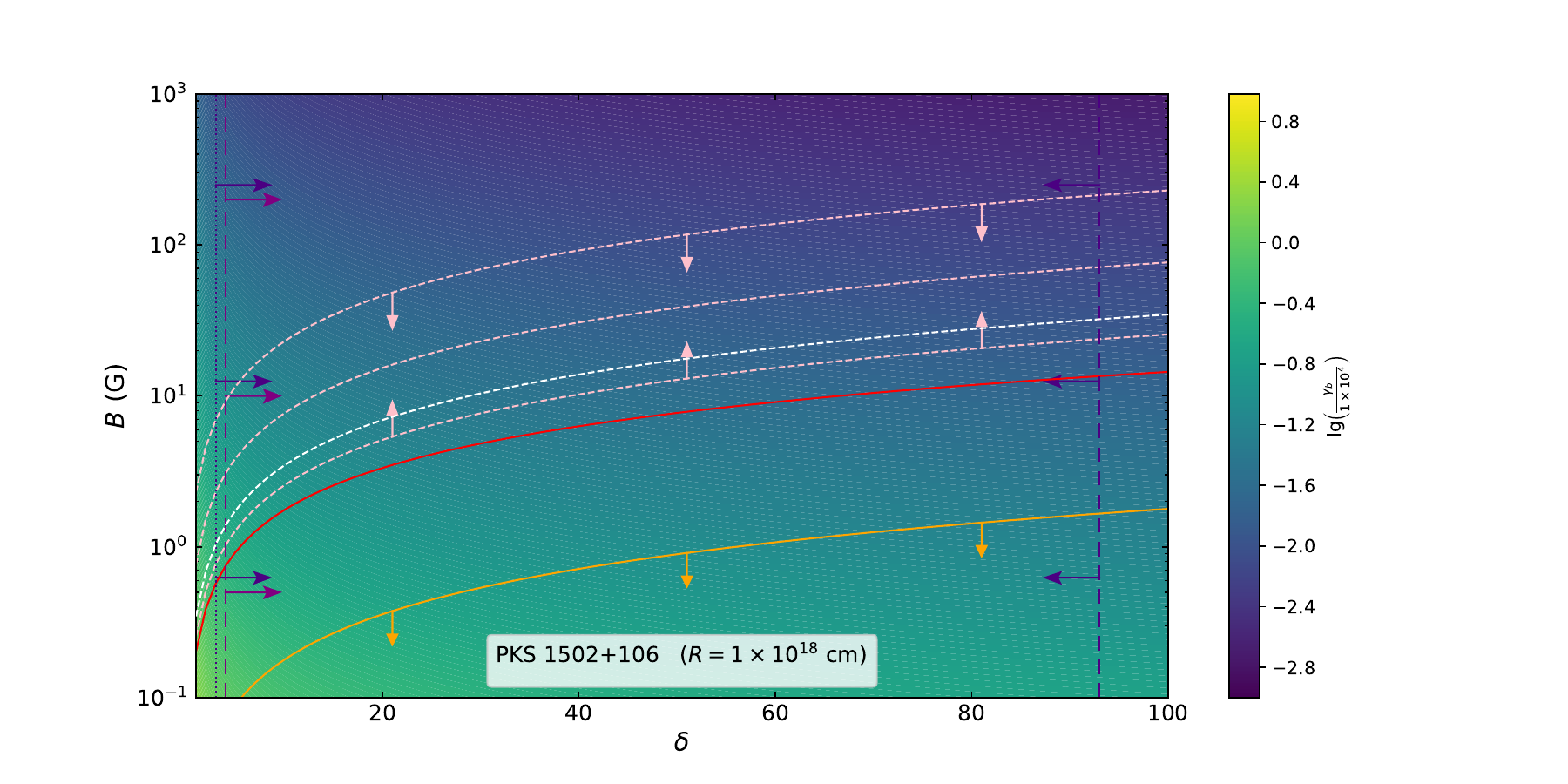}
        \label{PKS 1502+106 EC R=1e18 space}
    \end{minipage}

    \begin{minipage}{0.49\linewidth}
        \centering
        \includegraphics[width=\linewidth, trim=10 15 25 10,clip]{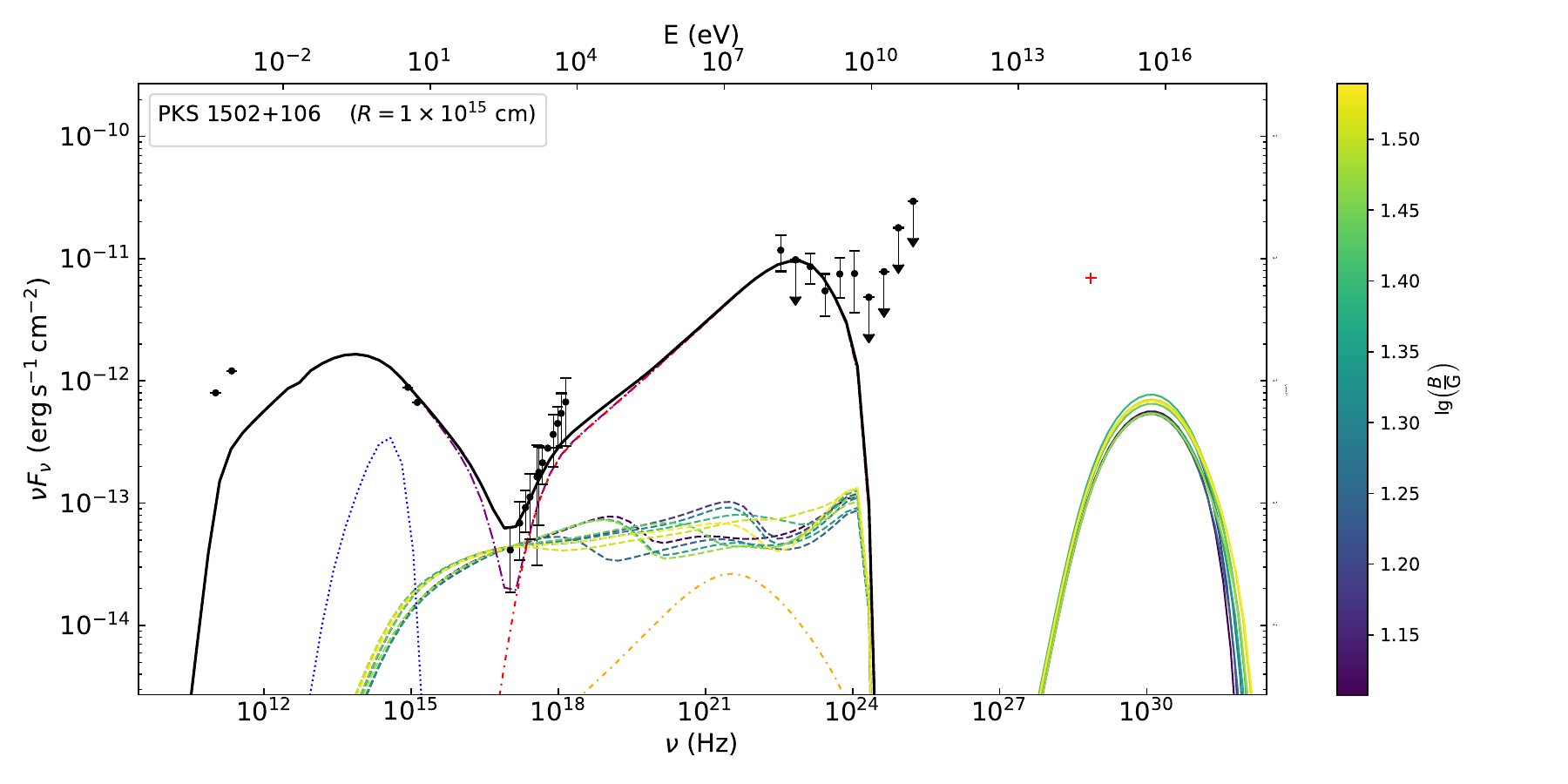}
        \label{PKS 1502+106 EC R=1e15}
    \end{minipage}\hspace{-4mm}
    \begin{minipage}{0.49\linewidth}
        \centering
        \includegraphics[width=\linewidth, trim=10 15 25 10,clip]{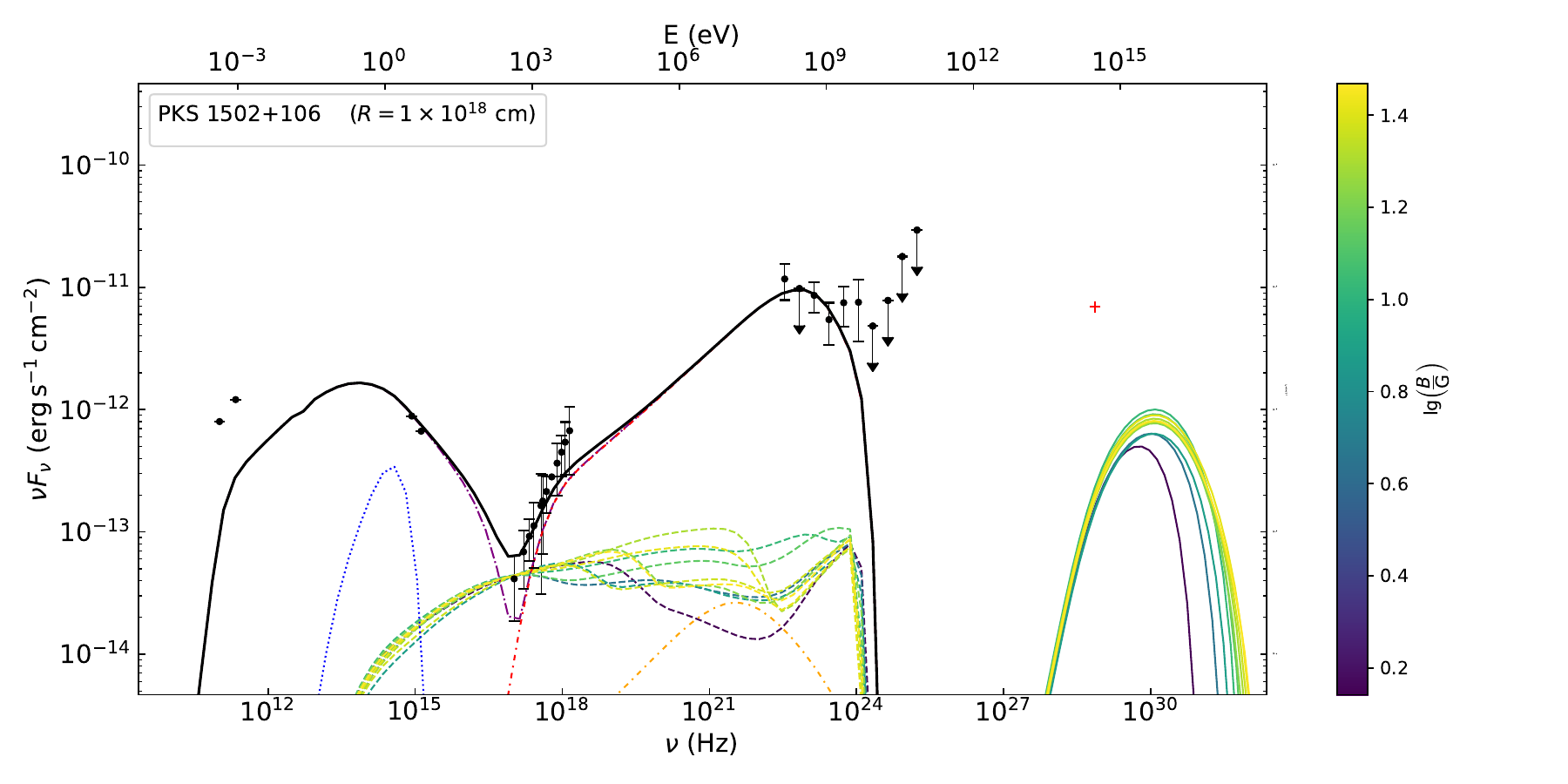}
        \label{PKS 1502+106 EC R=1e18}
    \end{minipage}
    \caption{PKS 1502+106 associated with IC-190730A. Upper panels: the fitting results of the SED for $R=1\times10^{18}~{\rm cm}$ (left panel) and $R=1\times10^{19}~{\rm cm}$ (right panel) under the SSC-dominated case. Middle panels: the parameter space for $R=1\times10^{15}~{\rm cm}$ (left panel) and $R=1\times10^{18}~{\rm cm}$ (right panel) under the EC-dominated case. Lower panels: the fitting results of the SED for $R=1\times10^{15}~{\rm cm}$ (left panel) and $R=1\times10^{18}~{\rm cm}$ (right panel) under the EC-dominated case. In upper and lower panels, the colored dashed and solid lines respectively represent the secondary pair cascade emission and the neutrino spectrum for different parameter combinations, which correspond to the color bar. The quasi-simultaneous data, neutrino data and other line styles in all panels have the same meaning as in Fig.~\ref{PKS 1502+106}.}
    \label{PKS 1502+106 appendix}
\end{figure*}

\begin{figure*}[htbp]
    \centering
    \begin{minipage}{0.49\linewidth}
        \centering
        \includegraphics[width=\linewidth, trim=10 15 30 10,clip]{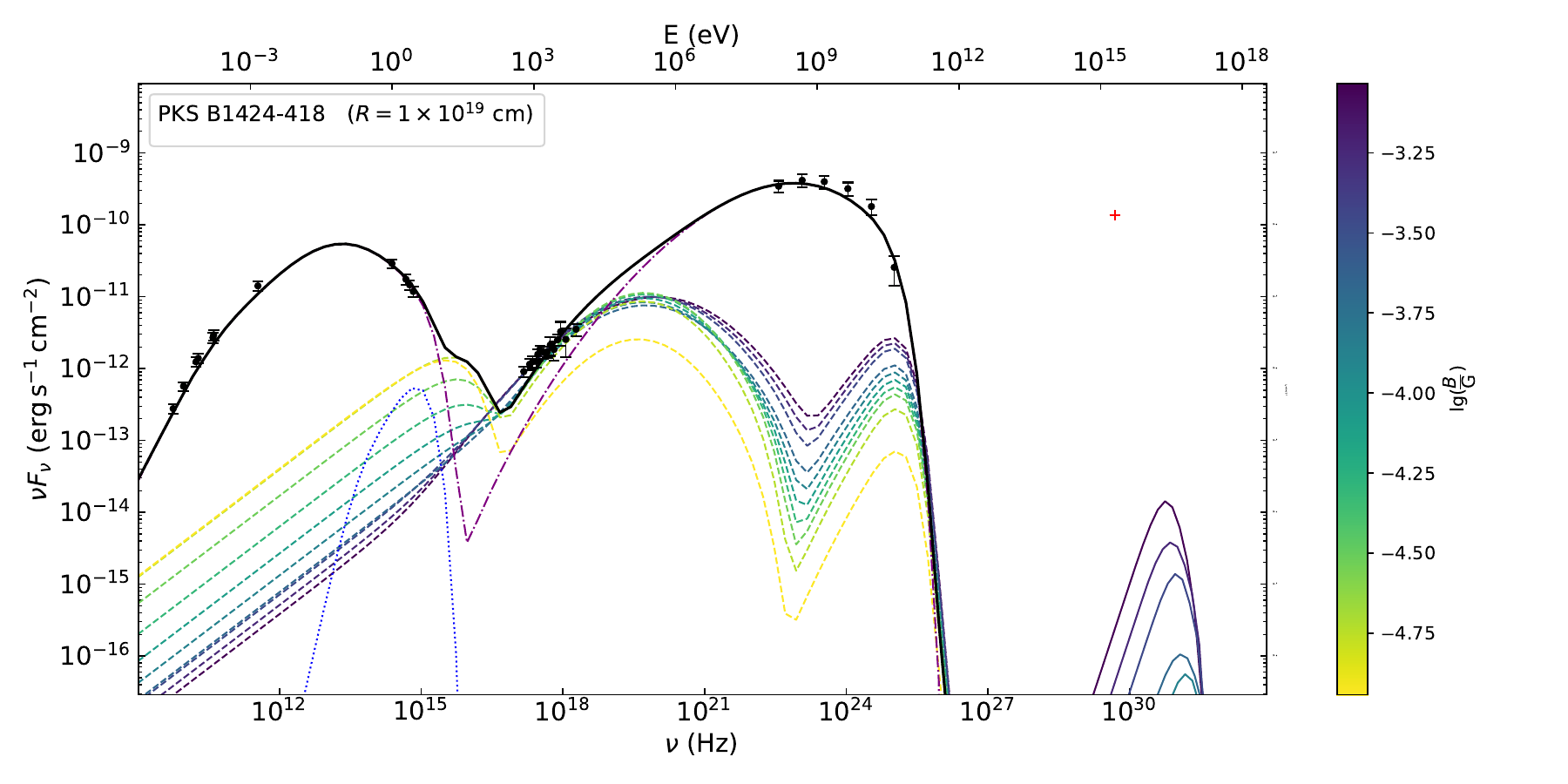}
        \label{PKS B1424-418 SSC R=1e19}
    \end{minipage}\hspace{-4mm}
    \begin{minipage}{0.49\linewidth}
        \centering
        \includegraphics[width=\linewidth, trim=10 15 30 10,clip]{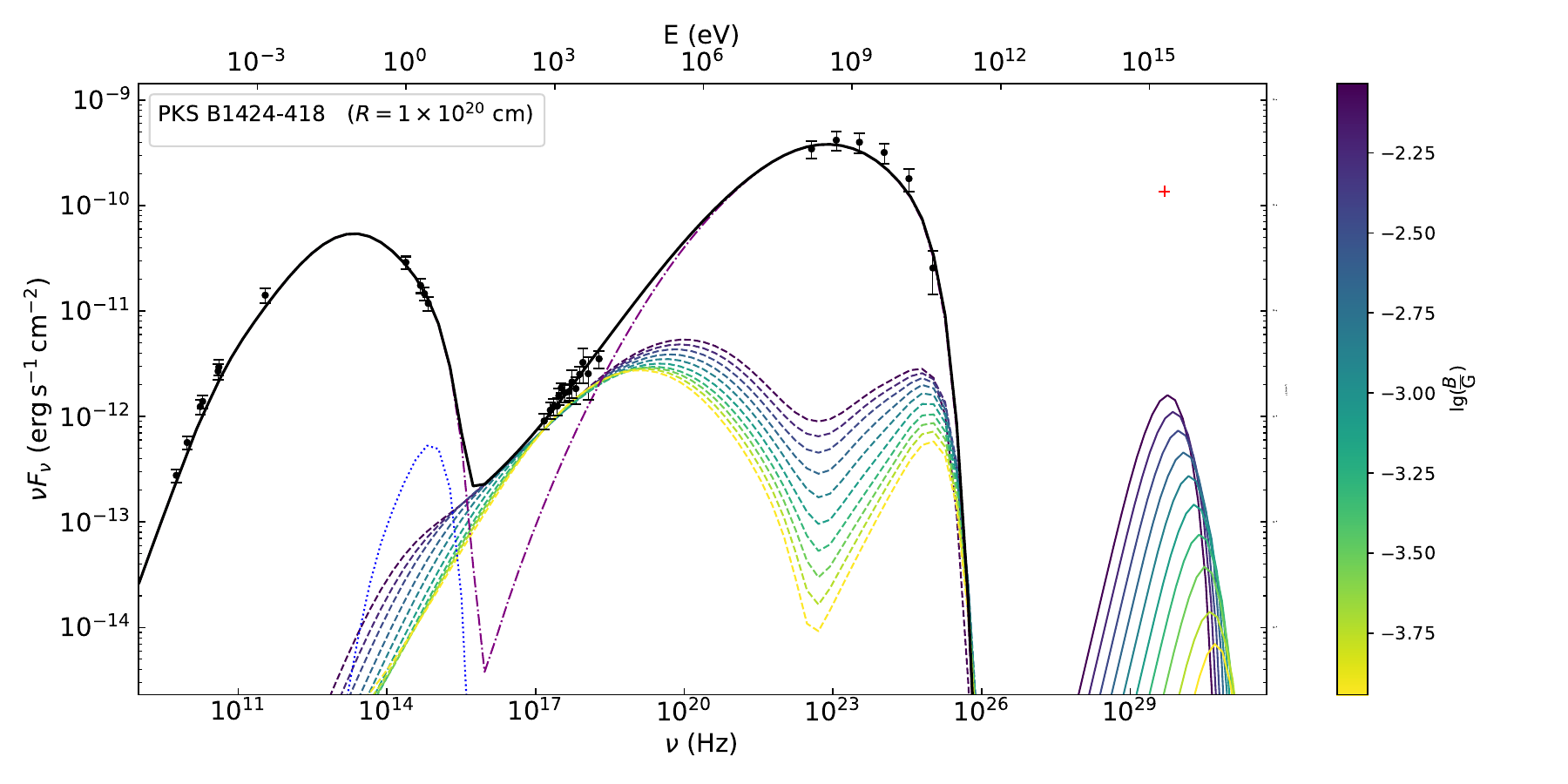}
        \label{PKS B1424-418 SSC R=1e20}
    \end{minipage}

    \begin{minipage}{0.49\linewidth}
        \centering
        \includegraphics[width=\linewidth, trim=35 15 90 30,clip]{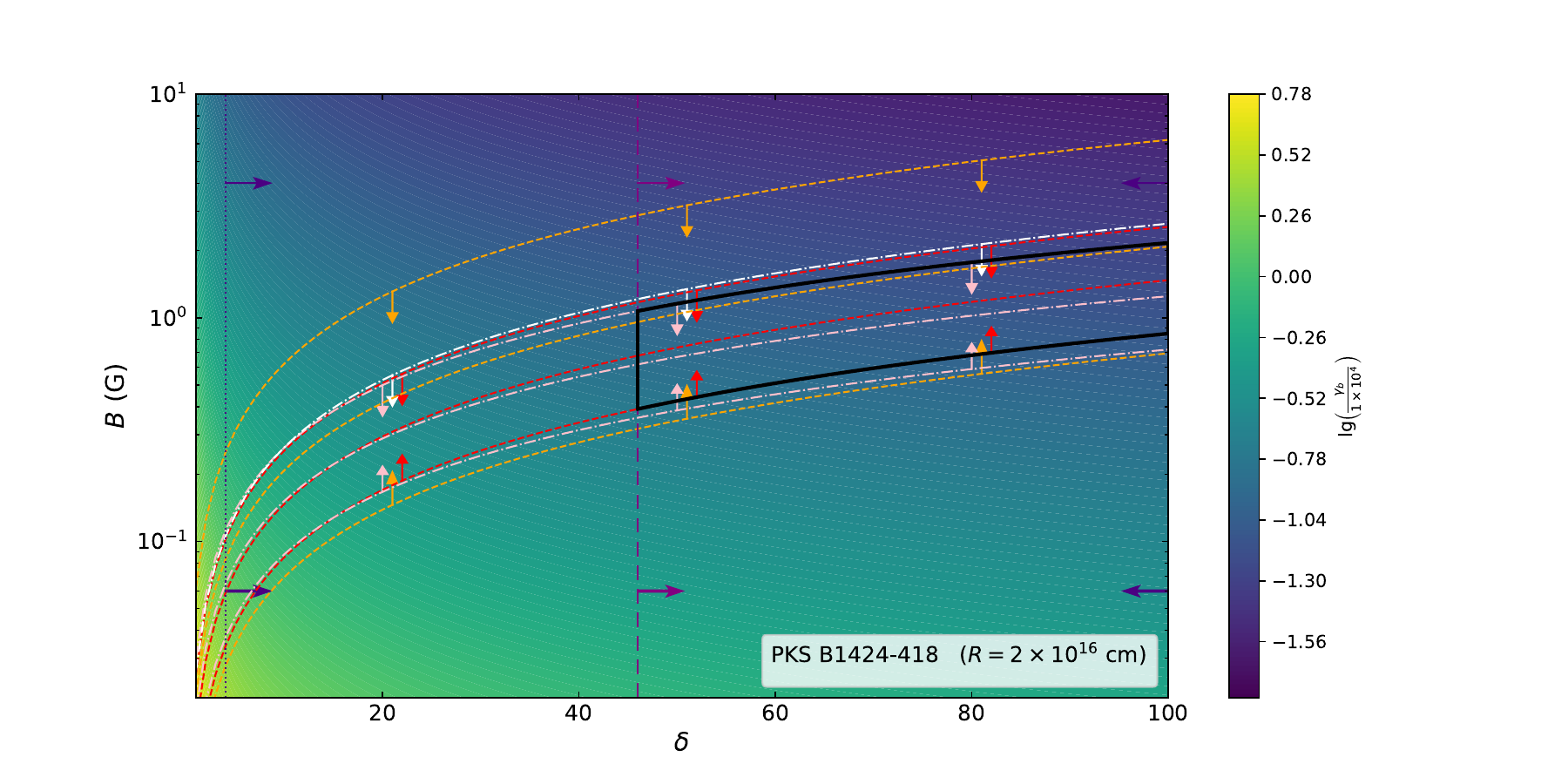}
        \label{PKS B1424-418 EC R=2e16 space}
    \end{minipage}\hspace{-4mm}
    \begin{minipage}{0.49\linewidth}
        \centering
        \includegraphics[width=\linewidth, trim=35 15 90 30,clip]{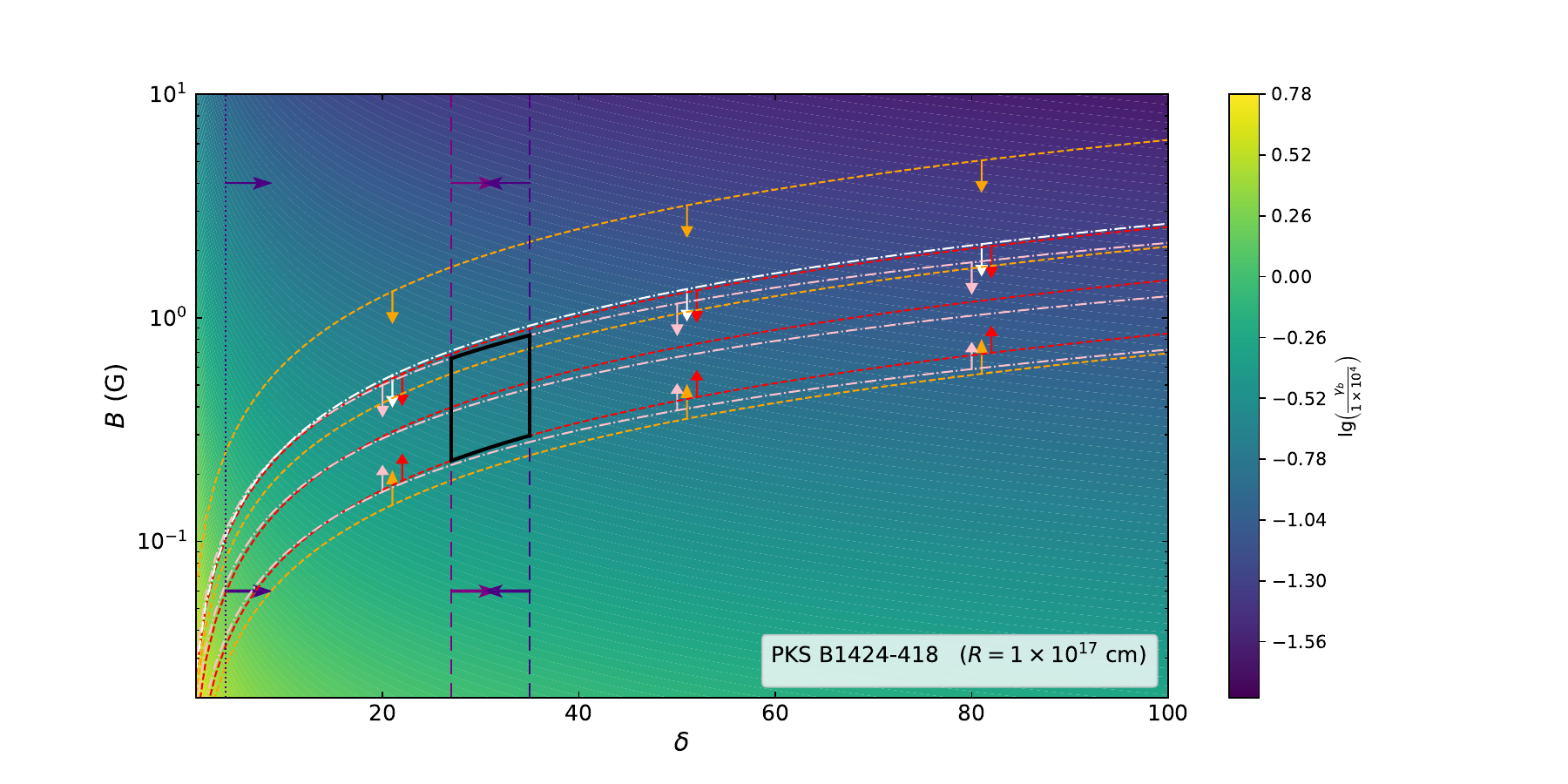}
        \label{PKS B1424-418 EC R=1e17 space}
    \end{minipage}

    \begin{minipage}{0.49\linewidth}
        \centering
        \includegraphics[width=\linewidth, trim=10 15 25 10,clip]{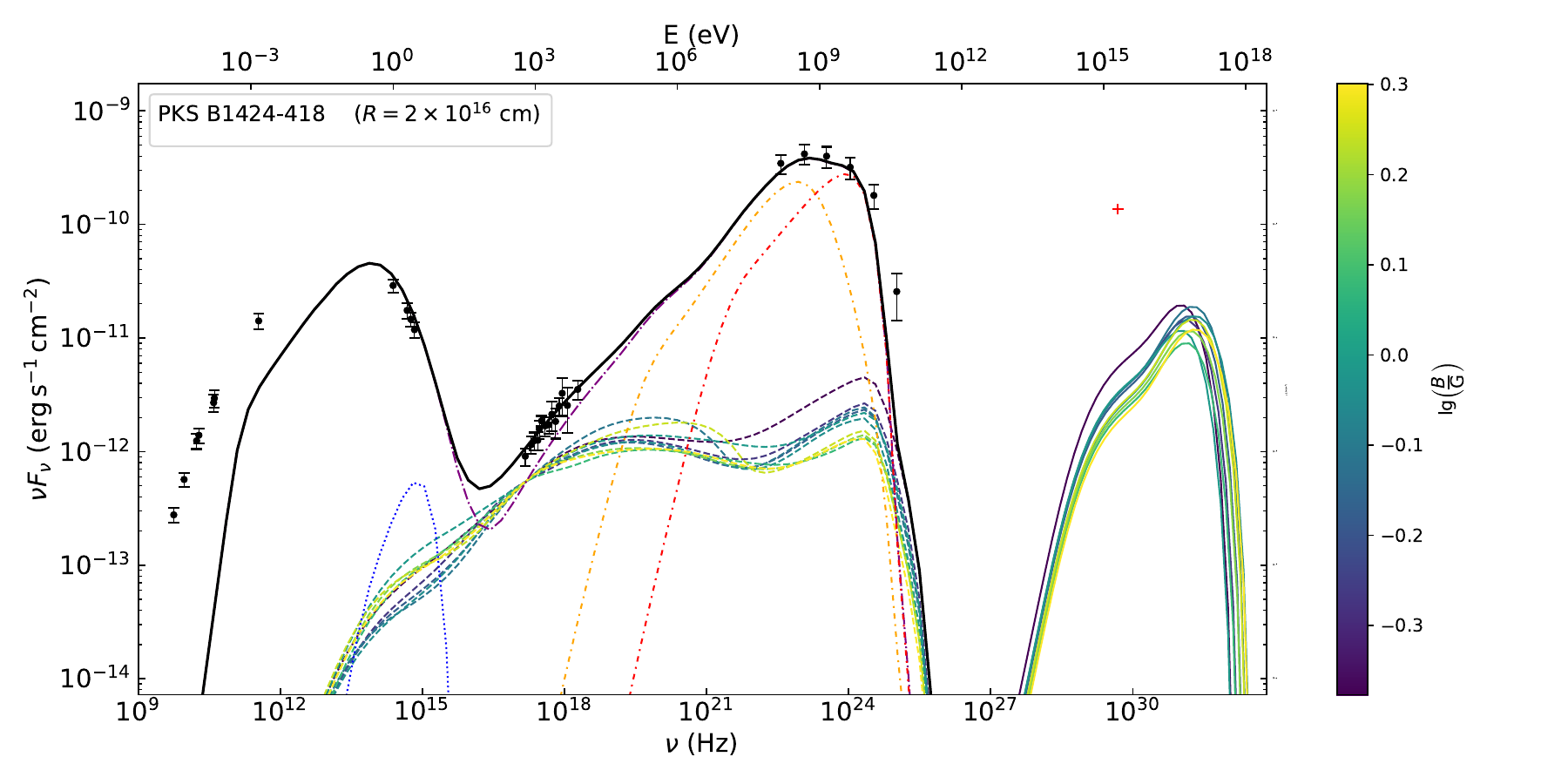}
        \label{PKS B1424-418 EC R=2e16}
    \end{minipage}\hspace{-4mm}
    \begin{minipage}{0.49\linewidth}
        \centering
        \includegraphics[width=\linewidth, trim=10 15 25 10,clip]{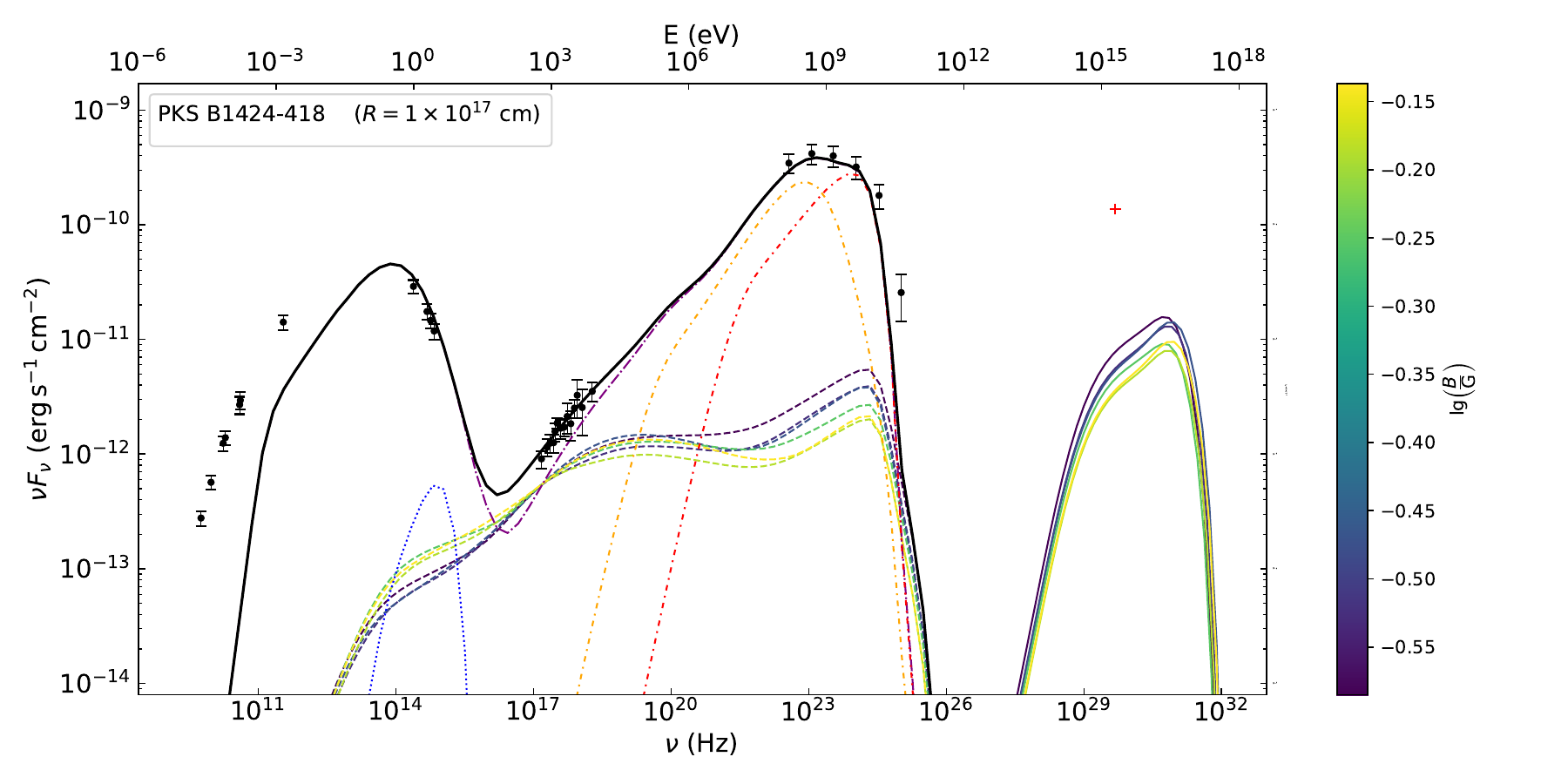}
        \label{PKS B1424-418 EC R=1e17}
    \end{minipage}
    \caption{PKS B1424-418 associated with Big Bird. Upper panels: the fitting results of the SED for $R=1\times10^{19}~{\rm cm}$ (left panel) and $R=1\times10^{20}~{\rm cm}$ (right panel) under the SSC-dominated case. Middle panels: the parameter space for $R=2\times10^{16}~{\rm cm}$ (left panel) and $R=1\times10^{17}~{\rm cm}$ (right panel) under the EC-dominated case. Lower panels: the fitting results of the SED for $R=2\times10^{16}~{\rm cm}$ (left panel) and $R=1\times10^{17}~{\rm cm}$ (right panel) under the EC-dominated case. In upper and lower panels, the colored dashed and solid lines respectively represent the secondary pair cascade emission and the neutrino spectrum for different parameter combinations, which correspond to the color bar. The quasi-simultaneous data, neutrino data and other line styles in all panels have the same meaning as in Fig.~\ref{PKS B1424-418}.}
    \label{PKS B1424-418 appendix}
\end{figure*}

\begin{figure*}[htbp]
    \centering
    \begin{minipage}{0.49\linewidth}
        \centering
        \includegraphics[width=\linewidth, trim=10 15 30 10,clip]{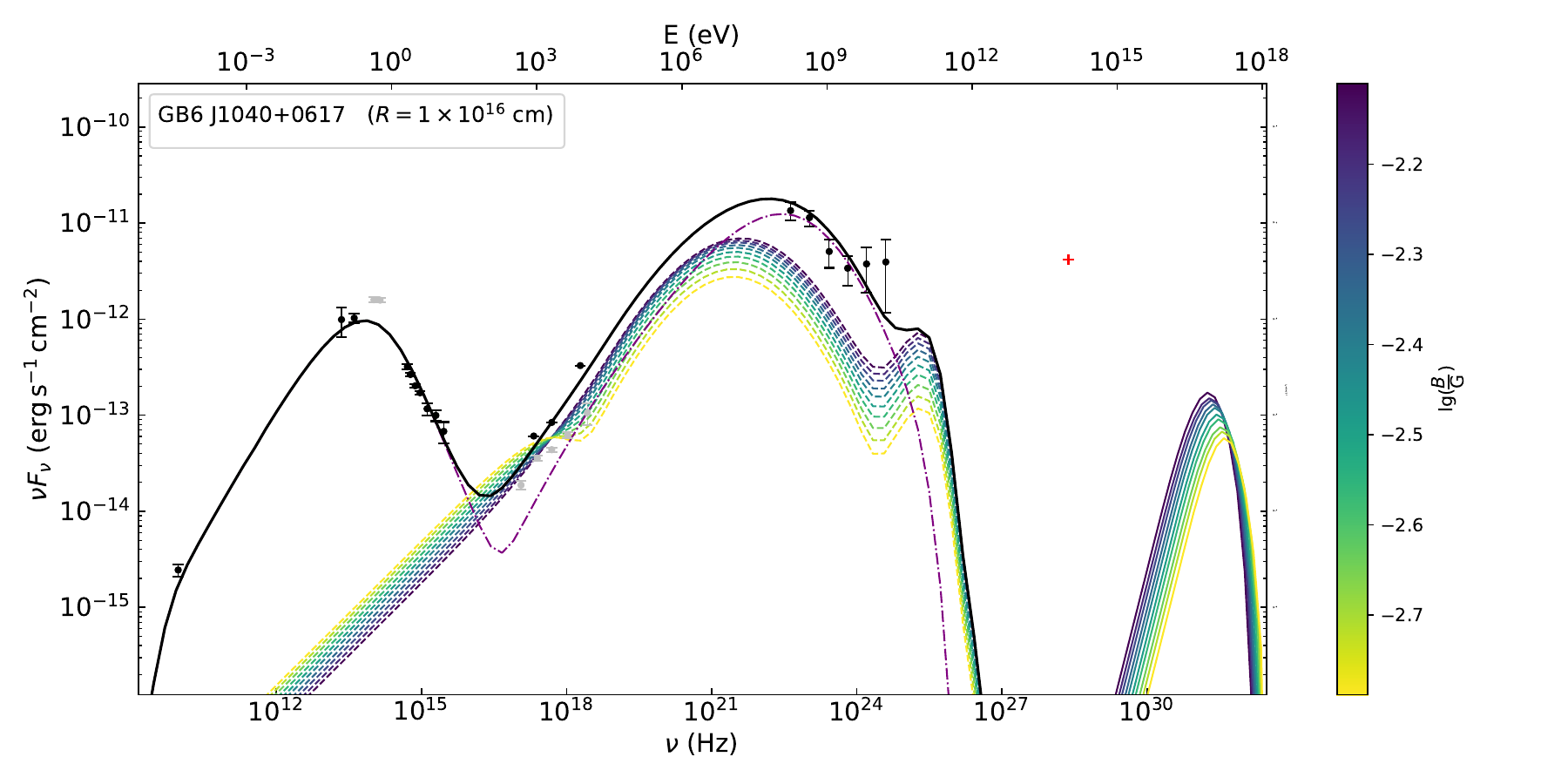}
        \label{GB6 J1040+0617 SSC R=1e16}
    \end{minipage}\hspace{-4mm}
    \begin{minipage}{0.49\linewidth}
        \centering
        \includegraphics[width=\linewidth, trim=10 15 30 10,clip]{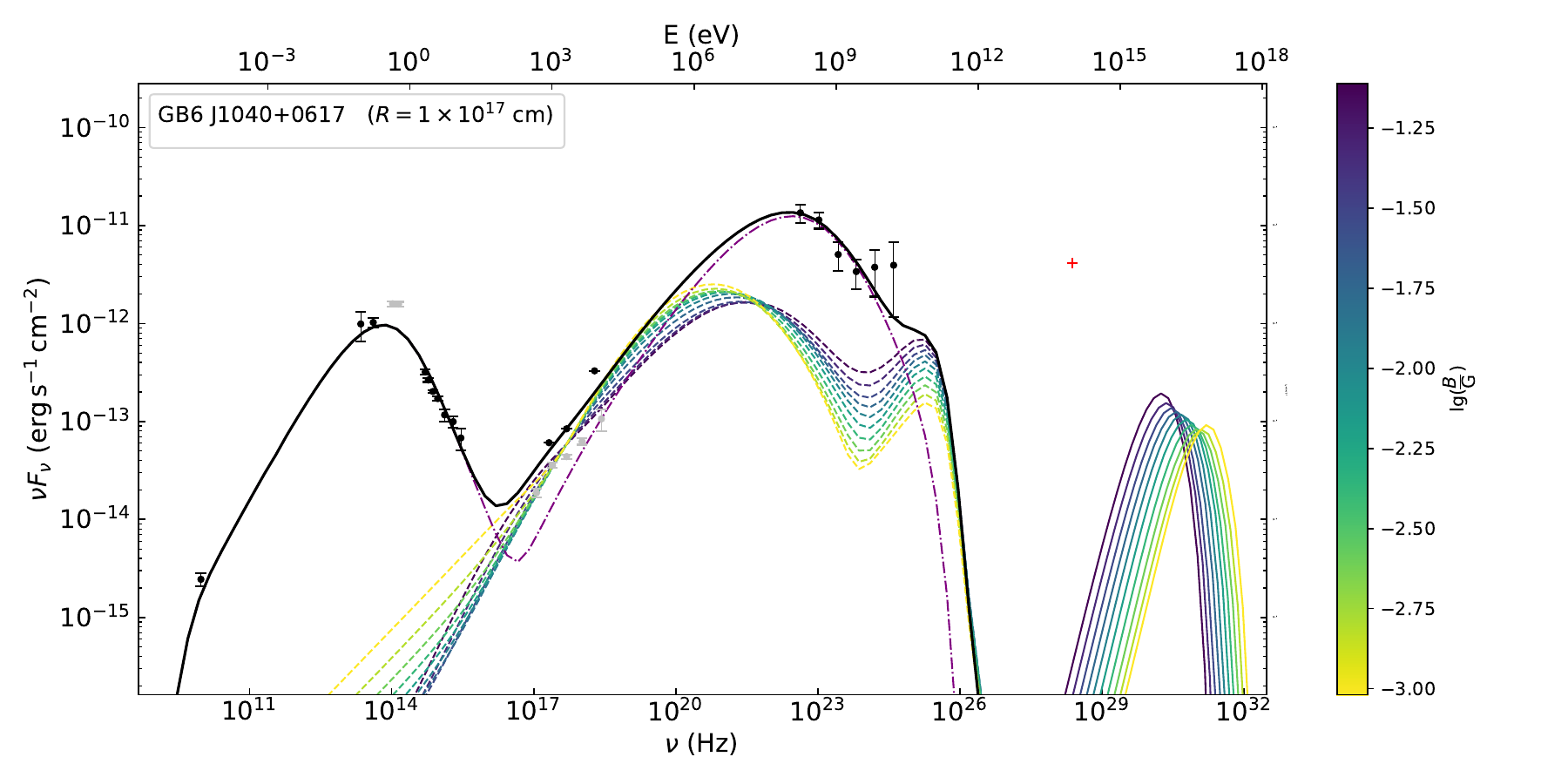}
        \label{GB6 J1040+0617 SSC R=1e17}
    \end{minipage}
    \caption{GB6 J1040+0617 associated with IC-141209A. The fitting results of the SED for $R=1\times10^{16}~{\rm cm}$ (left panel) and $R=1\times10^{17}~{\rm cm}$ (right panel) under the SSC-dominated case. The colored dashed and solid lines respectively represent the secondary pair cascade emission and the neutrino spectrum for different parameter combinations, which correspond to the color bar. The quasi-simultaneous data, neutrino data and other line styles have the same meaning as in Fig.~\ref{GB6 J1040+0617}.}
    \label{GB6 J1040+0617 appendix}
\end{figure*}

\begin{figure*}[htbp]
    \centering
    \begin{minipage}{0.49\linewidth}
        \centering
        \includegraphics[width=\linewidth, trim=10 15 30 10,clip]{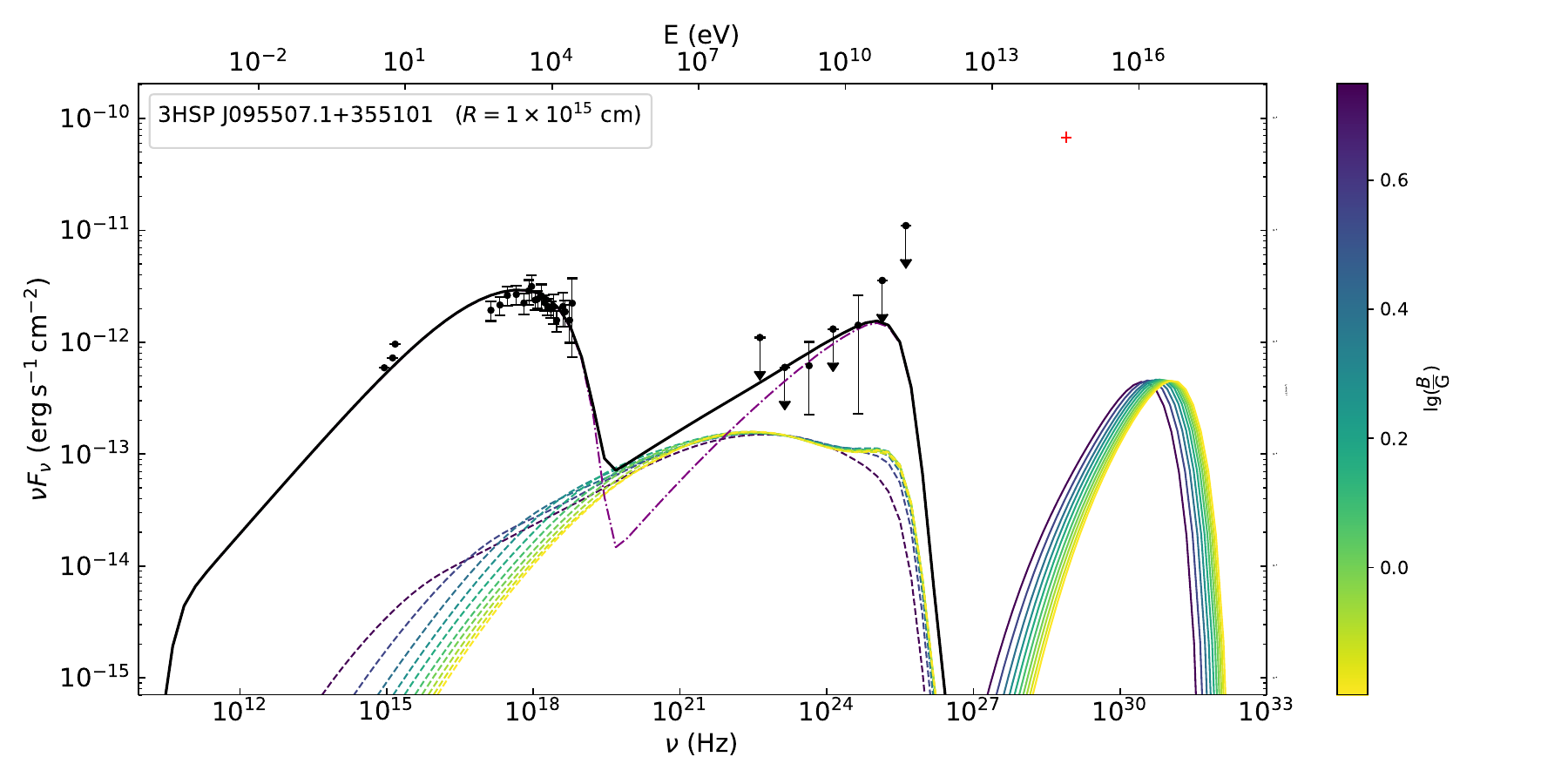}
        \label{3HSP J095507.1+355101 SSC R=1e15}
    \end{minipage}\hspace{-4mm}
    \begin{minipage}{0.49\linewidth}
        \centering
        \includegraphics[width=\linewidth, trim=10 15 30 10,clip]{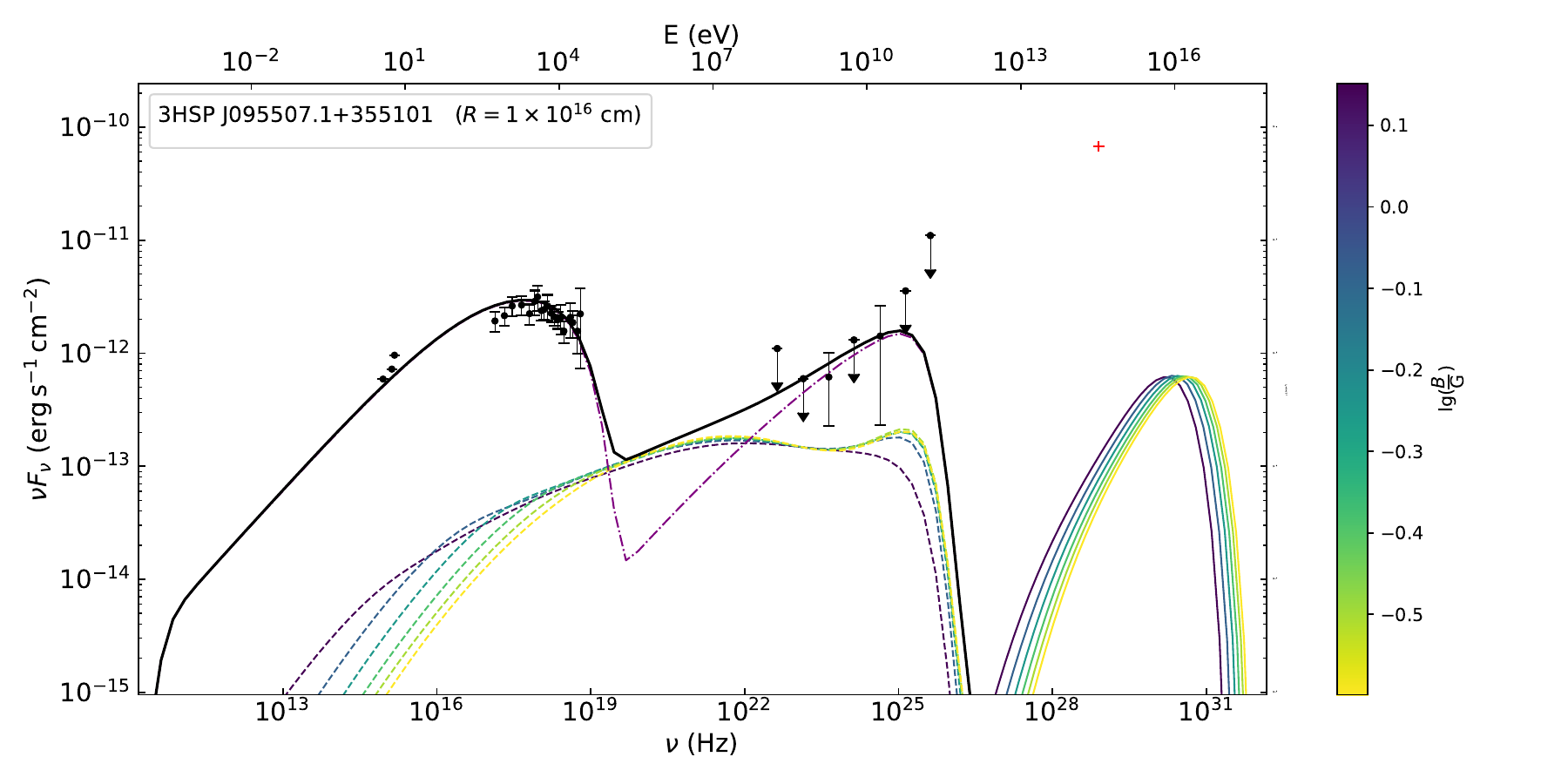}
        \label{3HSP J095507.1+355101 SSC R=1e16}
    \end{minipage}
    \caption{3HSP J095507.1+355101 associated with IC-200107A. The fitting results of the SED for $R=1\times10^{15}~{\rm cm}$ (left panel) and $R=1\times10^{16}~{\rm cm}$ (right panel) under the SSC-dominated case. The colored dashed and solid lines respectively represent the secondary pair cascade emission and the neutrino spectrum for different parameter combinations, which correspond to the color bar. The quasi-simultaneous data, neutrino data and other line styles have the same meaning as in Fig.~\ref{3HSP J095507.1+355101}.}
    \label{3HSP J095507.1+355101 appendix}
\end{figure*}

This appendix supplements the results not presented in the main text, as illustrated in \cref{TXS 17 appendix,TXS 21 appendix,TXS 22 appendix,PKS 0735+178 appendix,PKS 1502+106 appendix,PKS B1424-418 appendix,GB6 J1040+0617 appendix,GB6 J2113+1121 appendix,5BZB J0630-2406 appendix,3HSP J095507.1+355101 appendix}. These include EC parameter spaces at specified blob radii (middle panels in \cref{TXS 17 appendix,TXS 21 appendix,TXS 22 appendix,PKS 0735+178 appendix,PKS 1502+106 appendix,PKS B1424-418 appendix,GB6 J2113+1121 appendix,5BZB J0630-2406 appendix}) and the fitting results of the SED for specified radii in the SSC-dominated (\cref{GB6 J1040+0617 appendix}, \cref{3HSP J095507.1+355101 appendix} and upper panels in \cref{TXS 17 appendix,TXS 21 appendix,TXS 22 appendix,PKS 0735+178 appendix,PKS 1502+106 appendix,PKS B1424-418 appendix,GB6 J2113+1121 appendix,5BZB J0630-2406 appendix}) and EC-dominated (lower panels in \cref{TXS 17 appendix,TXS 21 appendix,TXS 22 appendix,PKS 0735+178 appendix,PKS 1502+106 appendix,PKS B1424-418 appendix,GB6 J2113+1121 appendix,5BZB J0630-2406 appendix}) cases, respectively.

\clearpage

\bibliography{sample701}{}
\bibliographystyle{aasjournalv7}



\end{document}